%% file: formatted.tex
\documentclass[runningheads]{llncs}

\usepackage{xspace} \usepackage{xcolor} \usepackage{graphicx} \usepackage[normalem]{ulem} %
\usepackage{amsmath} \usepackage{amssymb} \usepackage{amsfonts} \usepackage{enumitem} \usepackage{xstring}

\usepackage[hidelinks,colorlinks=true,linkcolor=blue,citecolor=blue,urlcolor=blue]{hyperref} \usepackage{url} \usepackage{cite} \usepackage[numbers]{natbib}

\usepackage{booktabs} \usepackage{multirow} \usepackage{makecell} \usepackage{ragged2e} \usepackage{colortbl} \usepackage{tabu}

\usepackage{caption} \usepackage{subcaption} \usepackage{wrapfig}

\usepackage{listings} \fboxsep=0.8pt 

\usepackage{tikz} \usetikzlibrary{calc} \usetikzlibrary{shapes.geometric} \usetikzlibrary{decorations.pathreplacing} \usetikzlibrary{positioning} \usetikzlibrary{automata} \usetikzlibrary{arrows}

\usepackage[keeplastbox]{flushend} %
\usepackage{pgf} \usepackage{datetime} %

\newcommand{\XSpace}[1]{} \newcommand{\XComment}[1]{} \newcommand{\DefMacro}[2]{\expandafter\newcommand\csname rmk-#1\endcsname{#2}} \newcommand{\UseMacro}[1]{\csname rmk-#1\endcsname}

\newcommand{\FormattingTask}[1]{} \newcommand{\MyPara}[1]{\noindent\textbf{#1}:} \newcommand{\MyFinding}[1]{\noindent\textbf{Finding \##1}:}  \newcommand{\reducedstrut}{\vrule width 0pt height .9\ht\strutbox depth .9\dp\strutbox\relax} \newcommand{\InputWithSpace}[1]{\bgroup\def\arraystretch{1.1}\input{#1}\egroup} \newcommand{\Code}[1]{{\ifmmode{\mathtt{#1}}\else$\mathtt{#1}$\fi}} \newcommand{\CodeIn}[1]{{\ifmmode{\mathtt{#1}}\else$\mathtt{#1}$\fi}} \newcommand{\CoqIn}[1]{\lstinline[language=Coq,basicstyle=\normalsize\ttfamily]{#1}} \newcommand{\CoqInSmall}[1]{\lstinline[language=Coq,basicstyle=\small\ttfamily]{#1}} \newcommand{\ColorBack}[1]{%
  \begingroup   \setlength{\fboxsep}{0pt}%
  \colorbox{purple!20}{\reducedstrut#1\/}%
  \endgroup }

\newcommand{\specialcell}[2][c]{%
  \begin{tabular}[#1]{@{}c@{}}#2\end{tabular}} \newcolumntype{R}[1]{>{\RaggedLeft\arraybackslash}p{#1}} \newcolumntype{L}[1]{>{\RaggedRight\arraybackslash}p{#1}}

\newcommand{\numtoword}[1]{%
\IfStrEqCase{#1}{{0}{zero}{1}{one}{2}{two}{3}{three}{4}{four}{5}{five}{6}{six}{7}{seven}{8}{eight}{9}{nine}{10}{ten}}[#1]}

\newcommand{\ltrue}{\top} %
\newcommand{\lfalse}{\bot} %

\newcommand{\limply}{\rightarrow} %
\newcommand{\liff}{\leftrightarrow} %
\newcommand{\lcofactor}{\downarrow} %

\newcommand{\llimply}{\Rightarrow} %
\newcommand{\lliff}{\Leftrightarrow} %
\newcommand{\lsat}{\vDash} %
\newcommand{\lnsat}{\nvDash} %

\newcommand{\lqall}[1]{\forall #1.~} %
\newcommand{\lqexist}[1]{\exists #1.~} %
\newcommand{\lqallOnly}[1]{\forall #1 } %
\newcommand{\lqexistOnly}[1]{\exists #1 } %

\newcommand{\lAnd}[1]{\bigwedge_{\substack{#1}}} %
\newcommand{\lOr}[1]{\bigvee_{\substack{#1}}} %

\newcommand{\set}[1]{\{#1\}} %
\newcommand{\Set}[1]{\left\{#1\right\}} %
\newcommand{\tuple}[1]{\langle#1\rangle} %
\newcommand{\Tuple}[1]{\left\langle#1\right\rangle} %

\newcommand{\sunion}{\cup} %
\newcommand{\sinter}{\cap} %

\definecolor{gray}{RGB}{211,211,211} \newcommand{\jbasicstyle}{\small\sffamily} %
\newcommand{\textcode}[1]{{#1}} \newcommand{\jnumberstyle}{\scriptsize} \newcommand{\Hilight}{\makebox[0pt][l]{\color{gray}\rule[-3pt]{0.80\linewidth}{9pt}}}

\lstdefinelanguage{pseudo} {   morekeywords={},   keywordstyle=\bfseries,   lineskip=-0.1em,   numbers=left, %
  numberstyle=\jnumberstyle,   numbersep=4pt,   basicstyle=\jbasicstyle,   breaklines=true,   breakautoindent=true,   tabsize=2,   columns=fullflexible,   morecomment=*[l][\textsl]{//},   mathescape=true,   xleftmargin=10pt, }

\lstdefinelanguage{todo-comment} {   morekeywords={},   keywordstyle=\bfseries,   lineskip=-0.1em,   numbers=none,   basicstyle=\jbasicstyle,   breaklines=true,   breakautoindent=true,   tabsize=2,   columns=fullflexible,   morecomment=*[l][\textsl]{//},   mathescape=true,   xleftmargin=-10pt, }

\lstdefinelanguage{java-pretty} {   language=java,   numbers=left,   basicstyle=\scriptsize\ttfamily,   numberstyle=\scriptsize,   breaklines=true,   columns=fullflexible,   xleftmargin=16pt,   showstringspaces=false, }

 \definecolor{shadecolor}{gray}{1.00} \definecolor{darkgray}{gray}{0.30} \definecolor{violet}{rgb}{0.56, 0.0, 1.0} \definecolor{forestgreen}{rgb}{0.13, 0.55, 0.13}

\lstdefinelanguage{Coq} { mathescape=true,						 texcl=false, morekeywords=[1]{   Add,   All,   Arguments,   Axiom,   Bind,   Canonical,   Check,   Close,   CoFixpoint,   CoInductive,   Coercion,   Contextual,   Corollary,   Defined,   Definition,   Delimit,   End,   Example,   Export,   Fact,   Fixpoint,   Goal,   Graph,   Hint,   Hypotheses,   Hypothesis,   Implicit,   Implicits,   Import,   Inductive,   Lemma,   Let,   Local,   Locate,   Ltac,   Maximal   Module,   Morphism,   Next,   Notation,   Obligation,   Open,   Parameter,   Parameters,   Prenex,   Print,   Printing,   Program,   Projections,   Proof,   Proposition,   Qed,   Record,   Relation,   Remark,   Require,   Reserved,   Resolve,   Rewrite,   Save,   Scope,   Search,   Section,   Show,   Strict,   Structure,   Tactic,   Theorem,   Unset,   Variable,   Variables,   View,   inside,   outside }, morekeywords=[2]{   as,   cofix,   else,   end,   exists,   exists2,   fix,   for,   forall,   fun,   if,   in,   is,   let,   match,   nosimpl,   of,   return,   struct,   then,   vfun,   with }, morekeywords=[3]{Type, Prop, Set, True, False}, morekeywords=[4]{   after,   apply,   assert,   auto,   bool_congr,   case,   change,   clear,   compute,   congr,   cut,   cutrewrite,   destruct,   elim,   field,   fold,   generalize,   have,   heval,    hnf,   induction,   injection,   intro,   intros,   intuition,   inversion,   left,   loss,   move,   nat_congr,   nat_norm,   pattern,   pose,   refine,   rename,   replace,   revert,   rewrite,   right,   ring,   set,   simpl,   split,   subst,   suff,   suffices,   symmetry,   transitivity,   trivial,   unfold,   unlock,   using,   without,   wlog,   autorewrite },         morekeywords=[5]{   assumption,   by,   contradiction,   congruence,   done,   exact,   lia,   gappa,   omega,   reflexivity,   romega,   solve,   tauto,   discriminate,   unsat }, morecomment=[s]{(*}{*)}, morekeywords=[6]{do, first, try, idtac, repeat}, showstringspaces=false, morestring=[b]", tabsize=3,							 extendedchars=true,  		 		 sensitive=true,  breaklines=false, basicstyle=\footnotesize\ttfamily, captionpos=b,							 columns=[l]fullflexible, identifierstyle={\color{black}}, keywordstyle=[1]{\color{violet}}, keywordstyle=[2]{\color{forestgreen}}, keywordstyle=[3]{\color{forestgreen}}, keywordstyle=[4]{\color{blue}}, keywordstyle=[5]{\color{red}}, keywordstyle=[6]{\color{violet}}, stringstyle=, commentstyle=\it\ttfamily\color{brown}, numberstyle=\tiny%
}

\lstdefinestyle{Coq}{language=Coq} 

\newcommand{\urlsymbol}{\includegraphics[width=.8em]{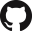}}

\newcommand{\RevisionInfo}{\Fix{Time: \today{} at \currenttime{.}}}

\newcommand{\cc}{coding conventions\xspace} \newcommand{\CC}{Coding Conventions\xspace} \newcommand{\Cc}{Coding conventions\xspace}

\newcommand{\LemmaN}{Lemma Names\xspace} \newcommand{\lemman}{lemma names\xspace} \newcommand{\Lemman}{Lemma names\xspace}

\newcommand{\MClibrary}{family of projects\xspace} \newcommand{\Coq}{Coq\xspace} \newcommand{\CoqConvTool}{\textsc{Roosterize}\xspace} \newcommand{\tooltype}{tool\-chain\xspace} \newcommand{\Tooltype}{Tool\-chain\xspace} \newcommand{\Title}{Deep Generation of Coq Lemma Names \\ Using Elaborated Terms} \newcommand{\ShortTitle}{Deep Generation of Coq Lemma Names Using Elaborated Terms}

\newcommand{\xcut}{\trimmed{}\xspace} \newcommand{\bidirectional}{bi-di\-rec\-tion\-al\xspace} \newcommand{\dsta}{all tiers\xspace} \newcommand{\Dsta}{All tiers\xspace} \newcommand{\DSta}{All Tiers\xspace} \newcommand{\dsti}{tier 1\xspace} \newcommand{\Dsti}{Tier 1\xspace} \newcommand{\DSti}{Tier 1\xspace} \newcommand{\dstii}{tier 2\xspace} \newcommand{\Dstii}{Tier 2\xspace} \newcommand{\DStii}{Tier 2\xspace} \newcommand{\dstiii}{tier 3\xspace} \newcommand{\Dstiii}{Tier 3\xspace} \newcommand{\DStiii}{Tier 3\xspace} \newcommand{\encdec}{encoder-decoder\xspace} \newcommand{\flatten}{flatten\xspace} \newcommand{\flattened}{flattened\xspace} \newcommand{\flattening}{flattening\xspace} \newcommand{\formatting}{formatting\xspace} \newcommand{\Formatting}{Formatting\xspace} \newcommand{\Ktree}{Kernel tree\xspace} \newcommand{\ktree}{kernel tree\xspace} \newcommand{\ktrees}{kernel trees\xspace} \newcommand{\etree}{elaborated tree\xspace} \newcommand{\etrees}{elaborated trees\xspace} \newcommand{\eterms}{elaborated terms\xspace} \newcommand{\eterm}{elaborated term\xspace} \newcommand{\KTreeAcro}{KnlTree\xspace} \newcommand{\kindcom}{\CodeIn{COM}\xspace} \newcommand{\kindid}{\CodeIn{ID}\xspace} \newcommand{\lemmanaming}{lemma naming\xspace} \newcommand{\Lemmanaming}{Lemma naming\xspace} \newcommand{\LemmaNaming}{Lemma Naming\xspace} \newcommand{\lstmt}{\lemmastmt{}\xspace} \newcommand{\Lstmt}{\Lemmastmt{}\xspace} \newcommand{\lstmts}{\lemmastmts{}\xspace} \newcommand{\leftoutcorpus}{left-out\xspace} \newcommand{\LeftOutCorpus}{Left-out\xspace} \newcommand{\lemmastmt}{lemma statement\xspace} \newcommand{\lemmastmts}{lemma statements\xspace} \newcommand{\Lemmastmt}{Lemma statement\xspace} \newcommand{\LemmaStmt}{Lemma Statement\xspace} \newcommand{\LStmtAcro}{Stmt\xspace} \newcommand{\maincorpus}{main\xspace} \newcommand{\MainCorpus}{Main\xspace} \newcommand{\notation}{notation\xspace} \newcommand{\Notation}{Notation\xspace} \newcommand{\ngram}{N-gram\xspace} \newcommand{\Ngram}{N-gram\xspace} \newcommand{\retrievalbased}{retrieval-based\xspace} \newcommand{\RetrievalBased}{Retrieval-based\xspace} \newcommand{\rnn}{RNN\xspace} \newcommand{\rnns}{RNNs\xspace} \newcommand{\rnnlm}{RNNLM\xspace} \newcommand{\simplification}{chopping\xspace} \newcommand{\Simplification}{\Trimming\xspace} \newcommand{\sexp}{sexp\xspace} \newcommand{\sexps}{sexps\xspace} \newcommand{\stree}{syntax tree\xspace} \newcommand{\strees}{syntax trees\xspace} \newcommand{\STreeAcro}{SynTree\xspace} \newcommand{\sktree}{syntax and kernel tree\xspace} \newcommand{\sktrees}{syntax and kernel trees\xspace} \newcommand{\SKTrees}{Syntax and Kernel Trees\xspace} \newcommand{\subtok}{sub-token\xspace} \newcommand{\SubTok}{Sub-token\xspace} \newcommand{\subtokenize}{sub-tokenize\xspace} \newcommand{\subtokenized}{sub-tokenized\xspace} \newcommand{\subtokenizer}{sub-tokenizer\xspace} \newcommand{\subtokenizertool}{\texttt{subtokenizer}\xspace} \newcommand{\subtokenization}{sub-tok\-eniza\-tion\xspace} \newcommand{\SubTokenization}{Sub-tok\-eniza\-tion\xspace} \newcommand{\taskformatting}{\formatting task\xspace} \newcommand{\Taskformatting}{\Formatting task\xspace} \newcommand{\TaskFormatting}{\Formatting Task\xspace} \newcommand{\tasklemmanaming}{\lemmanaming task\xspace} \newcommand{\Tasklemmanaming}{\Lemmanaming task\xspace} \newcommand{\TaskLemmaNaming}{\LemmaNaming Task\xspace} \newcommand{\test}{testing\xspace} \newcommand{\Test}{Testing\xspace} \newcommand{\train}{training\xspace} \newcommand{\Train}{Training\xspace} \newcommand{\trim}{chop\xspace} \newcommand{\trims}{chops\xspace} \newcommand{\trimmed}{chopped\xspace} %
\newcommand{\TrimmedAcro}{Chop\xspace} \newcommand{\trimmedktree}{\trimmed \ktree} \newcommand{\trimmedktrees}{\trimmed \ktrees} \newcommand{\TrimmedKTreeAcro}{\TrimmedAcro{}\KTreeAcro{}\xspace} \newcommand{\trimmedstree}{\trimmed \stree} \newcommand{\trimmedstrees}{\trimmed \strees} \newcommand{\TrimmedSTreeAcro}{\TrimmedAcro{}\STreeAcro{}\xspace} \newcommand{\trimming}{chopping\xspace} \newcommand{\Trimming}{Chopping\xspace} \newcommand{\val}{validation\xspace} \newcommand{\Val}{Validation\xspace} \newcommand{\vernacular}{vernacular\xspace} \newcommand{\keepcategory}{keep-category\xspace} \newcommand{\KeepCategory}{Keep-category\xspace}

\newcommand{\Gallina}{Gallina\xspace} \newcommand{\Ltac}{Ltac\xspace} \newcommand{\Vernac}{Vernacular\xspace} \newcommand{\MathematicalComponents}{Mathematical Components\xspace} \newcommand{\MathComp}{MathComp\xspace} \newcommand{\Naturalize}{Naturalize\xspace} \newcommand{\seqtoseq}{\textsc{seq2seq}\xspace} \newcommand{\SerAPI}{SerAPI\xspace} \newcommand{\ssreflect}{SSReflect\xspace} \newcommand{\sertok}{\texttt{sertok}\xspace} \newcommand{\sercomp}{\texttt{sercomp}\xspace} \newcommand{\sername}{\texttt{sername}\xspace} \newcommand{\postproc}{\texttt{postproc}\xspace} \newcommand{\learner}{\texttt{learner}\xspace} \newcommand{\DotV}{\texttt{.v}\xspace} \newcommand{\coqwc}{\texttt{coqwc}\xspace} \newcommand{\NLP}{NLP\xspace} \newcommand{\doc}{file\xspace} %
\newcommand{\docs}{files\xspace} %
\newcommand{\CNNs}{convolutional neural networks\xspace}

\newcommand{\bleu}{BLEU\xspace} \newcommand{\Bleu}{BLEU\xspace} \newcommand{\fragacc}{fragment accuracy\xspace} \newcommand{\Fragacc}{Fragment accuracy\xspace} \newcommand{\toponeacc}{top-1 accuracy\xspace} \newcommand{\Toponeacc}{Top-1 accuracy\xspace} \newcommand{\topthreeacc}{top-3 accuracy\xspace} \newcommand{\Topthreeacc}{Top-3 accuracy\xspace} \newcommand{\topfiveacc}{top-5 accuracy\xspace} \newcommand{\Topfiveacc}{Top-5 accuracy\xspace}

\newcommand{\NGNLMF}{\Ngram LM\xspace} \newcommand{\NGNLMFfull}{\Ngram Language Model\xspace} \newcommand{\NGLMN}{\Ngram LM\xspace} \newcommand{\NGLMNfull}{\Ngram Language Model\xspace}

\DefMacro{f-brnn}{BiRNNLM\xspace} \DefMacro{f-brnn-CK}{BiRNNLM(C+K)\xspace} \DefMacro{f-brnn-CS}{BiRNNLM(C+S)\xspace} \DefMacro{f-brnn-SK}{BiRNNLM(S+K)\xspace} \DefMacro{f-brnn-C}{BiRNNLM(C)\xspace} \DefMacro{f-brnn-S}{BiRNNLM(S)\xspace} \DefMacro{f-brnn-K}{BiRNNLM(K)\xspace} \DefMacro{f-rrnn}{Reversed RNNLM\xspace} \DefMacro{f-rnn}{RNNLM\xspace} \DefMacro{f-ngram}{\NGNLMF{}\xspace}

\DefMacro{f-brnn+cl+rr}{BiRNNLM+CL+Reranking\xspace} \DefMacro{f-brnn+masking+rr}{BiRNNLM+Masking+Reranking\xspace} \DefMacro{f-ngram+rr}{\NGNLMF{}+Reranking\xspace} \DefMacro{f-brnn+cl}{BiRNNLM+CL\xspace} \DefMacro{f-brnn+rr}{BiRNNLM+Reranking\xspace} \DefMacro{f-brnn+masking}{BiRNNLM+Masking\xspace}

\DefMacro{ln-s+bsexpl1+fsexpl1+attn+copy}{\LStmtAcro{}\allowbreak+\TrimmedKTreeAcro{}\allowbreak+\TrimmedSTreeAcro{}\allowbreak+attn\allowbreak+copy\xspace} \DefMacro{ln-s+bsexpl1+attn+copy}{\LStmtAcro{}\allowbreak+\TrimmedKTreeAcro{}\allowbreak+attn\allowbreak+copy\xspace} \DefMacro{ln-s+fsexpl1+attn+copy}{\LStmtAcro{}\allowbreak+\TrimmedSTreeAcro{}\allowbreak+attn\allowbreak+copy\xspace} \DefMacro{ln-bsexpl1+fsexpl1+attn+copy}{\TrimmedKTreeAcro{}\allowbreak+\TrimmedSTreeAcro{}\allowbreak+attn\allowbreak+copy\xspace} \DefMacro{ln-s+attn+copy}{\LStmtAcro{}\allowbreak+attn\allowbreak+copy\xspace} \DefMacro{ln-bsexpl1+attn+copy}{\TrimmedKTreeAcro{}\allowbreak+attn\allowbreak+copy\xspace} \DefMacro{ln-fsexpl1+attn+copy}{\TrimmedSTreeAcro{}\allowbreak+attn\allowbreak+copy\xspace} \DefMacro{ln-bsexpl0+attn+copy}{\KTreeAcro{}\allowbreak+attn\allowbreak+copy\xspace} \DefMacro{ln-fsexpl0+attn+copy}{\STreeAcro{}\allowbreak+attn\allowbreak+copy\xspace}

\DefMacro{ln-s+bsexpl1+fsexpl1+attn}{\LStmtAcro{}\allowbreak+\TrimmedKTreeAcro{}\allowbreak+\TrimmedSTreeAcro{}\allowbreak+attn\xspace} \DefMacro{ln-s+bsexpl1+attn}{\LStmtAcro{}\allowbreak+\TrimmedKTreeAcro{}\allowbreak+attn\xspace} \DefMacro{ln-s+fsexpl1+attn}{\LStmtAcro{}\allowbreak+\TrimmedSTreeAcro{}\allowbreak+attn\xspace} \DefMacro{ln-bsexpl1+fsexpl1+attn}{\TrimmedKTreeAcro{}\allowbreak+\TrimmedSTreeAcro{}\allowbreak+attn\xspace} \DefMacro{ln-s+attn}{\LStmtAcro{}\allowbreak+attn\xspace} \DefMacro{ln-bsexpl1+attn}{\TrimmedKTreeAcro{}\allowbreak+attn\xspace} \DefMacro{ln-fsexpl1+attn}{\TrimmedSTreeAcro{}\allowbreak+attn\xspace} \DefMacro{ln-bsexpl0+attn}{\KTreeAcro{}\allowbreak+attn\xspace} \DefMacro{ln-fsexpl0+attn}{\STreeAcro{}\allowbreak+attn\xspace}

\DefMacro{ln-s+bsexpl1+fsexpl1}{\LStmtAcro{}\allowbreak+\TrimmedKTreeAcro{}\allowbreak+\TrimmedSTreeAcro{}\xspace} \DefMacro{ln-s+bsexpl1}{\LStmtAcro{}\allowbreak+\TrimmedKTreeAcro{}\xspace} \DefMacro{ln-s+fsexpl1}{\LStmtAcro{}\allowbreak+\TrimmedSTreeAcro{}\xspace} \DefMacro{ln-bsexpl1+fsexpl1}{\TrimmedKTreeAcro{}\allowbreak+\TrimmedSTreeAcro{}\xspace} \DefMacro{ln-s}{\LStmtAcro{}\xspace} \DefMacro{ln-bsexpl1}{\TrimmedKTreeAcro{}\xspace} \DefMacro{ln-fsexpl1}{\TrimmedSTreeAcro{}\xspace} \DefMacro{ln-bsexpl0}{\KTreeAcro{}\xspace} \DefMacro{ln-fsexpl0}{\STreeAcro{}\xspace}

\DefMacro{ln-rb}{\RetrievalBased{}\xspace}

\DefMacro{ln-s+bsexpl1x+fsexpl1x+attn+copy}{\LStmtAcro{}\allowbreak+\TrimmedKTreeAcro{}\allowbreak+\TrimmedSTreeAcro{}\allowbreak+attn\allowbreak+copy\xspace} \DefMacro{ln-s+bsexpl1x+attn+copy}{\LStmtAcro{}\allowbreak+\TrimmedKTreeAcro{}\allowbreak+attn\allowbreak+copy\xspace} \DefMacro{ln-s+fsexpl1x+attn+copy}{\LStmtAcro{}\allowbreak+\TrimmedSTreeAcro{}\allowbreak+attn\allowbreak+copy\xspace} \DefMacro{ln-bsexpl1x+fsexpl1x+attn+copy}{\TrimmedKTreeAcro{}\allowbreak+\TrimmedSTreeAcro{}\allowbreak+attn\allowbreak+copy\xspace}

\DefMacro{ln-s+bsexpd10+fsexpd10+attn+copy}{\LStmtAcro{}\allowbreak+\TrimmedKTreeAcro{}\allowbreak+\TrimmedSTreeAcro{}\allowbreak+attn\allowbreak+copy\xspace} \DefMacro{ln-s+bsexpd10+attn+copy}{\LStmtAcro{}\allowbreak+\TrimmedKTreeAcro{}\allowbreak+attn\allowbreak+copy\xspace} \DefMacro{ln-s+fsexpd10+attn+copy}{\LStmtAcro{}\allowbreak+\TrimmedSTreeAcro{}\allowbreak+attn\allowbreak+copy\xspace} \DefMacro{ln-bsexpd10+fsexpd10+attn+copy}{\TrimmedKTreeAcro{}\allowbreak+\TrimmedSTreeAcro{}\allowbreak+attn\allowbreak+copy\xspace}

\DefMacro{ln-s+bsexprnd+fsexprnd+attn+copy}{\LStmtAcro{}\allowbreak+\TrimmedKTreeAcro{}\allowbreak+\TrimmedSTreeAcro{}\allowbreak+attn\allowbreak+copy\xspace} \DefMacro{ln-s+bsexprnd+attn+copy}{\LStmtAcro{}\allowbreak+\TrimmedKTreeAcro{}\allowbreak+attn\allowbreak+copy\xspace} \DefMacro{ln-s+fsexprnd+attn+copy}{\LStmtAcro{}\allowbreak+\TrimmedSTreeAcro{}\allowbreak+attn\allowbreak+copy\xspace} \DefMacro{ln-bsexprnd+fsexprnd+attn+copy}{\TrimmedKTreeAcro{}\allowbreak+\TrimmedSTreeAcro{}\allowbreak+attn\allowbreak+copy\xspace}

\DefMacro{group-t1}{\DSti} \DefMacro{group-t2}{\DStii} \DefMacro{group-t3}{\DStiii} \DefMacro{group-ta}{\DSta} \DefMacro{group-lo}{\LeftOutCorpus} \DefMacro{group-lo-small}{LO} \DefMacro{group-allgroup}{All} \DefMacro{lo-train-t1}{0} \DefMacro{lo-train-ta}{0}

\DefMacro{table-caption-corpus}{     \label{tbl:corpus}Projects from the \MathComp Family Used in Our     Corpus.} \DefMacro{table-caption-corpus-allgroup}{     \label{tbl:corpus-allgroup}Projects Used in Our Expanded Corpus.} \DefMacro{table-caption-experiments}{     \label{tbl:experiments}Links to Tables with Results for Various     Combinations of \Train, \Val, and \Test Sets.} \DefMacro{table-caption-dataset-split-allgroup}{     \label{tbl:dataset-split-allgroup}Split of Corpus into Training,     Validation, and Testing Sets.\vspace{-2pt}} \DefMacro{table-caption-dataset-lemma}{     \label{tbl:dataset-lemma} Statistics on the Lemmas     in the \Train, \Val, and \Test Sets.} \DefMacro{table-caption-dataset-lemma-big}{     \label{tbl:dataset-lemma-big} Additional Statistics on the Lemmas     in the \Train, \Val, and \Test Sets.} \DefMacro{table-caption-dataset-lemma-allgroup}{     \label{tbl:dataset-lemma-allgroup} Statistics on the Lemmas     in the \Train, \Val, and \Test Sets of the Expanded     Corpus in Each Tier.} \newcommand{\TableCaptionDatasetNL}{\label{tbl:corpus-stats}LOC, Sentence,     and Token Statistics (Averages) for the Corpus.\vspace{-2pt}} \newcommand{\TableCaptionDataset}{\label{tbl:dataset}Statistics     on the Datasets Used in Our Experiments.} \newcommand{\TableCaptionDatasetF}{\label{tbl:dataset-f}Split of     Dataset into Training, Validation, and Testing Sets.\vspace{-2pt}} \newcommand{\TableCaptionDatasetLid}{\label{tbl:dataset-lid}Statistics     with Language Identification Categories of the Datasets Used in Our Experiments.\vspace{-2pt}}

\DefMacro{table-caption-results-ln-ta--ta--ta-main}{     \label{tbl:results-ln-ta--ta--ta-main}Results of \CoqConvTool     Models with \Train and \Val Sets from \UseMacro{group-ta}     and \Test Set from \UseMacro{group-ta}.} \DefMacro{table-caption-results-ln-ta--ta--t1-main}{     \label{tbl:results-ln-ta--ta--t1-main}Results of \CoqConvTool     Models with \Train and \Val Sets from \UseMacro{group-ta}     and \Test Set from \UseMacro{group-t1}.} \DefMacro{table-caption-results-ln-ta--ta--t2-main}{     \label{tbl:results-ln-ta--ta--t2-main}Results of \CoqConvTool     Models with \Train and \Val Sets from \UseMacro{group-ta}     and \Test Set from \UseMacro{group-t2}.} \DefMacro{table-caption-results-ln-ta--ta--t3-main}{     \label{tbl:results-ln-ta--ta--t3-main}Results of \CoqConvTool     Models with \Train and \Val Sets from \UseMacro{group-ta}     and \Test Set from \UseMacro{group-t3}.} \DefMacro{table-caption-results-ln-t1--t1--t1-main}{     \label{tbl:results-ln-t1--t1--t1-main}Results of \CoqConvTool     Models.} \DefMacro{table-caption-results-ln-t1--t1--t2-main}{     \label{tbl:results-ln-t1--t1--t2-main}Results of \CoqConvTool     Models with \Train and \Val Sets from \UseMacro{group-t1}     and \Test Set from \UseMacro{group-t2}.} \DefMacro{table-caption-results-ln-t1--t1--t3-main}{     \label{tbl:results-ln-t1--t1--t3-main}Results of \CoqConvTool     Models with \Train and \Val Sets from \UseMacro{group-t1}     and \Test Set from \UseMacro{group-t3}.} \DefMacro{table-caption-results-ln-t1--t1--ta-main}{     \label{tbl:results-ln-t1--t1--ta-main}Results of \CoqConvTool     Models with \Train and \Val Sets from \UseMacro{group-t1}     and \Test Set from \UseMacro{group-ta}.} \DefMacro{table-caption-results-ln-t2--t2--t2-main}{     \label{tbl:results-ln-t2--t2--t2-main}Results of \CoqConvTool     Models with \Train and \Val Sets from \UseMacro{group-t2}     and \Test Set from \UseMacro{group-t2}.} \DefMacro{table-caption-results-ln-t3--t3--t3-main}{     \label{tbl:results-ln-t3--t3--t3-main}Results of \CoqConvTool     Models with \Train and \Val Sets from \UseMacro{group-t3}     and \Test Set from \UseMacro{group-t3}.}

\DefMacro{table-caption-results-ln-lo-t1-main}{     \label{tbl:results-ln-lo-t1-main}Results of the Generalization Study.} \DefMacro{table-caption-results-ln-lo-ta-main}{     \label{tbl:results-ln-lo-ta-main}Results of the Generalization Study     with \CoqConvTool Pre-trained on \DSta.} \DefMacro{table-caption-results-ln-chopping-chopping-main}{     \label{tbl:results-ln-chopping-chopping-main}Results of the     Ablation Study on Tree \Trimming{}.} \newcommand{\TableCaptionResultsF}{\label{tbl:results-f}Results     of Models for \TaskFormatting.\vspace{-2pt}} \newcommand{\TableCaptionSubTokenizationExamples}{\label{tbl:sub-tokenization-examples}\SubTokenization examples.\vspace{-2pt}} \DefMacro{table-caption-qualitative-ln}{     \label{tbl:qualitative-ln}Manual Quality Analysis Representative     Examples.} \newcommand{\TableCaptionQualitativeF}{\label{tbl:qualitative-f}Case Study Reports for \TaskFormatting.\vspace{-2pt}}

\newcommand{\TableHeadAVG}{\textbf{Avg.}\xspace} \newcommand{\TableHeadSUM}{\textbf{$\Sigma$}\xspace} \newcommand{\TableHeadNA}{N/A\xspace} \newcommand{\TableHeadProject}{\textbf{Project}\xspace} \newcommand{\TableHeadURL}{\textbf{URL} (https://github.com/)\xspace} \newcommand{\TableHeadSHA}{\textbf{SHA}\xspace} \newcommand{\TableHeadTextFiles}{\#Files\xspace} \newcommand{\TableHeadFiles}{\textbf{\TableHeadTextFiles}\xspace} \newcommand{\TableHeadTextLOC}{LOC\xspace} \newcommand{\TableHeadLOC}{\textbf{\TableHeadTextLOC}\xspace} \newcommand{\TableHeadTextSpecLOC}{Spec.\xspace} \newcommand{\TableHeadSpecLOC}{\textbf{\TableHeadTextSpecLOC}\xspace} \newcommand{\TableHeadTextProofLOC}{Proof\xspace} \newcommand{\TableHeadProofLOC}{\textbf{\TableHeadTextProofLOC}\xspace} \newcommand{\TableHeadTextCommentLOC}{Com.~LOC\xspace} \newcommand{\TableHeadCommentLOC}{\textbf{\TableHeadTextCommentLOC}\xspace} \newcommand{\TableHeadNumSpecSent}{\textbf{\#Spec.~sents.}\xspace} \newcommand{\TableHeadNumProofSent}{\textbf{\#Pr.~sents.}\xspace} \newcommand{\TableHeadTier}{\textbf{Tier}\xspace} \DefMacro{table-head-summary-group-t1}{\makecell[l]{\TableHeadAVG{}\\\TableHeadSUM{}}\xspace} \DefMacro{table-head-summary-group-ta}{\MainCorpus{}\hspace{0.5em}\makecell[l]{\TableHeadAVG{}\\\TableHeadSUM{}}\xspace} \DefMacro{table-head-summary-group-allgroup}{All\hspace{1.5em}\makecell[l]{\TableHeadAVG{}\\\TableHeadSUM{}}\xspace} \newcommand{\TableHeadTextLocPerDoc}{LOC/\doc{}\xspace} \newcommand{\TableHeadLocPerDoc}{\textbf{\TableHeadTextLocPerDoc}\xspace} \newcommand{\TableHeadTextSpecLocPerDoc}{Spec.\xspace} \newcommand{\TableHeadSpecLocPerDoc}{\textbf{\TableHeadTextSpecLocPerDoc}\xspace} \newcommand{\TableHeadTextProofLocPerDoc}{Proof\xspace} \newcommand{\TableHeadProofLocPerDoc}{\textbf{\TableHeadTextProofLocPerDoc}\xspace} \newcommand{\TableHeadSpecSentPerDoc}{\textbf{\#Spec.~sents./doc}\xspace} \newcommand{\TableHeadProofSentPerDoc}{\textbf{\#Pr.~sents./doc}\xspace} \newcommand{\TableHeadTokensPerSpecSent}{\textbf{\#Toks/Spec.~sent}\xspace} \newcommand{\TableHeadTokensPerProofSent}{\textbf{\#Toks/Pr.~sent}\xspace} \newcommand{\TableHeadDSTier}{\xspace} \newcommand{\TableHeadDSGroup}{\xspace} \newcommand{\TableHeadDSSet}{\xspace} \newcommand{\TableHeadTextNumPrj}{\#Prjs\xspace} \newcommand{\TableHeadNumPrj}{\textbf{\TableHeadTextNumPrj}\xspace} \newcommand{\TableHeadTextNumDoc}{\#Files\xspace} \newcommand{\TableHeadNumDoc}{\textbf{\TableHeadTextNumDoc}\xspace} \newcommand{\TableHeadTextNumSent}{\#Sents\xspace} \newcommand{\TableHeadNumSent}{\textbf{\TableHeadTextNumSent}\xspace} \newcommand{\TableHeadTextNumTok}{\#Toks\xspace} \newcommand{\TableHeadNumTok}{\textbf{\TableHeadTextNumTok}\xspace} \newcommand{\TableHeadTextLenOfDoc}{len(Files)\xspace} \newcommand{\TableHeadLenOfDoc}{\textbf{\TableHeadTextLenOfDoc}\xspace} \newcommand{\TableHeadTextLenOfSent}{len(Sents)\xspace} \newcommand{\TableHeadLenOfSent}{\textbf{\TableHeadTextLenOfSent}\xspace} \DefMacro{table-head-tier-alltier}{all\xspace} \DefMacro{table-head-tier-tier1}{\dsti\xspace} \DefMacro{table-head-tier-tier2}{\dstii\xspace} \DefMacro{table-head-tier-tier3}{\dstiii\xspace} \DefMacro{table-head-set-all}{all\xspace} \DefMacro{table-head-set-train}{\train\xspace} \DefMacro{table-head-set-val}{\val\xspace} \DefMacro{table-head-set-test}{\test\xspace} \newcommand{\TableHeadTextGallina}{G\xspace} \newcommand{\TableHeadGallina}{\textbf{\TableHeadTextGallina}\xspace} \newcommand{\TableHeadTextLtac}{L\xspace} \newcommand{\TableHeadLtac}{\textbf{\TableHeadTextLtac}\xspace} \newcommand{\TableHeadTextLtacWGallina}{L+G\xspace} \newcommand{\TableHeadLtacWGallina}{\textbf{\TableHeadTextLtacWGallina}\xspace} \newcommand{\TableHeadTextLtacWOGallina}{L\xspace} \newcommand{\TableHeadLtacWOGallina}{\textbf{\TableHeadTextLtacWOGallina}\xspace} \newcommand{\TableHeadTextVernacWGallina}{V+G\xspace} \newcommand{\TableHeadVernacWGallina}{\textbf{\TableHeadTextVernacWGallina}\xspace} \newcommand{\TableHeadTextVernacWOGallina}{V\xspace} \newcommand{\TableHeadVernacWOGallina}{\textbf{\TableHeadTextVernacWOGallina}\xspace} \newcommand{\TableHeadTextVernac}{V\xspace} \newcommand{\TableHeadVernac}{\textbf{\TableHeadTextVernac}\xspace} \newcommand{\TableHeadLMSet}{\xspace} \DefMacro{table-head-dslemma-all-t1-all}{before filtering\xspace} \DefMacro{table-head-dslemma-filtered-t1-all}{after filtering\xspace} \DefMacro{table-head-dslemma-filtered-t1-train}{\train{}\xspace} \DefMacro{table-head-dslemma-filtered-t1-val}{\val{}\xspace} \DefMacro{table-head-dslemma-filtered-t1-test}{\test{}\xspace} \DefMacro{table-head-dslemmaallgroup-all-ta-all}{before filtering\xspace} \DefMacro{table-head-dslemmaallgroup-filtered-ta-all}{after filtering\xspace} \DefMacro{table-head-dslemmaallgroup-filtered-ta-train}{\phantom{\DSta}\train{}\xspace} \DefMacro{table-head-dslemmaallgroup-filtered-ta-val}{\DSta\val{}\xspace} \DefMacro{table-head-dslemmaallgroup-filtered-ta-test}{\phantom{\DSta}\test{}\xspace} \DefMacro{table-head-dslemmaallgroup-filtered-t1-train}{\phantom{\DSti}\train{}\xspace} \DefMacro{table-head-dslemmaallgroup-filtered-t1-val}{\DSti\val{}\xspace} \DefMacro{table-head-dslemmaallgroup-filtered-t1-test}{\phantom{\DSti}\test{}\xspace} \DefMacro{table-head-dslemmaallgroup-filtered-t2-train}{\phantom{\DStii}\train{}\xspace} \DefMacro{table-head-dslemmaallgroup-filtered-t2-val}{\DStii\val{}\xspace} \DefMacro{table-head-dslemmaallgroup-filtered-t2-test}{\phantom{\DStii}\test{}\xspace} \DefMacro{table-head-dslemmaallgroup-filtered-t3-train}{\phantom{\DStiii}\train{}\xspace} \DefMacro{table-head-dslemmaallgroup-filtered-t3-val}{\DStiii\val{}\xspace} \DefMacro{table-head-dslemmaallgroup-filtered-t3-test}{\phantom{\DStiii}\test{}\xspace} \newcommand{\TableHeadTextNumLemma}{\#Lemmas\xspace} \newcommand{\TableHeadNumLemma}{\textbf{\TableHeadTextNumLemma}\xspace} \newcommand{\TableHeadNumLemmaFiltered}{\textbf{\TableHeadTextNumLemma}\xspace} \newcommand{\TableHeadLMName}{\textbf{Name}\xspace} \newcommand{\TableHeadLMStmt}{\textbf{\LStmtAcro}\xspace} \newcommand{\TableHeadLMKTree}{\textbf{\KTreeAcro}\xspace} \newcommand{\TableHeadLMKTreeTrimmed}{\textbf{\KTreeAcro{}\TrimmedAcro}\xspace} \newcommand{\TableHeadLMSTree}{\textbf{\STreeAcro}\xspace} \newcommand{\TableHeadLMSTreeTrimmed}{\textbf{\STreeAcro{}\TrimmedAcro}\xspace} \newcommand{\TableHeadNumChar}{\textbf{\#Char}\xspace} \newcommand{\TableHeadNumSubToken}{\textbf{\#SubToks}\xspace} \newcommand{\TableHeadDepth}{\textbf{Depth}\xspace} \newcommand{\TableHeadModel}{\textbf{Model}\xspace} \DefMacro{table-head-results-f-set-val}{\textbf{Val. Acc.}\xspace} \DefMacro{table-head-results-f-set-test}{\textbf{Test Acc.}\xspace} \DefMacro{table-head-results-f-acc-all}{\textbf{Top1 Acc.}\xspace} \DefMacro{table-head-results-f-acc-top-3}{\textbf{Top3 Acc.}\xspace} \DefMacro{table-head-results-f-acc-top-5}{\textbf{Top5 Acc.}\xspace} \newcommand{\TableHeadGroup}{\textbf{Group}\xspace} \DefMacro{table-head-bleu}{\bleu\xspace} \DefMacro{table-head-frag-acc}{Frag.Acc.\xspace} \DefMacro{table-head-full-acc-top-1}{Top1\xspace} \DefMacro{table-head-full-acc-top-5}{Top5\xspace} \DefMacro{table-head-results-ln-BLEU-4}{\textbf{\UseMacro{table-head-bleu}}\xspace} \DefMacro{table-head-results-ln-frag-acc}{\textbf{\UseMacro{table-head-frag-acc}}\xspace} \DefMacro{table-head-results-ln-full-acc-top-1}{\textbf{\UseMacro{table-head-full-acc-top-1}}\xspace} \DefMacro{table-head-results-ln-full-acc-top-5}{\textbf{\UseMacro{table-head-full-acc-top-5}}\xspace} \newcommand{\TablePartMultiAttnCopy}{\makecell{Multi-input\\+attn\\+copy}} \newcommand{\TablePartMonoAttnCopy}{\makecell{Single-input\\+attn\\+copy}} \newcommand{\TablePartMultiAttn}{\makecell{Multi-input\\+attn}} \newcommand{\TablePartMonoAttn}{\makecell{Single-input\\+attn}} \newcommand{\TablePartMulti}{\makecell{Multi-input}} \newcommand{\TablePartMono}{\makecell{Single-input}} \newcommand{\TablePartRB}{\makecell{-}} \newcommand{\TablePartChoppingRoosterize}{\makecell[l]{\CoqConvTool\\\Trimming}} \newcommand{\TablePartChoppingWithCategory}{\makecell[l]{\KeepCategory\\\Trimming}} \newcommand{\TablePartChoppingDepthX}{\makecell[l]{Rule-based\\\Trimming}} \newcommand{\TablePartChoppingRandom}{\makecell[l]{Random\\\Trimming}} \newcommand{\TableHeadLOTrainSet}{\textbf{\#Lemmas}} \newcommand{\QSStmt}[1]{\textbf{\Lstmt}: & \lstinline[language=Coq,basicstyle=\footnotesize\ttfamily]{#1}} \newcommand{\QSStmtCont}[1]{\multicolumn{2}{l}{\hspace{5.8em}\lstinline[language=Coq,basicstyle=\footnotesize\ttfamily]{#1}}} \newcommand{\QSTruth}[1]{\multicolumn{2}{l}{\textbf{Hand-written}: \lstinline[language=Coq,basicstyle=\footnotesize\ttfamily]{#1}}} \newcommand{\QSPred}[1]{\multicolumn{2}{l}{\textbf{\CoqConvTool}: \lstinline[language=Coq,basicstyle=\footnotesize\ttfamily]{#1}}} \newcommand{\QSTruthPred}[2]{\multicolumn{2}{l}{\textbf{Hand-written}: \lstinline[language=Coq,basicstyle=\footnotesize\ttfamily]{#1}\ \ \textbf{\CoqConvTool}: \lstinline[language=Coq,basicstyle=\footnotesize\ttfamily]{#2}}} \newcommand{\QSComment}[1]{\multicolumn{2}{l}{\makecell[{{p{\columnwidth}}}]{\textbf{Comment}: #1}}} \newcommand{\QSTruthOnly}{\textbf{Hand-written}} \newcommand{\QSPredOnly}{\textbf{\CoqConvTool}} \newcommand{\QSCode}[1]{\lstinline[language=Coq,basicstyle=\footnotesize\ttfamily,keepspaces=true]{#1}}

\DefMacro{vspace-corpus}{-18pt} \DefMacro{vspace-corpus-allgroup}{-10pt} \DefMacro{vspace-experiments}{-10pt} \DefMacro{vspace-dataset-nl}{0pt} \DefMacro{vspace-dataset}{-10pt} \DefMacro{vspace-dataset-f}{-10pt} \DefMacro{vspace-dataset-split-allgroup}{-10pt} \DefMacro{vspace-dataset-lid}{0pt} \DefMacro{vspace-dataset-lemma}{-15pt} \DefMacro{vspace-dataset-lemma-big}{-10pt} \DefMacro{vspace-dataset-lemma-allgroup}{-10pt} \DefMacro{vspace-results-f}{-10pt}

\DefMacro{vspace-results-ln-ta--ta--ta-main}{-5pt} \DefMacro{vspace-results-ln-ta--ta--t1-main}{-5pt} \DefMacro{vspace-results-ln-ta--ta--t2-main}{-5pt} \DefMacro{vspace-results-ln-ta--ta--t3-main}{-5pt} \DefMacro{vspace-results-ln-t1--t1--t1-main}{-20pt} \DefMacro{vspace-results-ln-t1--t1--t2-main}{-5pt} \DefMacro{vspace-results-ln-t1--t1--t3-main}{-5pt} \DefMacro{vspace-results-ln-t1--t1--ta-main}{-5pt} \DefMacro{vspace-results-ln-t2--t2--t2-main}{-5pt} \DefMacro{vspace-results-ln-t3--t3--t3-main}{-5pt}

\DefMacro{vspace-results-ln-lo-t1-main}{-5pt} \DefMacro{vspace-results-ln-lo-ta-main}{-5pt} \DefMacro{vspace-results-ln-chopping-chopping-main}{-5pt} \DefMacro{vspace-sub-tokenization-examples}{-10pt} \DefMacro{vspace-qualitative-ln}{-25pt} \DefMacro{vspace-qualitative-f}{0pt}

\input{tables/numbers-corpus} \input{tables/numbers-dataset} \input{tables/numbers-results-ln-t1--t1--t1} \input{tables/numbers-results-ln-t1--t1--t2} \input{tables/numbers-results-ln-t1--t1--t3} \input{tables/numbers-results-ln-t1--t1--ta} \input{tables/numbers-results-ln-ta--ta--t1} \input{tables/numbers-results-ln-ta--ta--t2} \input{tables/numbers-results-ln-ta--ta--t3} \input{tables/numbers-results-ln-ta--ta--ta} \input{tables/numbers-results-ln-t2--t2--t2} \input{tables/numbers-results-ln-t3--t3--t3} \input{tables/numbers-results-ln-lo} \input{tables/numbers-results-ln-lo-t1} \input{tables/numbers-results-ln-lo-ta} \input{tables/numbers-results-ln-chopping-chopping} \input{tables/numbers-lemma-metrics-filtered} \input{tables/numbers-lemma-metrics-all} \input{tables/numbers-misc}

\newcommand{\CorpusKLOC}{\UseMacro{corpus-t1-SUM-k-code-loc}k\xspace} \newcommand{\CorpusNumProjects}{\numtoword{\UseMacro{corpus-t1-num-projects}}\xspace}

\newcommand{\CoqVersion}{8.10.2\xspace} \newcommand{\ToolURL}{\url{https://github.com/EngineeringSoftware/roosterize}}

\newcommand{\NumExpTrials}{3\xspace} \newcommand{\NumExpTimeout}{12\xspace} \newcommand{\NumSubTokenInspected}{200\xspace} \newcommand{\NumSubTokenAcc}{79.5\%\xspace} \newcommand{\NumQSLemmaNameComments}{150\xspace} \newcommand{\NumQSLemmaNameGood}{17\xspace} \newcommand{\PerQSLemmaNameGood}{11.3\%\xspace} \newcommand{\NumQSLemmaNameNeutral}{13\xspace} \newcommand{\PerQSLemmaNameGoodOrNeutral}{20\%\xspace} \newcommand{\NumQSLemmaNameBad}{120\xspace} \newcommand{\NumQSFormattingomments}{56\xspace}

\newcommand{\NumVocabContent}{10,257\xspace} \newcommand{\NumVocabSpacing}{75\xspace} \newcommand{\NumVocabKind}{10\xspace}

\newcommand{\Pmathcomp}{math-comp\xspace} \newcommand{\Poddorder}{odd-order\xspace} \newcommand{\Pfourcolor}{fourcolor\xspace} \newcommand{\Pfinmap}{finmap\xspace}

\newcommand{\Pbigenough}{bigenough\xspace} \newcommand{\Panalysis}{analysis\xspace} \newcommand{\Prealclosed}{real-closed\xspace} \newcommand{\Probot}{robot\xspace} \newcommand{\Pmultinomials}{multinomials\xspace} \newcommand{\Pellipticcurves}{elliptic-curves\xspace} \newcommand{\Ptwosquare}{two-square\xspace} \newcommand{\Pgrobner}{grobner\xspace} \newcommand{\Pinfotheo}{infotheo\xspace}

\newcommand{\Pfcslpcm}{fcsl-pcm\xspace} \newcommand{\Pdisel}{disel\xspace} \newcommand{\Preglang}{reglang\xspace} \newcommand{\Pcompdecpdl}{comp-dec-pdl\xspace} \newcommand{\Ptoychain}{toychain\xspace} \newcommand{\Pbits}{bits\xspace} \newcommand{\Pmonae}{monae\xspace} \newcommand{\Pgames}{games\xspace}

\newcommand{\PTEXTfcslpcm}{the PCM library\xspace} \newcommand{\PTEXTFcslpcm}{The PCM library\xspace} \newcommand{\PTEXTinfotheo}{infotheo\xspace} \newcommand{\PTEXTInfotheo}{Infotheo\xspace}

\newcommand{\mycheckmark}{{\normalsize \checkmark}\xspace} \newcommand{\mycross}{$\mathbin{\tikz [x=1.4ex,y=1.4ex,line width=.2ex] \draw (0,0) -- (1,1) (0,1) -- (1,0);}$\xspace\xspace}

\newcommand{\citeappendix}[1]{}

\newcommand{\PaperVersion}{arxiv}

\renewcommand\UrlFont{\color{blue}\rmfamily}

\begin{document} \title{\Title} \titlerunning{\ShortTitle} \author{Pengyu Nie\inst{1} \and Karl Palmskog\inst{2} \and Junyi Jessy Li\inst{1} \and Milos Gligoric\inst{1}} \authorrunning{P. Nie et al.} \institute{The University of Texas at Austin, Austin, TX, USA\\ \email{pynie@utexas.edu,jessy@austin.utexas.edu,gligoric@utexas.edu} \and KTH Royal Institute of Technology, Stockholm, Sweden\\ \email{palmskog@kth.se}} \maketitle              %
\begin{abstract} \Cc for naming, spacing, and other essentially stylistic properties are necessary for developers to effectively understand, review, and modify source code in large software projects.   Consistent conventions in verification projects based on proof   assistants, such as \Coq, increase in importance as projects grow in   size and scope.   %
  While conventions can be documented and enforced manually at high cost, emerging approaches automatically learn and suggest idiomatic names in Java-like languages by applying statistical language models on large code corpora.   %
  However, due to its powerful language extension   facilities and fusion of type checking and computation, \Coq is a   challenging target for automated learning techniques.   We present novel generation models for learning and suggesting   \lemman for \Coq projects.   %
  Our models, based on multi-input neural networks,   are the first to leverage syntactic and   semantic information from \Coq's lexer (tokens in lemma statements), parser (\stree{s}), and kernel (\eterms)   for naming; the key insight is that learning from elaborated terms   can substantially boost model performance.   %
  We implemented our models in a \tooltype, dubbed \CoqConvTool, and   applied it on a large corpus of code derived   from the   \MathematicalComponents family of projects, known for its   stringent \cc. Our results show that \CoqConvTool substantially   outperforms baselines for suggesting \lemman, highlighting the   importance of using multi-input models and elaborated terms.

   \end{abstract}

\section{Introduction} \label{sec:intro}

Programming language source code with deficient \cc, such as misleading function and variable names and irregular spacing, is difficult for developers to effectively understand, review, and modify~\citep{Shneiderman1976,Miara1983,Avidan2017}. Code with haphazard adherence to conventions may also be more bug-prone~\citep{Boogerd2009}. The problem is exacerbated in large projects with many developers, where different source code files and components may have inconsistent and clashing conventions.

Many open source software projects manually document \cc that contributors are expected to follow, and maintainers willingly accept fixes of violations to such conventions~\cite{AllamanisETAL14Learning}. Enforcement of conventions can be performed by static analysis tools~\cite{GoogleJavaFormatWebPage,OguraETAL18Bring}. However, such tools require developers to write precise checks for conventions, which are tedious to define and often \emph{incomplete}. To address this problem, researchers have proposed techniques for automatically learning \cc for Java-like languages from code corpora by applying statistical language models~\cite{AllamanisETAL18Survey}. These models are applicable because code in these languages has high \emph{naturalness}~\cite{Hindle2012}, i.e., statistical regularities and repetitiveness. Learned conventions can then be used to, e.g., suggest names in code.

Proof assistants, such as \Coq~\cite{CoqArt}, are increasingly used to formalize results in advanced mathematics~\cite{Gonthier2008,Gonthier2013} and develop large trustworthy software systems, e.g., compilers, operating systems, file systems, and distributed systems~\cite{Leroy2009,Chen2015,Woos2016}. Such projects typically involve contributions of many participants over several years, and require considerable effort to maintain over time. \Cc are essential for evolution of large verification projects, and are thus highly emphasized in the Coq libraries HoTT~\cite{HoTTStyle} and Iris~\cite{IrisStyle}, in Lean's Mathlib~\cite{MathlibNaming}, and in particular in the influential \MathematicalComponents (\MathComp) \emph{family of \Coq projects}~\cite{MathCompContribGuide}. Extensive changes to adhere to conventions, e.g., on naming, are regularly requested by \MathComp maintainers for proposed external contributions~\cite{MathCompPR}, and its conventions have been adopted, to varying degrees, by a growing number of independent Coq projects~\cite{Bartzia2014,Sergey2015,Affeldt2015,Doczkal2018}.

We believe these properties make \Coq code related to \MathComp an attractive target for automated learning and suggesting of \cc, in particular, for suggesting \emph{lemma names}~\cite{Aspinall2016b}. However, serious challenges are posed by, on the one hand, \Coq's powerful language extension facilities and fusion of type checking and computation~\cite{Barendregt2002}, and on the other hand, the idiosyncratic conventions used by Coq practitioners compared to software engineers. Hence, although suggesting lemma names is similar in spirit to suggesting method names in Java-like languages~\cite{XuETAL19Method}, the former task is more challenging in that lemma names are typically much shorter than method names and tend to include heavily abbreviated terminology from logic and advanced mathematics; a single character can carry significant information about a lemma's meaning. For example, the \MathComp lemma names \texttt{card\_support\_normedTI} (``cardinality of support groups of a normed trivial intersection group'') and \CodeIn{extprod\_mulgA} (``associativity of multiplication operations in external product groups'') concisely convey information on \lemmastmt structure and meaning through both abbreviations and suffixes, as when the suffix \CodeIn{A} indicates an associative property.

In this paper, we present novel generation models for learning and suggesting \lemman for \Coq verification projects that address these challenges. Specifically, based on our knowledge of Coq and its implementation, we developed multi-input encoder-decoder neural networks for generating names that use information directly from Coq's internal data structures related to lexing, parsing, and type checking. In the context of naming, our models are the first to leverage the \emph{lemma \lstmt} as well as the corresponding \emph{\stree} and \emph{elaborated term} (which we call \emph{\ktree}) processed by Coq's kernel~\cite{DeMoura2015b}.

We implemented our models in a \tooltype, dubbed \CoqConvTool, which we used to learn from a high-quality Coq corpus derived from the \MathComp family. We then measured the performance of \CoqConvTool using automatic metrics, finding that it significantly outperforms baselines. Using our best model, we also suggested lemma names for \PTEXTfcslpcm~\cite{fcslpcm,Sergey2015}, which were manually reviewed by the project maintainer with encouraging results.

To allow \CoqConvTool to use information directly from \Coq's lexer, parser, and kernel, we extended the \SerAPI library~\cite{Gallego2016} to serialize \Coq tokens, \stree{s}, and \ktrees into a machine-readable format. This allowed us to achieve robustness against user-defined notations and other extensions to \Coq syntax. Thanks to our integration with \SerAPI and its use of metaprogramming, we expect our \tooltype to only require modest maintenance as \Coq evolves.

\vspace{3pt} \noindent We make the following key contributions in this work:

\begin{itemize}[leftmargin=*,itemsep=2pt,topsep=0pt] \item[$\bullet$] \textbf{Models}: We propose novel generation models based on   multi-input neural networks to learn and suggest \lemman for \Coq   verification projects. A~key property of our models is that they   combine data from several Coq phases, including lexing, parsing, and term elaboration. \item[$\bullet$] \textbf{Corpus}: Advised by \MathComp developers, we constructed   a corpus of high-quality \Coq code for learning \cc, totaling over   \CorpusKLOC{} LOC taken from \CorpusNumProjects core projects.  We   believe that our corpus can enable   development of many novel techniques for \Coq based on   statistical language models.   %
\item[$\bullet$] \textbf{\Tooltype}: We implemented a \tooltype, dubbed \CoqConvTool, which   suggests lemma names for a given \Coq project.   We envision \CoqConvTool being useful during the review process of   proposed contributions to a \Coq project. \item[$\bullet$] \textbf{Evaluation}: We performed several experiments with   \CoqConvTool to evaluate our models using our corpus.   %
  Our results show that \CoqConvTool performs significantly better than   several strong baselines, as measured by standard   automatic metrics~\cite{PapineniETAL02BLEU}. The results also reveal that   our novel multi-input models, as well as the incorporation of kernel   trees, are important for suggestion quality. Finally, we   performed a manual quality analysis by suggesting \lemman for a   medium sized Coq project~\cite{fcslpcm}, evaluated by its maintainer, who found   many of the suggestions useful for improving naming consistency. \end{itemize}

\noindent The appendix\IfStrEq{\PaperVersion}{main}{ of the extended version of   the paper~\cite{Nie2020}}{} describes more experiments, including an automatic evaluation on additional Coq projects.  We provide artifacts related to our \tooltype and corpus at: \ToolURL.

 \vspace{-3pt} \section{Background} \label{sec:background}

This section gives brief background related to \Coq and the Mathematical Components (\MathComp) family of projects, as well as the \SerAPI library.

\MyPara{Coq and Gallina} Coq is a proof assistant based on dependent types, implemented in the OCaml language~\cite{Coq810,CoqArt}. For our purposes, we view Coq as a programming language and type-checking toolchain. Specifically, Coq \emph{\docs} are sequences of \emph{sentences}, with each sentence ending with a period. Sentences are essentially either (a)~commands for printing and other output, (b)~definitions of functions, lemmas, and datatypes in the Gallina language~\cite{Gallina}, or (c)~expressions in the \Ltac tactic language~\cite{Delahaye2000}. We will refer to definitions of lemmas as in (b) as \emph{lemma sentences}. Coq internally represents a lemma sentence both as a sequence of tokens (lexing phase) and as a \stree (parsing phase).

\begin{figure}[t]
  \centering
\begin{lstlisting}[language=Coq,frame=single,keepspaces=true,xrightmargin=4pt,xleftmargin=10pt,numbers=left,numberstyle=\scriptsize,basicstyle=\scriptsize\ttfamily,escapechar=@]
Lemma mg_eq_proof L1 L2 (N1 : mgClassifier L1) : L1 =i L2 -> nerode L2 N1.@\label{mgeq:one}@
Proof. move => H0 u v. split => [/nerodeP H1 w|H1].@\label{mgeq:two}@
  - by rewrite -!H0.
  - apply/nerodeP => w. by rewrite !H0.@\label{mgeq:four}@
Qed.@\label{mgeq:five}@
\end{lstlisting}
\vspace{-8pt}
\caption{\Coq lemma on the theory of regular languages, including proof script.}
\vspace{-20pt}
\label{fig:reglang-before}
\end{figure}

In the typical workflow for a \Coq-based verification project, users write datatypes and functions and then interactively prove lemmas about them by executing different tactic expressions that may, e.g., discharge or split the current proof goal. Both statements to be proved and proofs are represented internally as \emph{terms} produced during an \emph{elaboration} phase~\cite{DeMoura2015b}; we refer to elaborated terms as \emph{\ktrees}. Hence, as tactics are successfully executed, they gradually build a \ktree. The \CoqIn{Qed} command sends the \ktree for a tentative proof to Coq's kernel for final certification. We call a collection of \Ltac tactic sentences that build a \ktree a \emph{proof script}.

\figurename~\ref{fig:reglang-before} shows a \Coq lemma and its proof script, taken verbatim from a development on the theory of regular languages~\cite{Doczkal2018}.  Line~\ref{mgeq:one} contains a lemma sentence with the lemma name \CoqIn{mg_eq_proof}, followed by a \emph{\lemmastmt} (on the same line) involving the arbitrary languages \CoqIn{L1} and \CoqIn{L2}, i.e., typed variables that are implicitly universally quantified. When \Coq processes line~\ref{mgeq:five}, the kernel certifies that the \ktree generated by the proof script (lines \ref{mgeq:two} to \ref{mgeq:four}) has the type (is a proof) of the \ktree for the \lemmastmt on line~\ref{mgeq:one}.

\MyPara{\MathComp and lemma naming} The \MathComp family of \Coq projects, including in particular the \MathComp library of general mathematical definitions and results~\cite{MCB}, grew out of Gonthier's proof of the four-color theorem~\citep{Gonthier2008}, with substantial developments in the context of the landmark proof of the odd order theorem in abstract algebra~\cite{Gonthier2013}. The \MathComp library is now used in many projects outside of the \MathComp family, such as in the project containing the lemma in \figurename~\ref{fig:reglang-before}~\cite{CoqRegLang}. \MathComp has documented naming conventions for two kinds of entities: (1)~variables and (2)~functions and lemmas~\cite{MathCompContribGuide}. Variable names tend to be short and simple, while function and lemma names can be long and consist of several \emph{name components}, typically separated by an underscore, but sometimes using CamelCase. Examples of definition and lemma names in \figurename~\ref{fig:reglang-before} include \CoqIn{mg\_eq\_proof}, \CoqIn{mgClassifier}, \CoqIn{nerode}, and \CoqIn{nerodeP}. Note that lemma names sometimes have \emph{suffixes} to indicate their meaning, such as \CoqIn{P} in \CoqIn{nerodeP} which says that the lemma is a \emph{characteristic property}. Coq functions tend to be named based on corresponding function definition bodies rather than just types (of the parameters and return value), analogously to methods in Java~\cite{LiuETAL19Learning}. In contrast, \MathComp lemma names tend to be based solely on the \lemmastmt. Hence, a more suitable name for the lemma in \figurename~\ref{fig:reglang-before} is \CoqIn{mg_eq_nerode}.

\MyPara{\SerAPI and Coq serialization} \SerAPI is an OCaml library and toolchain for machine interaction with Coq~\cite{Gallego2016}, which provides serialization and deserialization of Coq internal data structures to and from S-expressions (\sexps)~\cite{McCarthy1960}. \SerAPI is implemented using OCaml's PPX metaprogramming facilities~\cite{PPX}, which enable modifying OCaml program syntax trees at compilation time. \figurename~\ref{fig:lemma-sexp} shows the lemma sentence on line~\ref{mgeq:one} in \figurename~\ref{fig:reglang-before}, and below it, the corresponding (simplified) \sexps for its tokens, \stree, and \ktree, with the \lemmastmt highlighted in each representation. Note that the \stree omits the types of some quantified variables, e.g., for the types of \CoqIn{L1} and \CoqIn{L2}, as indicated by the \CoqIn{CHole} constructor. Note also that during elaboration of the \stree into the \ktree by Coq, an implicit variable \CoqIn{char} is added (all-quantified via \CoqIn{Prod}), and the extensional equality operator \CoqIn{=i} is translated to its globally unique \emph{kernel name}, \CoqIn{Coq.ssr.ssrbool.eq_mem}. Hence, a \ktree can be much larger and contain more information than the corresponding \stree.

\newsavebox\boxLemmaSexpSentence
\begin{lrbox}{\boxLemmaSexpSentence}
\begin{lstlisting}[language=Coq,escapechar=@,basicstyle=\scriptsize\ttfamily]
Lemma mg_eq_proof @\colorbox{lightgray}{L1 L2 (N1 : mgClassifier L1) : L1 =i L2 -> nerode L2 N1}@.
\end{lstlisting}
\end{lrbox}

\newsavebox\boxLemmaSexpTokens
\begin{lrbox}{\boxLemmaSexpTokens}
\begin{lstlisting}[language=LISP,escapechar=@,basicstyle=\scriptsize\ttfamily]
(Sentence((IDENT Lemma)(IDENT mg_eq_proof)@\colorbox{lightgray}{(IDENT L1)(IDENT L2)}@
  @\colorbox{lightgray}{(KEYWORD"(")(IDENT N1)(KEYWORD :)(IDENT mgClassifier)}@
  @\colorbox{lightgray}{(IDENT L1)(KEYWORD")")(KEYWORD :)(IDENT L1)(KEYWORD =i)(IDENT L2)}@
  @\colorbox{lightgray}{(KEYWORD ->)(IDENT nerode)(IDENT L2)(IDENT N1)}@(KEYWORD .)))
\end{lstlisting}
\end{lrbox}

\newsavebox\boxLemmaSexpSTree
\begin{lrbox}{\boxLemmaSexpSTree}
\begin{lstlisting}[language=LISP,escapechar=@,basicstyle=\scriptsize\ttfamily]
(VernacExpr()(VernacStartTheoremProof Lemma (Id mg_eq_proof)
 @\colorbox{lightgray}{(((CLocalAssum(Name(Id L1))(CHole()IntroAnonymous()))}@
   @\colorbox{lightgray}{(CLocalAssum(Name(Id L2))(CHole()IntroAnonymous()))}@
   @\colorbox{lightgray}{(CLocalAssum(Name(Id N1))}@
    @\colorbox{lightgray}{(CApp(CRef(Ser\_Qualid(DirPath())(Id mgClassifier)))(CRef(Ser\_Qualid(DirPath())(Id L1))))))}@
  @\colorbox{lightgray}{(CNotation(InConstrEntrySomeLevel"\_ -> \_")}@
   @\colorbox{lightgray}{(CNotation(InConstrEntrySomeLevel"\_ =i \_")}@
    @\colorbox{lightgray}{(CRef(Ser\_Qualid(DirPath())(Id L1)))(CRef(Ser\_Qualid(DirPath())(Id L2))))}@
   @\colorbox{lightgray}{(CApp(CRef(Ser\_Qualid(DirPath())(Id nerode)))}@
    @\colorbox{lightgray}{(CRef(Ser\_Qualid(DirPath())(Id L2)))(CRef(Ser\_Qualid(DirPath())(Id N1))))))}@))
\end{lstlisting}
\end{lrbox}

\newsavebox\boxLemmaSexpKTree
\begin{lrbox}{\boxLemmaSexpKTree}
\begin{lstlisting}[language=LISP,escapechar=@,basicstyle=\scriptsize\ttfamily]
(Prod (Name (Id char)) ... @\colorbox{lightgray}{(Prod (Name (Id L1))}@ ...
  @\colorbox{lightgray}{(Prod (Name (Id L2))}@ ... @\colorbox{lightgray}{(Prod (Name (Id N1))}@ ...
   @\colorbox{lightgray}{(Prod Anonymous (App (Ref (DirPath ((Id ssrbool) (Id ssr) (Id Coq))) (Id eq\_mem))}@ ...
     @\colorbox{lightgray}{(Var (Id L1))}@ ... @\colorbox{lightgray}{(Var (Id L2)))}@
    @\colorbox{lightgray}{(App (Ref (DirPath ((Id myhill\_nerode) (Id RegLang))) (Id nerode))}@ ...
     @\colorbox{lightgray}{(Var (Id L2))}@ ... @\colorbox{lightgray}{(Var (Id N1)))))))}@)
\end{lstlisting}
\end{lrbox}

\begin{figure}[t]
\centering
\begin{tikzpicture}[
  line width=0.4pt,
  node distance=0ex and 0em,
  every node/.style={scale=1},
  gridBox/.style={rectangle, opacity=0, draw=red},
  box/.style={rectangle, draw=black, inner sep=2pt, font=\small},
  rounded box/.style={rectangle, rounded corners, draw=black, inner sep=2pt, font=\small},
  anno/.style={font=\footnotesize},
]

  \DefMacro{wCodeBox}{\textwidth}

  \node (g-Sentence) at (0,0) [box, minimum width=\UseMacro{wCodeBox}, minimum height=3.5ex] {};
  \node (b-Sentence) [right = 0 of g-Sentence.west] [scale=0.97] {\usebox\boxLemmaSexpSentence};
  \node (b-SentenceText) [below left = 1pt and 1pt of g-Sentence.north east] [box, scale=0.9] {\bf sentence};

  \node (g-Tokens) [below = 0 of g-Sentence.south] [box, minimum width=\UseMacro{wCodeBox}, minimum height=11ex] {};
  \node (b-Tokens) [right = 0 of g-Tokens.west] [scale=0.97] {\usebox\boxLemmaSexpTokens};
  \node (b-TokensText) [below left = 1pt and 1pt of g-Tokens.north east] [box, scale=0.9] {\bf tokens};

  \node (g-STree) [below = 0 of g-Tokens.south] [box, minimum width=\UseMacro{wCodeBox}, minimum height=24ex] {};
  \node (b-STree) [right = 0 of g-STree.west] [scale=0.97] {\usebox\boxLemmaSexpSTree};
  \node (b-STreeText) [below left = 1pt and 1pt of g-STree.north east] [box, scale=0.9] {\bf \stree};

  \node (g-KTree) [below = 0 of g-STree.south] [box, minimum width=\UseMacro{wCodeBox}, minimum height=15ex] {};
  \node (b-KTree) [right = 0 of g-KTree.west] [scale=0.97] {\usebox\boxLemmaSexpKTree};
  \node (b-KTreeText) [below left = 1pt and 1pt of g-KTree.north east] [box, scale=0.9] {\bf \ktree};

\end{tikzpicture}
\vspace{-15pt}
\caption{Coq lemma sentence at the top, with \sexps for, from
just below to bottom: tokens, \stree, and \ktree; the \lemmastmt in each is highlighted.}
\vspace{-15pt}
\label{fig:lemma-sexp}
\end{figure}

\vspace{-5pt} \section{Models} \label{sec:technique} \vspace{-5pt}

In this section, we describe our multi-input generation models for suggesting Coq \lemman. Our models consider lemma name generation with an \emph{\encdec} mindset, i.e., we use neural architectures specifically designed for transduction tasks~\cite{sutskever2014sequence}. This family of architectures is commonly used for sequence generation, e.g., in machine translation~\cite{BahdanauETAL15Neural} and code summarization~\cite{LeClairETAL19Neural}, where it has been found to be much more effective than traditional probabilistic and retrieval-based approaches.

\vspace{-2pt} \subsection{Core Architecture}\label{sec:corearchitect}

\begin{figure}[t]   \centering   \includegraphics[width=.7\columnwidth]{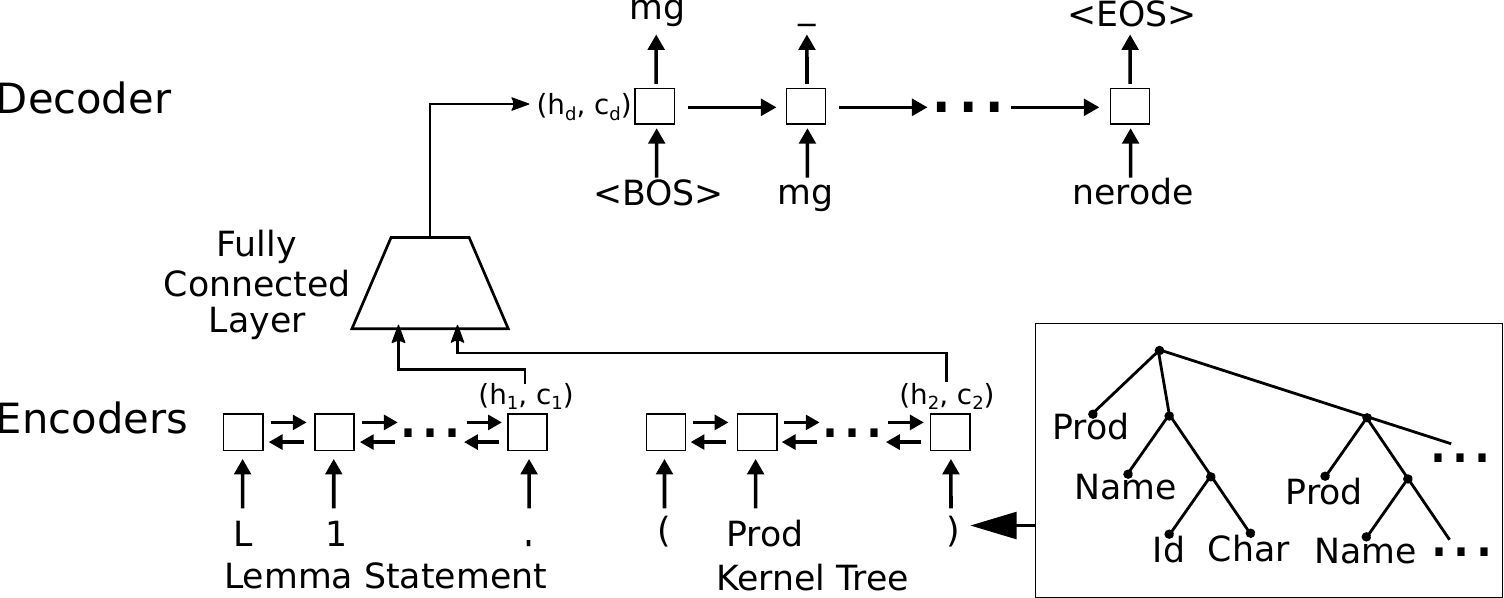}   \vspace{-5pt}   \caption{Core architecture of our   multi-input \encdec models.     \label{fig:seqtoseq-arch}   }   \vspace{-8pt} \end{figure}

Our encoders are Recurrent Neural Networks (RNNs) that learn a deep semantic representation of a given \lemmastmt from its tokens, \stree, and \ktree. The decoder---another RNN---generates the descriptive lemma name as a sequence. The model is trained end-to-end, maximizing the probability of the generated lemma name given the input. In contrast to prior work in language-code tasks that uses a single encoder~\cite{GaoETAL19Neural}, we design multi-input models that leverage both syntactic and semantic information from \Coq's lexer, parser, and kernel. A high-level visualization of our architecture is shown in \figurename~\ref{fig:seqtoseq-arch}.

\MyPara{Encoding} Our multi-input encoders combine different kinds of syntactic and semantic information in the encoding phase. We use a different encoder for each input, which are: \lstmt, \stree, and \ktree.

Coq data structure instances can be large, with \stree{s} having an average depth of \UseMacro{lemma-all-t1-all-AVG-depth_fsexp} and \ktree{s} \UseMacro{lemma-all-t1-all-AVG-depth_bsexp} in our corpus (we provide detailed statistics in Section~\ref{sec:corpus}). Therefore, we \flatten the trees into sequences, which can be trained more efficiently than tree encoders without performance loss~\cite{HuETAL18Deep}. We \flatten the trees with pre-order traversal, and we use ``('' and ``)'' as the boundaries of the children of a node.  In later parts of this paper, we use \sktrees to refer to their \flattened versions.  In Section~\ref{sec:treesimp}, we introduce \emph{tree \simplification{}} to reduce the length of the resulting sequences.

To encode \lstmts and \flattened tree sequences, we use \bidirectional Long-Short Term Memory (LSTM)~\cite{hochreiter1997long} networks. LSTMs are advanced RNNs good at capturing long-range dependencies in a sequence, and are widely used in encoders~\cite{HuETAL18Deep}. A \bidirectional LSTM learns stronger representations (than a uni-directional LSTM) by encoding a sequence from both left to right and right to left~\cite{ZhangETAL15Bidirectional}.

\MyPara{Decoding} We use an LSTM (left to right direction only) as our decoder. To obtain the initial hidden and cell states $(h_d, c_d)$ of the decoder, we learn a unified representation of these separate encoders by concatenating their final hidden and cell states $(h_i, c_i)$, and then applying a fully connected layer on the concatenated states: $   h_d = W_h \cdot \mathtt{concat}([h_i]) +b_h\; \mathrm{and}\;   c_d = W_c \cdot \mathtt{concat}([c_i]) +b_c, $ where $W_h$, $W_c$, $b_h$, and $b_c$ are learnable parameters.

During training, we maximize the log likelihood of the reference lemma name given all input sequences.  Standard beam search is used to reduce the search space for the optimal sequence of tokens. With regular decoding, at each time step the decoder generates a new token relying on the preceding \emph{generated} token, which can be error-prone and leads to slow convergence and instability. We mitigate this problem by performing decoding with teacher forcing~\cite{WilliamsAndZipser89Learning} such that the decoder relies on the preceding \emph{reference} token. At test time, the decoder still uses the proceeding generated token as input.

\MyPara{Attention}  With RNN encoders, the input  sequence is compressed into the RNN's final hidden states, which results in a loss of information, especially for longer sequences. The attention mechanism~\cite{LuongETAL15Effective} grants the decoder access to the encoder hidden and cell states for all previous tokens.  At each decoder time step, an attention vector is calculated as a distribution over all encoded tokens, indicating which token the decoder should ``pay attention to''. To make the attention mechanism work with multiple encoders, we concatenate the hidden states of the $n$ encoders $[h_1,...,h_n]$ and apply an attention layer on the result~\cite{UnanueETAL18Shared}.

\MyPara{Initialization} Since there are no pre-trained token embeddings for Coq, we initialize each unique token in the vocabulary with a random vector sampled from the uniform distribution $U(-0.1, 0.1)$. These embeddings are trained together with the model.  The hidden layer parameters of the encoders and decoders are also initialized with random vectors sampled from the same uniform distribution.

\begin{figure}[t]
\begin{lstlisting}[language=LISP,frame=single,escapechar=@,basicstyle=\scriptsize\ttfamily]
(Prod Anonymous (App @\colorbox{lightgray}{(Ref (DirPath ((Id ssrbool) (Id ssr) (Id Coq))) (Id}@ eq_mem@\colorbox{lightgray}{))}@ ...
  @\colorbox{lightgray}{(}@(App (Ref ... ))@\colorbox{lightgray}{)}@ ... ))
\end{lstlisting}
\vspace{-10pt}
\begin{lstlisting}[language=LISP,frame=single,escapechar=@,basicstyle=\scriptsize\ttfamily]
(Prod Anonymous (App eq_mem ... (App (Ref ... )) ... ))
\end{lstlisting}
\vspace{-8pt}
\caption{\Ktree \sexp before and after \trimming; \trimmed parts are highlighted.}
\vspace{-10pt}
\label{fig:ktree-trimming-example}
\end{figure}

\vspace{-2pt} \subsection{Tree \Simplification} \label{sec:treesimp}

While \sktrees for \lemmastmts can be large, not all parts of the trees are relevant for naming. For instance, each constant reference is expanded to its fully qualified form in the \ktree, but the added prefixes are usually related to directory paths and likely do not contain relevant information for generating the name of the lemma. Irrelevant information in long sequences can be detrimental to the model, since the model would have to reason about and encode all tokens in the sequence.

To this end, we implemented \emph{\trimming} heuristics for both \strees and \ktrees to remove irrelevant parts.  The heuristics essentially: (1)~replace the fully qualified name sub-trees with only the last component of the name; (2)~remove the location information from sub-trees; (3)~extract the singletons, i.e., non-leaf nodes that have only one child. \figurename~\ref{fig:ktree-trimming-example} illustrates the \trimming of a \ktree, with the upper box showing the tree before \trimming with the parts to be removed highlighted, and the lower box showing the tree after \trimming. In the example in the figure, we \trimmed a fully qualified name and extracted a singleton.  These heuristics greatly reduce the size of the tree: for \ktree{s}, they reduce the average depth from \UseMacro{lemma-filtered-t1-all-AVG-depth_bsexp} to \UseMacro{lemma-filtered-t1-all-AVG-depth_bsexp_trimmed}.

Our models use \trimmed trees as the inputs to the encoders. As we discuss in more detail in Section~\ref{sec:eval}, the \trimmed trees help the models to focus better on the relevant parts of the inputs. While the attention mechanism in principle could learn what the relevant parts of the trees are, our evaluation shows that it can easily be overwhelmed by  large amounts of irrelevant information.

\vspace{-2pt} \subsection{Copy Mechanism}\label{sec:copy}

We found it common for  lemma name tokens to only occur in a single Coq \doc, whence they are unlikely to appear in the vocabulary learned from the \train set, but can still appear in the respective \lemmastmt, \stree, or \ktree. For example, \CoqIn{mg} occurs in both the lemma name and \lstmt in \figurename~\ref{fig:reglang-before}, but not outside the file the lemma is in. To account for this, we adopt the copy mechanism~\cite{SeeETAL17Get} which improves the generalizability of our model  by allowing the decoder to \emph{copy} from inputs rather than always choosing one word from the fixed vocabulary from the training set. To handle multiple encoders, similar to what we did with the attention layer, we concatenate the hidden states of each encoder and apply a copy layer on the concatenated hidden states.

\vspace{-2pt} \subsection{\SubTokenization} \label{sec:technique:subtok}

We \subtokenize all inputs (\lstmts, \sktrees) and outputs (lemma names) in a pre-processing step.  Previous work on learning from software projects has shown that \subtokenization helps to reduce the sparsity of the vocabulary and improves the performance of the model~\cite{Babii2019}.  However, unlike Java-like languages where the method names (almost) always follow the CamelCase convention, lemma names in Coq use a mix of snake\_case, CamelCase, prefixes, and suffixes, thus making \subtokenization more complex.  For example, \CodeIn{extprod\_mulgA} should be \subtokenized to \CodeIn{extprod}, \CodeIn{\_}, \CodeIn{mul}, \CodeIn{g}, and \CodeIn{A}.

To perform \subtokenization, we implemented a set of heuristics based on the conventions outlined by \MathComp developers~\cite{MathCompContribGuide}. After \subtokenization, the vocabulary size of lemma names in our corpus was reduced from \UseMacro{lemma-filtered-t1-train-num-lemmas} to \UseMacro{vocab-size-sub-token}. When applying the \subtokenizer on the \lemmastmts and \sktrees, we \subtokenize the identifiers and not the keywords or operators. 

\vspace{-2pt} \subsection{Repetition Prevention}\label{sec:repetition}

We observed that decoders often generated repeated tokens, e.g., \CoqIn{mem_mem_mem}.  This issue also exists in natural language summarization~\cite{SuzukiAndNagata17Cutting-off}.  We further observed that it is very unlikely to have repeated \subtok{s} in lemma names used by proof engineers (only \UseMacro{sub-token-repeat-rate} of cases in our corpus).  Hence, we simply forbid the decoder from repeating a \subtok (modulo ``\_'') during beam search.

\vspace{-5pt} \section{Corpus} \label{sec:corpus} \vspace{-3pt}

We constructed a corpus of \CorpusNumProjects large Coq projects from the \MathComp family, totaling \CorpusKLOC{} lines of code (LOC). We selected these projects based on the recommendation of \MathComp developers, who emphasized their high quality and stringent adherence to \cc.  Our corpus is \emph{self-contained}: there are inter-project dependencies within the corpus, but no project depends on a project outside the corpus (except Coq's standard library).  All projects build with \Coq version \CoqVersion.  Note that we need to be able to build projects to be able to extract tokens, \strees, and \ktrees.

\begin{table*}[t]
\begin{scriptsize}
\begin{center}
\caption{\UseMacro{table-caption-corpus}}
\begin{tabular}{l c c r r r r r c r r}
\toprule
\multirow{2}{*}{\TableHeadProject} &  & \multirow{2}{*}{\TableHeadSHA} & \multirow{2}{*}{\TableHeadFiles} & \multirow{2}{*}{\TableHeadNumLemma} & \multirow{2}{*}{\TableHeadNumTok} & \multicolumn{2}{c}{\TableHeadLOC} &  & \multicolumn{2}{c}{\TableHeadLocPerDoc} \\ \cline{7-8}\cline{10-11}
 &  &  &  &  &  & \TableHeadSpecLOC & \TableHeadProofLOC & & \TableHeadSpecLocPerDoc & \TableHeadProofLocPerDoc \\
\midrule
\Pfinmap
& \href{\UseMacro{corpus-finmap-url}}{\urlsymbol}
& \texttt{\UseMacro{corpus-finmap-sha-pretty}}
& \UseMacro{corpus-finmap-num-docs}
& \UseMacro{lemma-filtered-project-finmap-num-lemmas}
& \UseMacro{dataset-project-finmap-num-tokens}
& \UseMacro{corpus-finmap-spec-loc}
& \UseMacro{corpus-finmap-proof-loc}
 & 
& \UseMacro{corpus-finmap-AVG-spec-loc-per-doc}
& \UseMacro{corpus-finmap-AVG-proof-loc-per-doc}
\\
\Pfourcolor
& \href{\UseMacro{corpus-fourcolor-url}}{\urlsymbol}
& \texttt{\UseMacro{corpus-fourcolor-sha-pretty}}
& \UseMacro{corpus-fourcolor-num-docs}
& \UseMacro{lemma-filtered-project-fourcolor-num-lemmas}
& \UseMacro{dataset-project-fourcolor-num-tokens}
& \UseMacro{corpus-fourcolor-spec-loc}
& \UseMacro{corpus-fourcolor-proof-loc}
 & 
& \UseMacro{corpus-fourcolor-AVG-spec-loc-per-doc}
& \UseMacro{corpus-fourcolor-AVG-proof-loc-per-doc}
\\
\Pmathcomp
& \href{\UseMacro{corpus-mathcomp-url}}{\urlsymbol}
& \texttt{\UseMacro{corpus-mathcomp-sha-pretty}}
& \UseMacro{corpus-mathcomp-num-docs}
& \UseMacro{lemma-filtered-project-mathcomp-num-lemmas}
& \UseMacro{dataset-project-mathcomp-num-tokens}
& \UseMacro{corpus-mathcomp-spec-loc}
& \UseMacro{corpus-mathcomp-proof-loc}
 & 
& \UseMacro{corpus-mathcomp-AVG-spec-loc-per-doc}
& \UseMacro{corpus-mathcomp-AVG-proof-loc-per-doc}
\\
\Poddorder
& \href{\UseMacro{corpus-oddorder-url}}{\urlsymbol}
& \texttt{\UseMacro{corpus-oddorder-sha-pretty}}
& \UseMacro{corpus-oddorder-num-docs}
& \UseMacro{lemma-filtered-project-oddorder-num-lemmas}
& \UseMacro{dataset-project-oddorder-num-tokens}
& \UseMacro{corpus-oddorder-spec-loc}
& \UseMacro{corpus-oddorder-proof-loc}
 & 
& \UseMacro{corpus-oddorder-AVG-spec-loc-per-doc}
& \UseMacro{corpus-oddorder-AVG-proof-loc-per-doc}
\\
\midrule
\multirow{2}{*}{\UseMacro{table-head-summary-group-t1}}
 & & \TableHeadNA 
& \UseMacro{corpus-t1-AVG-num-docs}
& \UseMacro{lemma-filtered-t1-project-AVG-num-lemmas}
& \UseMacro{dataset-t1-project-AVG-num-tokens}
& \UseMacro{corpus-t1-AVG-spec-loc}
& \UseMacro{corpus-t1-AVG-proof-loc}
 & 
& \UseMacro{corpus-t1-AVG-spec-loc-per-doc}
& \UseMacro{corpus-t1-AVG-proof-loc-per-doc}
\\
 & & \TableHeadNA 
& \UseMacro{corpus-t1-SUM-num-docs}
& \UseMacro{lemma-filtered-t1-project-SUM-num-lemmas}
& \UseMacro{dataset-t1-project-SUM-num-tokens}
& \UseMacro{corpus-t1-SUM-spec-loc}
& \UseMacro{corpus-t1-SUM-proof-loc}
 & 
& \UseMacro{corpus-t1-SUM-spec-loc-per-doc}
& \UseMacro{corpus-t1-SUM-proof-loc-per-doc}
\\
\bottomrule
\end{tabular}
\end{center}
\end{scriptsize}
\vspace{\UseMacro{vspace-corpus}}
\end{table*}

\MyPara{Constituent projects} Table~\ref{tbl:corpus} lists the projects in the corpus, along with basic information about each project. The table includes columns for the project identifier, revision SHA, number of \docs (\TableHeadTextFiles), number of lemmas (\TableHeadTextNumLemma), number of tokens (\TableHeadTextNumTok), LOC for specifications (\TableHeadTextSpecLOC) and proof scripts (\TableHeadTextProofLOC), and average LOC per \doc for specifications and proof scripts. The \Pmathcomp SHA corresponds to version 1.9.0 of the library.  The LOC numbers are computed with Coq's bundled \coqwc tool. The last two rows of the table show the averages and sums across all projects.

\begin{table}[t]
\begin{scriptsize}
\begin{center}
\caption{\UseMacro{table-caption-dataset-lemma}}
\begin{tabular}{l | r r c rr c rr}
\toprule
\multirow{2}{*}{\TableHeadLMSet}  & \multirow{2}{*}{\TableHeadNumDoc}  & \multirow{2}{*}{\TableHeadNumLemma}  & & \multicolumn{2}{c}{\TableHeadLMName}  & & \multicolumn{2}{c}{\TableHeadLMStmt}  \\ \cline{5-6} \cline{8-9}
 & &  & & \TableHeadNumChar & \TableHeadNumSubToken & & \TableHeadNumChar & \TableHeadNumSubToken \\
\midrule
\UseMacro{table-head-dslemma-filtered-t1-train}
& \UseMacro{dataset-t1-train-num-documents}
& \UseMacro{lemma-filtered-t1-train-num-lemmas}
& 
& \UseMacro{lemma-filtered-t1-train-AVG-num_char_name}
& \UseMacro{lemma-filtered-t1-train-AVG-num_subtok_name}
& 
& \UseMacro{lemma-filtered-t1-train-AVG-num_char_statement}
& \UseMacro{lemma-filtered-t1-train-AVG-num_subtok_statement}
\\
\UseMacro{table-head-dslemma-filtered-t1-val}
& \UseMacro{dataset-t1-val-num-documents}
& \UseMacro{lemma-filtered-t1-val-num-lemmas}
& 
& \UseMacro{lemma-filtered-t1-val-AVG-num_char_name}
& \UseMacro{lemma-filtered-t1-val-AVG-num_subtok_name}
& 
& \UseMacro{lemma-filtered-t1-val-AVG-num_char_statement}
& \UseMacro{lemma-filtered-t1-val-AVG-num_subtok_statement}
\\
\UseMacro{table-head-dslemma-filtered-t1-test}
& \UseMacro{dataset-t1-test-num-documents}
& \UseMacro{lemma-filtered-t1-test-num-lemmas}
& 
& \UseMacro{lemma-filtered-t1-test-AVG-num_char_name}
& \UseMacro{lemma-filtered-t1-test-AVG-num_subtok_name}
& 
& \UseMacro{lemma-filtered-t1-test-AVG-num_char_statement}
& \UseMacro{lemma-filtered-t1-test-AVG-num_subtok_statement}
\\
\bottomrule
\end{tabular}
\end{center}
\end{scriptsize}
\vspace{\UseMacro{vspace-dataset-lemma}}
\end{table}

\MyPara{Corpus statistics} We extracted all lemmas from the corpus, and initially we obtained \UseMacro{lemma-all-t1-all-num-lemmas} lemmas in total.  However, we found several outlier lemmas where the \lstmt, \stree and \ktree were very large. To ensure stable training, and similar to prior work on generating method names for Java~\cite{LiuETAL19Learning}, we excluded the lemmas with the deepest 25\% \ktrees. This left us with \UseMacro{lemma-filtered-t1-all-num-lemmas} lemmas. Column~4 of Table~\ref{tbl:corpus} shows the number of lemmas after filtering.

We randomly split corpus \docs into \train, \val, and \test sets which contain 80\%, 10\%, 10\% of the \docs, respectively. Table~\ref{tbl:dataset-lemma} shows statistics on the lemmas in each set, which includes columns for the number of \docs, the number of lemmas, the average number of characters and \subtok{s} in lemma names, and the average number of characters and \subtok{s} in \lstmt{s}.

\begin{figure}[t]   \centering   \includegraphics[scale=0.65]{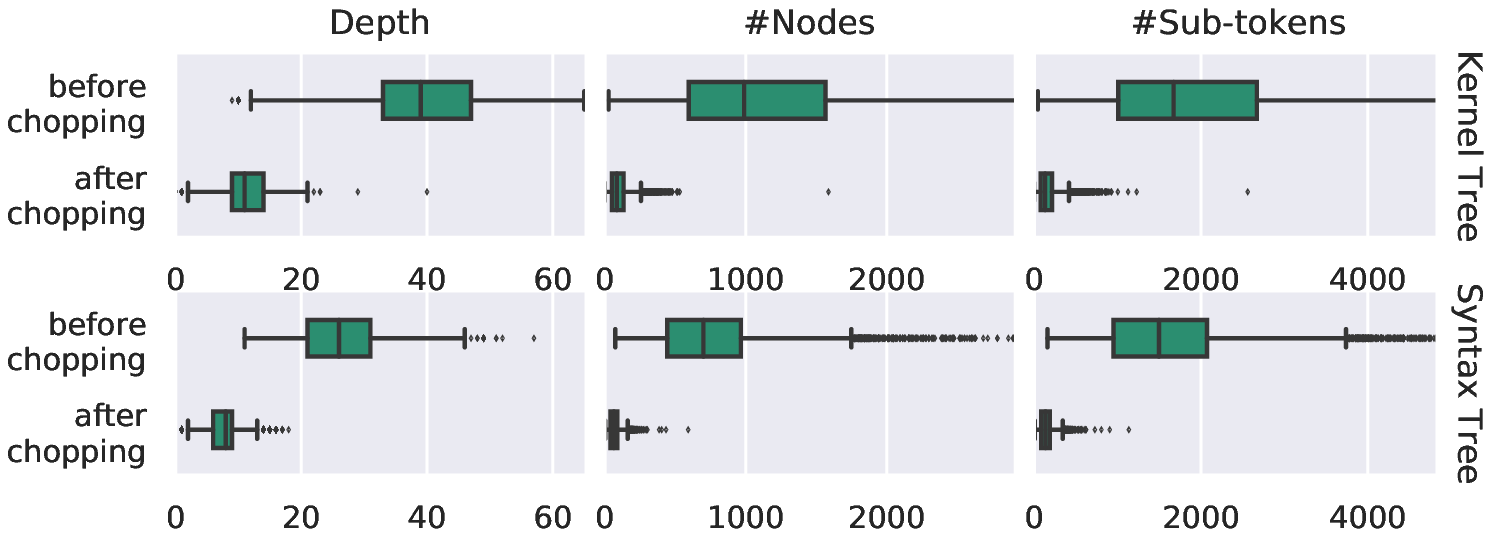}   \vspace{-10pt}   \caption{Statistics on \sktrees.}\label{fig:lemma-metrics}   \vspace{-10pt} \end{figure}

\figurename~\ref{fig:lemma-metrics} illustrates the changes of the depth, number of nodes and number of \subtok{s} (after \flattening) of the \ktrees (first row) and \strees (second row) before and after \simplification{}.  Our \trimming process reduced tree depth by \UseMacro{lemma-filtered-t1-all-depth_bsexp_reduction} for \ktrees and \UseMacro{lemma-filtered-t1-all-depth_fsexp_reduction} for \strees, and reduced the number of nodes by \UseMacro{lemma-filtered-t1-all-num_nodes_bsexp_reduction} for \ktrees and \UseMacro{lemma-filtered-t1-all-num_nodes_fsexp_reduction} for \strees; after \flattening, the resulting average sequence length is, for \ktrees \UseMacro{lemma-filtered-t1-all-AVG-num_subtok_bsexp_trimmed-int} comparing to the original \UseMacro{lemma-filtered-t1-all-AVG-num_subtok_bsexp-int}, and for \strees \UseMacro{lemma-filtered-t1-all-AVG-num_subtok_fsexp_trimmed-int} comparing to the original \UseMacro{lemma-filtered-t1-all-AVG-num_subtok_fsexp-int}. We provide additional statistics on lemmas before filtering in the appendix\IfStrEq{\PaperVersion}{main}{ of the extended paper~\cite{Nie2020}}{}. %

\vspace{-5pt} \section{Implementation} \label{sec:impl} \vspace{-3pt}

In this section, we briefly describe our \tooltype which implements the models in Section~\ref{sec:technique} and processes and learns from the corpus in Section~\ref{sec:corpus}; we dub this \tooltype \CoqConvTool. The components of the \tooltype can be divided into two categories: (1)~components that interact with \Coq or directly process information extracted from Coq, and (2)~components concerned with machine learning and name generation.

The first category includes several OCaml-based tools integrated with \SerAPI~\cite{Gallego2016} (and thus \Coq itself), and Python-based tools for processing of data obtained via \SerAPI from \Coq. All OCaml tools have either already been included in, or accepted for inclusion into, \SerAPI itself. The tools are as follows:

\MyPara{\sercomp} We integrated the existing program \sercomp distributed with \SerAPI into \CoqConvTool to serialize Coq \docs to lists of \sexps for \strees.

\MyPara{\sertok} We developed an OCaml program dubbed \sertok on top of \SerAPI. The program takes a Coq \doc as input and produces \sexps of all tokens found by Coq's lexer in the \doc, organized at the sentence level.

\MyPara{\sername} We developed an OCaml program dubbed \sername on top of \SerAPI. The program takes a list of fully qualified (kernel) lemma names and produces \sexps for the \ktrees of the corresponding \lemmastmts.

\MyPara{\postproc \& \subtokenizertool} We created two small independent tools in Python to post-process \Coq \sexps and perform \subtokenization, respectively.

For the second category, we implemented our machine learning models in Python using two widely-used deep learning libraries: PyTorch~\cite{PaszkeETAL17Automatic} and OpenNMT~\cite{KleinETAL17OpenNMT}. More specifically, we extended the sequence-to-sequence models in OpenNMT to use multi-input encoders, and extended attention and copy layers to use multiple inputs. Source code for the components of \CoqConvTool is available from: \ToolURL.

 \vspace{-3pt} \section{Evaluation} \label{sec:eval} \vspace{-3pt}

\pgfmathsetmacro{\varNumQSLemmaNameEvaled}{\UseMacro{qs-fcslpcm-num-lemmas}-\UseMacro{qs-fcslpcm-ln-full-correct-top-1}} \newcommand{\NumQSLemmaNameEvaled}{\pgfmathprintnumber[fixed, precision=0]{\varNumQSLemmaNameEvaled}\xspace}

This section presents an extensive evaluation of our models as implemented in \CoqConvTool. Our automatic evaluation (Section~\ref{sec:eval:auto}) compares \CoqConvTool with a series of strong baselines and reports on ablation experiments; additional experiments, e.g., on \trimming heuristics, are described in the appendix\IfStrEq{\PaperVersion}{main}{ of the extended version of the paper~\cite{Nie2020}}{}. Our manual quality assessment (Section~\ref{sec:eval:human}) analyzes \NumQSLemmaNameComments comments we received from the maintainer of \PTEXTfcslpcm on names suggested by \CoqConvTool for that project using our best model.

\vspace{-5pt} \subsection{Models and Baselines} \label{sec:eval:models-and-baselines}

We study the combinations of: (1)~using individual input (\lstmt and trees) in a single encoder, or multi-input encoders with different mixture of these inputs; and (2)~using the attention and copy mechanisms. Our inputs include: \lstmt ({\em \LStmtAcro}), \stree ({\em \STreeAcro}), \trimmedstree ({\em \TrimmedSTreeAcro}), \ktree ({\em \KTreeAcro}), and \trimmedktree ({\em \TrimmedKTreeAcro}).  For multiple inputs, the models are named by concatenating inputs with ``+''; a ``+'' is also used to denote the presence of attention ({\em attn}) or copy ({\em copy}).  For example, \UseMacro{ln-s+bsexpl1+attn+copy} refers to a model that uses two encoders---one for \lstmt and one for \trimmedktree---and uses attention and copy mechanisms.

We consider the vanilla \encdec models with only one input (\lstmt, \ktree, or \stree) as baseline models.  We also compare with a \retrievalbased baseline model implemented using Lucene~\cite{LuceneWebpage}: a k-nearest neighbors classifier using the tf-idf of the tokens in \lstmt as features.

Hyperparameters are tuned on the \val set within the following options: embedding dimensions from \{200, 500, 1000\}, number of hidden units in each LSTM from \{200, 500, 1000\}, number of stacked LSTM layers from \{1, 2, 3\}. We set the dropout rate between LSTM layers to 0.5.  We set the output dimension of the fully connected layer for combining encoders to the same number as the number of hidden units in each LSTM. We checked the validation loss every 200 training steps (as defined in OpenNMT~\cite{KleinETAL17OpenNMT}, which is similar to one training epoch on our dataset), and set an early stopping threshold of 3.  We used the Adam~\cite{KingmaAndBa15Adam} optimizer with a learning rate of 0.001.  We used a beam size of 5 in beam search.  All the experiments were run with one NVIDIA 1080-TI GPU and Intel Xeon E5-2620 v4 CPU.

\vspace{-3pt} \subsection{Automatic Evaluation} \label{sec:eval:auto}

\MyPara{Metrics} We use four automatic metrics which evaluate generated lemma names against the reference lemma name (as written by developers) in the \test set. Each metric captures a different level of granularity of the generation quality.  \emph{\Bleu}~\cite{PapineniETAL02BLEU} is a standard metric used in transduction tasks including language $\leftrightarrow$ code transduction. It calculates the number of n-grams in a generated sequence that also appear in the reference sequence, where one ``n-gram'' is n consecutive items in a sequence (in our case, one ``n-gram'' is n consecutive characters in the sequence of characters of the lemma name). We use it to compute the $1\!\sim\!4$-grams overlap between the characters in generated name and characters in the reference name, averaged between $1\!\sim\!4$-grams with smoothing method proposed by Lin and Och~\cite{LinAndOch04ORANGE}. \emph{\Fragacc} computes the accuracy of generated names on the fragment level, which is defined by splitting the name by underscores (``\_''). For example, \CodeIn{map\_determinant\_mx} has a \fragacc of 66.7\%
when evaluated against \CodeIn{det\_map\_mx}.  Unlike \Bleu, \fragacc ignores the ordering of the fragments. Finally, \emph{\toponeacc} and \emph{\topfiveacc} compute how often the true name fully matches the generated name or is one of the top-5 generated names.

\begin{table*}[t]
\begin{scriptsize}
\begin{center}
\caption{\UseMacro{table-caption-results-ln-t1--t1--t1-main}}
\begin{tabular}{c l r r r r}
\toprule
\TableHeadGroup & \TableHeadModel
& \UseMacro{table-head-results-ln-BLEU-4}
& \UseMacro{table-head-results-ln-frag-acc}
& \UseMacro{table-head-results-ln-full-acc-top-1}
& \UseMacro{table-head-results-ln-full-acc-top-5}
 \\
\midrule
 & 
\UseMacro{ln-s+bsexpl1+fsexpl1+attn+copy}
& 
\UseMacro{results-ln-t1--t1--t1-ln-s+bsexpl1+fsexpl1+attn+copy-test-AVG-BLEU-4}
& 
\UseMacro{results-ln-t1--t1--t1-ln-s+bsexpl1+fsexpl1+attn+copy-test-AVG-frag-acc}
& 
\UseMacro{results-ln-t1--t1--t1-ln-s+bsexpl1+fsexpl1+attn+copy-test-AVG-full-acc-top-1}
& 
\UseMacro{results-ln-t1--t1--t1-ln-s+bsexpl1+fsexpl1+attn+copy-test-AVG-full-acc-top-5}
\\
 & 
\UseMacro{ln-s+bsexpl1+attn+copy}
& 
\textbf{\UseMacro{results-ln-t1--t1--t1-ln-s+bsexpl1+attn+copy-test-AVG-BLEU-4}}
& 
\textbf{\UseMacro{results-ln-t1--t1--t1-ln-s+bsexpl1+attn+copy-test-AVG-frag-acc}}
& 
\textbf{\UseMacro{results-ln-t1--t1--t1-ln-s+bsexpl1+attn+copy-test-AVG-full-acc-top-1}}
& 
\textbf{\UseMacro{results-ln-t1--t1--t1-ln-s+bsexpl1+attn+copy-test-AVG-full-acc-top-5}}
\\
 & 
\UseMacro{ln-s+fsexpl1+attn+copy}
& 
\UseMacro{results-ln-t1--t1--t1-ln-s+fsexpl1+attn+copy-test-AVG-BLEU-4}
& 
\UseMacro{results-ln-t1--t1--t1-ln-s+fsexpl1+attn+copy-test-AVG-frag-acc}
& 
\UseMacro{results-ln-t1--t1--t1-ln-s+fsexpl1+attn+copy-test-AVG-full-acc-top-1}
& 
\UseMacro{results-ln-t1--t1--t1-ln-s+fsexpl1+attn+copy-test-AVG-full-acc-top-5}
\\
\multirow{-4}{*}{\TablePartMultiAttnCopy}
 & 
\UseMacro{ln-bsexpl1+fsexpl1+attn+copy}
& 
\UseMacro{results-ln-t1--t1--t1-ln-bsexpl1+fsexpl1+attn+copy-test-AVG-BLEU-4}
& 
\UseMacro{results-ln-t1--t1--t1-ln-bsexpl1+fsexpl1+attn+copy-test-AVG-frag-acc}
& 
\UseMacro{results-ln-t1--t1--t1-ln-bsexpl1+fsexpl1+attn+copy-test-AVG-full-acc-top-1}
& 
\UseMacro{results-ln-t1--t1--t1-ln-bsexpl1+fsexpl1+attn+copy-test-AVG-full-acc-top-5}
\\
\midrule
 & 
\UseMacro{ln-bsexpl1+attn+copy}
& 
\UseMacro{results-ln-t1--t1--t1-ln-bsexpl1+attn+copy-test-AVG-BLEU-4}
& 
\UseMacro{results-ln-t1--t1--t1-ln-bsexpl1+attn+copy-test-AVG-frag-acc}
& 
\UseMacro{results-ln-t1--t1--t1-ln-bsexpl1+attn+copy-test-AVG-full-acc-top-1}
& 
\UseMacro{results-ln-t1--t1--t1-ln-bsexpl1+attn+copy-test-AVG-full-acc-top-5}
\\
 & 
\UseMacro{ln-fsexpl1+attn+copy}
& 
\UseMacro{results-ln-t1--t1--t1-ln-fsexpl1+attn+copy-test-AVG-BLEU-4}
& 
\UseMacro{results-ln-t1--t1--t1-ln-fsexpl1+attn+copy-test-AVG-frag-acc}
& 
\UseMacro{results-ln-t1--t1--t1-ln-fsexpl1+attn+copy-test-AVG-full-acc-top-1}
& 
\UseMacro{results-ln-t1--t1--t1-ln-fsexpl1+attn+copy-test-AVG-full-acc-top-5}
\\
 & 
\UseMacro{ln-bsexpl0+attn+copy}
& 
\UseMacro{results-ln-t1--t1--t1-ln-bsexpl0+attn+copy-test-AVG-BLEU-4}
& 
\UseMacro{results-ln-t1--t1--t1-ln-bsexpl0+attn+copy-test-AVG-frag-acc}
& 
\UseMacro{results-ln-t1--t1--t1-ln-bsexpl0+attn+copy-test-AVG-full-acc-top-1}
& 
\UseMacro{results-ln-t1--t1--t1-ln-bsexpl0+attn+copy-test-AVG-full-acc-top-5}
\\
 & 
\UseMacro{ln-fsexpl0+attn+copy}
& 
\UseMacro{results-ln-t1--t1--t1-ln-fsexpl0+attn+copy-test-AVG-BLEU-4}
& 
\UseMacro{results-ln-t1--t1--t1-ln-fsexpl0+attn+copy-test-AVG-frag-acc}
& 
\UseMacro{results-ln-t1--t1--t1-ln-fsexpl0+attn+copy-test-AVG-full-acc-top-1}
& 
\UseMacro{results-ln-t1--t1--t1-ln-fsexpl0+attn+copy-test-AVG-full-acc-top-5}
\\
\multirow{-5}{*}{\TablePartMonoAttnCopy}
 & 
\UseMacro{ln-s+attn+copy}
& 
\UseMacro{results-ln-t1--t1--t1-ln-s+attn+copy-test-AVG-BLEU-4}
& 
\UseMacro{results-ln-t1--t1--t1-ln-s+attn+copy-test-AVG-frag-acc}
& 
\UseMacro{results-ln-t1--t1--t1-ln-s+attn+copy-test-AVG-full-acc-top-1}
& 
\UseMacro{results-ln-t1--t1--t1-ln-s+attn+copy-test-AVG-full-acc-top-5}
\\
\midrule
 & 
\UseMacro{ln-s+bsexpl1+fsexpl1+attn}
& 
\UseMacro{results-ln-t1--t1--t1-ln-s+bsexpl1+fsexpl1+attn-test-AVG-BLEU-4}
& 
\UseMacro{results-ln-t1--t1--t1-ln-s+bsexpl1+fsexpl1+attn-test-AVG-frag-acc}
& 
\UseMacro{results-ln-t1--t1--t1-ln-s+bsexpl1+fsexpl1+attn-test-AVG-full-acc-top-1}
& 
\UseMacro{results-ln-t1--t1--t1-ln-s+bsexpl1+fsexpl1+attn-test-AVG-full-acc-top-5}
\\
 & 
\UseMacro{ln-s+bsexpl1+attn}
& 
\UseMacro{results-ln-t1--t1--t1-ln-s+bsexpl1+attn-test-AVG-BLEU-4}
& 
\UseMacro{results-ln-t1--t1--t1-ln-s+bsexpl1+attn-test-AVG-frag-acc}
& 
\UseMacro{results-ln-t1--t1--t1-ln-s+bsexpl1+attn-test-AVG-full-acc-top-1}
& 
\UseMacro{results-ln-t1--t1--t1-ln-s+bsexpl1+attn-test-AVG-full-acc-top-5}
\\
 & 
\UseMacro{ln-s+fsexpl1+attn}
& 
\UseMacro{results-ln-t1--t1--t1-ln-s+fsexpl1+attn-test-AVG-BLEU-4}
& 
\UseMacro{results-ln-t1--t1--t1-ln-s+fsexpl1+attn-test-AVG-frag-acc}
& 
\UseMacro{results-ln-t1--t1--t1-ln-s+fsexpl1+attn-test-AVG-full-acc-top-1}
& 
\UseMacro{results-ln-t1--t1--t1-ln-s+fsexpl1+attn-test-AVG-full-acc-top-5}
\\
\multirow{-4}{*}{\TablePartMultiAttn}
 & 
\UseMacro{ln-bsexpl1+fsexpl1+attn}
& 
\UseMacro{results-ln-t1--t1--t1-ln-bsexpl1+fsexpl1+attn-test-AVG-BLEU-4}
& 
\UseMacro{results-ln-t1--t1--t1-ln-bsexpl1+fsexpl1+attn-test-AVG-frag-acc}
& 
\UseMacro{results-ln-t1--t1--t1-ln-bsexpl1+fsexpl1+attn-test-AVG-full-acc-top-1}
& 
\UseMacro{results-ln-t1--t1--t1-ln-bsexpl1+fsexpl1+attn-test-AVG-full-acc-top-5}
\\
\midrule
 & 
\UseMacro{ln-bsexpl1+attn}
& 
\UseMacro{results-ln-t1--t1--t1-ln-bsexpl1+attn-test-AVG-BLEU-4}
& 
\UseMacro{results-ln-t1--t1--t1-ln-bsexpl1+attn-test-AVG-frag-acc}
& 
\UseMacro{results-ln-t1--t1--t1-ln-bsexpl1+attn-test-AVG-full-acc-top-1}
& 
\UseMacro{results-ln-t1--t1--t1-ln-bsexpl1+attn-test-AVG-full-acc-top-5}
\\
 & 
\UseMacro{ln-fsexpl1+attn}
& 
\UseMacro{results-ln-t1--t1--t1-ln-fsexpl1+attn-test-AVG-BLEU-4}
& 
\UseMacro{results-ln-t1--t1--t1-ln-fsexpl1+attn-test-AVG-frag-acc}
& 
\UseMacro{results-ln-t1--t1--t1-ln-fsexpl1+attn-test-AVG-full-acc-top-1}
& 
\UseMacro{results-ln-t1--t1--t1-ln-fsexpl1+attn-test-AVG-full-acc-top-5}
\\
 & 
\UseMacro{ln-bsexpl0+attn}
& 
\UseMacro{results-ln-t1--t1--t1-ln-bsexpl0+attn-test-AVG-BLEU-4}
& 
\UseMacro{results-ln-t1--t1--t1-ln-bsexpl0+attn-test-AVG-frag-acc}
& 
\UseMacro{results-ln-t1--t1--t1-ln-bsexpl0+attn-test-AVG-full-acc-top-1}
& 
\UseMacro{results-ln-t1--t1--t1-ln-bsexpl0+attn-test-AVG-full-acc-top-5}
\\
 & 
\UseMacro{ln-fsexpl0+attn}
& 
\UseMacro{results-ln-t1--t1--t1-ln-fsexpl0+attn-test-AVG-BLEU-4}
& 
\UseMacro{results-ln-t1--t1--t1-ln-fsexpl0+attn-test-AVG-frag-acc}
& 
\UseMacro{results-ln-t1--t1--t1-ln-fsexpl0+attn-test-AVG-full-acc-top-1}
& 
\UseMacro{results-ln-t1--t1--t1-ln-fsexpl0+attn-test-AVG-full-acc-top-5}
\\
\multirow{-5}{*}{\TablePartMonoAttn}
 & 
\UseMacro{ln-s+attn}
& 
\UseMacro{results-ln-t1--t1--t1-ln-s+attn-test-AVG-BLEU-4}
& 
\UseMacro{results-ln-t1--t1--t1-ln-s+attn-test-AVG-frag-acc}
& 
\UseMacro{results-ln-t1--t1--t1-ln-s+attn-test-AVG-full-acc-top-1}
& 
\UseMacro{results-ln-t1--t1--t1-ln-s+attn-test-AVG-full-acc-top-5}
\\
\midrule
 & 
\UseMacro{ln-s+bsexpl1+fsexpl1}
& 
\UseMacro{results-ln-t1--t1--t1-ln-s+bsexpl1+fsexpl1-test-AVG-BLEU-4}
& 
\UseMacro{results-ln-t1--t1--t1-ln-s+bsexpl1+fsexpl1-test-AVG-frag-acc}
& 
\UseMacro{results-ln-t1--t1--t1-ln-s+bsexpl1+fsexpl1-test-AVG-full-acc-top-1}
& 
\UseMacro{results-ln-t1--t1--t1-ln-s+bsexpl1+fsexpl1-test-AVG-full-acc-top-5}
\\
 & 
\UseMacro{ln-s+bsexpl1}
& 
\UseMacro{results-ln-t1--t1--t1-ln-s+bsexpl1-test-AVG-BLEU-4}
& 
\UseMacro{results-ln-t1--t1--t1-ln-s+bsexpl1-test-AVG-frag-acc}
& 
\UseMacro{results-ln-t1--t1--t1-ln-s+bsexpl1-test-AVG-full-acc-top-1}
& 
\UseMacro{results-ln-t1--t1--t1-ln-s+bsexpl1-test-AVG-full-acc-top-5}
\\
 & 
\UseMacro{ln-s+fsexpl1}
& 
\UseMacro{results-ln-t1--t1--t1-ln-s+fsexpl1-test-AVG-BLEU-4}
& 
\UseMacro{results-ln-t1--t1--t1-ln-s+fsexpl1-test-AVG-frag-acc}
& 
\UseMacro{results-ln-t1--t1--t1-ln-s+fsexpl1-test-AVG-full-acc-top-1}
& 
\UseMacro{results-ln-t1--t1--t1-ln-s+fsexpl1-test-AVG-full-acc-top-5}
\\
\multirow{-4}{*}{\TablePartMulti}
 & 
\UseMacro{ln-bsexpl1+fsexpl1}
& 
\UseMacro{results-ln-t1--t1--t1-ln-bsexpl1+fsexpl1-test-AVG-BLEU-4}
& 
\UseMacro{results-ln-t1--t1--t1-ln-bsexpl1+fsexpl1-test-AVG-frag-acc}
& 
\UseMacro{results-ln-t1--t1--t1-ln-bsexpl1+fsexpl1-test-AVG-full-acc-top-1}
& 
\UseMacro{results-ln-t1--t1--t1-ln-bsexpl1+fsexpl1-test-AVG-full-acc-top-5}
\\
\midrule
 & 
\UseMacro{ln-bsexpl1}
& 
\UseMacro{results-ln-t1--t1--t1-ln-bsexpl1-test-AVG-BLEU-4}
& 
\UseMacro{results-ln-t1--t1--t1-ln-bsexpl1-test-AVG-frag-acc}
& 
\UseMacro{results-ln-t1--t1--t1-ln-bsexpl1-test-AVG-full-acc-top-1}
& 
\UseMacro{results-ln-t1--t1--t1-ln-bsexpl1-test-AVG-full-acc-top-5}
\\
 & 
\UseMacro{ln-fsexpl1}
& 
\UseMacro{results-ln-t1--t1--t1-ln-fsexpl1-test-AVG-BLEU-4}
& 
\UseMacro{results-ln-t1--t1--t1-ln-fsexpl1-test-AVG-frag-acc}
& 
\UseMacro{results-ln-t1--t1--t1-ln-fsexpl1-test-AVG-full-acc-top-1}
& 
\UseMacro{results-ln-t1--t1--t1-ln-fsexpl1-test-AVG-full-acc-top-5}
\\
\rowcolor{lightgray}
 & 
\UseMacro{ln-bsexpl0}
& 
\UseMacro{results-ln-t1--t1--t1-ln-bsexpl0-test-AVG-BLEU-4}
& 
\UseMacro{results-ln-t1--t1--t1-ln-bsexpl0-test-AVG-frag-acc}
& 
\UseMacro{results-ln-t1--t1--t1-ln-bsexpl0-test-AVG-full-acc-top-1}
& 
\UseMacro{results-ln-t1--t1--t1-ln-bsexpl0-test-AVG-full-acc-top-5}
\\
\rowcolor{lightgray}
 & 
\UseMacro{ln-fsexpl0}
& 
\UseMacro{results-ln-t1--t1--t1-ln-fsexpl0-test-AVG-BLEU-4}
& 
\UseMacro{results-ln-t1--t1--t1-ln-fsexpl0-test-AVG-frag-acc}
& 
\UseMacro{results-ln-t1--t1--t1-ln-fsexpl0-test-AVG-full-acc-top-1}
& 
\UseMacro{results-ln-t1--t1--t1-ln-fsexpl0-test-AVG-full-acc-top-5}
\\
\rowcolor{lightgray}
\multirow{-5}{*}{\TablePartMono}
 & 
\UseMacro{ln-s}
& 
\UseMacro{results-ln-t1--t1--t1-ln-s-test-AVG-BLEU-4}
& 
\UseMacro{results-ln-t1--t1--t1-ln-s-test-AVG-frag-acc}
& 
\UseMacro{results-ln-t1--t1--t1-ln-s-test-AVG-full-acc-top-1}
& 
\UseMacro{results-ln-t1--t1--t1-ln-s-test-AVG-full-acc-top-5}
\\
\midrule
\rowcolor{lightgray}
\multirow{-1}{*}{\TablePartRB}
 & 
\UseMacro{ln-rb}
& 
\UseMacro{results-ln-t1--t1--t1-ln-rb-test-AVG-BLEU-4}
& 
\UseMacro{results-ln-t1--t1--t1-ln-rb-test-AVG-frag-acc}
& 
\UseMacro{results-ln-t1--t1--t1-ln-rb-test-AVG-full-acc-top-1}
& 
\UseMacro{results-ln-t1--t1--t1-ln-rb-test-AVG-full-acc-top-5}
\\
\bottomrule
\end{tabular}
\end{center}
\end{scriptsize}
\vspace{\UseMacro{vspace-results-ln-t1--t1--t1-main}}
\end{table*}

\MyPara{Results} Table~\ref{tbl:results-ln-t1--t1--t1-main} shows the performance of the models.  Similar models are grouped together. The first column shows the names of the model groups and the second column shows the names of the models. For each model, we show values for the four automatic metrics, \bleu, \fragacc (\UseMacro{table-head-frag-acc}), \toponeacc (\UseMacro{table-head-full-acc-top-1}), and \topfiveacc (\UseMacro{table-head-full-acc-top-5}).   We repeated each experiment \NumExpTrials{} times, with different random initialization each time, and computed the averages of each automated metric.  We performed statistical significance tests---under significance level $p<0.05$ using the bootstrap method~\cite{Berg-KirkpatrickETAL12Empirical}---to compare each pair of models.  We use bold text to highlight the best value for each automatic metric, and gray background for baseline models.  We make several observations:

\pgfmathsetmacro{\varbleudiffktreetrim}{\UseMacro{results-ln-t1--t1--t1-ln-bsexpl1+attn+copy-test-AVG-BLEU-4}-\UseMacro{results-ln-t1--t1--t1-ln-bsexpl0+attn+copy-test-AVG-BLEU-4}} \newcommand{\bleudiffktreetrim}{\pgfmathprintnumber[fixed, precision=1]{\varbleudiffktreetrim}}

\MyFinding{1} The best overall performance (\bleu =   \UseMacro{results-ln-t1--t1--t1-ln-s+bsexpl1+attn+copy-test-AVG-BLEU-4})   is obtained using the multi-input model with \lstmt   and \trimmedktree as inputs, which also includes copy and attention   mechanisms (\UseMacro{ln-s+bsexpl1+attn+copy}). The improvements   over all other models are statistically significant and all   automatic metrics are consistent in identifying the best model.   This shows the importance of using Coq's internal structures and   focusing only on certain parts of those structures.   %

\MyFinding{2} The copy mechanism brings statistically   significant improvements to all models.  This can be clearly   observed by comparing groups 1 and 3 in the table, as well as groups 2   and 4.  For example, \bleu for \UseMacro{ln-s+attn} and   \UseMacro{ln-s+attn+copy} are   \UseMacro{results-ln-t1--t1--t1-ln-s+attn-test-AVG-BLEU-4} and   \UseMacro{results-ln-t1--t1--t1-ln-s+attn+copy-test-AVG-BLEU-4},   respectively.  We believe that the copy mechanism plays an important   role because many \subtok{s} are specific to the file context and do   not appear in the fixed vocabulary learned on the \docs in \train   set.   %

\MyFinding{3} Using \trimmed trees greatly improves performance of models and   the improvements brought by upgrading \KTreeAcro   to \TrimmedKTreeAcro or \STreeAcro to \TrimmedSTreeAcro are   statistically significant.  For example, this can be clearly seen in   the second group: \bleu   for \UseMacro{ln-bsexpl0+attn+copy}   and \UseMacro{ln-bsexpl1+attn+copy}   are \UseMacro{results-ln-t1--t1--t1-ln-bsexpl0+attn+copy-test-AVG-BLEU-4}   and \UseMacro{results-ln-t1--t1--t1-ln-bsexpl1+attn+copy-test-AVG-BLEU-4},   respectively.  We believe that the size of the original trees, and a   lot of irrelevant data in those trees, hurt the performance.   The fact that \TrimmedKTreeAcro and \TrimmedSTreeAcro both perform   much better than using \KTreeAcro or \STreeAcro across all groups   indicate that the \trimmed trees could be viewed as a form of   supervised attention with flat values that helps later attention   layers to focus better.

\MyFinding{4} Although \trimmedstree with attention outperforms (statistically   significant) \trimmedktree with attention (\bleu   \UseMacro{results-ln-t1--t1--t1-ln-fsexpl1+attn-test-AVG-BLEU-4}   vs. \UseMacro{results-ln-t1--t1--t1-ln-bsexpl1+attn-test-AVG-BLEU-4}),   \trimmedktree with attention and copy by far outperforms   (statistically significant) \trimmedstree with attention and copy   (\bleu   \UseMacro{results-ln-t1--t1--t1-ln-bsexpl1+attn+copy-test-AVG-BLEU-4}   vs. \UseMacro{results-ln-t1--t1--t1-ln-fsexpl1+attn+copy-test-AVG-BLEU-4}).   The copy mechanism helps \ktrees much more than the \strees, because   the mathematical notations and symbols in the \strees get expanded   to their names in the \ktrees, and some of them are needed as a part   of the lemma names.

\MyFinding{5} \Lstmt and \stree do not work well together, primarily because the   two representations contain mostly the same information.  In which   case, a model taking both as inputs may not work as well as using   only one of the inputs, because more parameters need to be trained.

\MyFinding{6} The \retrievalbased baseline, which is the strongest among   baselines, outperforms several \encdec models without attention and   copy or with only attention, but is worse than (statistically   significant) all models with both attention and copy mechanisms   enabled.

\subsection{Manual Quality Analysis} \label{sec:eval:human}

While generated lemma names may not always match the manually written ones in the training set, they can still be semantically valid and conform to prevailing conventions. However, such name properties are not reflected in our automatic evaluation metrics, since these metrics only consider exactly matched tokens as correct. To obtain a more complete evaluation, we therefore performed a manual quality analysis of generated lemma names from \CoqConvTool by applying it to a \Coq project outside of our corpus, \PTEXTfcslpcm~\cite{fcslpcm}. This project depends on \MathComp, and follows, to a degree, many of the \MathComp \cc. \PTEXTFcslpcm consists of \UseMacro{corpus-fcslpcm-num-docs} \Coq \docs, and contains \UseMacro{qs-fcslpcm-num-lemmas} lemmas.

We ran \CoqConvTool with the best model (\UseMacro{ln-s+bsexpl1+attn+copy}) on \PTEXTfcslpcm to get the top-1 suggestions for all lemma names. Overall, the \CoqConvTool suggestions achieved a \bleu score of \UseMacro{qs-fcslpcm-ln-BLEU-4} and a fragment accuracy of \UseMacro{qs-fcslpcm-ln-frag-acc}, and \UseMacro{qs-fcslpcm-ln-full-correct-top-1} suggestions (\UseMacro{qs-fcslpcm-ln-full-correct-top-1-percent}) exactly match the existing lemma names. Next, we asked the maintainer of \PTEXTfcslpcm to evaluate the remaining \NumQSLemmaNameEvaled lemma names (those that do not match exactly) and send us feedback.

{\renewcommand{\arraystretch}{1.1} 
\begin{table}[t]
\begin{scriptsize}
\begin{center}
\caption{\UseMacro{table-caption-qualitative-ln}}
\begin{tabular}{l@{\hspace{4pt}}l}
  \toprule

  \QSStmt{p s : supp (kfilter p s) = filter p (supp s)} \\
  \QSTruthPred{supp_kfilt}{supp_kfilter}\\
  \QSComment{\mycheckmark Using only \CoqInSmall{kfilt} has cognitive overhead.} \\

  \midrule
  
  \QSStmt{g e k v f : path ord k (supp f) ->}\\
  \QSStmtCont{foldfmap g e (ins k v f) = g (k, v) (foldfmap g e f)}\\
  \QSTruthPred{foldf_ins}{foldfmap_ins}\\
  \QSComment{\mycheckmark The whole function name is used in the suggested name.} \\

  \midrule

  \QSStmt{: transitive (@ord T)}\\
  \QSTruthPred{trans}{ord_trans}\\
  \QSComment{\mycheckmark Useful to add the \CoqInSmall{ord} prefix to the name.}\\

  \midrule

  \QSStmt{s : sorted (@ord T) s -> sorted (@oleq T) s}\\
  \QSTruthPred{sorted_oleq}{ord_sorted}\\
  \QSComment{\mycross The conclusion content should have greater priority.}\\

  \midrule

  \QSStmt{x y : total_spec x y (ord x y) (x == y) (ord y x)}\\
  \QSTruthPred{totalP}{ordP}\\
  \QSComment{\mycross Maybe this lemma should be named \CoqInSmall{ord_totalP}?}\\  

  \midrule

  \QSStmt{p1 p2 s : kfilter (predI p1 p2) s =}\\
  \QSStmtCont{kfilter p1 (kfilter p2 s)}\\
  \QSTruthPred{kfilter_predI}{eq_kfilter}\\
  \QSComment{\mycross The suggested name is too generic.}\\

  \bottomrule
\end{tabular}
\end{center}
\end{scriptsize}
\vspace{\UseMacro{vspace-qualitative-ln}}
\end{table}

 }

The maintainer spent one day on the task and provided comments on \NumQSLemmaNameComments suggested names. We analyzed these comments to identify patterns and trends. He found that 20\% of the suggested names he inspected were of good quality, out of which more than half were of high quality. Considering that the analysis was of top-1 suggestions excluding exact matches, we find these figures encouraging. For low-quality names, a clear trend was that they were often ``too generic''. Similar observations have been made about the results from \encdec models in dialog generation~\cite{serban2016building,li2016diversity}.  In contrast, useful suggestions were typically able to expand or elaborate on name components that are intuitively too concise, e.g., replacing \CoqIn{kfilt} with \CoqIn{kfilter}. Table~\ref{tbl:qualitative-ln} lists examples that are representative of these trends; checkmarks indicate useful suggestions, while crosses indicate unsuitability.  We also include comments from the maintainer. As illustrated by the comments, even suggestions considered unsuitable may contain useful parts.

\vspace{-5pt} \section{Discussion} \vspace{-5pt}

Our toolchain builds on \Coq \CoqVersion, and thus we only used projects that support this version. However, we do not expect any fundamental obstacles in supporting future \Coq releases. Thanks to the use of OCaml metaprogramming via PPX, which allowed eliding explicit references to the internal structure of Coq datatypes, \SerAPI itself and our extensions to it are expected to require only modest effort to maintain as Coq evolves.

Our models and \tooltype may not be applicable to \Coq projects unrelated to the \MathComp family of projects, i.e., projects which do not follow any \MathComp conventions. To the best of our knowledge, \MathComp's \cc are the most recognizable and well-documented in the \Coq community; suggesting \cc based on learning from projects unrelated to \MathComp are likely to give more ambiguous results that are difficult to validate manually. Our case study also included generating suggestions for a project outside the \MathComp family, \PTEXTfcslpcm, with encouraging results.

Our models are in principle applicable to proof assistants with similar foundations, such as Lean~\cite{DeMoura2015}. However, the current version of Lean, Lean 3, does not provide serialization of internal data structures as \SerAPI does for \Coq, which prevents direct application of our \tooltype. Application of our models to proof assistants with different foundations and proof-checking toolchains, such as Isabelle/HOL, is even less straightforward, although the Archive of Formal Proofs (AFP) contains many projects with high-quality lemma names~\cite{AFP}.

 \vspace{-5pt} \section{Related Work} \vspace{-5pt}

\MyPara{Naturalness and \cc} Hindle et al.~\cite{Hindle2012} first applied the concept of naturalness to Java-like languages, noting that program statement regularities and repetitiveness make statistical language models applicable for performing software engineering tasks~\cite{AllamanisETAL18Survey}. Rahman et al.~\citep{Rahman2019} validated the naturalness of other similar programming languages, and Hellendoorn et al.~\cite{HellendoornETAL18Naturalness} found high naturalness in \Coq code, providing motivation for our application of statistical language models to \Coq. Allamanis et al.~\cite{AllamanisETAL14Learning} used the concept of naturalness and statistical language models to learn and suggest \cc, including names, for Java, and Raychev et al.~\cite{Raychev2015} used conditional random fields to learn and suggest \cc for JavaScript. To our knowledge, no previous work has developed \emph{applications} of naturalness for proof assistants; Hellendorn et al.~\cite{HellendoornETAL18Naturalness} only measured naturalness for their Coq corpus.

\MyPara{Suggesting names} Prior work on suggesting names mostly concerns Java method names.  Liu et al.~\cite{LiuETAL19Learning} used a similarity matching algorithm, based on deep representations of Java method names and bodies learned with Paragraph Vector and \CNNs, to detect and fix inconsistent Java method names. Allamanis et al.~\cite{AllamanisETAL15Suggesting} used logbilinear neural language models supplemented by additional manual features to predict Java method and class names. Java method names have also been treated as short, descriptive ``summaries'' of its body; in this view, prior work has augmented attention mechanisms in convolutional networks~\cite{AllamanisETAL16Convolutional}, used sequence-to-sequence models to learn from descriptions (e.g., Javadoc comments)~\cite{GaoETAL19Neural}, and utilized the tree-structure of the code in a hierarchical attention network~\cite{XuETAL19Method}. Unlike Java syntax trees, Coq \sktrees contain considerable semantic information useful for naming. In the work closest to our domain, Aspinall and Kaliszyk used a k-nearest neighbors multi-label classifier on a corpus for the HOL Light proof assistant to suggest names of lemmas~\cite{Aspinall2016b}. However, their technique only suggests names that exist in the training data and therefore does not generalize. To our knowledge, ours is the first neural generation model for suggesting names in a proof assistant context.

\MyPara{Mining and learning for proof assistants} M\"uller et al.~\cite{Mueller2019} exported Coq \ktrees as XML strings to translate 49 Coq projects to the OMDoc theory graph format. Rather than translating documents to an independently specified format, we produce lightweight machine-readable representations of Coq's internal data structures. Wiedijk~\cite{Wiedijk2009} collected early basic statistics on the core libraries of several proof assistants, including Coq and Isabelle/HOL. Blanchette et al.~\cite{Blanchette2015} mined the AFP to gather statistics such as the average number of lines of Isabelle/HOL specifications and proof scripts. However, these corpora were not used to perform learning. Komendantskaya et al.~\cite{Komendantskaya2012,Heras2013b,Heras2014,Heras2014b} used machine learning without neural networks to identify patterns in Coq tactic sequences and proof \ktrees, e.g., to find structural similarities between lemmas and simplify proof development. In contrast, our models capture similarity among several different representations of lemma \emph{statements} to generate lemma names.

 \vspace{-7pt} \section{Conclusion} \label{sec:conclusion} \vspace{-5pt}

We presented novel techniques, based on neural networks, for learning and suggesting \lemman in \Coq verification projects. We designed and implemented multi-input \encdec models that use \Coq's internal data structures, including (\trimmed) \strees and \ktrees. Additionally, we constructed a large corpus of high quality \Coq code that will enable development and evaluation of future techniques for \Coq.  We performed an extensive evaluation of our models using the corpus.  Our results show that the multi-input models, which use internal data structures, substantially outperform several baselines; the model that uses the \lemmastmt tokens and the \trimmed \ktree with attention and copy mechanism performs the best. Based on our findings, we believe that multi-input models leveraging key parts of internal data structures can play a critical role in producing high-quality lemma name suggestions.

 \vspace{-5pt} \section*{Acknowledgments} \vspace{-5pt} The authors thank Yves Bertot, Cyril Cohen, Emilio Jes{\'u}s Gallego Arias, Ga\"etan Gilbert, Hugo Herbelin, Anton Trunov, Th{\'e}o Zimmermann, and the anonymous reviewers for their comments and feedback. This work was partially supported by the US National Science Foundation under Grant Nos. CCF-1652517 and IIS-1850153, and by the Swedish Foundation for Strategic Research under the TrustFull project.

\bibliographystyle{splncs04} \bibliography{bib}

\newpage \clearpage \appendix \section{Explanatory Notes on \CoqConvTool Models} \label{sec:ex-technique}

In this section, we explain some key terminology and concepts used to describe our generation models; these explanations were omitted from the main text to conserve space and avoid distracting the reader with excessive detail.

\begin{figure}[t]   \centering   \includegraphics[width=.8\columnwidth]{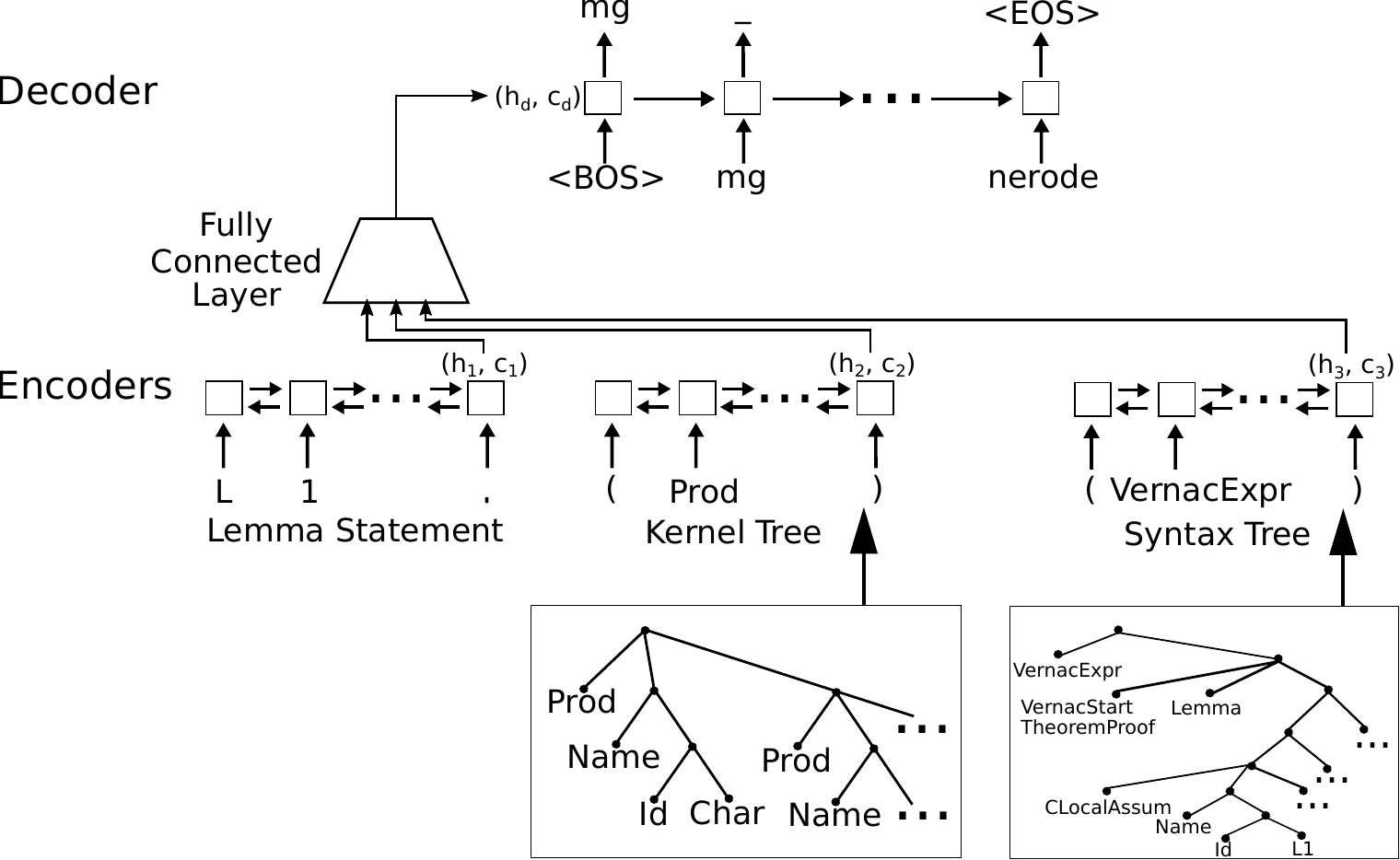}   \vspace{-10pt}   \caption{Core architecture of our multi-input \encdec models, with     \lemmastmts, \strees and \ktrees as inputs.     \label{fig:seqtoseq-arch-3inputs}   }   \vspace{-10pt} \end{figure}

\MyPara{Architecture of our multi-input \encdec models with   \lemmastmt, \stree, and \ktree} \figurename~\ref{fig:seqtoseq-arch-3inputs} illustrates our architecture when all three inputs are used, in contrast to \figurename~\ref{fig:seqtoseq-arch}, which only uses two inputs (\lemmastmt and \ktree).

\MyPara{Repetition of decoders' generated tokens {\normalfont     (cf. Section~\ref{sec:repetition})}} This is a common problem in \encdec models; it is largely because the attention mechanism (while helping the model for the most part) does not store information on how much information the model has ``covered'' in the encoded sequence. See also Sutskever~et~al.~\cite{sutskever2014sequence}.

\MyPara{Tf-idf {\normalfont (cf. the \retrievalbased baseline model in     Section~\ref{sec:eval:models-and-baselines})}} This is a numerical metric reflecting the importance of a token to a document in a corpus, calculated as the product of term frequency (proportional to the frequency of the token in the document) and inverse document frequency (inversely proportional to the number of documents containing the token).  In our \retrievalbased baseline model, we used Lucene's implementation of tf-idf~\cite{LuceneWebpage}. %

\MyPara{Early stopping {\normalfont (cf. hyperparameters in     Section~\ref{sec:eval:models-and-baselines})}} This is a common strategy to mitigate overfitting in training a machine learning model by monitoring the model's performance on both the \train set and the \val set and halting the training if the model stops improving on the \val set even if it improves on the \train set.  If early stopping is not used, the model is fully trained to maximize its performance on the \train set, but may have bad performance on a separate set (e.g., \test set).  In our experiments, we set an early stopping threshold of 3, which means the training is halted if the model does not obtain smaller loss on the \val set for 3 consecutive checkpoints.

\MyPara{Learning rate {\normalfont (cf. hyperparameters in     Section~\ref{sec:eval:models-and-baselines})}} This controls the speed of adjusting models' learnable parameters based on the loss at each iteration of the training.  An excessively large learning rate makes training faster, but may result in ``overshooting'': adjusting so much that it results in jumping over the minima. A too low learning rate means training is unnecessarily slow to complete, and may result in the training getting stuck in a local minimum. Guided by our previous experience, we used a value of 0.001 paired with the Adam~\cite{KingmaAndBa15Adam} optimizer (an algorithm for adjusting models' learnable parameters).

\MyPara{Dropout {\normalfont (cf. hyperparameters in     Section~\ref{sec:eval:models-and-baselines})}} This is a regularization technique for reducing overfitting, by randomly resetting a fraction of neural connections between two layers during training (and during training only). In our experiments, a dropout rate of 0.5 between the LSTM layers means that 50\% of the bits of the hidden and cell states are set to 0 when they are passed from a previous layer to its next layer in the LSTM during training.

\section{Additional Statistics} \label{sec:appendix:ex-statistics}

\begin{table}[t]
\begin{scriptsize}
\begin{center}
\caption{\UseMacro{table-caption-dataset-lemma-big}}
\begin{tabular}{l | r r c rr c rr}
\toprule
\multirow{2}{*}{\TableHeadLMSet}  & \multirow{2}{*}{\TableHeadNumDoc}  & \multirow{2}{*}{\TableHeadNumLemma}  & & \multicolumn{2}{c}{\TableHeadLMName}  & & \multicolumn{2}{c}{\TableHeadLMStmt}  \\ \cline{5-6} \cline{8-9}
 & &  & & \TableHeadNumChar & \TableHeadNumSubToken & & \TableHeadNumChar & \TableHeadNumSubToken \\
\midrule
\UseMacro{table-head-dslemma-all-t1-all}
& \multirow{2}{*}{\UseMacro{dataset-t1-all-num-documents}}
& \UseMacro{lemma-all-t1-all-num-lemmas}
& 
& \UseMacro{lemma-all-t1-all-AVG-num_char_name}
& \UseMacro{lemma-all-t1-all-AVG-num_subtok_name}
& 
& \UseMacro{lemma-all-t1-all-AVG-num_char_statement}
& \UseMacro{lemma-all-t1-all-AVG-num_subtok_statement}
\\
\UseMacro{table-head-dslemma-filtered-t1-all}
& 
& \UseMacro{lemma-filtered-t1-all-num-lemmas}
& 
& \UseMacro{lemma-filtered-t1-all-AVG-num_char_name}
& \UseMacro{lemma-filtered-t1-all-AVG-num_subtok_name}
& 
& \UseMacro{lemma-filtered-t1-all-AVG-num_char_statement}
& \UseMacro{lemma-filtered-t1-all-AVG-num_subtok_statement}
\\
\midrule
\UseMacro{table-head-dslemma-filtered-t1-train}
& \UseMacro{dataset-t1-train-num-documents}
& \UseMacro{lemma-filtered-t1-train-num-lemmas}
& 
& \UseMacro{lemma-filtered-t1-train-AVG-num_char_name}
& \UseMacro{lemma-filtered-t1-train-AVG-num_subtok_name}
& 
& \UseMacro{lemma-filtered-t1-train-AVG-num_char_statement}
& \UseMacro{lemma-filtered-t1-train-AVG-num_subtok_statement}
\\
\UseMacro{table-head-dslemma-filtered-t1-val}
& \UseMacro{dataset-t1-val-num-documents}
& \UseMacro{lemma-filtered-t1-val-num-lemmas}
& 
& \UseMacro{lemma-filtered-t1-val-AVG-num_char_name}
& \UseMacro{lemma-filtered-t1-val-AVG-num_subtok_name}
& 
& \UseMacro{lemma-filtered-t1-val-AVG-num_char_statement}
& \UseMacro{lemma-filtered-t1-val-AVG-num_subtok_statement}
\\
\UseMacro{table-head-dslemma-filtered-t1-test}
& \UseMacro{dataset-t1-test-num-documents}
& \UseMacro{lemma-filtered-t1-test-num-lemmas}
& 
& \UseMacro{lemma-filtered-t1-test-AVG-num_char_name}
& \UseMacro{lemma-filtered-t1-test-AVG-num_subtok_name}
& 
& \UseMacro{lemma-filtered-t1-test-AVG-num_char_statement}
& \UseMacro{lemma-filtered-t1-test-AVG-num_subtok_statement}
\\
\bottomrule
\end{tabular}
\end{center}
\end{scriptsize}
\vspace{\UseMacro{vspace-dataset-lemma-big}}
\end{table}

\begin{figure}[t]   \centering   \includegraphics[scale=0.65]{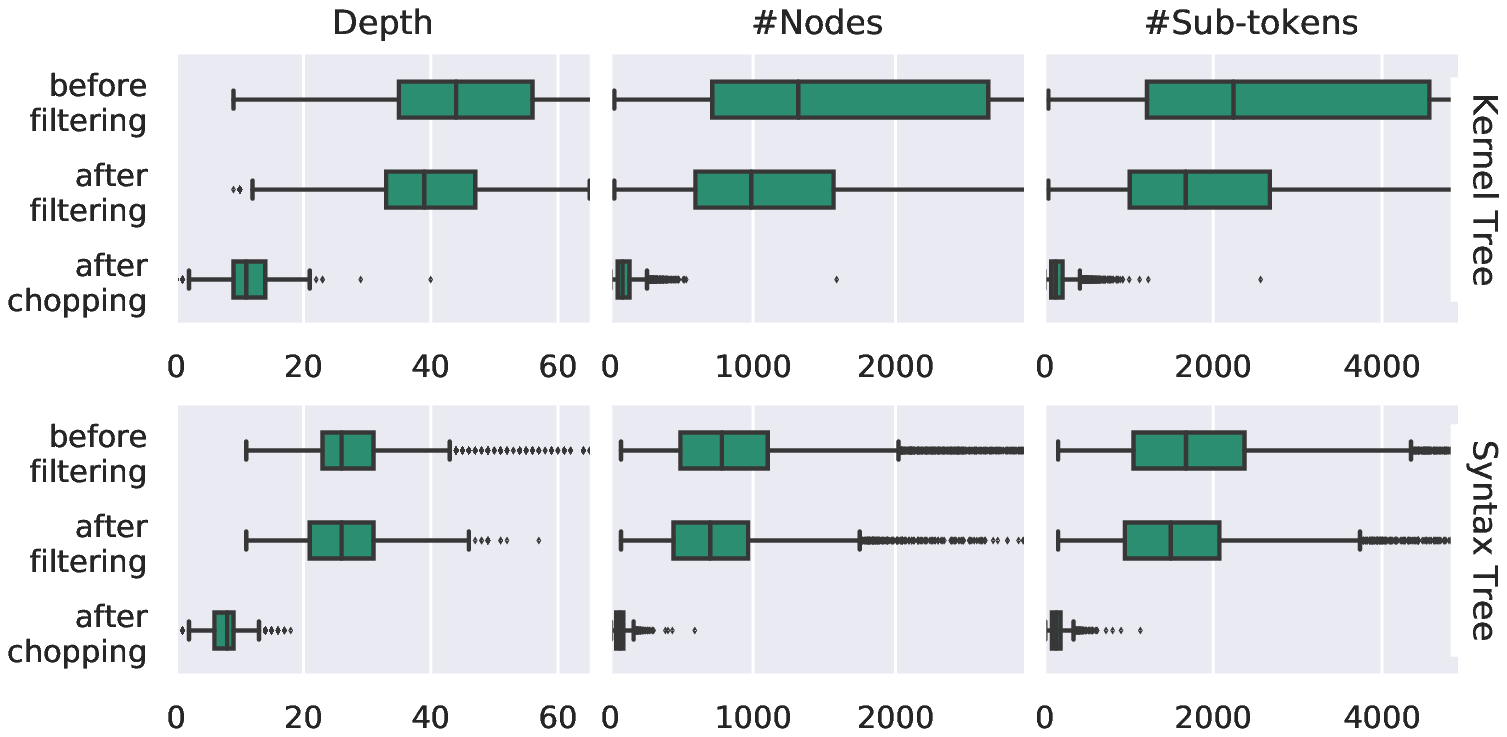}   \vspace{-10pt}   \caption{Additional statistics on \sktrees.}\label{fig:lemma-metrics-big}   \vspace{-10pt} \end{figure}

Table~\ref{tbl:dataset-lemma-big} shows additional statistics on the lemmas we used. The first row is for the lemmas before filtering the outliers, the second row is for the lemmas after the filtering, and the last three rows are for the \train, \val, and \test sets, respectively. \figurename~\ref{fig:lemma-metrics-big} illustrates the changes in the depth, number of nodes and number of \subtok{s} (after \flattening) of the \ktrees (first row) and \strees (second row) before filtering, after filtering, and after \simplification{}.

\section{Ablation Study on Tree \Trimming Heuristics} \label{Sec:appendix:ex-ablation-chopping}

In order to corroborate the effectiveness of \CoqConvTool's tree \trimming heuristics, we designed an ablation study that applies three different sets of \trimming heuristics and compares them with the one in \CoqConvTool (Section~\ref{sec:treesimp}).  The three sets of \trimming heuristics are: \begin{itemize} \item \MyPara{\KeepCategory \trimming} This set of heuristics is   almost the same as \CoqConvTool \trimming, except that it keeps the   category of a referenced name in \ktrees (e.g., whether it is a constant or inductive type),   since that semantic information could be relevant for naming. \item \MyPara{Rule-based \trimming} Removes all nodes after depth 10   for \sktrees. This is similar to the proof \ktree processing heuristics used   in ML4PG~\cite{Heras2014b}. \item \MyPara{Random \trimming} Randomly removes a subset of nodes   from \sktrees so that the resulting trees have the same average   number of nodes compared to \CoqConvTool's \trimmed trees, i.e.,   the heuristic   removes \UseMacro{lemma-filtered-t1-train-num_nodes_fsexp_reduction}   nodes from \strees and   \UseMacro{lemma-filtered-t1-train-num_nodes_bsexp_reduction} nodes   from \ktrees. \end{itemize} We performed the ablation study using the \encdec models with multi-input encoders and with both the attention and copy mechanisms. The other hyperparameters and experimental settings are the same as in Section~\ref{sec:eval}.

\begin{table*}[t]
\begin{scriptsize}
\begin{center}
\caption{\UseMacro{table-caption-results-ln-chopping-chopping-main}}
\begin{tabular}{c l r r r r}
\toprule
\TableHeadGroup & \TableHeadModel
& \UseMacro{table-head-results-ln-BLEU-4}
& \UseMacro{table-head-results-ln-frag-acc}
& \UseMacro{table-head-results-ln-full-acc-top-1}
& \UseMacro{table-head-results-ln-full-acc-top-5}
 \\
\midrule
 & 
\UseMacro{ln-s+bsexpl1+fsexpl1+attn+copy}
& 
\UseMacro{results-ln-chopping-chopping-ln-s+bsexpl1+fsexpl1+attn+copy-test-AVG-BLEU-4}
& 
\UseMacro{results-ln-chopping-chopping-ln-s+bsexpl1+fsexpl1+attn+copy-test-AVG-frag-acc}
& 
\UseMacro{results-ln-chopping-chopping-ln-s+bsexpl1+fsexpl1+attn+copy-test-AVG-full-acc-top-1}
& 
\UseMacro{results-ln-chopping-chopping-ln-s+bsexpl1+fsexpl1+attn+copy-test-AVG-full-acc-top-5}
\\
 & 
\UseMacro{ln-s+bsexpl1+attn+copy}
& 
\textbf{\UseMacro{results-ln-chopping-chopping-ln-s+bsexpl1+attn+copy-test-AVG-BLEU-4}}
& 
\UseMacro{results-ln-chopping-chopping-ln-s+bsexpl1+attn+copy-test-AVG-frag-acc}
& 
\textbf{\UseMacro{results-ln-chopping-chopping-ln-s+bsexpl1+attn+copy-test-AVG-full-acc-top-1}}
& 
\UseMacro{results-ln-chopping-chopping-ln-s+bsexpl1+attn+copy-test-AVG-full-acc-top-5}
\\
 & 
\UseMacro{ln-s+fsexpl1+attn+copy}
& 
\UseMacro{results-ln-chopping-chopping-ln-s+fsexpl1+attn+copy-test-AVG-BLEU-4}
& 
\UseMacro{results-ln-chopping-chopping-ln-s+fsexpl1+attn+copy-test-AVG-frag-acc}
& 
\UseMacro{results-ln-chopping-chopping-ln-s+fsexpl1+attn+copy-test-AVG-full-acc-top-1}
& 
\UseMacro{results-ln-chopping-chopping-ln-s+fsexpl1+attn+copy-test-AVG-full-acc-top-5}
\\
\multirow{-4}{*}{\TablePartChoppingRoosterize}
 & 
\UseMacro{ln-bsexpl1+fsexpl1+attn+copy}
& 
\UseMacro{results-ln-chopping-chopping-ln-bsexpl1+fsexpl1+attn+copy-test-AVG-BLEU-4}
& 
\UseMacro{results-ln-chopping-chopping-ln-bsexpl1+fsexpl1+attn+copy-test-AVG-frag-acc}
& 
\UseMacro{results-ln-chopping-chopping-ln-bsexpl1+fsexpl1+attn+copy-test-AVG-full-acc-top-1}
& 
\UseMacro{results-ln-chopping-chopping-ln-bsexpl1+fsexpl1+attn+copy-test-AVG-full-acc-top-5}
\\
\midrule
 & 
\UseMacro{ln-s+bsexpl1x+fsexpl1x+attn+copy}
& 
\UseMacro{results-ln-chopping-chopping-ln-s+bsexpl1x+fsexpl1x+attn+copy-test-AVG-BLEU-4}
& 
\textbf{\UseMacro{results-ln-chopping-chopping-ln-s+bsexpl1x+fsexpl1x+attn+copy-test-AVG-frag-acc}}
& 
\UseMacro{results-ln-chopping-chopping-ln-s+bsexpl1x+fsexpl1x+attn+copy-test-AVG-full-acc-top-1}
& 
\textbf{\UseMacro{results-ln-chopping-chopping-ln-s+bsexpl1x+fsexpl1x+attn+copy-test-AVG-full-acc-top-5}}
\\
 & 
\UseMacro{ln-s+bsexpl1x+attn+copy}
& 
\textbf{\UseMacro{results-ln-chopping-chopping-ln-s+bsexpl1x+attn+copy-test-AVG-BLEU-4}}
& 
\UseMacro{results-ln-chopping-chopping-ln-s+bsexpl1x+attn+copy-test-AVG-frag-acc}
& 
\UseMacro{results-ln-chopping-chopping-ln-s+bsexpl1x+attn+copy-test-AVG-full-acc-top-1}
& 
\UseMacro{results-ln-chopping-chopping-ln-s+bsexpl1x+attn+copy-test-AVG-full-acc-top-5}
\\
 & 
\UseMacro{ln-s+fsexpl1x+attn+copy}
& 
\UseMacro{results-ln-chopping-chopping-ln-s+fsexpl1x+attn+copy-test-AVG-BLEU-4}
& 
\UseMacro{results-ln-chopping-chopping-ln-s+fsexpl1x+attn+copy-test-AVG-frag-acc}
& 
\UseMacro{results-ln-chopping-chopping-ln-s+fsexpl1x+attn+copy-test-AVG-full-acc-top-1}
& 
\UseMacro{results-ln-chopping-chopping-ln-s+fsexpl1x+attn+copy-test-AVG-full-acc-top-5}
\\
\multirow{-4}{*}{\TablePartChoppingWithCategory}
 & 
\UseMacro{ln-bsexpl1x+fsexpl1x+attn+copy}
& 
\UseMacro{results-ln-chopping-chopping-ln-bsexpl1x+fsexpl1x+attn+copy-test-AVG-BLEU-4}
& 
\UseMacro{results-ln-chopping-chopping-ln-bsexpl1x+fsexpl1x+attn+copy-test-AVG-frag-acc}
& 
\UseMacro{results-ln-chopping-chopping-ln-bsexpl1x+fsexpl1x+attn+copy-test-AVG-full-acc-top-1}
& 
\UseMacro{results-ln-chopping-chopping-ln-bsexpl1x+fsexpl1x+attn+copy-test-AVG-full-acc-top-5}
\\
\midrule
 & 
\UseMacro{ln-s+bsexpd10+fsexpd10+attn+copy}
& 
\UseMacro{results-ln-chopping-chopping-ln-s+bsexpd10+fsexpd10+attn+copy-test-AVG-BLEU-4}
& 
\UseMacro{results-ln-chopping-chopping-ln-s+bsexpd10+fsexpd10+attn+copy-test-AVG-frag-acc}
& 
\UseMacro{results-ln-chopping-chopping-ln-s+bsexpd10+fsexpd10+attn+copy-test-AVG-full-acc-top-1}
& 
\UseMacro{results-ln-chopping-chopping-ln-s+bsexpd10+fsexpd10+attn+copy-test-AVG-full-acc-top-5}
\\
 & 
\UseMacro{ln-s+bsexpd10+attn+copy}
& 
\UseMacro{results-ln-chopping-chopping-ln-s+bsexpd10+attn+copy-test-AVG-BLEU-4}
& 
\UseMacro{results-ln-chopping-chopping-ln-s+bsexpd10+attn+copy-test-AVG-frag-acc}
& 
\UseMacro{results-ln-chopping-chopping-ln-s+bsexpd10+attn+copy-test-AVG-full-acc-top-1}
& 
\UseMacro{results-ln-chopping-chopping-ln-s+bsexpd10+attn+copy-test-AVG-full-acc-top-5}
\\
 & 
\UseMacro{ln-s+fsexpd10+attn+copy}
& 
\UseMacro{results-ln-chopping-chopping-ln-s+fsexpd10+attn+copy-test-AVG-BLEU-4}
& 
\UseMacro{results-ln-chopping-chopping-ln-s+fsexpd10+attn+copy-test-AVG-frag-acc}
& 
\UseMacro{results-ln-chopping-chopping-ln-s+fsexpd10+attn+copy-test-AVG-full-acc-top-1}
& 
\UseMacro{results-ln-chopping-chopping-ln-s+fsexpd10+attn+copy-test-AVG-full-acc-top-5}
\\
\multirow{-4}{*}{\TablePartChoppingDepthX}
 & 
\UseMacro{ln-bsexpd10+fsexpd10+attn+copy}
& 
\UseMacro{results-ln-chopping-chopping-ln-bsexpd10+fsexpd10+attn+copy-test-AVG-BLEU-4}
& 
\UseMacro{results-ln-chopping-chopping-ln-bsexpd10+fsexpd10+attn+copy-test-AVG-frag-acc}
& 
\UseMacro{results-ln-chopping-chopping-ln-bsexpd10+fsexpd10+attn+copy-test-AVG-full-acc-top-1}
& 
\UseMacro{results-ln-chopping-chopping-ln-bsexpd10+fsexpd10+attn+copy-test-AVG-full-acc-top-5}
\\
\midrule
 & 
\UseMacro{ln-s+bsexprnd+fsexprnd+attn+copy}
& 
\UseMacro{results-ln-chopping-chopping-ln-s+bsexprnd+fsexprnd+attn+copy-test-AVG-BLEU-4}
& 
\UseMacro{results-ln-chopping-chopping-ln-s+bsexprnd+fsexprnd+attn+copy-test-AVG-frag-acc}
& 
\UseMacro{results-ln-chopping-chopping-ln-s+bsexprnd+fsexprnd+attn+copy-test-AVG-full-acc-top-1}
& 
\UseMacro{results-ln-chopping-chopping-ln-s+bsexprnd+fsexprnd+attn+copy-test-AVG-full-acc-top-5}
\\
 & 
\UseMacro{ln-s+bsexprnd+attn+copy}
& 
\UseMacro{results-ln-chopping-chopping-ln-s+bsexprnd+attn+copy-test-AVG-BLEU-4}
& 
\UseMacro{results-ln-chopping-chopping-ln-s+bsexprnd+attn+copy-test-AVG-frag-acc}
& 
\UseMacro{results-ln-chopping-chopping-ln-s+bsexprnd+attn+copy-test-AVG-full-acc-top-1}
& 
\UseMacro{results-ln-chopping-chopping-ln-s+bsexprnd+attn+copy-test-AVG-full-acc-top-5}
\\
 & 
\UseMacro{ln-s+fsexprnd+attn+copy}
& 
\UseMacro{results-ln-chopping-chopping-ln-s+fsexprnd+attn+copy-test-AVG-BLEU-4}
& 
\UseMacro{results-ln-chopping-chopping-ln-s+fsexprnd+attn+copy-test-AVG-frag-acc}
& 
\UseMacro{results-ln-chopping-chopping-ln-s+fsexprnd+attn+copy-test-AVG-full-acc-top-1}
& 
\UseMacro{results-ln-chopping-chopping-ln-s+fsexprnd+attn+copy-test-AVG-full-acc-top-5}
\\
\multirow{-4}{*}{\TablePartChoppingRandom}
 & 
\UseMacro{ln-bsexprnd+fsexprnd+attn+copy}
& 
\UseMacro{results-ln-chopping-chopping-ln-bsexprnd+fsexprnd+attn+copy-test-AVG-BLEU-4}
& 
\UseMacro{results-ln-chopping-chopping-ln-bsexprnd+fsexprnd+attn+copy-test-AVG-frag-acc}
& 
\UseMacro{results-ln-chopping-chopping-ln-bsexprnd+fsexprnd+attn+copy-test-AVG-full-acc-top-1}
& 
\UseMacro{results-ln-chopping-chopping-ln-bsexprnd+fsexprnd+attn+copy-test-AVG-full-acc-top-5}
\\
\bottomrule
\end{tabular}
\end{center}
\end{scriptsize}
\vspace{\UseMacro{vspace-results-ln-chopping-chopping-main}}
\end{table*}

Table~\ref{tbl:results-ln-chopping-chopping-main} shows the results of the ablation study.  The models using the same \trimming heuristics are grouped together.  We make several observations:

\begin{itemize} \item Among \keepcategory \trimming models, \UseMacro{ln-s+bsexpl1+fsexpl1+attn+copy} and   \UseMacro{ln-s+bsexpl1+attn+copy} perform the best, and they have performance similar   to \UseMacro{ln-s+bsexpl1+attn+copy} using \CoqConvTool   \trimming (\CoqConvTool's best model).  The measured differences between these three   models are not statistically significant, under significance level $p<0.05$ using the bootstrap   method~\cite{Berg-KirkpatrickETAL12Empirical}. This indicates that   although the category of a referenced name may contain some relevant semantic   information, the most relevant information is already preserved   by \CoqConvTool \trimming heuristics. \item The models using rule-based \trimming and random \trimming have   poor performance.  This indicates that the performance gain achieved   by \CoqConvTool through \trimming is not only due to the size reduction   of the input trees, but also due to the relevant information   retained by our \trimming heuristics. \end{itemize}

\section{Expanded Corpus and Evaluation} \label{sec:appendix:ex-study}

In addition to evaluating \CoqConvTool using the high-quality corpus consisting of \CorpusNumProjects \MathComp projects (Section~\ref{sec:corpus}), we also performed an evaluation on an expanded corpus that includes \UseMacro{corpus-allgroup-num-projects} \Coq projects related to the \MathComp family which follow (to various degrees) the same \cc, totaling over \UseMacro{corpus-allgroup-SUM-k-code-loc}k LOC. All projects depend, directly or indirectly, on the \MathComp library, but not on projects outside the corpus itself except for \Coq's standard library. We introduce the expanded corpus in Section~\ref{sec:appendix:ex-corpus} and describe our additional evaluation on this corpus in Section~\ref{sec:appendix:ex-eval}.

\subsection{Expanded Corpus} \label{sec:appendix:ex-corpus}

\begin{table*}[t]
\begin{scriptsize}
\begin{center}
\caption{\UseMacro{table-caption-corpus-allgroup}}
\begin{tabular}{c | l c c r r r r r c r r}
\toprule
\multicolumn{1}{c}{} & \multirow{2}{*}{\TableHeadProject} &  & \multirow{2}{*}{\TableHeadSHA} & \multirow{2}{*}{\TableHeadFiles} & \multirow{2}{*}{\TableHeadNumLemma} & \multirow{2}{*}{\TableHeadNumTok} & \multicolumn{2}{c}{\TableHeadLOC} &  & \multicolumn{2}{c}{\TableHeadLocPerDoc} \\ \cline{8-9}\cline{11-12}
\multicolumn{1}{c}{} &  &  &  &  &  &  & \TableHeadSpecLOC & \TableHeadProofLOC & & \TableHeadSpecLocPerDoc & \TableHeadProofLocPerDoc \\
\midrule
\multirow{4}{*}{\rotatebox[origin=c]{270}{\UseMacro{group-t1}}} & 
\Pfinmap
& \href{\UseMacro{corpus-finmap-url}}{\urlsymbol}
& \texttt{\UseMacro{corpus-finmap-sha-pretty}}
& \UseMacro{corpus-finmap-num-docs}
& \UseMacro{lemma-filtered-project-finmap-num-lemmas}
& \UseMacro{dataset-project-finmap-num-tokens}
& \UseMacro{corpus-finmap-spec-loc}
& \UseMacro{corpus-finmap-proof-loc}
 & 
& \UseMacro{corpus-finmap-AVG-spec-loc-per-doc}
& \UseMacro{corpus-finmap-AVG-proof-loc-per-doc}
\\
 & 
\Pfourcolor
& \href{\UseMacro{corpus-fourcolor-url}}{\urlsymbol}
& \texttt{\UseMacro{corpus-fourcolor-sha-pretty}}
& \UseMacro{corpus-fourcolor-num-docs}
& \UseMacro{lemma-filtered-project-fourcolor-num-lemmas}
& \UseMacro{dataset-project-fourcolor-num-tokens}
& \UseMacro{corpus-fourcolor-spec-loc}
& \UseMacro{corpus-fourcolor-proof-loc}
 & 
& \UseMacro{corpus-fourcolor-AVG-spec-loc-per-doc}
& \UseMacro{corpus-fourcolor-AVG-proof-loc-per-doc}
\\
 & 
\Pmathcomp
& \href{\UseMacro{corpus-mathcomp-url}}{\urlsymbol}
& \texttt{\UseMacro{corpus-mathcomp-sha-pretty}}
& \UseMacro{corpus-mathcomp-num-docs}
& \UseMacro{lemma-filtered-project-mathcomp-num-lemmas}
& \UseMacro{dataset-project-mathcomp-num-tokens}
& \UseMacro{corpus-mathcomp-spec-loc}
& \UseMacro{corpus-mathcomp-proof-loc}
 & 
& \UseMacro{corpus-mathcomp-AVG-spec-loc-per-doc}
& \UseMacro{corpus-mathcomp-AVG-proof-loc-per-doc}
\\
 & 
\Poddorder
& \href{\UseMacro{corpus-oddorder-url}}{\urlsymbol}
& \texttt{\UseMacro{corpus-oddorder-sha-pretty}}
& \UseMacro{corpus-oddorder-num-docs}
& \UseMacro{lemma-filtered-project-oddorder-num-lemmas}
& \UseMacro{dataset-project-oddorder-num-tokens}
& \UseMacro{corpus-oddorder-spec-loc}
& \UseMacro{corpus-oddorder-proof-loc}
 & 
& \UseMacro{corpus-oddorder-AVG-spec-loc-per-doc}
& \UseMacro{corpus-oddorder-AVG-proof-loc-per-doc}
\\
\midrule
\multirow{8}{*}{\rotatebox[origin=c]{270}{\UseMacro{group-t2}}} & 
\Panalysis
& \href{\UseMacro{corpus-analysis-url}}{\urlsymbol}
& \texttt{\UseMacro{corpus-analysis-sha-pretty}}
& \UseMacro{corpus-analysis-num-docs}
& \UseMacro{lemma-filtered-project-analysis-num-lemmas}
& \UseMacro{dataset-project-analysis-num-tokens}
& \UseMacro{corpus-analysis-spec-loc}
& \UseMacro{corpus-analysis-proof-loc}
 & 
& \UseMacro{corpus-analysis-AVG-spec-loc-per-doc}
& \UseMacro{corpus-analysis-AVG-proof-loc-per-doc}
\\
 & 
\Pbigenough
& \href{\UseMacro{corpus-bigenough-url}}{\urlsymbol}
& \texttt{\UseMacro{corpus-bigenough-sha-pretty}}
& \UseMacro{corpus-bigenough-num-docs}
& \UseMacro{lemma-filtered-project-bigenough-num-lemmas}
& \UseMacro{dataset-project-bigenough-num-tokens}
& \UseMacro{corpus-bigenough-spec-loc}
& \UseMacro{corpus-bigenough-proof-loc}
 & 
& \UseMacro{corpus-bigenough-AVG-spec-loc-per-doc}
& \UseMacro{corpus-bigenough-AVG-proof-loc-per-doc}
\\
 & 
\Pellipticcurves
& \href{\UseMacro{corpus-ellipticcurves-url}}{\urlsymbol}
& \texttt{\UseMacro{corpus-ellipticcurves-sha-pretty}}
& \UseMacro{corpus-ellipticcurves-num-docs}
& \UseMacro{lemma-filtered-project-ellipticcurves-num-lemmas}
& \UseMacro{dataset-project-ellipticcurves-num-tokens}
& \UseMacro{corpus-ellipticcurves-spec-loc}
& \UseMacro{corpus-ellipticcurves-proof-loc}
 & 
& \UseMacro{corpus-ellipticcurves-AVG-spec-loc-per-doc}
& \UseMacro{corpus-ellipticcurves-AVG-proof-loc-per-doc}
\\
 & 
\Pgrobner
& \href{\UseMacro{corpus-grobner-url}}{\urlsymbol}
& \texttt{\UseMacro{corpus-grobner-sha-pretty}}
& \UseMacro{corpus-grobner-num-docs}
& \UseMacro{lemma-filtered-project-grobner-num-lemmas}
& \UseMacro{dataset-project-grobner-num-tokens}
& \UseMacro{corpus-grobner-spec-loc}
& \UseMacro{corpus-grobner-proof-loc}
 & 
& \UseMacro{corpus-grobner-AVG-spec-loc-per-doc}
& \UseMacro{corpus-grobner-AVG-proof-loc-per-doc}
\\
 & 
\Pmultinomials
& \href{\UseMacro{corpus-multinomials-url}}{\urlsymbol}
& \texttt{\UseMacro{corpus-multinomials-sha-pretty}}
& \UseMacro{corpus-multinomials-num-docs}
& \UseMacro{lemma-filtered-project-multinomials-num-lemmas}
& \UseMacro{dataset-project-multinomials-num-tokens}
& \UseMacro{corpus-multinomials-spec-loc}
& \UseMacro{corpus-multinomials-proof-loc}
 & 
& \UseMacro{corpus-multinomials-AVG-spec-loc-per-doc}
& \UseMacro{corpus-multinomials-AVG-proof-loc-per-doc}
\\
 & 
\Prealclosed
& \href{\UseMacro{corpus-realclosed-url}}{\urlsymbol}
& \texttt{\UseMacro{corpus-realclosed-sha-pretty}}
& \UseMacro{corpus-realclosed-num-docs}
& \UseMacro{lemma-filtered-project-realclosed-num-lemmas}
& \UseMacro{dataset-project-realclosed-num-tokens}
& \UseMacro{corpus-realclosed-spec-loc}
& \UseMacro{corpus-realclosed-proof-loc}
 & 
& \UseMacro{corpus-realclosed-AVG-spec-loc-per-doc}
& \UseMacro{corpus-realclosed-AVG-proof-loc-per-doc}
\\
 & 
\Probot
& \href{\UseMacro{corpus-robot-url}}{\urlsymbol}
& \texttt{\UseMacro{corpus-robot-sha-pretty}}
& \UseMacro{corpus-robot-num-docs}
& \UseMacro{lemma-filtered-project-robot-num-lemmas}
& \UseMacro{dataset-project-robot-num-tokens}
& \UseMacro{corpus-robot-spec-loc}
& \UseMacro{corpus-robot-proof-loc}
 & 
& \UseMacro{corpus-robot-AVG-spec-loc-per-doc}
& \UseMacro{corpus-robot-AVG-proof-loc-per-doc}
\\
 & 
\Ptwosquare
& \href{\UseMacro{corpus-twosquare-url}}{\urlsymbol}
& \texttt{\UseMacro{corpus-twosquare-sha-pretty}}
& \UseMacro{corpus-twosquare-num-docs}
& \UseMacro{lemma-filtered-project-twosquare-num-lemmas}
& \UseMacro{dataset-project-twosquare-num-tokens}
& \UseMacro{corpus-twosquare-spec-loc}
& \UseMacro{corpus-twosquare-proof-loc}
 & 
& \UseMacro{corpus-twosquare-AVG-spec-loc-per-doc}
& \UseMacro{corpus-twosquare-AVG-proof-loc-per-doc}
\\
\midrule
\multirow{8}{*}{\rotatebox[origin=c]{270}{\UseMacro{group-t3}}} & 
\Pbits
& \href{\UseMacro{corpus-bits-url}}{\urlsymbol}
& \texttt{\UseMacro{corpus-bits-sha-pretty}}
& \UseMacro{corpus-bits-num-docs}
& \UseMacro{lemma-filtered-project-bits-num-lemmas}
& \UseMacro{dataset-project-bits-num-tokens}
& \UseMacro{corpus-bits-spec-loc}
& \UseMacro{corpus-bits-proof-loc}
 & 
& \UseMacro{corpus-bits-AVG-spec-loc-per-doc}
& \UseMacro{corpus-bits-AVG-proof-loc-per-doc}
\\
 & 
\Pcompdecpdl
& \href{\UseMacro{corpus-compdecpdl-url}}{\urlsymbol}
& \texttt{\UseMacro{corpus-compdecpdl-sha-pretty}}
& \UseMacro{corpus-compdecpdl-num-docs}
& \UseMacro{lemma-filtered-project-compdecpdl-num-lemmas}
& \UseMacro{dataset-project-compdecpdl-num-tokens}
& \UseMacro{corpus-compdecpdl-spec-loc}
& \UseMacro{corpus-compdecpdl-proof-loc}
 & 
& \UseMacro{corpus-compdecpdl-AVG-spec-loc-per-doc}
& \UseMacro{corpus-compdecpdl-AVG-proof-loc-per-doc}
\\
 & 
\Pdisel
& \href{\UseMacro{corpus-disel-url}}{\urlsymbol}
& \texttt{\UseMacro{corpus-disel-sha-pretty}}
& \UseMacro{corpus-disel-num-docs}
& \UseMacro{lemma-filtered-project-disel-num-lemmas}
& \UseMacro{dataset-project-disel-num-tokens}
& \UseMacro{corpus-disel-spec-loc}
& \UseMacro{corpus-disel-proof-loc}
 & 
& \UseMacro{corpus-disel-AVG-spec-loc-per-doc}
& \UseMacro{corpus-disel-AVG-proof-loc-per-doc}
\\
 & 
\Pfcslpcm
& \href{\UseMacro{corpus-fcslpcm-url}}{\urlsymbol}
& \texttt{\UseMacro{corpus-fcslpcm-sha-pretty}}
& \UseMacro{corpus-fcslpcm-num-docs}
& \UseMacro{lemma-filtered-project-fcslpcm-num-lemmas}
& \UseMacro{dataset-project-fcslpcm-num-tokens}
& \UseMacro{corpus-fcslpcm-spec-loc}
& \UseMacro{corpus-fcslpcm-proof-loc}
 & 
& \UseMacro{corpus-fcslpcm-AVG-spec-loc-per-doc}
& \UseMacro{corpus-fcslpcm-AVG-proof-loc-per-doc}
\\
 & 
\Pgames
& \href{\UseMacro{corpus-games-url}}{\urlsymbol}
& \texttt{\UseMacro{corpus-games-sha-pretty}}
& \UseMacro{corpus-games-num-docs}
& \UseMacro{lemma-filtered-project-games-num-lemmas}
& \UseMacro{dataset-project-games-num-tokens}
& \UseMacro{corpus-games-spec-loc}
& \UseMacro{corpus-games-proof-loc}
 & 
& \UseMacro{corpus-games-AVG-spec-loc-per-doc}
& \UseMacro{corpus-games-AVG-proof-loc-per-doc}
\\
 & 
\Pmonae
& \href{\UseMacro{corpus-monae-url}}{\urlsymbol}
& \texttt{\UseMacro{corpus-monae-sha-pretty}}
& \UseMacro{corpus-monae-num-docs}
& \UseMacro{lemma-filtered-project-monae-num-lemmas}
& \UseMacro{dataset-project-monae-num-tokens}
& \UseMacro{corpus-monae-spec-loc}
& \UseMacro{corpus-monae-proof-loc}
 & 
& \UseMacro{corpus-monae-AVG-spec-loc-per-doc}
& \UseMacro{corpus-monae-AVG-proof-loc-per-doc}
\\
 & 
\Preglang
& \href{\UseMacro{corpus-reglang-url}}{\urlsymbol}
& \texttt{\UseMacro{corpus-reglang-sha-pretty}}
& \UseMacro{corpus-reglang-num-docs}
& \UseMacro{lemma-filtered-project-reglang-num-lemmas}
& \UseMacro{dataset-project-reglang-num-tokens}
& \UseMacro{corpus-reglang-spec-loc}
& \UseMacro{corpus-reglang-proof-loc}
 & 
& \UseMacro{corpus-reglang-AVG-spec-loc-per-doc}
& \UseMacro{corpus-reglang-AVG-proof-loc-per-doc}
\\
 & 
\Ptoychain
& \href{\UseMacro{corpus-toychain-url}}{\urlsymbol}
& \texttt{\UseMacro{corpus-toychain-sha-pretty}}
& \UseMacro{corpus-toychain-num-docs}
& \UseMacro{lemma-filtered-project-toychain-num-lemmas}
& \UseMacro{dataset-project-toychain-num-tokens}
& \UseMacro{corpus-toychain-spec-loc}
& \UseMacro{corpus-toychain-proof-loc}
 & 
& \UseMacro{corpus-toychain-AVG-spec-loc-per-doc}
& \UseMacro{corpus-toychain-AVG-proof-loc-per-doc}
\\
\midrule
\multicolumn{1}{c}{} & 
\multirow{2}{*}{\UseMacro{table-head-summary-group-ta}}
 & & \TableHeadNA 
& \UseMacro{corpus-ta-AVG-num-docs}
& \UseMacro{lemma-filtered-ta-project-AVG-num-lemmas}
& \UseMacro{dataset-ta-project-AVG-num-tokens}
& \UseMacro{corpus-ta-AVG-spec-loc}
& \UseMacro{corpus-ta-AVG-proof-loc}
 & 
& \UseMacro{corpus-ta-AVG-spec-loc-per-doc}
& \UseMacro{corpus-ta-AVG-proof-loc-per-doc}
\\
\multicolumn{1}{c}{} & 
 & & \TableHeadNA 
& \UseMacro{corpus-ta-SUM-num-docs}
& \UseMacro{lemma-filtered-ta-project-SUM-num-lemmas}
& \UseMacro{dataset-ta-project-SUM-num-tokens}
& \UseMacro{corpus-ta-SUM-spec-loc}
& \UseMacro{corpus-ta-SUM-proof-loc}
 & 
& \UseMacro{corpus-ta-SUM-spec-loc-per-doc}
& \UseMacro{corpus-ta-SUM-proof-loc-per-doc}
\\
\midrule
\multirow{1}{*}{\rotatebox[origin=c]{270}{\tiny \UseMacro{group-lo-small}}} & 
\Pinfotheo
& \href{\UseMacro{corpus-infotheo-url}}{\urlsymbol}
& \texttt{\UseMacro{corpus-infotheo-sha-pretty}}
& \UseMacro{corpus-infotheo-num-docs}
& \UseMacro{lemma-filtered-project-infotheo-num-lemmas}
& \UseMacro{dataset-project-infotheo-num-tokens}
& \UseMacro{corpus-infotheo-spec-loc}
& \UseMacro{corpus-infotheo-proof-loc}
 & 
& \UseMacro{corpus-infotheo-AVG-spec-loc-per-doc}
& \UseMacro{corpus-infotheo-AVG-proof-loc-per-doc}
\\
\midrule
\multicolumn{1}{c}{} & 
\multirow{2}{*}{\UseMacro{table-head-summary-group-allgroup}}
 & & \TableHeadNA 
& \UseMacro{corpus-allgroup-AVG-num-docs}
& \UseMacro{lemma-filtered-allgroup-project-AVG-num-lemmas}
& \UseMacro{dataset-allgroup-project-AVG-num-tokens}
& \UseMacro{corpus-allgroup-AVG-spec-loc}
& \UseMacro{corpus-allgroup-AVG-proof-loc}
 & 
& \UseMacro{corpus-allgroup-AVG-spec-loc-per-doc}
& \UseMacro{corpus-allgroup-AVG-proof-loc-per-doc}
\\
\multicolumn{1}{c}{} & 
 & & \TableHeadNA 
& \UseMacro{corpus-allgroup-SUM-num-docs}
& \UseMacro{lemma-filtered-allgroup-project-SUM-num-lemmas}
& \UseMacro{dataset-allgroup-project-SUM-num-tokens}
& \UseMacro{corpus-allgroup-SUM-spec-loc}
& \UseMacro{corpus-allgroup-SUM-proof-loc}
 & 
& \UseMacro{corpus-allgroup-SUM-spec-loc-per-doc}
& \UseMacro{corpus-allgroup-SUM-proof-loc-per-doc}
\\
\bottomrule
\end{tabular}
\end{center}
\end{scriptsize}
\vspace{\UseMacro{vspace-corpus-allgroup}}
\end{table*}

\pgfmathsetmacro{\varActualTiiNumProjects}{\UseMacro{corpus-t2-num-projects}+\UseMacro{corpus-lo-num-projects}} \newcommand{\ActualTiiNumProjects}{\pgfmathprintnumber[fixed, precision=0]{\varActualTiiNumProjects}\xspace}

Table~\ref{tbl:corpus-allgroup} lists the projects in the expanded corpus, along with basic information about each project. The expanded corpus consists of two parts: the \maincorpus part consists of \UseMacro{corpus-ta-num-projects} projects and is used for  training and evaluating \CoqConvTool; the \leftoutcorpus (\UseMacro{group-lo-small}) part is one project, \PTEXTinfotheo, which is used to study the generalizability of \CoqConvTool on an unseen project.

We constructed and organized the corpus based on recommendations from \MathComp developers. The \UseMacro{corpus-t1-num-projects} core \MathComp projects used in the original corpus are included as the \emph{\dsti} set. We selected \ActualTiiNumProjects projects for the \emph{\dstii} set, such that each included project (1) has a main contributor who is also a significant contributor to one of the \dsti projects, and (2) follows to a significant degree the \cc specified for \MathComp. (\PTEXTinfotheo would be in this set had we not added it to the \leftoutcorpus part.)  Finally, we selected \UseMacro{corpus-t3-num-projects} projects which follow \MathComp \cc but do not fullfil the \dstii criteria, for inclusion in the \emph{\dstiii} set.

We briefly describe each project in our corpus:

\vspace{5pt}

\MyPara{\Panalysis} A library for general real analysis.

\MyPara{\Pbigenough} A small library for $\epsilon - N$ reasoning.

\MyPara{\Pbits} A library for reasoning about bit-level operations\citeappendix{Blot2016}.

\MyPara{\Pcompdecpdl} Formal proofs of completeness and decidability of converse propositional dynamic logic\citeappendix{Doczkal2018b}.

\MyPara{\Pdisel} A framework for distributed separation logic, useful   for verifying implementations of distributed   systems\citeappendix{Sergey2018}.

\MyPara{\Pellipticcurves} A formalization of the algebraic theory of elliptic curves\citeappendix{Bartzia2014}.

\MyPara{\Pfcslpcm} A library formalizing partial commutative monoids,   which are useful for reasoning about pointer-based   programs\citeappendix{Sergey2015}.

\MyPara{\Pfinmap} A library with definitions and results about finite maps and sets with finitely many members.

\MyPara{\Pfourcolor} An updated version of the formal proof of the four-color theorem in graph theory\citeappendix{Gonthier2008}, which states that in all planar graphs, four colors suffice for coloring all vertices such that no two adjacent vertices have the same color.

\MyPara{\Pgames} Definitions and formal proofs of theorems in algorithmic game theory\citeappendix{Bagnall2017}.

\MyPara{\Pgrobner} A formalization of Gr\"obner bases.

\MyPara{\Pmathcomp} The \MathComp library itself\citeappendix{MCB}.

\MyPara{\Pmonae} A library for monadic equational reasoning\citeappendix{Affeldt2019}.

\MyPara{\Pmultinomials} A library formalizing monoidal rings and   multinomials, and related results.

\MyPara{\Poddorder} The formal proof of the odd order (Feit-Thompson) theorem in abstract algebra\citeappendix{Gonthier2013}, which states that all groups of odd order are solvable.

\MyPara{\Prealclosed} Theorems on real closed fields in algebra.

\MyPara{\Preglang} A formalization of the theory of regular languages\citeappendix{Doczkal2018}.

\MyPara{\Probot} A formalization of the mathematics of rigid body   transformations to enable proofs about robot   manipulators\citeappendix{Affeldt2017}.

\MyPara{\Ptoychain} Formalization and verification of a blockchain network protocol\citeappendix{Pirlea2018}.

\MyPara{\Ptwosquare} A proof of Fermat's theorem on the sum of two   squares, including a definition of Gaussian integers.

\MyPara{\Pinfotheo} Formalizations of notions and results from   information theory and probability   theory\citeappendix{Affeldt2015}.

\begin{table}[t]
\begin{scriptsize}
\begin{center}
\caption{\UseMacro{table-caption-dataset-lemma-allgroup}}
\begin{tabular}{l | r c rr c rr}
\toprule
\multirow{2}{*}{\TableHeadLMSet}  & \multirow{2}{*}{\TableHeadNumLemma}  & & \multicolumn{2}{c}{\TableHeadLMName}  & & \multicolumn{2}{c}{\TableHeadLMStmt}  \\ \cline{4-5} \cline{7-8}
 &  & & \TableHeadNumChar & \TableHeadNumSubToken & & \TableHeadNumChar & \TableHeadNumSubToken \\
\midrule
\UseMacro{table-head-dslemmaallgroup-all-ta-all}
& \UseMacro{lemma-all-ta-all-num-lemmas}
& 
& \UseMacro{lemma-all-ta-all-AVG-num_char_name}
& \UseMacro{lemma-all-ta-all-AVG-num_subtok_name}
& 
& \UseMacro{lemma-all-ta-all-AVG-num_char_statement}
& \UseMacro{lemma-all-ta-all-AVG-num_subtok_statement}
\\
\UseMacro{table-head-dslemmaallgroup-filtered-ta-all}
& \UseMacro{lemma-filtered-ta-all-num-lemmas}
& 
& \UseMacro{lemma-filtered-ta-all-AVG-num_char_name}
& \UseMacro{lemma-filtered-ta-all-AVG-num_subtok_name}
& 
& \UseMacro{lemma-filtered-ta-all-AVG-num_char_statement}
& \UseMacro{lemma-filtered-ta-all-AVG-num_subtok_statement}
\\
\midrule
\UseMacro{table-head-dslemmaallgroup-filtered-ta-train}
& \UseMacro{lemma-filtered-ta-train-num-lemmas}
& 
& \UseMacro{lemma-filtered-ta-train-AVG-num_char_name}
& \UseMacro{lemma-filtered-ta-train-AVG-num_subtok_name}
& 
& \UseMacro{lemma-filtered-ta-train-AVG-num_char_statement}
& \UseMacro{lemma-filtered-ta-train-AVG-num_subtok_statement}
\\
\UseMacro{table-head-dslemmaallgroup-filtered-ta-val}
& \UseMacro{lemma-filtered-ta-val-num-lemmas}
& 
& \UseMacro{lemma-filtered-ta-val-AVG-num_char_name}
& \UseMacro{lemma-filtered-ta-val-AVG-num_subtok_name}
& 
& \UseMacro{lemma-filtered-ta-val-AVG-num_char_statement}
& \UseMacro{lemma-filtered-ta-val-AVG-num_subtok_statement}
\\
\UseMacro{table-head-dslemmaallgroup-filtered-ta-test}
& \UseMacro{lemma-filtered-ta-test-num-lemmas}
& 
& \UseMacro{lemma-filtered-ta-test-AVG-num_char_name}
& \UseMacro{lemma-filtered-ta-test-AVG-num_subtok_name}
& 
& \UseMacro{lemma-filtered-ta-test-AVG-num_char_statement}
& \UseMacro{lemma-filtered-ta-test-AVG-num_subtok_statement}
\\
\midrule
\UseMacro{table-head-dslemmaallgroup-filtered-t1-train}
& \UseMacro{lemma-filtered-t1-train-num-lemmas}
& 
& \UseMacro{lemma-filtered-t1-train-AVG-num_char_name}
& \UseMacro{lemma-filtered-t1-train-AVG-num_subtok_name}
& 
& \UseMacro{lemma-filtered-t1-train-AVG-num_char_statement}
& \UseMacro{lemma-filtered-t1-train-AVG-num_subtok_statement}
\\
\UseMacro{table-head-dslemmaallgroup-filtered-t1-val}
& \UseMacro{lemma-filtered-t1-val-num-lemmas}
& 
& \UseMacro{lemma-filtered-t1-val-AVG-num_char_name}
& \UseMacro{lemma-filtered-t1-val-AVG-num_subtok_name}
& 
& \UseMacro{lemma-filtered-t1-val-AVG-num_char_statement}
& \UseMacro{lemma-filtered-t1-val-AVG-num_subtok_statement}
\\
\UseMacro{table-head-dslemmaallgroup-filtered-t1-test}
& \UseMacro{lemma-filtered-t1-test-num-lemmas}
& 
& \UseMacro{lemma-filtered-t1-test-AVG-num_char_name}
& \UseMacro{lemma-filtered-t1-test-AVG-num_subtok_name}
& 
& \UseMacro{lemma-filtered-t1-test-AVG-num_char_statement}
& \UseMacro{lemma-filtered-t1-test-AVG-num_subtok_statement}
\\
\midrule
\UseMacro{table-head-dslemmaallgroup-filtered-t2-train}
& \UseMacro{lemma-filtered-t2-train-num-lemmas}
& 
& \UseMacro{lemma-filtered-t2-train-AVG-num_char_name}
& \UseMacro{lemma-filtered-t2-train-AVG-num_subtok_name}
& 
& \UseMacro{lemma-filtered-t2-train-AVG-num_char_statement}
& \UseMacro{lemma-filtered-t2-train-AVG-num_subtok_statement}
\\
\UseMacro{table-head-dslemmaallgroup-filtered-t2-val}
& \UseMacro{lemma-filtered-t2-val-num-lemmas}
& 
& \UseMacro{lemma-filtered-t2-val-AVG-num_char_name}
& \UseMacro{lemma-filtered-t2-val-AVG-num_subtok_name}
& 
& \UseMacro{lemma-filtered-t2-val-AVG-num_char_statement}
& \UseMacro{lemma-filtered-t2-val-AVG-num_subtok_statement}
\\
\UseMacro{table-head-dslemmaallgroup-filtered-t2-test}
& \UseMacro{lemma-filtered-t2-test-num-lemmas}
& 
& \UseMacro{lemma-filtered-t2-test-AVG-num_char_name}
& \UseMacro{lemma-filtered-t2-test-AVG-num_subtok_name}
& 
& \UseMacro{lemma-filtered-t2-test-AVG-num_char_statement}
& \UseMacro{lemma-filtered-t2-test-AVG-num_subtok_statement}
\\
\midrule
\UseMacro{table-head-dslemmaallgroup-filtered-t3-train}
& \UseMacro{lemma-filtered-t3-train-num-lemmas}
& 
& \UseMacro{lemma-filtered-t3-train-AVG-num_char_name}
& \UseMacro{lemma-filtered-t3-train-AVG-num_subtok_name}
& 
& \UseMacro{lemma-filtered-t3-train-AVG-num_char_statement}
& \UseMacro{lemma-filtered-t3-train-AVG-num_subtok_statement}
\\
\UseMacro{table-head-dslemmaallgroup-filtered-t3-val}
& \UseMacro{lemma-filtered-t3-val-num-lemmas}
& 
& \UseMacro{lemma-filtered-t3-val-AVG-num_char_name}
& \UseMacro{lemma-filtered-t3-val-AVG-num_subtok_name}
& 
& \UseMacro{lemma-filtered-t3-val-AVG-num_char_statement}
& \UseMacro{lemma-filtered-t3-val-AVG-num_subtok_statement}
\\
\UseMacro{table-head-dslemmaallgroup-filtered-t3-test}
& \UseMacro{lemma-filtered-t3-test-num-lemmas}
& 
& \UseMacro{lemma-filtered-t3-test-AVG-num_char_name}
& \UseMacro{lemma-filtered-t3-test-AVG-num_subtok_name}
& 
& \UseMacro{lemma-filtered-t3-test-AVG-num_char_statement}
& \UseMacro{lemma-filtered-t3-test-AVG-num_subtok_statement}
\\
\bottomrule
\end{tabular}
\end{center}
\end{scriptsize}
\vspace{\UseMacro{vspace-dataset-lemma-allgroup}}
\end{table}

\begin{figure}[t]   \centering   \includegraphics[scale=0.65]{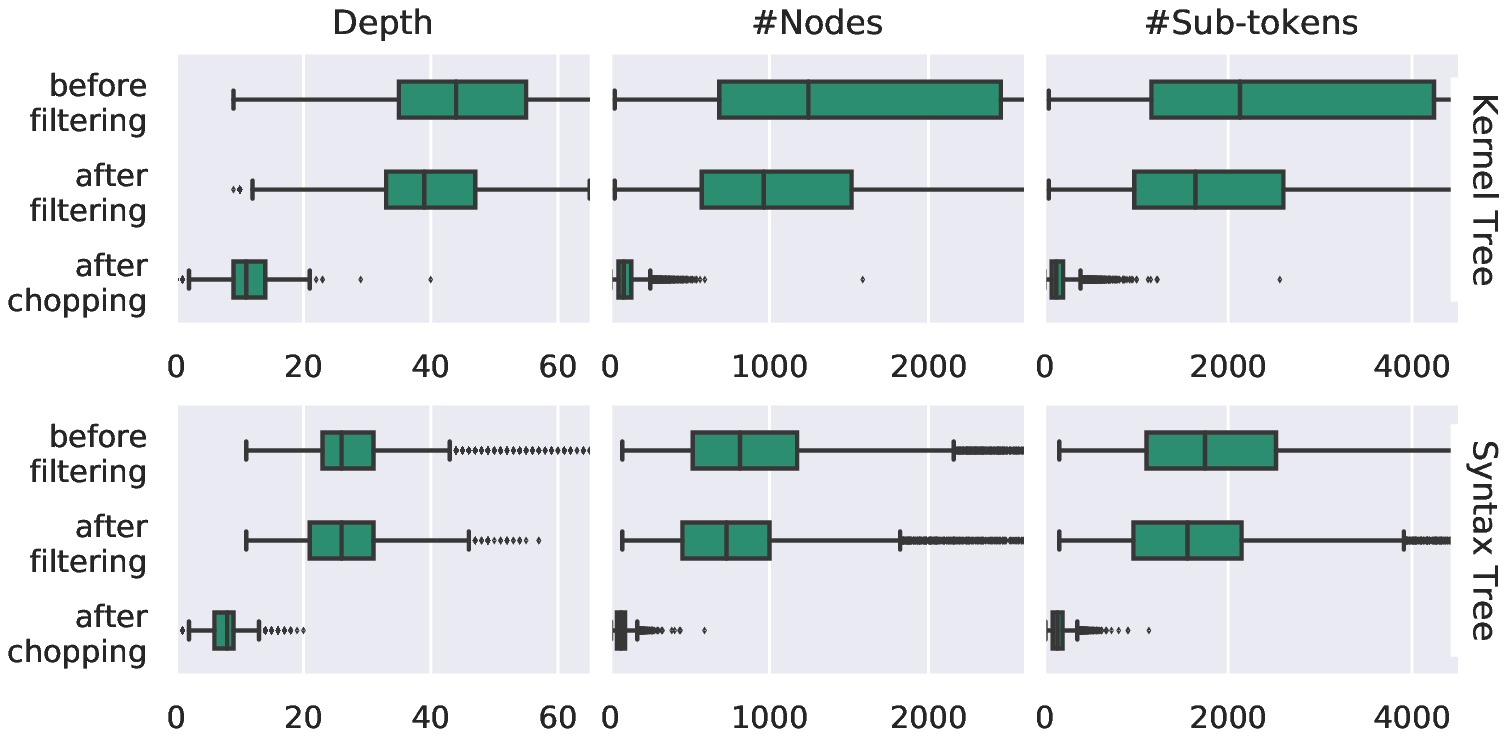}   \vspace{-10pt}   \caption{Statistics of \sktrees in the expanded corpus.}\label{fig:lemma-metrics-allgroup}   \vspace{-10pt} \end{figure}

We follow the same procedure as for the original corpus to extract the lemmas and filter out the lemmas with the deepest 25\% of the \ktrees. Table~\ref{tbl:dataset-lemma-allgroup} shows statistics on the lemmas obtained from each set and each tier. \figurename~\ref{fig:lemma-metrics-allgroup} illustrates the changes of depth, number of nodes, and number of \subtok{s} (after serialization) of the \ktrees (first row) and \strees (second row) before filtering, after filtering, and after \simplification.

\subsection{Automatic Evaluation on the Expanded Corpus} \label{sec:appendix:ex-eval}

\begin{table}[t] 
\begin{scriptsize}
\begin{center}
\caption{\UseMacro{table-caption-experiments}}
\begin{tabular}{c c | c}

  \toprule
  \makecell[l]{\Train \& \Val} & \Test & Results Table \\
  \midrule

  \UseMacro{group-ta} & \UseMacro{group-ta} & Table~\ref{tbl:results-ln-ta--ta--ta-main} \\
  \UseMacro{group-ta} & \UseMacro{group-t1} & Table~\ref{tbl:results-ln-ta--ta--t1-main} \\
  \UseMacro{group-ta} & \UseMacro{group-t2} & Table~\ref{tbl:results-ln-ta--ta--t2-main} \\
  \UseMacro{group-ta} & \UseMacro{group-t3} & Table~\ref{tbl:results-ln-ta--ta--t3-main} \\
  \UseMacro{group-t1} & \UseMacro{group-ta} & Table~\ref{tbl:results-ln-t1--t1--ta-main} \\
  \UseMacro{group-t1} & \UseMacro{group-t1} & Table~\ref{tbl:results-ln-t1--t1--t1-main} \\
  \UseMacro{group-t1} & \UseMacro{group-t2} & Table~\ref{tbl:results-ln-t1--t1--t2-main} \\
  \UseMacro{group-t1} & \UseMacro{group-t3} & Table~\ref{tbl:results-ln-t1--t1--t3-main} \\
  \UseMacro{group-t2} & \UseMacro{group-t2} & Table~\ref{tbl:results-ln-t2--t2--t2-main} \\
  \UseMacro{group-t3} & \UseMacro{group-t3} & Table~\ref{tbl:results-ln-t3--t3--t3-main} \\

  \bottomrule

\end{tabular}
\end{center}
\end{scriptsize}
\vspace{\UseMacro{vspace-experiments}}

 \end{table}

To investigate whether \CoqConvTool can benefit from learning from a larger but less focused corpus than in our original automatic evaluation, we experimented with different combinations of \train, \val, and \test sets.  Table~\ref{tbl:experiments} lists the combinations we used; the first column shows the corpus that \train and \val sets are from, the second column shows the corpus that \test set is from, and the third column indicates the results table for each combination.

\begin{table*}[t]
\begin{scriptsize}
\begin{center}
\caption{\UseMacro{table-caption-results-ln-ta--ta--ta-main}}
\begin{tabular}{c l r r r r}
\toprule
\TableHeadGroup & \TableHeadModel
& \UseMacro{table-head-results-ln-BLEU-4}
& \UseMacro{table-head-results-ln-frag-acc}
& \UseMacro{table-head-results-ln-full-acc-top-1}
& \UseMacro{table-head-results-ln-full-acc-top-5}
 \\
\midrule
 & 
\UseMacro{ln-s+bsexpl1+fsexpl1+attn+copy}
& 
\UseMacro{results-ln-ta--ta--ta-ln-s+bsexpl1+fsexpl1+attn+copy-test-AVG-BLEU-4}
& 
\UseMacro{results-ln-ta--ta--ta-ln-s+bsexpl1+fsexpl1+attn+copy-test-AVG-frag-acc}
& 
\UseMacro{results-ln-ta--ta--ta-ln-s+bsexpl1+fsexpl1+attn+copy-test-AVG-full-acc-top-1}
& 
\UseMacro{results-ln-ta--ta--ta-ln-s+bsexpl1+fsexpl1+attn+copy-test-AVG-full-acc-top-5}
\\
 & 
\UseMacro{ln-s+bsexpl1+attn+copy}
& 
\textbf{\UseMacro{results-ln-ta--ta--ta-ln-s+bsexpl1+attn+copy-test-AVG-BLEU-4}}
& 
\textbf{\UseMacro{results-ln-ta--ta--ta-ln-s+bsexpl1+attn+copy-test-AVG-frag-acc}}
& 
\textbf{\UseMacro{results-ln-ta--ta--ta-ln-s+bsexpl1+attn+copy-test-AVG-full-acc-top-1}}
& 
\textbf{\UseMacro{results-ln-ta--ta--ta-ln-s+bsexpl1+attn+copy-test-AVG-full-acc-top-5}}
\\
 & 
\UseMacro{ln-s+fsexpl1+attn+copy}
& 
\UseMacro{results-ln-ta--ta--ta-ln-s+fsexpl1+attn+copy-test-AVG-BLEU-4}
& 
\UseMacro{results-ln-ta--ta--ta-ln-s+fsexpl1+attn+copy-test-AVG-frag-acc}
& 
\UseMacro{results-ln-ta--ta--ta-ln-s+fsexpl1+attn+copy-test-AVG-full-acc-top-1}
& 
\UseMacro{results-ln-ta--ta--ta-ln-s+fsexpl1+attn+copy-test-AVG-full-acc-top-5}
\\
\multirow{-4}{*}{\TablePartMultiAttnCopy}
 & 
\UseMacro{ln-bsexpl1+fsexpl1+attn+copy}
& 
\UseMacro{results-ln-ta--ta--ta-ln-bsexpl1+fsexpl1+attn+copy-test-AVG-BLEU-4}
& 
\UseMacro{results-ln-ta--ta--ta-ln-bsexpl1+fsexpl1+attn+copy-test-AVG-frag-acc}
& 
\UseMacro{results-ln-ta--ta--ta-ln-bsexpl1+fsexpl1+attn+copy-test-AVG-full-acc-top-1}
& 
\UseMacro{results-ln-ta--ta--ta-ln-bsexpl1+fsexpl1+attn+copy-test-AVG-full-acc-top-5}
\\
\bottomrule
\end{tabular}
\end{center}
\end{scriptsize}
\vspace{\UseMacro{vspace-results-ln-ta--ta--ta-main}}
\end{table*}

\begin{table*}[t]
\begin{scriptsize}
\begin{center}
\caption{\UseMacro{table-caption-results-ln-ta--ta--t1-main}}
\begin{tabular}{c l r r r r}
\toprule
\TableHeadGroup & \TableHeadModel
& \UseMacro{table-head-results-ln-BLEU-4}
& \UseMacro{table-head-results-ln-frag-acc}
& \UseMacro{table-head-results-ln-full-acc-top-1}
& \UseMacro{table-head-results-ln-full-acc-top-5}
 \\
\midrule
 & 
\UseMacro{ln-s+bsexpl1+fsexpl1+attn+copy}
& 
\UseMacro{results-ln-ta--ta--t1-ln-s+bsexpl1+fsexpl1+attn+copy-test-AVG-BLEU-4}
& 
\UseMacro{results-ln-ta--ta--t1-ln-s+bsexpl1+fsexpl1+attn+copy-test-AVG-frag-acc}
& 
\UseMacro{results-ln-ta--ta--t1-ln-s+bsexpl1+fsexpl1+attn+copy-test-AVG-full-acc-top-1}
& 
\UseMacro{results-ln-ta--ta--t1-ln-s+bsexpl1+fsexpl1+attn+copy-test-AVG-full-acc-top-5}
\\
 & 
\UseMacro{ln-s+bsexpl1+attn+copy}
& 
\textbf{\UseMacro{results-ln-ta--ta--t1-ln-s+bsexpl1+attn+copy-test-AVG-BLEU-4}}
& 
\textbf{\UseMacro{results-ln-ta--ta--t1-ln-s+bsexpl1+attn+copy-test-AVG-frag-acc}}
& 
\textbf{\UseMacro{results-ln-ta--ta--t1-ln-s+bsexpl1+attn+copy-test-AVG-full-acc-top-1}}
& 
\textbf{\UseMacro{results-ln-ta--ta--t1-ln-s+bsexpl1+attn+copy-test-AVG-full-acc-top-5}}
\\
 & 
\UseMacro{ln-s+fsexpl1+attn+copy}
& 
\UseMacro{results-ln-ta--ta--t1-ln-s+fsexpl1+attn+copy-test-AVG-BLEU-4}
& 
\UseMacro{results-ln-ta--ta--t1-ln-s+fsexpl1+attn+copy-test-AVG-frag-acc}
& 
\UseMacro{results-ln-ta--ta--t1-ln-s+fsexpl1+attn+copy-test-AVG-full-acc-top-1}
& 
\UseMacro{results-ln-ta--ta--t1-ln-s+fsexpl1+attn+copy-test-AVG-full-acc-top-5}
\\
\multirow{-4}{*}{\TablePartMultiAttnCopy}
 & 
\UseMacro{ln-bsexpl1+fsexpl1+attn+copy}
& 
\UseMacro{results-ln-ta--ta--t1-ln-bsexpl1+fsexpl1+attn+copy-test-AVG-BLEU-4}
& 
\UseMacro{results-ln-ta--ta--t1-ln-bsexpl1+fsexpl1+attn+copy-test-AVG-frag-acc}
& 
\UseMacro{results-ln-ta--ta--t1-ln-bsexpl1+fsexpl1+attn+copy-test-AVG-full-acc-top-1}
& 
\UseMacro{results-ln-ta--ta--t1-ln-bsexpl1+fsexpl1+attn+copy-test-AVG-full-acc-top-5}
\\
\bottomrule
\end{tabular}
\end{center}
\end{scriptsize}
\vspace{\UseMacro{vspace-results-ln-ta--ta--t1-main}}
\end{table*}

\begin{table*}[t]
\begin{scriptsize}
\begin{center}
\caption{\UseMacro{table-caption-results-ln-ta--ta--t2-main}}
\begin{tabular}{c l r r r r}
\toprule
\TableHeadGroup & \TableHeadModel
& \UseMacro{table-head-results-ln-BLEU-4}
& \UseMacro{table-head-results-ln-frag-acc}
& \UseMacro{table-head-results-ln-full-acc-top-1}
& \UseMacro{table-head-results-ln-full-acc-top-5}
 \\
\midrule
 & 
\UseMacro{ln-s+bsexpl1+fsexpl1+attn+copy}
& 
\UseMacro{results-ln-ta--ta--t2-ln-s+bsexpl1+fsexpl1+attn+copy-test-AVG-BLEU-4}
& 
\UseMacro{results-ln-ta--ta--t2-ln-s+bsexpl1+fsexpl1+attn+copy-test-AVG-frag-acc}
& 
\UseMacro{results-ln-ta--ta--t2-ln-s+bsexpl1+fsexpl1+attn+copy-test-AVG-full-acc-top-1}
& 
\UseMacro{results-ln-ta--ta--t2-ln-s+bsexpl1+fsexpl1+attn+copy-test-AVG-full-acc-top-5}
\\
 & 
\UseMacro{ln-s+bsexpl1+attn+copy}
& 
\UseMacro{results-ln-ta--ta--t2-ln-s+bsexpl1+attn+copy-test-AVG-BLEU-4}
& 
\UseMacro{results-ln-ta--ta--t2-ln-s+bsexpl1+attn+copy-test-AVG-frag-acc}
& 
\textbf{\UseMacro{results-ln-ta--ta--t2-ln-s+bsexpl1+attn+copy-test-AVG-full-acc-top-1}}
& 
\textbf{\UseMacro{results-ln-ta--ta--t2-ln-s+bsexpl1+attn+copy-test-AVG-full-acc-top-5}}
\\
 & 
\UseMacro{ln-s+fsexpl1+attn+copy}
& 
\UseMacro{results-ln-ta--ta--t2-ln-s+fsexpl1+attn+copy-test-AVG-BLEU-4}
& 
\UseMacro{results-ln-ta--ta--t2-ln-s+fsexpl1+attn+copy-test-AVG-frag-acc}
& 
\UseMacro{results-ln-ta--ta--t2-ln-s+fsexpl1+attn+copy-test-AVG-full-acc-top-1}
& 
\UseMacro{results-ln-ta--ta--t2-ln-s+fsexpl1+attn+copy-test-AVG-full-acc-top-5}
\\
\multirow{-4}{*}{\TablePartMultiAttnCopy}
 & 
\UseMacro{ln-bsexpl1+fsexpl1+attn+copy}
& 
\UseMacro{results-ln-ta--ta--t2-ln-bsexpl1+fsexpl1+attn+copy-test-AVG-BLEU-4}
& 
\textbf{\UseMacro{results-ln-ta--ta--t2-ln-bsexpl1+fsexpl1+attn+copy-test-AVG-frag-acc}}
& 
\UseMacro{results-ln-ta--ta--t2-ln-bsexpl1+fsexpl1+attn+copy-test-AVG-full-acc-top-1}
& 
\UseMacro{results-ln-ta--ta--t2-ln-bsexpl1+fsexpl1+attn+copy-test-AVG-full-acc-top-5}
\\
\bottomrule
\end{tabular}
\end{center}
\end{scriptsize}
\vspace{\UseMacro{vspace-results-ln-ta--ta--t2-main}}
\end{table*}

\begin{table*}[t]
\begin{scriptsize}
\begin{center}
\caption{\UseMacro{table-caption-results-ln-ta--ta--t3-main}}
\begin{tabular}{c l r r r r}
\toprule
\TableHeadGroup & \TableHeadModel
& \UseMacro{table-head-results-ln-BLEU-4}
& \UseMacro{table-head-results-ln-frag-acc}
& \UseMacro{table-head-results-ln-full-acc-top-1}
& \UseMacro{table-head-results-ln-full-acc-top-5}
 \\
\midrule
 & 
\UseMacro{ln-s+bsexpl1+fsexpl1+attn+copy}
& 
\UseMacro{results-ln-ta--ta--t3-ln-s+bsexpl1+fsexpl1+attn+copy-test-AVG-BLEU-4}
& 
\UseMacro{results-ln-ta--ta--t3-ln-s+bsexpl1+fsexpl1+attn+copy-test-AVG-frag-acc}
& 
\UseMacro{results-ln-ta--ta--t3-ln-s+bsexpl1+fsexpl1+attn+copy-test-AVG-full-acc-top-1}
& 
\UseMacro{results-ln-ta--ta--t3-ln-s+bsexpl1+fsexpl1+attn+copy-test-AVG-full-acc-top-5}
\\
 & 
\UseMacro{ln-s+bsexpl1+attn+copy}
& 
\UseMacro{results-ln-ta--ta--t3-ln-s+bsexpl1+attn+copy-test-AVG-BLEU-4}
& 
\UseMacro{results-ln-ta--ta--t3-ln-s+bsexpl1+attn+copy-test-AVG-frag-acc}
& 
\UseMacro{results-ln-ta--ta--t3-ln-s+bsexpl1+attn+copy-test-AVG-full-acc-top-1}
& 
\UseMacro{results-ln-ta--ta--t3-ln-s+bsexpl1+attn+copy-test-AVG-full-acc-top-5}
\\
 & 
\UseMacro{ln-s+fsexpl1+attn+copy}
& 
\UseMacro{results-ln-ta--ta--t3-ln-s+fsexpl1+attn+copy-test-AVG-BLEU-4}
& 
\UseMacro{results-ln-ta--ta--t3-ln-s+fsexpl1+attn+copy-test-AVG-frag-acc}
& 
\UseMacro{results-ln-ta--ta--t3-ln-s+fsexpl1+attn+copy-test-AVG-full-acc-top-1}
& 
\textbf{\UseMacro{results-ln-ta--ta--t3-ln-s+fsexpl1+attn+copy-test-AVG-full-acc-top-5}}
\\
\multirow{-4}{*}{\TablePartMultiAttnCopy}
 & 
\UseMacro{ln-bsexpl1+fsexpl1+attn+copy}
& 
\UseMacro{results-ln-ta--ta--t3-ln-bsexpl1+fsexpl1+attn+copy-test-AVG-BLEU-4}
& 
\UseMacro{results-ln-ta--ta--t3-ln-bsexpl1+fsexpl1+attn+copy-test-AVG-frag-acc}
& 
\UseMacro{results-ln-ta--ta--t3-ln-bsexpl1+fsexpl1+attn+copy-test-AVG-full-acc-top-1}
& 
\UseMacro{results-ln-ta--ta--t3-ln-bsexpl1+fsexpl1+attn+copy-test-AVG-full-acc-top-5}
\\
\bottomrule
\end{tabular}
\end{center}
\end{scriptsize}
\vspace{\UseMacro{vspace-results-ln-ta--ta--t3-main}}
\end{table*}

\begin{table*}[t]
\begin{scriptsize}
\begin{center}
\caption{\UseMacro{table-caption-results-ln-t1--t1--ta-main}}
\begin{tabular}{c l r r r r}
\toprule
\TableHeadGroup & \TableHeadModel
& \UseMacro{table-head-results-ln-BLEU-4}
& \UseMacro{table-head-results-ln-frag-acc}
& \UseMacro{table-head-results-ln-full-acc-top-1}
& \UseMacro{table-head-results-ln-full-acc-top-5}
 \\
\midrule
 & 
\UseMacro{ln-s+bsexpl1+fsexpl1+attn+copy}
& 
\UseMacro{results-ln-t1--t1--ta-ln-s+bsexpl1+fsexpl1+attn+copy-test-AVG-BLEU-4}
& 
\UseMacro{results-ln-t1--t1--ta-ln-s+bsexpl1+fsexpl1+attn+copy-test-AVG-frag-acc}
& 
\UseMacro{results-ln-t1--t1--ta-ln-s+bsexpl1+fsexpl1+attn+copy-test-AVG-full-acc-top-1}
& 
\UseMacro{results-ln-t1--t1--ta-ln-s+bsexpl1+fsexpl1+attn+copy-test-AVG-full-acc-top-5}
\\
 & 
\UseMacro{ln-s+bsexpl1+attn+copy}
& 
\textbf{\UseMacro{results-ln-t1--t1--ta-ln-s+bsexpl1+attn+copy-test-AVG-BLEU-4}}
& 
\textbf{\UseMacro{results-ln-t1--t1--ta-ln-s+bsexpl1+attn+copy-test-AVG-frag-acc}}
& 
\textbf{\UseMacro{results-ln-t1--t1--ta-ln-s+bsexpl1+attn+copy-test-AVG-full-acc-top-1}}
& 
\textbf{\UseMacro{results-ln-t1--t1--ta-ln-s+bsexpl1+attn+copy-test-AVG-full-acc-top-5}}
\\
 & 
\UseMacro{ln-s+fsexpl1+attn+copy}
& 
\UseMacro{results-ln-t1--t1--ta-ln-s+fsexpl1+attn+copy-test-AVG-BLEU-4}
& 
\UseMacro{results-ln-t1--t1--ta-ln-s+fsexpl1+attn+copy-test-AVG-frag-acc}
& 
\UseMacro{results-ln-t1--t1--ta-ln-s+fsexpl1+attn+copy-test-AVG-full-acc-top-1}
& 
\UseMacro{results-ln-t1--t1--ta-ln-s+fsexpl1+attn+copy-test-AVG-full-acc-top-5}
\\
\multirow{-4}{*}{\TablePartMultiAttnCopy}
 & 
\UseMacro{ln-bsexpl1+fsexpl1+attn+copy}
& 
\UseMacro{results-ln-t1--t1--ta-ln-bsexpl1+fsexpl1+attn+copy-test-AVG-BLEU-4}
& 
\UseMacro{results-ln-t1--t1--ta-ln-bsexpl1+fsexpl1+attn+copy-test-AVG-frag-acc}
& 
\UseMacro{results-ln-t1--t1--ta-ln-bsexpl1+fsexpl1+attn+copy-test-AVG-full-acc-top-1}
& 
\UseMacro{results-ln-t1--t1--ta-ln-bsexpl1+fsexpl1+attn+copy-test-AVG-full-acc-top-5}
\\
\bottomrule
\end{tabular}
\end{center}
\end{scriptsize}
\vspace{\UseMacro{vspace-results-ln-t1--t1--ta-main}}
\end{table*}

\begin{table*}[t]
\begin{scriptsize}
\begin{center}
\caption{\UseMacro{table-caption-results-ln-t1--t1--t2-main}}
\begin{tabular}{c l r r r r}
\toprule
\TableHeadGroup & \TableHeadModel
& \UseMacro{table-head-results-ln-BLEU-4}
& \UseMacro{table-head-results-ln-frag-acc}
& \UseMacro{table-head-results-ln-full-acc-top-1}
& \UseMacro{table-head-results-ln-full-acc-top-5}
 \\
\midrule
 & 
\UseMacro{ln-s+bsexpl1+fsexpl1+attn+copy}
& 
\UseMacro{results-ln-t1--t1--t2-ln-s+bsexpl1+fsexpl1+attn+copy-test-AVG-BLEU-4}
& 
\UseMacro{results-ln-t1--t1--t2-ln-s+bsexpl1+fsexpl1+attn+copy-test-AVG-frag-acc}
& 
\textbf{\UseMacro{results-ln-t1--t1--t2-ln-s+bsexpl1+fsexpl1+attn+copy-test-AVG-full-acc-top-1}}
& 
\UseMacro{results-ln-t1--t1--t2-ln-s+bsexpl1+fsexpl1+attn+copy-test-AVG-full-acc-top-5}
\\
 & 
\UseMacro{ln-s+bsexpl1+attn+copy}
& 
\UseMacro{results-ln-t1--t1--t2-ln-s+bsexpl1+attn+copy-test-AVG-BLEU-4}
& 
\textbf{\UseMacro{results-ln-t1--t1--t2-ln-s+bsexpl1+attn+copy-test-AVG-frag-acc}}
& 
\UseMacro{results-ln-t1--t1--t2-ln-s+bsexpl1+attn+copy-test-AVG-full-acc-top-1}
& 
\UseMacro{results-ln-t1--t1--t2-ln-s+bsexpl1+attn+copy-test-AVG-full-acc-top-5}
\\
 & 
\UseMacro{ln-s+fsexpl1+attn+copy}
& 
\UseMacro{results-ln-t1--t1--t2-ln-s+fsexpl1+attn+copy-test-AVG-BLEU-4}
& 
\UseMacro{results-ln-t1--t1--t2-ln-s+fsexpl1+attn+copy-test-AVG-frag-acc}
& 
\UseMacro{results-ln-t1--t1--t2-ln-s+fsexpl1+attn+copy-test-AVG-full-acc-top-1}
& 
\UseMacro{results-ln-t1--t1--t2-ln-s+fsexpl1+attn+copy-test-AVG-full-acc-top-5}
\\
\multirow{-4}{*}{\TablePartMultiAttnCopy}
 & 
\UseMacro{ln-bsexpl1+fsexpl1+attn+copy}
& 
\UseMacro{results-ln-t1--t1--t2-ln-bsexpl1+fsexpl1+attn+copy-test-AVG-BLEU-4}
& 
\UseMacro{results-ln-t1--t1--t2-ln-bsexpl1+fsexpl1+attn+copy-test-AVG-frag-acc}
& 
\textbf{\UseMacro{results-ln-t1--t1--t2-ln-bsexpl1+fsexpl1+attn+copy-test-AVG-full-acc-top-1}}
& 
\textbf{\UseMacro{results-ln-t1--t1--t2-ln-bsexpl1+fsexpl1+attn+copy-test-AVG-full-acc-top-5}}
\\
\bottomrule
\end{tabular}
\end{center}
\end{scriptsize}
\vspace{\UseMacro{vspace-results-ln-t1--t1--t2-main}}
\end{table*}

\begin{table*}[t]
\begin{scriptsize}
\begin{center}
\caption{\UseMacro{table-caption-results-ln-t1--t1--t3-main}}
\begin{tabular}{c l r r r r}
\toprule
\TableHeadGroup & \TableHeadModel
& \UseMacro{table-head-results-ln-BLEU-4}
& \UseMacro{table-head-results-ln-frag-acc}
& \UseMacro{table-head-results-ln-full-acc-top-1}
& \UseMacro{table-head-results-ln-full-acc-top-5}
 \\
\midrule
 & 
\UseMacro{ln-s+bsexpl1+fsexpl1+attn+copy}
& 
\UseMacro{results-ln-t1--t1--t3-ln-s+bsexpl1+fsexpl1+attn+copy-test-AVG-BLEU-4}
& 
\UseMacro{results-ln-t1--t1--t3-ln-s+bsexpl1+fsexpl1+attn+copy-test-AVG-frag-acc}
& 
\UseMacro{results-ln-t1--t1--t3-ln-s+bsexpl1+fsexpl1+attn+copy-test-AVG-full-acc-top-1}
& 
\UseMacro{results-ln-t1--t1--t3-ln-s+bsexpl1+fsexpl1+attn+copy-test-AVG-full-acc-top-5}
\\
 & 
\UseMacro{ln-s+bsexpl1+attn+copy}
& 
\UseMacro{results-ln-t1--t1--t3-ln-s+bsexpl1+attn+copy-test-AVG-BLEU-4}
& 
\UseMacro{results-ln-t1--t1--t3-ln-s+bsexpl1+attn+copy-test-AVG-frag-acc}
& 
\UseMacro{results-ln-t1--t1--t3-ln-s+bsexpl1+attn+copy-test-AVG-full-acc-top-1}
& 
\UseMacro{results-ln-t1--t1--t3-ln-s+bsexpl1+attn+copy-test-AVG-full-acc-top-5}
\\
 & 
\UseMacro{ln-s+fsexpl1+attn+copy}
& 
\UseMacro{results-ln-t1--t1--t3-ln-s+fsexpl1+attn+copy-test-AVG-BLEU-4}
& 
\UseMacro{results-ln-t1--t1--t3-ln-s+fsexpl1+attn+copy-test-AVG-frag-acc}
& 
\UseMacro{results-ln-t1--t1--t3-ln-s+fsexpl1+attn+copy-test-AVG-full-acc-top-1}
& 
\UseMacro{results-ln-t1--t1--t3-ln-s+fsexpl1+attn+copy-test-AVG-full-acc-top-5}
\\
\multirow{-4}{*}{\TablePartMultiAttnCopy}
 & 
\UseMacro{ln-bsexpl1+fsexpl1+attn+copy}
& 
\UseMacro{results-ln-t1--t1--t3-ln-bsexpl1+fsexpl1+attn+copy-test-AVG-BLEU-4}
& 
\textbf{\UseMacro{results-ln-t1--t1--t3-ln-bsexpl1+fsexpl1+attn+copy-test-AVG-frag-acc}}
& 
\UseMacro{results-ln-t1--t1--t3-ln-bsexpl1+fsexpl1+attn+copy-test-AVG-full-acc-top-1}
& 
\UseMacro{results-ln-t1--t1--t3-ln-bsexpl1+fsexpl1+attn+copy-test-AVG-full-acc-top-5}
\\
\bottomrule
\end{tabular}
\end{center}
\end{scriptsize}
\vspace{\UseMacro{vspace-results-ln-t1--t1--t3-main}}
\end{table*}

\begin{table*}[t]
\begin{scriptsize}
\begin{center}
\caption{\UseMacro{table-caption-results-ln-t2--t2--t2-main}}
\begin{tabular}{c l r r r r}
\toprule
\TableHeadGroup & \TableHeadModel
& \UseMacro{table-head-results-ln-BLEU-4}
& \UseMacro{table-head-results-ln-frag-acc}
& \UseMacro{table-head-results-ln-full-acc-top-1}
& \UseMacro{table-head-results-ln-full-acc-top-5}
 \\
\midrule
 & 
\UseMacro{ln-s+bsexpl1+fsexpl1+attn+copy}
& 
\UseMacro{results-ln-t2--t2--t2-ln-s+bsexpl1+fsexpl1+attn+copy-test-AVG-BLEU-4}
& 
\UseMacro{results-ln-t2--t2--t2-ln-s+bsexpl1+fsexpl1+attn+copy-test-AVG-frag-acc}
& 
\textbf{\UseMacro{results-ln-t2--t2--t2-ln-s+bsexpl1+fsexpl1+attn+copy-test-AVG-full-acc-top-1}}
& 
\UseMacro{results-ln-t2--t2--t2-ln-s+bsexpl1+fsexpl1+attn+copy-test-AVG-full-acc-top-5}
\\
 & 
\UseMacro{ln-s+bsexpl1+attn+copy}
& 
\UseMacro{results-ln-t2--t2--t2-ln-s+bsexpl1+attn+copy-test-AVG-BLEU-4}
& 
\UseMacro{results-ln-t2--t2--t2-ln-s+bsexpl1+attn+copy-test-AVG-frag-acc}
& 
\UseMacro{results-ln-t2--t2--t2-ln-s+bsexpl1+attn+copy-test-AVG-full-acc-top-1}
& 
\UseMacro{results-ln-t2--t2--t2-ln-s+bsexpl1+attn+copy-test-AVG-full-acc-top-5}
\\
 & 
\UseMacro{ln-s+fsexpl1+attn+copy}
& 
\UseMacro{results-ln-t2--t2--t2-ln-s+fsexpl1+attn+copy-test-AVG-BLEU-4}
& 
\UseMacro{results-ln-t2--t2--t2-ln-s+fsexpl1+attn+copy-test-AVG-frag-acc}
& 
\UseMacro{results-ln-t2--t2--t2-ln-s+fsexpl1+attn+copy-test-AVG-full-acc-top-1}
& 
\UseMacro{results-ln-t2--t2--t2-ln-s+fsexpl1+attn+copy-test-AVG-full-acc-top-5}
\\
\multirow{-4}{*}{\TablePartMultiAttnCopy}
 & 
\UseMacro{ln-bsexpl1+fsexpl1+attn+copy}
& 
\UseMacro{results-ln-t2--t2--t2-ln-bsexpl1+fsexpl1+attn+copy-test-AVG-BLEU-4}
& 
\UseMacro{results-ln-t2--t2--t2-ln-bsexpl1+fsexpl1+attn+copy-test-AVG-frag-acc}
& 
\UseMacro{results-ln-t2--t2--t2-ln-bsexpl1+fsexpl1+attn+copy-test-AVG-full-acc-top-1}
& 
\textbf{\UseMacro{results-ln-t2--t2--t2-ln-bsexpl1+fsexpl1+attn+copy-test-AVG-full-acc-top-5}}
\\
\bottomrule
\end{tabular}
\end{center}
\end{scriptsize}
\vspace{\UseMacro{vspace-results-ln-t2--t2--t2-main}}
\end{table*}

\begin{table*}[t]
\begin{scriptsize}
\begin{center}
\caption{\UseMacro{table-caption-results-ln-t3--t3--t3-main}}
\begin{tabular}{c l r r r r}
\toprule
\TableHeadGroup & \TableHeadModel
& \UseMacro{table-head-results-ln-BLEU-4}
& \UseMacro{table-head-results-ln-frag-acc}
& \UseMacro{table-head-results-ln-full-acc-top-1}
& \UseMacro{table-head-results-ln-full-acc-top-5}
 \\
\midrule
 & 
\UseMacro{ln-s+bsexpl1+fsexpl1+attn+copy}
& 
\UseMacro{results-ln-t3--t3--t3-ln-s+bsexpl1+fsexpl1+attn+copy-test-AVG-BLEU-4}
& 
\UseMacro{results-ln-t3--t3--t3-ln-s+bsexpl1+fsexpl1+attn+copy-test-AVG-frag-acc}
& 
\UseMacro{results-ln-t3--t3--t3-ln-s+bsexpl1+fsexpl1+attn+copy-test-AVG-full-acc-top-1}
& 
\UseMacro{results-ln-t3--t3--t3-ln-s+bsexpl1+fsexpl1+attn+copy-test-AVG-full-acc-top-5}
\\
 & 
\UseMacro{ln-s+bsexpl1+attn+copy}
& 
\UseMacro{results-ln-t3--t3--t3-ln-s+bsexpl1+attn+copy-test-AVG-BLEU-4}
& 
\UseMacro{results-ln-t3--t3--t3-ln-s+bsexpl1+attn+copy-test-AVG-frag-acc}
& 
\UseMacro{results-ln-t3--t3--t3-ln-s+bsexpl1+attn+copy-test-AVG-full-acc-top-1}
& 
\UseMacro{results-ln-t3--t3--t3-ln-s+bsexpl1+attn+copy-test-AVG-full-acc-top-5}
\\
 & 
\UseMacro{ln-s+fsexpl1+attn+copy}
& 
\UseMacro{results-ln-t3--t3--t3-ln-s+fsexpl1+attn+copy-test-AVG-BLEU-4}
& 
\UseMacro{results-ln-t3--t3--t3-ln-s+fsexpl1+attn+copy-test-AVG-frag-acc}
& 
\UseMacro{results-ln-t3--t3--t3-ln-s+fsexpl1+attn+copy-test-AVG-full-acc-top-1}
& 
\textbf{\UseMacro{results-ln-t3--t3--t3-ln-s+fsexpl1+attn+copy-test-AVG-full-acc-top-5}}
\\
\multirow{-4}{*}{\TablePartMultiAttnCopy}
 & 
\UseMacro{ln-bsexpl1+fsexpl1+attn+copy}
& 
\UseMacro{results-ln-t3--t3--t3-ln-bsexpl1+fsexpl1+attn+copy-test-AVG-BLEU-4}
& 
\UseMacro{results-ln-t3--t3--t3-ln-bsexpl1+fsexpl1+attn+copy-test-AVG-frag-acc}
& 
\UseMacro{results-ln-t3--t3--t3-ln-bsexpl1+fsexpl1+attn+copy-test-AVG-full-acc-top-1}
& 
\UseMacro{results-ln-t3--t3--t3-ln-bsexpl1+fsexpl1+attn+copy-test-AVG-full-acc-top-5}
\\
\bottomrule
\end{tabular}
\end{center}
\end{scriptsize}
\vspace{\UseMacro{vspace-results-ln-t3--t3--t3-main}}
\end{table*}

We conclude that all our observations in Section~\ref{sec:eval:auto} on training and testing on our original corpus (here, \dsti) hold when training and testing on \dsta. Additionally, we make the following observations based on the results of models using different combinations of \train, \val, and \test sets:

\begin{itemize}[leftmargin=1em,itemsep=1pt,topsep=5pt] \item Training on \dsta helps \CoqConvTool obtain better   performance, although the corpus includes some noise from \dstii and   \dstiii projects.  This observation is based on comparing the   results of training on different sets and testing on the same   set.  For example, when testing on \dsta, the best \bleu score among   models trained on \dsta    (\UseMacro{results-ln-ta--ta--ta-ln-s+bsexpl1+attn+copy-test-AVG-BLEU-4}, cf. Table~\ref{tbl:results-ln-ta--ta--ta-main})   is higher than the best score for models trained on \dsti   (\UseMacro{results-ln-t1--t1--ta-ln-s+bsexpl1+attn+copy-test-AVG-BLEU-4},   cf. Table~\ref{tbl:results-ln-t1--t1--ta-main}). As another   example, when testing on \dstii, the best \bleu score among models trained   on \dsta is   \UseMacro{results-ln-ta--ta--t2-ln-fsexpl1+attn+copy-test-AVG-BLEU-4}   (cf. Table~\ref{tbl:results-ln-ta--ta--t2-main}), which is higher than   the best score among models trained on \dstii, namely, \UseMacro{results-ln-t2--t2--t2-ln-bsexpl1+attn+copy-test-AVG-BLEU-4}   (cf. Table~\ref{tbl:results-ln-t2--t2--t2-main}). \item \Dstii and \dstiii projects are indeed less conforming to   \MathComp naming conventions than \dsti projects, confirming the judgment of domain experts.   With the same models trained on \dsta, \CoqConvTool's   best \bleu score on the \dsti \test set   (\UseMacro{results-ln-ta--ta--t1-ln-s+bsexpl1+attn+copy-test-AVG-BLEU-4})   is greater than the best \bleu score on the \dstii \test set   (\UseMacro{results-ln-ta--ta--t2-ln-fsexpl1+attn+copy-test-AVG-BLEU-4}),   and the latter is greater than the best \bleu score on the \dstiii \test set   (\UseMacro{results-ln-ta--ta--t3-ln-s+attn+copy-test-AVG-BLEU-4}).   The same relationships hold for the models trained only on \dsti. \end{itemize}

\subsection{Generalizability Case Study} \label{sec:eval:generalizability} \PTEXTInfotheo consists of \UseMacro{corpus-infotheo-num-docs} \Coq \docs, and contains \UseMacro{lemma-filtered-lo-all-num-lemmas} lemmas.  We randomly split the \docs into \train, \val, and \test sets which contain 40\%, 10\%, 50\% of the \docs, respectively.  After splitting, there were \UseMacro{lemma-filtered-lo-train-num-lemmas} lemmas in the \train set, \UseMacro{lemma-filtered-lo-val-num-lemmas} lemmas in the \val set, and \UseMacro{lemma-filtered-lo-test-num-lemmas} lemmas in the \test set.

\begin{table*}[t]
\begin{scriptsize}
\begin{center}
\caption{\UseMacro{table-caption-results-ln-lo-ta-main}}
\begin{tabular}{r r r r r}
\toprule
\TableHeadLOTrainSet
& \UseMacro{table-head-results-ln-BLEU-4}
& \UseMacro{table-head-results-ln-frag-acc}
& \UseMacro{table-head-results-ln-full-acc-top-1}
& \UseMacro{table-head-results-ln-full-acc-top-5}
 \\
\midrule
\UseMacro{lo-train-ta}
& 
\UseMacro{results-ln-lo-ta-ln-s+bsexpl1+attn+copy_ta-test-AVG-BLEU-4}
& 
\UseMacro{results-ln-lo-ta-ln-s+bsexpl1+attn+copy_ta-test-AVG-frag-acc}
& 
\UseMacro{results-ln-lo-ta-ln-s+bsexpl1+attn+copy_ta-test-AVG-full-acc-top-1}
& 
\UseMacro{results-ln-lo-ta-ln-s+bsexpl1+attn+copy_ta-test-AVG-full-acc-top-5}
\\
\UseMacro{lo-train-lo-25}
& 
\UseMacro{results-ln-lo-lo-25-ln-s+bsexpl1+attn+copy_ta-test-AVG-BLEU-4}
& 
\UseMacro{results-ln-lo-lo-25-ln-s+bsexpl1+attn+copy_ta-test-AVG-frag-acc}
& 
\UseMacro{results-ln-lo-lo-25-ln-s+bsexpl1+attn+copy_ta-test-AVG-full-acc-top-1}
& 
\UseMacro{results-ln-lo-lo-25-ln-s+bsexpl1+attn+copy_ta-test-AVG-full-acc-top-5}
\\
\UseMacro{lo-train-lo-5}
& 
\UseMacro{results-ln-lo-lo-5-ln-s+bsexpl1+attn+copy_ta-test-AVG-BLEU-4}
& 
\UseMacro{results-ln-lo-lo-5-ln-s+bsexpl1+attn+copy_ta-test-AVG-frag-acc}
& 
\UseMacro{results-ln-lo-lo-5-ln-s+bsexpl1+attn+copy_ta-test-AVG-full-acc-top-1}
& 
\UseMacro{results-ln-lo-lo-5-ln-s+bsexpl1+attn+copy_ta-test-AVG-full-acc-top-5}
\\
\UseMacro{lo-train-lo-75}
& 
\UseMacro{results-ln-lo-lo-75-ln-s+bsexpl1+attn+copy_ta-test-AVG-BLEU-4}
& 
\UseMacro{results-ln-lo-lo-75-ln-s+bsexpl1+attn+copy_ta-test-AVG-frag-acc}
& 
\UseMacro{results-ln-lo-lo-75-ln-s+bsexpl1+attn+copy_ta-test-AVG-full-acc-top-1}
& 
\UseMacro{results-ln-lo-lo-75-ln-s+bsexpl1+attn+copy_ta-test-AVG-full-acc-top-5}
\\
\UseMacro{lo-train-lo}
& 
\textbf{\UseMacro{results-ln-lo-lo-ln-s+bsexpl1+attn+copy_ta-test-AVG-BLEU-4}}
& 
\textbf{\UseMacro{results-ln-lo-lo-ln-s+bsexpl1+attn+copy_ta-test-AVG-frag-acc}}
& 
\textbf{\UseMacro{results-ln-lo-lo-ln-s+bsexpl1+attn+copy_ta-test-AVG-full-acc-top-1}}
& 
\textbf{\UseMacro{results-ln-lo-lo-ln-s+bsexpl1+attn+copy_ta-test-AVG-full-acc-top-5}}
\\
\bottomrule
\end{tabular}
\end{center}
\end{scriptsize}
\vspace{\UseMacro{vspace-results-ln-lo-ta-main}}
\end{table*}

Table~\ref{tbl:results-ln-lo-ta-main} shows the results of applying \CoqConvTool with the best model on \PTEXTinfotheo without and with additional training.  The first column shows the number of lemmas from the \PTEXTinfotheo \train set used for additional training.  The rest of the columns show the four automatic metrics. We can observe that applying \CoqConvTool without additional training achieves moderate performance (\bleu = \UseMacro{results-ln-lo-ta-ln-s+bsexpl1+attn+copy_ta-test-AVG-BLEU-4}). With some additional training, performance can be markedly improved (up to a \bleu score of \UseMacro{results-ln-lo-lo-ln-s+bsexpl1+attn+copy_ta-test-AVG-BLEU-4} when training on all \UseMacro{lo-train-lo} lemmas).

\end{document}

%% file: tables/numbers-corpus.tex
\DefMacro{corpus-mathcomp-full-name}{math-comp\_math-comp}
\DefMacro{corpus-mathcomp-short-name}{math-comp}
\DefMacro{corpus-mathcomp-url}{https://github.com/math-comp/math-comp}
\DefMacro{corpus-mathcomp-url-pretty}{math-comp/math-comp}
\DefMacro{corpus-mathcomp-sha}{748d716efb2f2f75946c8386e441ce1789806a39}
\DefMacro{corpus-mathcomp-sha-pretty}{748d716}
\DefMacro{corpus-mathcomp-tag}{748d716efb2f2f75946c8386e441ce1789806a39}
\DefMacro{corpus-mathcomp-tier}{1}
\DefMacro{corpus-fcslpcm-full-name}{imdea-software\_fcsl-pcm}
\DefMacro{corpus-fcslpcm-short-name}{fcsl-pcm}
\DefMacro{corpus-fcslpcm-url}{https://github.com/imdea-software/fcsl-pcm}
\DefMacro{corpus-fcslpcm-url-pretty}{imdea-software/fcsl-pcm}
\DefMacro{corpus-fcslpcm-sha}{eef45034808487fa3fa7dc360514c6734efc8d07}
\DefMacro{corpus-fcslpcm-sha-pretty}{eef4503}
\DefMacro{corpus-fcslpcm-tag}{eef45034808487fa3fa7dc360514c6734efc8d07}
\DefMacro{corpus-fcslpcm-tier}{3}
\DefMacro{corpus-disel-full-name}{DistributedComponents\_disel}
\DefMacro{corpus-disel-short-name}{disel}
\DefMacro{corpus-disel-url}{https://github.com/DistributedComponents/disel}
\DefMacro{corpus-disel-url-pretty}{DistributedComponents/disel}
\DefMacro{corpus-disel-sha}{e8aa80c486ec618888b6d0c801da7f61b6046daa}
\DefMacro{corpus-disel-sha-pretty}{e8aa80c}
\DefMacro{corpus-disel-tag}{e8aa80c486ec618888b6d0c801da7f61b6046daa}
\DefMacro{corpus-disel-tier}{3}
\DefMacro{corpus-reglang-full-name}{palmskog\_coq-reglang}
\DefMacro{corpus-reglang-short-name}{reglang}
\DefMacro{corpus-reglang-url}{https://github.com/palmskog/coq-reglang}
\DefMacro{corpus-reglang-url-pretty}{palmskog/coq-reglang}
\DefMacro{corpus-reglang-sha}{da333e01771f41debc3086a06e45d1393533d5c4}
\DefMacro{corpus-reglang-sha-pretty}{da333e0}
\DefMacro{corpus-reglang-tag}{da333e01771f41debc3086a06e45d1393533d5c4}
\DefMacro{corpus-reglang-tier}{3}
\DefMacro{corpus-compdecpdl-full-name}{palmskog\_comp-dec-pdl}
\DefMacro{corpus-compdecpdl-short-name}{comp-dec-pdl}
\DefMacro{corpus-compdecpdl-url}{https://github.com/palmskog/comp-dec-pdl}
\DefMacro{corpus-compdecpdl-url-pretty}{palmskog/comp-dec-pdl}
\DefMacro{corpus-compdecpdl-sha}{c1f813b8c97805e3c676cb559d0eab8381988a8a}
\DefMacro{corpus-compdecpdl-sha-pretty}{c1f813b}
\DefMacro{corpus-compdecpdl-tag}{c1f813b8c97805e3c676cb559d0eab8381988a8a}
\DefMacro{corpus-compdecpdl-tier}{3}
\DefMacro{corpus-oddorder-full-name}{math-comp\_odd-order}
\DefMacro{corpus-oddorder-short-name}{odd-order}
\DefMacro{corpus-oddorder-url}{https://github.com/math-comp/odd-order}
\DefMacro{corpus-oddorder-url-pretty}{math-comp/odd-order}
\DefMacro{corpus-oddorder-sha}{ca602a4638a9fe8ac30780095543d861f60fbfa0}
\DefMacro{corpus-oddorder-sha-pretty}{ca602a4}
\DefMacro{corpus-oddorder-tag}{ca602a4638a9fe8ac30780095543d861f60fbfa0}
\DefMacro{corpus-oddorder-tier}{1}
\DefMacro{corpus-fourcolor-full-name}{math-comp\_fourcolor}
\DefMacro{corpus-fourcolor-short-name}{fourcolor}
\DefMacro{corpus-fourcolor-url}{https://github.com/math-comp/fourcolor}
\DefMacro{corpus-fourcolor-url-pretty}{math-comp/fourcolor}
\DefMacro{corpus-fourcolor-sha}{0851d498d9a06ab81cf978494b6d41fbeb5f6631}
\DefMacro{corpus-fourcolor-sha-pretty}{0851d49}
\DefMacro{corpus-fourcolor-tag}{0851d498d9a06ab81cf978494b6d41fbeb5f6631}
\DefMacro{corpus-fourcolor-tier}{1}
\DefMacro{corpus-finmap-full-name}{math-comp\_finmap}
\DefMacro{corpus-finmap-short-name}{finmap}
\DefMacro{corpus-finmap-url}{https://github.com/math-comp/finmap}
\DefMacro{corpus-finmap-url-pretty}{math-comp/finmap}
\DefMacro{corpus-finmap-sha}{27642a81f796def99f050f2b3af41c275eac889f}
\DefMacro{corpus-finmap-sha-pretty}{27642a8}
\DefMacro{corpus-finmap-tag}{27642a81f796def99f050f2b3af41c275eac889f}
\DefMacro{corpus-finmap-tier}{1}
\DefMacro{corpus-bigenough-full-name}{math-comp\_bigenough}
\DefMacro{corpus-bigenough-short-name}{bigenough}
\DefMacro{corpus-bigenough-url}{https://github.com/math-comp/bigenough}
\DefMacro{corpus-bigenough-url-pretty}{math-comp/bigenough}
\DefMacro{corpus-bigenough-sha}{5f79a32ab69e255b9313b1e0d91577066d02bee6}
\DefMacro{corpus-bigenough-sha-pretty}{5f79a32}
\DefMacro{corpus-bigenough-tag}{5f79a32ab69e255b9313b1e0d91577066d02bee6}
\DefMacro{corpus-bigenough-tier}{2}
\DefMacro{corpus-analysis-full-name}{math-comp\_analysis}
\DefMacro{corpus-analysis-short-name}{analysis}
\DefMacro{corpus-analysis-url}{https://github.com/math-comp/analysis}
\DefMacro{corpus-analysis-url-pretty}{math-comp/analysis}
\DefMacro{corpus-analysis-sha}{9e5fe1dfa8faca954ba1c74ed301276d0e9d8339}
\DefMacro{corpus-analysis-sha-pretty}{9e5fe1d}
\DefMacro{corpus-analysis-tag}{9e5fe1dfa8faca954ba1c74ed301276d0e9d8339}
\DefMacro{corpus-analysis-tier}{2}
\DefMacro{corpus-realclosed-full-name}{math-comp\_real-closed}
\DefMacro{corpus-realclosed-short-name}{real-closed}
\DefMacro{corpus-realclosed-url}{https://github.com/math-comp/real-closed}
\DefMacro{corpus-realclosed-url-pretty}{math-comp/real-closed}
\DefMacro{corpus-realclosed-sha}{495a1faf3a13f5a5587646b6ba46061cfdeca33b}
\DefMacro{corpus-realclosed-sha-pretty}{495a1fa}
\DefMacro{corpus-realclosed-tag}{495a1faf3a13f5a5587646b6ba46061cfdeca33b}
\DefMacro{corpus-realclosed-tier}{2}
\DefMacro{corpus-robot-full-name}{affeldt-aist\_coq-robot}
\DefMacro{corpus-robot-short-name}{robot}
\DefMacro{corpus-robot-url}{https://github.com/affeldt-aist/coq-robot}
\DefMacro{corpus-robot-url-pretty}{affeldt-aist/coq-robot}
\DefMacro{corpus-robot-sha}{b341ad1dbbd7d693a1672084c97876b145792683}
\DefMacro{corpus-robot-sha-pretty}{b341ad1}
\DefMacro{corpus-robot-tag}{b341ad1dbbd7d693a1672084c97876b145792683}
\DefMacro{corpus-robot-tier}{2}
\DefMacro{corpus-multinomials-full-name}{math-comp\_multinomials}
\DefMacro{corpus-multinomials-short-name}{multinomials}
\DefMacro{corpus-multinomials-url}{https://github.com/math-comp/multinomials}
\DefMacro{corpus-multinomials-url-pretty}{math-comp/multinomials}
\DefMacro{corpus-multinomials-sha}{691d7954d3e08db172784f3e4fffe2367cd08c95}
\DefMacro{corpus-multinomials-sha-pretty}{691d795}
\DefMacro{corpus-multinomials-tag}{691d7954d3e08db172784f3e4fffe2367cd08c95}
\DefMacro{corpus-multinomials-tier}{2}
\DefMacro{corpus-ellipticcurves-full-name}{strub\_elliptic-curves-ssr}
\DefMacro{corpus-ellipticcurves-short-name}{elliptic-curves}
\DefMacro{corpus-ellipticcurves-url}{https://github.com/strub/elliptic-curves-ssr}
\DefMacro{corpus-ellipticcurves-url-pretty}{strub/elliptic-curves-ssr}
\DefMacro{corpus-ellipticcurves-sha}{631af893e591466207929714c45b5f7476d579d0}
\DefMacro{corpus-ellipticcurves-sha-pretty}{631af89}
\DefMacro{corpus-ellipticcurves-tag}{631af893e591466207929714c45b5f7476d579d0}
\DefMacro{corpus-ellipticcurves-tier}{2}
\DefMacro{corpus-toychain-full-name}{certichain\_toychain}
\DefMacro{corpus-toychain-short-name}{toychain}
\DefMacro{corpus-toychain-url}{https://github.com/certichain/toychain}
\DefMacro{corpus-toychain-url-pretty}{certichain/toychain}
\DefMacro{corpus-toychain-sha}{97bd6972f8d59eb1a56a429d59bfe10a0de6a281}
\DefMacro{corpus-toychain-sha-pretty}{97bd697}
\DefMacro{corpus-toychain-tag}{97bd6972f8d59eb1a56a429d59bfe10a0de6a281}
\DefMacro{corpus-toychain-tier}{3}
\DefMacro{corpus-bits-full-name}{coq-community\_coq-bits}
\DefMacro{corpus-bits-short-name}{bits}
\DefMacro{corpus-bits-url}{https://github.com/coq-community/coq-bits}
\DefMacro{corpus-bits-url-pretty}{coq-community/coq-bits}
\DefMacro{corpus-bits-sha}{3cd298ac7718305da6d22eaa5872b7e63651dbd0}
\DefMacro{corpus-bits-sha-pretty}{3cd298a}
\DefMacro{corpus-bits-tag}{3cd298ac7718305da6d22eaa5872b7e63651dbd0}
\DefMacro{corpus-bits-tier}{3}
\DefMacro{corpus-twosquare-full-name}{thery\_twoSquare}
\DefMacro{corpus-twosquare-short-name}{two-square}
\DefMacro{corpus-twosquare-url}{https://github.com/thery/twoSquare}
\DefMacro{corpus-twosquare-url-pretty}{thery/twoSquare}
\DefMacro{corpus-twosquare-sha}{1c09aca1080f36ec24276a2025cf87a25efdd29e}
\DefMacro{corpus-twosquare-sha-pretty}{1c09aca}
\DefMacro{corpus-twosquare-tag}{1c09aca1080f36ec24276a2025cf87a25efdd29e}
\DefMacro{corpus-twosquare-tier}{2}
\DefMacro{corpus-grobner-full-name}{thery\_grobner}
\DefMacro{corpus-grobner-short-name}{grobner}
\DefMacro{corpus-grobner-url}{https://github.com/thery/grobner}
\DefMacro{corpus-grobner-url-pretty}{thery/grobner}
\DefMacro{corpus-grobner-sha}{dfa54f9c2cd2a3913cfa4705efd9307b340197a4}
\DefMacro{corpus-grobner-sha-pretty}{dfa54f9}
\DefMacro{corpus-grobner-tag}{dfa54f9c2cd2a3913cfa4705efd9307b340197a4}
\DefMacro{corpus-grobner-tier}{2}
\DefMacro{corpus-infotheo-full-name}{palmskog\_infotheo}
\DefMacro{corpus-infotheo-short-name}{infotheo}
\DefMacro{corpus-infotheo-url}{https://github.com/palmskog/infotheo}
\DefMacro{corpus-infotheo-url-pretty}{palmskog/infotheo}
\DefMacro{corpus-infotheo-sha}{6c1724247583d084b05b092e79921465a9510e2e}
\DefMacro{corpus-infotheo-sha-pretty}{6c17242}
\DefMacro{corpus-infotheo-tag}{6c1724247583d084b05b092e79921465a9510e2e}
\DefMacro{corpus-infotheo-tier}{2}
\DefMacro{corpus-monae-full-name}{palmskog\_monae}
\DefMacro{corpus-monae-short-name}{monae}
\DefMacro{corpus-monae-url}{https://github.com/palmskog/monae}
\DefMacro{corpus-monae-url-pretty}{palmskog/monae}
\DefMacro{corpus-monae-sha}{9d0e461695e27059022271b304b75ac1bab70d3f}
\DefMacro{corpus-monae-sha-pretty}{9d0e461}
\DefMacro{corpus-monae-tag}{9d0e461695e27059022271b304b75ac1bab70d3f}
\DefMacro{corpus-monae-tier}{3}
\DefMacro{corpus-games-full-name}{gstew5\_games}
\DefMacro{corpus-games-short-name}{games}
\DefMacro{corpus-games-url}{https://github.com/gstew5/games}
\DefMacro{corpus-games-url-pretty}{gstew5/games}
\DefMacro{corpus-games-sha}{3d3bd31c7e4d1ea6f4b5a6815bca5c0a039b204f}
\DefMacro{corpus-games-sha-pretty}{3d3bd31}
\DefMacro{corpus-games-tag}{3d3bd31c7e4d1ea6f4b5a6815bca5c0a039b204f}
\DefMacro{corpus-games-tier}{3}
\DefMacro{corpus-mathcomp-num-docs}{89}
\DefMacro{corpus-mathcomp-spec-loc}{38,243}
\DefMacro{corpus-mathcomp-k-spec-loc}{38}
\DefMacro{corpus-mathcomp-proof-loc}{46,470}
\DefMacro{corpus-mathcomp-k-proof-loc}{46}
\DefMacro{corpus-mathcomp-comment-loc}{6,131}
\DefMacro{corpus-mathcomp-k-comment-loc}{6}
\DefMacro{corpus-mathcomp-total-loc}{90,844}
\DefMacro{corpus-mathcomp-k-total-loc}{90}
\DefMacro{corpus-mathcomp-code-loc}{84,713}
\DefMacro{corpus-mathcomp-k-code-loc}{84}
\DefMacro{corpus-mathcomp-AVG-spec-loc-per-doc}{429.70}
\DefMacro{corpus-mathcomp-SUM-spec-loc-per-doc}{38,243}
\DefMacro{corpus-mathcomp-MAX-spec-loc-per-doc}{4,331}
\DefMacro{corpus-mathcomp-MIN-spec-loc-per-doc}{1}
\DefMacro{corpus-mathcomp-MEDIAN-spec-loc-per-doc}{268.00}
\DefMacro{corpus-mathcomp-STDEV-spec-loc-per-doc}{567.03}
\DefMacro{corpus-mathcomp-AVG-proof-loc-per-doc}{522.13}
\DefMacro{corpus-mathcomp-SUM-proof-loc-per-doc}{46,470}
\DefMacro{corpus-mathcomp-MAX-proof-loc-per-doc}{3,084}
\DefMacro{corpus-mathcomp-MIN-proof-loc-per-doc}{0}
\DefMacro{corpus-mathcomp-MEDIAN-proof-loc-per-doc}{449.00}
\DefMacro{corpus-mathcomp-STDEV-proof-loc-per-doc}{514.98}
\DefMacro{corpus-mathcomp-AVG-comment-loc-per-doc}{68.89}
\DefMacro{corpus-mathcomp-SUM-comment-loc-per-doc}{6,131}
\DefMacro{corpus-mathcomp-MAX-comment-loc-per-doc}{588}
\DefMacro{corpus-mathcomp-MIN-comment-loc-per-doc}{0}
\DefMacro{corpus-mathcomp-MEDIAN-comment-loc-per-doc}{42.00}
\DefMacro{corpus-mathcomp-STDEV-comment-loc-per-doc}{84.63}
\DefMacro{corpus-mathcomp-AVG-total-loc-per-doc}{1,020.72}
\DefMacro{corpus-mathcomp-SUM-total-loc-per-doc}{90,844}
\DefMacro{corpus-mathcomp-MAX-total-loc-per-doc}{5,619}
\DefMacro{corpus-mathcomp-MIN-total-loc-per-doc}{1}
\DefMacro{corpus-mathcomp-MEDIAN-total-loc-per-doc}{729.00}
\DefMacro{corpus-mathcomp-STDEV-total-loc-per-doc}{1,008.01}
\DefMacro{corpus-mathcomp-AVG-code-loc-per-doc}{951.83}
\DefMacro{corpus-mathcomp-SUM-code-loc-per-doc}{84,713}
\DefMacro{corpus-mathcomp-MAX-code-loc-per-doc}{5,031}
\DefMacro{corpus-mathcomp-MIN-code-loc-per-doc}{1}
\DefMacro{corpus-mathcomp-MEDIAN-code-loc-per-doc}{675.00}
\DefMacro{corpus-mathcomp-STDEV-code-loc-per-doc}{936.58}
\DefMacro{corpus-fcslpcm-num-docs}{12}
\DefMacro{corpus-fcslpcm-spec-loc}{2,937}
\DefMacro{corpus-fcslpcm-k-spec-loc}{2}
\DefMacro{corpus-fcslpcm-proof-loc}{2,852}
\DefMacro{corpus-fcslpcm-k-proof-loc}{2}
\DefMacro{corpus-fcslpcm-comment-loc}{466}
\DefMacro{corpus-fcslpcm-k-comment-loc}{0}
\DefMacro{corpus-fcslpcm-total-loc}{6,255}
\DefMacro{corpus-fcslpcm-k-total-loc}{6}
\DefMacro{corpus-fcslpcm-code-loc}{5,789}
\DefMacro{corpus-fcslpcm-k-code-loc}{5}
\DefMacro{corpus-fcslpcm-AVG-spec-loc-per-doc}{244.75}
\DefMacro{corpus-fcslpcm-SUM-spec-loc-per-doc}{2,937}
\DefMacro{corpus-fcslpcm-MAX-spec-loc-per-doc}{722}
\DefMacro{corpus-fcslpcm-MIN-spec-loc-per-doc}{38}
\DefMacro{corpus-fcslpcm-MEDIAN-spec-loc-per-doc}{224.00}
\DefMacro{corpus-fcslpcm-STDEV-spec-loc-per-doc}{172.58}
\DefMacro{corpus-fcslpcm-AVG-proof-loc-per-doc}{237.67}
\DefMacro{corpus-fcslpcm-SUM-proof-loc-per-doc}{2,852}
\DefMacro{corpus-fcslpcm-MAX-proof-loc-per-doc}{981}
\DefMacro{corpus-fcslpcm-MIN-proof-loc-per-doc}{25}
\DefMacro{corpus-fcslpcm-MEDIAN-proof-loc-per-doc}{119.50}
\DefMacro{corpus-fcslpcm-STDEV-proof-loc-per-doc}{287.34}
\DefMacro{corpus-fcslpcm-AVG-comment-loc-per-doc}{38.83}
\DefMacro{corpus-fcslpcm-SUM-comment-loc-per-doc}{466}
\DefMacro{corpus-fcslpcm-MAX-comment-loc-per-doc}{126}
\DefMacro{corpus-fcslpcm-MIN-comment-loc-per-doc}{7}
\DefMacro{corpus-fcslpcm-MEDIAN-comment-loc-per-doc}{19.50}
\DefMacro{corpus-fcslpcm-STDEV-comment-loc-per-doc}{39.02}
\DefMacro{corpus-fcslpcm-AVG-total-loc-per-doc}{521.25}
\DefMacro{corpus-fcslpcm-SUM-total-loc-per-doc}{6,255}
\DefMacro{corpus-fcslpcm-MAX-total-loc-per-doc}{1,821}
\DefMacro{corpus-fcslpcm-MIN-total-loc-per-doc}{77}
\DefMacro{corpus-fcslpcm-MEDIAN-total-loc-per-doc}{427.00}
\DefMacro{corpus-fcslpcm-STDEV-total-loc-per-doc}{470.64}
\DefMacro{corpus-fcslpcm-AVG-code-loc-per-doc}{482.42}
\DefMacro{corpus-fcslpcm-SUM-code-loc-per-doc}{5,789}
\DefMacro{corpus-fcslpcm-MAX-code-loc-per-doc}{1,703}
\DefMacro{corpus-fcslpcm-MIN-code-loc-per-doc}{63}
\DefMacro{corpus-fcslpcm-MEDIAN-code-loc-per-doc}{383.00}
\DefMacro{corpus-fcslpcm-STDEV-code-loc-per-doc}{447.18}
\DefMacro{corpus-disel-num-docs}{20}
\DefMacro{corpus-disel-spec-loc}{2,575}
\DefMacro{corpus-disel-k-spec-loc}{2}
\DefMacro{corpus-disel-proof-loc}{1,898}
\DefMacro{corpus-disel-k-proof-loc}{1}
\DefMacro{corpus-disel-comment-loc}{505}
\DefMacro{corpus-disel-k-comment-loc}{0}
\DefMacro{corpus-disel-total-loc}{4,978}
\DefMacro{corpus-disel-k-total-loc}{4}
\DefMacro{corpus-disel-code-loc}{4,473}
\DefMacro{corpus-disel-k-code-loc}{4}
\DefMacro{corpus-disel-AVG-spec-loc-per-doc}{128.75}
\DefMacro{corpus-disel-SUM-spec-loc-per-doc}{2,575}
\DefMacro{corpus-disel-MAX-spec-loc-per-doc}{610}
\DefMacro{corpus-disel-MIN-spec-loc-per-doc}{17}
\DefMacro{corpus-disel-MEDIAN-spec-loc-per-doc}{89.00}
\DefMacro{corpus-disel-STDEV-spec-loc-per-doc}{125.21}
\DefMacro{corpus-disel-AVG-proof-loc-per-doc}{94.90}
\DefMacro{corpus-disel-SUM-proof-loc-per-doc}{1,898}
\DefMacro{corpus-disel-MAX-proof-loc-per-doc}{315}
\DefMacro{corpus-disel-MIN-proof-loc-per-doc}{0}
\DefMacro{corpus-disel-MEDIAN-proof-loc-per-doc}{60.00}
\DefMacro{corpus-disel-STDEV-proof-loc-per-doc}{89.67}
\DefMacro{corpus-disel-AVG-comment-loc-per-doc}{25.25}
\DefMacro{corpus-disel-SUM-comment-loc-per-doc}{505}
\DefMacro{corpus-disel-MAX-comment-loc-per-doc}{84}
\DefMacro{corpus-disel-MIN-comment-loc-per-doc}{0}
\DefMacro{corpus-disel-MEDIAN-comment-loc-per-doc}{23.00}
\DefMacro{corpus-disel-STDEV-comment-loc-per-doc}{22.87}
\DefMacro{corpus-disel-AVG-total-loc-per-doc}{248.90}
\DefMacro{corpus-disel-SUM-total-loc-per-doc}{4,978}
\DefMacro{corpus-disel-MAX-total-loc-per-doc}{939}
\DefMacro{corpus-disel-MIN-total-loc-per-doc}{17}
\DefMacro{corpus-disel-MEDIAN-total-loc-per-doc}{204.50}
\DefMacro{corpus-disel-STDEV-total-loc-per-doc}{205.41}
\DefMacro{corpus-disel-AVG-code-loc-per-doc}{223.65}
\DefMacro{corpus-disel-SUM-code-loc-per-doc}{4,473}
\DefMacro{corpus-disel-MAX-code-loc-per-doc}{855}
\DefMacro{corpus-disel-MIN-code-loc-per-doc}{17}
\DefMacro{corpus-disel-MEDIAN-code-loc-per-doc}{170.50}
\DefMacro{corpus-disel-STDEV-code-loc-per-doc}{190.65}
\DefMacro{corpus-reglang-num-docs}{12}
\DefMacro{corpus-reglang-spec-loc}{1,299}
\DefMacro{corpus-reglang-k-spec-loc}{1}
\DefMacro{corpus-reglang-proof-loc}{1,734}
\DefMacro{corpus-reglang-k-proof-loc}{1}
\DefMacro{corpus-reglang-comment-loc}{250}
\DefMacro{corpus-reglang-k-comment-loc}{0}
\DefMacro{corpus-reglang-total-loc}{3,283}
\DefMacro{corpus-reglang-k-total-loc}{3}
\DefMacro{corpus-reglang-code-loc}{3,033}
\DefMacro{corpus-reglang-k-code-loc}{3}
\DefMacro{corpus-reglang-AVG-spec-loc-per-doc}{108.25}
\DefMacro{corpus-reglang-SUM-spec-loc-per-doc}{1,299}
\DefMacro{corpus-reglang-MAX-spec-loc-per-doc}{273}
\DefMacro{corpus-reglang-MIN-spec-loc-per-doc}{20}
\DefMacro{corpus-reglang-MEDIAN-spec-loc-per-doc}{90.00}
\DefMacro{corpus-reglang-STDEV-spec-loc-per-doc}{67.77}
\DefMacro{corpus-reglang-AVG-proof-loc-per-doc}{144.50}
\DefMacro{corpus-reglang-SUM-proof-loc-per-doc}{1,734}
\DefMacro{corpus-reglang-MAX-proof-loc-per-doc}{506}
\DefMacro{corpus-reglang-MIN-proof-loc-per-doc}{13}
\DefMacro{corpus-reglang-MEDIAN-proof-loc-per-doc}{101.00}
\DefMacro{corpus-reglang-STDEV-proof-loc-per-doc}{126.67}
\DefMacro{corpus-reglang-AVG-comment-loc-per-doc}{20.83}
\DefMacro{corpus-reglang-SUM-comment-loc-per-doc}{250}
\DefMacro{corpus-reglang-MAX-comment-loc-per-doc}{38}
\DefMacro{corpus-reglang-MIN-comment-loc-per-doc}{2}
\DefMacro{corpus-reglang-MEDIAN-comment-loc-per-doc}{19.50}
\DefMacro{corpus-reglang-STDEV-comment-loc-per-doc}{10.50}
\DefMacro{corpus-reglang-AVG-total-loc-per-doc}{273.58}
\DefMacro{corpus-reglang-SUM-total-loc-per-doc}{3,283}
\DefMacro{corpus-reglang-MAX-total-loc-per-doc}{816}
\DefMacro{corpus-reglang-MIN-total-loc-per-doc}{39}
\DefMacro{corpus-reglang-MEDIAN-total-loc-per-doc}{211.50}
\DefMacro{corpus-reglang-STDEV-total-loc-per-doc}{199.27}
\DefMacro{corpus-reglang-AVG-code-loc-per-doc}{252.75}
\DefMacro{corpus-reglang-SUM-code-loc-per-doc}{3,033}
\DefMacro{corpus-reglang-MAX-code-loc-per-doc}{779}
\DefMacro{corpus-reglang-MIN-code-loc-per-doc}{33}
\DefMacro{corpus-reglang-MEDIAN-code-loc-per-doc}{192.50}
\DefMacro{corpus-reglang-STDEV-code-loc-per-doc}{191.22}
\DefMacro{corpus-compdecpdl-num-docs}{16}
\DefMacro{corpus-compdecpdl-spec-loc}{2,305}
\DefMacro{corpus-compdecpdl-k-spec-loc}{2}
\DefMacro{corpus-compdecpdl-proof-loc}{2,114}
\DefMacro{corpus-compdecpdl-k-proof-loc}{2}
\DefMacro{corpus-compdecpdl-comment-loc}{394}
\DefMacro{corpus-compdecpdl-k-comment-loc}{0}
\DefMacro{corpus-compdecpdl-total-loc}{4,813}
\DefMacro{corpus-compdecpdl-k-total-loc}{4}
\DefMacro{corpus-compdecpdl-code-loc}{4,419}
\DefMacro{corpus-compdecpdl-k-code-loc}{4}
\DefMacro{corpus-compdecpdl-AVG-spec-loc-per-doc}{144.06}
\DefMacro{corpus-compdecpdl-SUM-spec-loc-per-doc}{2,305}
\DefMacro{corpus-compdecpdl-MAX-spec-loc-per-doc}{464}
\DefMacro{corpus-compdecpdl-MIN-spec-loc-per-doc}{18}
\DefMacro{corpus-compdecpdl-MEDIAN-spec-loc-per-doc}{78.50}
\DefMacro{corpus-compdecpdl-STDEV-spec-loc-per-doc}{145.99}
\DefMacro{corpus-compdecpdl-AVG-proof-loc-per-doc}{132.12}
\DefMacro{corpus-compdecpdl-SUM-proof-loc-per-doc}{2,114}
\DefMacro{corpus-compdecpdl-MAX-proof-loc-per-doc}{415}
\DefMacro{corpus-compdecpdl-MIN-proof-loc-per-doc}{0}
\DefMacro{corpus-compdecpdl-MEDIAN-proof-loc-per-doc}{82.50}
\DefMacro{corpus-compdecpdl-STDEV-proof-loc-per-doc}{128.70}
\DefMacro{corpus-compdecpdl-AVG-comment-loc-per-doc}{24.62}
\DefMacro{corpus-compdecpdl-SUM-comment-loc-per-doc}{394}
\DefMacro{corpus-compdecpdl-MAX-comment-loc-per-doc}{89}
\DefMacro{corpus-compdecpdl-MIN-comment-loc-per-doc}{1}
\DefMacro{corpus-compdecpdl-MEDIAN-comment-loc-per-doc}{12.00}
\DefMacro{corpus-compdecpdl-STDEV-comment-loc-per-doc}{25.94}
\DefMacro{corpus-compdecpdl-AVG-total-loc-per-doc}{300.81}
\DefMacro{corpus-compdecpdl-SUM-total-loc-per-doc}{4,813}
\DefMacro{corpus-compdecpdl-MAX-total-loc-per-doc}{891}
\DefMacro{corpus-compdecpdl-MIN-total-loc-per-doc}{22}
\DefMacro{corpus-compdecpdl-MEDIAN-total-loc-per-doc}{211.50}
\DefMacro{corpus-compdecpdl-STDEV-total-loc-per-doc}{283.14}
\DefMacro{corpus-compdecpdl-AVG-code-loc-per-doc}{276.19}
\DefMacro{corpus-compdecpdl-SUM-code-loc-per-doc}{4,419}
\DefMacro{corpus-compdecpdl-MAX-code-loc-per-doc}{802}
\DefMacro{corpus-compdecpdl-MIN-code-loc-per-doc}{19}
\DefMacro{corpus-compdecpdl-MEDIAN-code-loc-per-doc}{199.50}
\DefMacro{corpus-compdecpdl-STDEV-code-loc-per-doc}{259.82}
\DefMacro{corpus-oddorder-num-docs}{34}
\DefMacro{corpus-oddorder-spec-loc}{11,882}
\DefMacro{corpus-oddorder-k-spec-loc}{11}
\DefMacro{corpus-oddorder-proof-loc}{24,243}
\DefMacro{corpus-oddorder-k-proof-loc}{24}
\DefMacro{corpus-oddorder-comment-loc}{2,313}
\DefMacro{corpus-oddorder-k-comment-loc}{2}
\DefMacro{corpus-oddorder-total-loc}{38,438}
\DefMacro{corpus-oddorder-k-total-loc}{38}
\DefMacro{corpus-oddorder-code-loc}{36,125}
\DefMacro{corpus-oddorder-k-code-loc}{36}
\DefMacro{corpus-oddorder-AVG-spec-loc-per-doc}{349.47}
\DefMacro{corpus-oddorder-SUM-spec-loc-per-doc}{11,882}
\DefMacro{corpus-oddorder-MAX-spec-loc-per-doc}{889}
\DefMacro{corpus-oddorder-MIN-spec-loc-per-doc}{30}
\DefMacro{corpus-oddorder-MEDIAN-spec-loc-per-doc}{285.50}
\DefMacro{corpus-oddorder-STDEV-spec-loc-per-doc}{247.22}
\DefMacro{corpus-oddorder-AVG-proof-loc-per-doc}{713.03}
\DefMacro{corpus-oddorder-SUM-proof-loc-per-doc}{24,243}
\DefMacro{corpus-oddorder-MAX-proof-loc-per-doc}{1,829}
\DefMacro{corpus-oddorder-MIN-proof-loc-per-doc}{104}
\DefMacro{corpus-oddorder-MEDIAN-proof-loc-per-doc}{600.50}
\DefMacro{corpus-oddorder-STDEV-proof-loc-per-doc}{433.11}
\DefMacro{corpus-oddorder-AVG-comment-loc-per-doc}{68.03}
\DefMacro{corpus-oddorder-SUM-comment-loc-per-doc}{2,313}
\DefMacro{corpus-oddorder-MAX-comment-loc-per-doc}{191}
\DefMacro{corpus-oddorder-MIN-comment-loc-per-doc}{2}
\DefMacro{corpus-oddorder-MEDIAN-comment-loc-per-doc}{55.50}
\DefMacro{corpus-oddorder-STDEV-comment-loc-per-doc}{46.93}
\DefMacro{corpus-oddorder-AVG-total-loc-per-doc}{1,130.53}
\DefMacro{corpus-oddorder-SUM-total-loc-per-doc}{38,438}
\DefMacro{corpus-oddorder-MAX-total-loc-per-doc}{2,662}
\DefMacro{corpus-oddorder-MIN-total-loc-per-doc}{176}
\DefMacro{corpus-oddorder-MEDIAN-total-loc-per-doc}{1,124.00}
\DefMacro{corpus-oddorder-STDEV-total-loc-per-doc}{600.21}
\DefMacro{corpus-oddorder-AVG-code-loc-per-doc}{1,062.50}
\DefMacro{corpus-oddorder-SUM-code-loc-per-doc}{36,125}
\DefMacro{corpus-oddorder-MAX-code-loc-per-doc}{2,526}
\DefMacro{corpus-oddorder-MIN-code-loc-per-doc}{164}
\DefMacro{corpus-oddorder-MEDIAN-code-loc-per-doc}{1,060.50}
\DefMacro{corpus-oddorder-STDEV-code-loc-per-doc}{569.58}
\DefMacro{corpus-fourcolor-num-docs}{60}
\DefMacro{corpus-fourcolor-spec-loc}{9,175}
\DefMacro{corpus-fourcolor-k-spec-loc}{9}
\DefMacro{corpus-fourcolor-proof-loc}{27,963}
\DefMacro{corpus-fourcolor-k-proof-loc}{27}
\DefMacro{corpus-fourcolor-comment-loc}{2,816}
\DefMacro{corpus-fourcolor-k-comment-loc}{2}
\DefMacro{corpus-fourcolor-total-loc}{39,954}
\DefMacro{corpus-fourcolor-k-total-loc}{39}
\DefMacro{corpus-fourcolor-code-loc}{37,138}
\DefMacro{corpus-fourcolor-k-code-loc}{37}
\DefMacro{corpus-fourcolor-AVG-spec-loc-per-doc}{152.92}
\DefMacro{corpus-fourcolor-SUM-spec-loc-per-doc}{9,175}
\DefMacro{corpus-fourcolor-MAX-spec-loc-per-doc}{1,486}
\DefMacro{corpus-fourcolor-MIN-spec-loc-per-doc}{6}
\DefMacro{corpus-fourcolor-MEDIAN-spec-loc-per-doc}{115.00}
\DefMacro{corpus-fourcolor-STDEV-spec-loc-per-doc}{214.81}
\DefMacro{corpus-fourcolor-AVG-proof-loc-per-doc}{466.05}
\DefMacro{corpus-fourcolor-SUM-proof-loc-per-doc}{27,963}
\DefMacro{corpus-fourcolor-MAX-proof-loc-per-doc}{5,551}
\DefMacro{corpus-fourcolor-MIN-proof-loc-per-doc}{0}
\DefMacro{corpus-fourcolor-MEDIAN-proof-loc-per-doc}{232.00}
\DefMacro{corpus-fourcolor-STDEV-proof-loc-per-doc}{1,006.70}
\DefMacro{corpus-fourcolor-AVG-comment-loc-per-doc}{46.93}
\DefMacro{corpus-fourcolor-SUM-comment-loc-per-doc}{2,816}
\DefMacro{corpus-fourcolor-MAX-comment-loc-per-doc}{178}
\DefMacro{corpus-fourcolor-MIN-comment-loc-per-doc}{2}
\DefMacro{corpus-fourcolor-MEDIAN-comment-loc-per-doc}{41.50}
\DefMacro{corpus-fourcolor-STDEV-comment-loc-per-doc}{39.47}
\DefMacro{corpus-fourcolor-AVG-total-loc-per-doc}{665.90}
\DefMacro{corpus-fourcolor-SUM-total-loc-per-doc}{39,954}
\DefMacro{corpus-fourcolor-MAX-total-loc-per-doc}{5,563}
\DefMacro{corpus-fourcolor-MIN-total-loc-per-doc}{17}
\DefMacro{corpus-fourcolor-MEDIAN-total-loc-per-doc}{398.50}
\DefMacro{corpus-fourcolor-STDEV-total-loc-per-doc}{994.17}
\DefMacro{corpus-fourcolor-AVG-code-loc-per-doc}{618.97}
\DefMacro{corpus-fourcolor-SUM-code-loc-per-doc}{37,138}
\DefMacro{corpus-fourcolor-MAX-code-loc-per-doc}{5,560}
\DefMacro{corpus-fourcolor-MIN-code-loc-per-doc}{15}
\DefMacro{corpus-fourcolor-MEDIAN-code-loc-per-doc}{343.00}
\DefMacro{corpus-fourcolor-STDEV-code-loc-per-doc}{999.58}
\DefMacro{corpus-finmap-num-docs}{4}
\DefMacro{corpus-finmap-spec-loc}{4,260}
\DefMacro{corpus-finmap-k-spec-loc}{4}
\DefMacro{corpus-finmap-proof-loc}{2,191}
\DefMacro{corpus-finmap-k-proof-loc}{2}
\DefMacro{corpus-finmap-comment-loc}{238}
\DefMacro{corpus-finmap-k-comment-loc}{0}
\DefMacro{corpus-finmap-total-loc}{6,689}
\DefMacro{corpus-finmap-k-total-loc}{6}
\DefMacro{corpus-finmap-code-loc}{6,451}
\DefMacro{corpus-finmap-k-code-loc}{6}
\DefMacro{corpus-finmap-AVG-spec-loc-per-doc}{1,065.00}
\DefMacro{corpus-finmap-SUM-spec-loc-per-doc}{4,260}
\DefMacro{corpus-finmap-MAX-spec-loc-per-doc}{1,831}
\DefMacro{corpus-finmap-MIN-spec-loc-per-doc}{338}
\DefMacro{corpus-finmap-MEDIAN-spec-loc-per-doc}{1,045.50}
\DefMacro{corpus-finmap-STDEV-spec-loc-per-doc}{651.57}
\DefMacro{corpus-finmap-AVG-proof-loc-per-doc}{547.75}
\DefMacro{corpus-finmap-SUM-proof-loc-per-doc}{2,191}
\DefMacro{corpus-finmap-MAX-proof-loc-per-doc}{1,184}
\DefMacro{corpus-finmap-MIN-proof-loc-per-doc}{264}
\DefMacro{corpus-finmap-MEDIAN-proof-loc-per-doc}{371.50}
\DefMacro{corpus-finmap-STDEV-proof-loc-per-doc}{370.29}
\DefMacro{corpus-finmap-AVG-comment-loc-per-doc}{59.50}
\DefMacro{corpus-finmap-SUM-comment-loc-per-doc}{238}
\DefMacro{corpus-finmap-MAX-comment-loc-per-doc}{136}
\DefMacro{corpus-finmap-MIN-comment-loc-per-doc}{18}
\DefMacro{corpus-finmap-MEDIAN-comment-loc-per-doc}{42.00}
\DefMacro{corpus-finmap-STDEV-comment-loc-per-doc}{45.90}
\DefMacro{corpus-finmap-AVG-total-loc-per-doc}{1,672.25}
\DefMacro{corpus-finmap-SUM-total-loc-per-doc}{6,689}
\DefMacro{corpus-finmap-MAX-total-loc-per-doc}{2,905}
\DefMacro{corpus-finmap-MIN-total-loc-per-doc}{718}
\DefMacro{corpus-finmap-MEDIAN-total-loc-per-doc}{1,533.00}
\DefMacro{corpus-finmap-STDEV-total-loc-per-doc}{945.93}
\DefMacro{corpus-finmap-AVG-code-loc-per-doc}{1,612.75}
\DefMacro{corpus-finmap-SUM-code-loc-per-doc}{6,451}
\DefMacro{corpus-finmap-MAX-code-loc-per-doc}{2,769}
\DefMacro{corpus-finmap-MIN-code-loc-per-doc}{687}
\DefMacro{corpus-finmap-MEDIAN-code-loc-per-doc}{1,497.50}
\DefMacro{corpus-finmap-STDEV-code-loc-per-doc}{905.40}
\DefMacro{corpus-bigenough-num-docs}{1}
\DefMacro{corpus-bigenough-spec-loc}{70}
\DefMacro{corpus-bigenough-k-spec-loc}{0}
\DefMacro{corpus-bigenough-proof-loc}{8}
\DefMacro{corpus-bigenough-k-proof-loc}{0}
\DefMacro{corpus-bigenough-comment-loc}{16}
\DefMacro{corpus-bigenough-k-comment-loc}{0}
\DefMacro{corpus-bigenough-total-loc}{94}
\DefMacro{corpus-bigenough-k-total-loc}{0}
\DefMacro{corpus-bigenough-code-loc}{78}
\DefMacro{corpus-bigenough-k-code-loc}{0}
\DefMacro{corpus-bigenough-AVG-spec-loc-per-doc}{70.00}
\DefMacro{corpus-bigenough-SUM-spec-loc-per-doc}{70}
\DefMacro{corpus-bigenough-MAX-spec-loc-per-doc}{70}
\DefMacro{corpus-bigenough-MIN-spec-loc-per-doc}{70}
\DefMacro{corpus-bigenough-MEDIAN-spec-loc-per-doc}{70.00}
\DefMacro{corpus-bigenough-STDEV-spec-loc-per-doc}{0.00}
\DefMacro{corpus-bigenough-AVG-proof-loc-per-doc}{8.00}
\DefMacro{corpus-bigenough-SUM-proof-loc-per-doc}{8}
\DefMacro{corpus-bigenough-MAX-proof-loc-per-doc}{8}
\DefMacro{corpus-bigenough-MIN-proof-loc-per-doc}{8}
\DefMacro{corpus-bigenough-MEDIAN-proof-loc-per-doc}{8.00}
\DefMacro{corpus-bigenough-STDEV-proof-loc-per-doc}{0.00}
\DefMacro{corpus-bigenough-AVG-comment-loc-per-doc}{16.00}
\DefMacro{corpus-bigenough-SUM-comment-loc-per-doc}{16}
\DefMacro{corpus-bigenough-MAX-comment-loc-per-doc}{16}
\DefMacro{corpus-bigenough-MIN-comment-loc-per-doc}{16}
\DefMacro{corpus-bigenough-MEDIAN-comment-loc-per-doc}{16.00}
\DefMacro{corpus-bigenough-STDEV-comment-loc-per-doc}{0.00}
\DefMacro{corpus-bigenough-AVG-total-loc-per-doc}{94.00}
\DefMacro{corpus-bigenough-SUM-total-loc-per-doc}{94}
\DefMacro{corpus-bigenough-MAX-total-loc-per-doc}{94}
\DefMacro{corpus-bigenough-MIN-total-loc-per-doc}{94}
\DefMacro{corpus-bigenough-MEDIAN-total-loc-per-doc}{94.00}
\DefMacro{corpus-bigenough-STDEV-total-loc-per-doc}{0.00}
\DefMacro{corpus-bigenough-AVG-code-loc-per-doc}{78.00}
\DefMacro{corpus-bigenough-SUM-code-loc-per-doc}{78}
\DefMacro{corpus-bigenough-MAX-code-loc-per-doc}{78}
\DefMacro{corpus-bigenough-MIN-code-loc-per-doc}{78}
\DefMacro{corpus-bigenough-MEDIAN-code-loc-per-doc}{78.00}
\DefMacro{corpus-bigenough-STDEV-code-loc-per-doc}{0.00}
\DefMacro{corpus-analysis-num-docs}{17}
\DefMacro{corpus-analysis-spec-loc}{5,553}
\DefMacro{corpus-analysis-k-spec-loc}{5}
\DefMacro{corpus-analysis-proof-loc}{6,186}
\DefMacro{corpus-analysis-k-proof-loc}{6}
\DefMacro{corpus-analysis-comment-loc}{1,352}
\DefMacro{corpus-analysis-k-comment-loc}{1}
\DefMacro{corpus-analysis-total-loc}{13,091}
\DefMacro{corpus-analysis-k-total-loc}{13}
\DefMacro{corpus-analysis-code-loc}{11,739}
\DefMacro{corpus-analysis-k-code-loc}{11}
\DefMacro{corpus-analysis-AVG-spec-loc-per-doc}{326.65}
\DefMacro{corpus-analysis-SUM-spec-loc-per-doc}{5,553}
\DefMacro{corpus-analysis-MAX-spec-loc-per-doc}{1,196}
\DefMacro{corpus-analysis-MIN-spec-loc-per-doc}{29}
\DefMacro{corpus-analysis-MEDIAN-spec-loc-per-doc}{256.00}
\DefMacro{corpus-analysis-STDEV-spec-loc-per-doc}{291.79}
\DefMacro{corpus-analysis-AVG-proof-loc-per-doc}{363.88}
\DefMacro{corpus-analysis-SUM-proof-loc-per-doc}{6,186}
\DefMacro{corpus-analysis-MAX-proof-loc-per-doc}{986}
\DefMacro{corpus-analysis-MIN-proof-loc-per-doc}{5}
\DefMacro{corpus-analysis-MEDIAN-proof-loc-per-doc}{282.00}
\DefMacro{corpus-analysis-STDEV-proof-loc-per-doc}{307.69}
\DefMacro{corpus-analysis-AVG-comment-loc-per-doc}{79.53}
\DefMacro{corpus-analysis-SUM-comment-loc-per-doc}{1,352}
\DefMacro{corpus-analysis-MAX-comment-loc-per-doc}{311}
\DefMacro{corpus-analysis-MIN-comment-loc-per-doc}{3}
\DefMacro{corpus-analysis-MEDIAN-comment-loc-per-doc}{36.00}
\DefMacro{corpus-analysis-STDEV-comment-loc-per-doc}{92.70}
\DefMacro{corpus-analysis-AVG-total-loc-per-doc}{770.06}
\DefMacro{corpus-analysis-SUM-total-loc-per-doc}{13,091}
\DefMacro{corpus-analysis-MAX-total-loc-per-doc}{2,493}
\DefMacro{corpus-analysis-MIN-total-loc-per-doc}{37}
\DefMacro{corpus-analysis-MEDIAN-total-loc-per-doc}{503.00}
\DefMacro{corpus-analysis-STDEV-total-loc-per-doc}{632.56}
\DefMacro{corpus-analysis-AVG-code-loc-per-doc}{690.53}
\DefMacro{corpus-analysis-SUM-code-loc-per-doc}{11,739}
\DefMacro{corpus-analysis-MAX-code-loc-per-doc}{2,182}
\DefMacro{corpus-analysis-MIN-code-loc-per-doc}{34}
\DefMacro{corpus-analysis-MEDIAN-code-loc-per-doc}{451.00}
\DefMacro{corpus-analysis-STDEV-code-loc-per-doc}{560.05}
\DefMacro{corpus-realclosed-num-docs}{10}
\DefMacro{corpus-realclosed-spec-loc}{4,348}
\DefMacro{corpus-realclosed-k-spec-loc}{4}
\DefMacro{corpus-realclosed-proof-loc}{4,577}
\DefMacro{corpus-realclosed-k-proof-loc}{4}
\DefMacro{corpus-realclosed-comment-loc}{356}
\DefMacro{corpus-realclosed-k-comment-loc}{0}
\DefMacro{corpus-realclosed-total-loc}{9,281}
\DefMacro{corpus-realclosed-k-total-loc}{9}
\DefMacro{corpus-realclosed-code-loc}{8,925}
\DefMacro{corpus-realclosed-k-code-loc}{8}
\DefMacro{corpus-realclosed-AVG-spec-loc-per-doc}{434.80}
\DefMacro{corpus-realclosed-SUM-spec-loc-per-doc}{4,348}
\DefMacro{corpus-realclosed-MAX-spec-loc-per-doc}{1,148}
\DefMacro{corpus-realclosed-MIN-spec-loc-per-doc}{8}
\DefMacro{corpus-realclosed-MEDIAN-spec-loc-per-doc}{382.00}
\DefMacro{corpus-realclosed-STDEV-spec-loc-per-doc}{355.87}
\DefMacro{corpus-realclosed-AVG-proof-loc-per-doc}{457.70}
\DefMacro{corpus-realclosed-SUM-proof-loc-per-doc}{4,577}
\DefMacro{corpus-realclosed-MAX-proof-loc-per-doc}{1,073}
\DefMacro{corpus-realclosed-MIN-proof-loc-per-doc}{0}
\DefMacro{corpus-realclosed-MEDIAN-proof-loc-per-doc}{420.50}
\DefMacro{corpus-realclosed-STDEV-proof-loc-per-doc}{421.37}
\DefMacro{corpus-realclosed-AVG-comment-loc-per-doc}{35.60}
\DefMacro{corpus-realclosed-SUM-comment-loc-per-doc}{356}
\DefMacro{corpus-realclosed-MAX-comment-loc-per-doc}{111}
\DefMacro{corpus-realclosed-MIN-comment-loc-per-doc}{0}
\DefMacro{corpus-realclosed-MEDIAN-comment-loc-per-doc}{34.50}
\DefMacro{corpus-realclosed-STDEV-comment-loc-per-doc}{35.96}
\DefMacro{corpus-realclosed-AVG-total-loc-per-doc}{928.10}
\DefMacro{corpus-realclosed-SUM-total-loc-per-doc}{9,281}
\DefMacro{corpus-realclosed-MAX-total-loc-per-doc}{1,607}
\DefMacro{corpus-realclosed-MIN-total-loc-per-doc}{8}
\DefMacro{corpus-realclosed-MEDIAN-total-loc-per-doc}{1,148.50}
\DefMacro{corpus-realclosed-STDEV-total-loc-per-doc}{530.94}
\DefMacro{corpus-realclosed-AVG-code-loc-per-doc}{892.50}
\DefMacro{corpus-realclosed-SUM-code-loc-per-doc}{8,925}
\DefMacro{corpus-realclosed-MAX-code-loc-per-doc}{1,496}
\DefMacro{corpus-realclosed-MIN-code-loc-per-doc}{8}
\DefMacro{corpus-realclosed-MEDIAN-code-loc-per-doc}{1,091.00}
\DefMacro{corpus-realclosed-STDEV-code-loc-per-doc}{509.72}
\DefMacro{corpus-robot-num-docs}{13}
\DefMacro{corpus-robot-spec-loc}{3,881}
\DefMacro{corpus-robot-k-spec-loc}{3}
\DefMacro{corpus-robot-proof-loc}{7,249}
\DefMacro{corpus-robot-k-proof-loc}{7}
\DefMacro{corpus-robot-comment-loc}{648}
\DefMacro{corpus-robot-k-comment-loc}{0}
\DefMacro{corpus-robot-total-loc}{11,778}
\DefMacro{corpus-robot-k-total-loc}{11}
\DefMacro{corpus-robot-code-loc}{11,130}
\DefMacro{corpus-robot-k-code-loc}{11}
\DefMacro{corpus-robot-AVG-spec-loc-per-doc}{298.54}
\DefMacro{corpus-robot-SUM-spec-loc-per-doc}{3,881}
\DefMacro{corpus-robot-MAX-spec-loc-per-doc}{544}
\DefMacro{corpus-robot-MIN-spec-loc-per-doc}{72}
\DefMacro{corpus-robot-MEDIAN-spec-loc-per-doc}{314.00}
\DefMacro{corpus-robot-STDEV-spec-loc-per-doc}{130.85}
\DefMacro{corpus-robot-AVG-proof-loc-per-doc}{557.62}
\DefMacro{corpus-robot-SUM-proof-loc-per-doc}{7,249}
\DefMacro{corpus-robot-MAX-proof-loc-per-doc}{1,466}
\DefMacro{corpus-robot-MIN-proof-loc-per-doc}{53}
\DefMacro{corpus-robot-MEDIAN-proof-loc-per-doc}{532.00}
\DefMacro{corpus-robot-STDEV-proof-loc-per-doc}{351.84}
\DefMacro{corpus-robot-AVG-comment-loc-per-doc}{49.85}
\DefMacro{corpus-robot-SUM-comment-loc-per-doc}{648}
\DefMacro{corpus-robot-MAX-comment-loc-per-doc}{110}
\DefMacro{corpus-robot-MIN-comment-loc-per-doc}{6}
\DefMacro{corpus-robot-MEDIAN-comment-loc-per-doc}{52.00}
\DefMacro{corpus-robot-STDEV-comment-loc-per-doc}{31.56}
\DefMacro{corpus-robot-AVG-total-loc-per-doc}{906.00}
\DefMacro{corpus-robot-SUM-total-loc-per-doc}{11,778}
\DefMacro{corpus-robot-MAX-total-loc-per-doc}{1,912}
\DefMacro{corpus-robot-MIN-total-loc-per-doc}{131}
\DefMacro{corpus-robot-MEDIAN-total-loc-per-doc}{856.00}
\DefMacro{corpus-robot-STDEV-total-loc-per-doc}{474.47}
\DefMacro{corpus-robot-AVG-code-loc-per-doc}{856.15}
\DefMacro{corpus-robot-SUM-code-loc-per-doc}{11,130}
\DefMacro{corpus-robot-MAX-code-loc-per-doc}{1,858}
\DefMacro{corpus-robot-MIN-code-loc-per-doc}{125}
\DefMacro{corpus-robot-MEDIAN-code-loc-per-doc}{819.00}
\DefMacro{corpus-robot-STDEV-code-loc-per-doc}{455.07}
\DefMacro{corpus-multinomials-num-docs}{5}
\DefMacro{corpus-multinomials-spec-loc}{3,699}
\DefMacro{corpus-multinomials-k-spec-loc}{3}
\DefMacro{corpus-multinomials-proof-loc}{3,664}
\DefMacro{corpus-multinomials-k-proof-loc}{3}
\DefMacro{corpus-multinomials-comment-loc}{308}
\DefMacro{corpus-multinomials-k-comment-loc}{0}
\DefMacro{corpus-multinomials-total-loc}{7,671}
\DefMacro{corpus-multinomials-k-total-loc}{7}
\DefMacro{corpus-multinomials-code-loc}{7,363}
\DefMacro{corpus-multinomials-k-code-loc}{7}
\DefMacro{corpus-multinomials-AVG-spec-loc-per-doc}{739.80}
\DefMacro{corpus-multinomials-SUM-spec-loc-per-doc}{3,699}
\DefMacro{corpus-multinomials-MAX-spec-loc-per-doc}{2,069}
\DefMacro{corpus-multinomials-MIN-spec-loc-per-doc}{23}
\DefMacro{corpus-multinomials-MEDIAN-spec-loc-per-doc}{502.00}
\DefMacro{corpus-multinomials-STDEV-spec-loc-per-doc}{737.37}
\DefMacro{corpus-multinomials-AVG-proof-loc-per-doc}{732.80}
\DefMacro{corpus-multinomials-SUM-proof-loc-per-doc}{3,664}
\DefMacro{corpus-multinomials-MAX-proof-loc-per-doc}{2,387}
\DefMacro{corpus-multinomials-MIN-proof-loc-per-doc}{5}
\DefMacro{corpus-multinomials-MEDIAN-proof-loc-per-doc}{521.00}
\DefMacro{corpus-multinomials-STDEV-proof-loc-per-doc}{855.37}
\DefMacro{corpus-multinomials-AVG-comment-loc-per-doc}{61.60}
\DefMacro{corpus-multinomials-SUM-comment-loc-per-doc}{308}
\DefMacro{corpus-multinomials-MAX-comment-loc-per-doc}{155}
\DefMacro{corpus-multinomials-MIN-comment-loc-per-doc}{2}
\DefMacro{corpus-multinomials-MEDIAN-comment-loc-per-doc}{51.00}
\DefMacro{corpus-multinomials-STDEV-comment-loc-per-doc}{53.55}
\DefMacro{corpus-multinomials-AVG-total-loc-per-doc}{1,534.20}
\DefMacro{corpus-multinomials-SUM-total-loc-per-doc}{7,671}
\DefMacro{corpus-multinomials-MAX-total-loc-per-doc}{4,611}
\DefMacro{corpus-multinomials-MIN-total-loc-per-doc}{30}
\DefMacro{corpus-multinomials-MEDIAN-total-loc-per-doc}{1,143.00}
\DefMacro{corpus-multinomials-STDEV-total-loc-per-doc}{1,631.48}
\DefMacro{corpus-multinomials-AVG-code-loc-per-doc}{1,472.60}
\DefMacro{corpus-multinomials-SUM-code-loc-per-doc}{7,363}
\DefMacro{corpus-multinomials-MAX-code-loc-per-doc}{4,456}
\DefMacro{corpus-multinomials-MIN-code-loc-per-doc}{28}
\DefMacro{corpus-multinomials-MEDIAN-code-loc-per-doc}{1,092.00}
\DefMacro{corpus-multinomials-STDEV-code-loc-per-doc}{1,578.81}
\DefMacro{corpus-ellipticcurves-num-docs}{18}
\DefMacro{corpus-ellipticcurves-spec-loc}{3,298}
\DefMacro{corpus-ellipticcurves-k-spec-loc}{3}
\DefMacro{corpus-ellipticcurves-proof-loc}{6,298}
\DefMacro{corpus-ellipticcurves-k-proof-loc}{6}
\DefMacro{corpus-ellipticcurves-comment-loc}{279}
\DefMacro{corpus-ellipticcurves-k-comment-loc}{0}
\DefMacro{corpus-ellipticcurves-total-loc}{9,875}
\DefMacro{corpus-ellipticcurves-k-total-loc}{9}
\DefMacro{corpus-ellipticcurves-code-loc}{9,596}
\DefMacro{corpus-ellipticcurves-k-code-loc}{9}
\DefMacro{corpus-ellipticcurves-AVG-spec-loc-per-doc}{183.22}
\DefMacro{corpus-ellipticcurves-SUM-spec-loc-per-doc}{3,298}
\DefMacro{corpus-ellipticcurves-MAX-spec-loc-per-doc}{503}
\DefMacro{corpus-ellipticcurves-MIN-spec-loc-per-doc}{38}
\DefMacro{corpus-ellipticcurves-MEDIAN-spec-loc-per-doc}{150.50}
\DefMacro{corpus-ellipticcurves-STDEV-spec-loc-per-doc}{136.29}
\DefMacro{corpus-ellipticcurves-AVG-proof-loc-per-doc}{349.89}
\DefMacro{corpus-ellipticcurves-SUM-proof-loc-per-doc}{6,298}
\DefMacro{corpus-ellipticcurves-MAX-proof-loc-per-doc}{1,115}
\DefMacro{corpus-ellipticcurves-MIN-proof-loc-per-doc}{48}
\DefMacro{corpus-ellipticcurves-MEDIAN-proof-loc-per-doc}{271.00}
\DefMacro{corpus-ellipticcurves-STDEV-proof-loc-per-doc}{278.67}
\DefMacro{corpus-ellipticcurves-AVG-comment-loc-per-doc}{15.50}
\DefMacro{corpus-ellipticcurves-SUM-comment-loc-per-doc}{279}
\DefMacro{corpus-ellipticcurves-MAX-comment-loc-per-doc}{36}
\DefMacro{corpus-ellipticcurves-MIN-comment-loc-per-doc}{2}
\DefMacro{corpus-ellipticcurves-MEDIAN-comment-loc-per-doc}{14.50}
\DefMacro{corpus-ellipticcurves-STDEV-comment-loc-per-doc}{9.12}
\DefMacro{corpus-ellipticcurves-AVG-total-loc-per-doc}{548.61}
\DefMacro{corpus-ellipticcurves-SUM-total-loc-per-doc}{9,875}
\DefMacro{corpus-ellipticcurves-MAX-total-loc-per-doc}{1,649}
\DefMacro{corpus-ellipticcurves-MIN-total-loc-per-doc}{100}
\DefMacro{corpus-ellipticcurves-MEDIAN-total-loc-per-doc}{400.50}
\DefMacro{corpus-ellipticcurves-STDEV-total-loc-per-doc}{388.61}
\DefMacro{corpus-ellipticcurves-AVG-code-loc-per-doc}{533.11}
\DefMacro{corpus-ellipticcurves-SUM-code-loc-per-doc}{9,596}
\DefMacro{corpus-ellipticcurves-MAX-code-loc-per-doc}{1,615}
\DefMacro{corpus-ellipticcurves-MIN-code-loc-per-doc}{94}
\DefMacro{corpus-ellipticcurves-MEDIAN-code-loc-per-doc}{390.00}
\DefMacro{corpus-ellipticcurves-STDEV-code-loc-per-doc}{384.65}
\DefMacro{corpus-toychain-num-docs}{14}
\DefMacro{corpus-toychain-spec-loc}{1,747}
\DefMacro{corpus-toychain-k-spec-loc}{1}
\DefMacro{corpus-toychain-proof-loc}{3,528}
\DefMacro{corpus-toychain-k-proof-loc}{3}
\DefMacro{corpus-toychain-comment-loc}{263}
\DefMacro{corpus-toychain-k-comment-loc}{0}
\DefMacro{corpus-toychain-total-loc}{5,538}
\DefMacro{corpus-toychain-k-total-loc}{5}
\DefMacro{corpus-toychain-code-loc}{5,275}
\DefMacro{corpus-toychain-k-code-loc}{5}
\DefMacro{corpus-toychain-AVG-spec-loc-per-doc}{124.79}
\DefMacro{corpus-toychain-SUM-spec-loc-per-doc}{1,747}
\DefMacro{corpus-toychain-MAX-spec-loc-per-doc}{567}
\DefMacro{corpus-toychain-MIN-spec-loc-per-doc}{10}
\DefMacro{corpus-toychain-MEDIAN-spec-loc-per-doc}{57.00}
\DefMacro{corpus-toychain-STDEV-spec-loc-per-doc}{149.06}
\DefMacro{corpus-toychain-AVG-proof-loc-per-doc}{252.00}
\DefMacro{corpus-toychain-SUM-proof-loc-per-doc}{3,528}
\DefMacro{corpus-toychain-MAX-proof-loc-per-doc}{2,176}
\DefMacro{corpus-toychain-MIN-proof-loc-per-doc}{0}
\DefMacro{corpus-toychain-MEDIAN-proof-loc-per-doc}{107.50}
\DefMacro{corpus-toychain-STDEV-proof-loc-per-doc}{546.07}
\DefMacro{corpus-toychain-AVG-comment-loc-per-doc}{18.79}
\DefMacro{corpus-toychain-SUM-comment-loc-per-doc}{263}
\DefMacro{corpus-toychain-MAX-comment-loc-per-doc}{78}
\DefMacro{corpus-toychain-MIN-comment-loc-per-doc}{1}
\DefMacro{corpus-toychain-MEDIAN-comment-loc-per-doc}{6.00}
\DefMacro{corpus-toychain-STDEV-comment-loc-per-doc}{25.46}
\DefMacro{corpus-toychain-AVG-total-loc-per-doc}{395.57}
\DefMacro{corpus-toychain-SUM-total-loc-per-doc}{5,538}
\DefMacro{corpus-toychain-MAX-total-loc-per-doc}{2,821}
\DefMacro{corpus-toychain-MIN-total-loc-per-doc}{21}
\DefMacro{corpus-toychain-MEDIAN-total-loc-per-doc}{167.00}
\DefMacro{corpus-toychain-STDEV-total-loc-per-doc}{705.51}
\DefMacro{corpus-toychain-AVG-code-loc-per-doc}{376.79}
\DefMacro{corpus-toychain-SUM-code-loc-per-doc}{5,275}
\DefMacro{corpus-toychain-MAX-code-loc-per-doc}{2,743}
\DefMacro{corpus-toychain-MIN-code-loc-per-doc}{14}
\DefMacro{corpus-toychain-MEDIAN-code-loc-per-doc}{164.00}
\DefMacro{corpus-toychain-STDEV-code-loc-per-doc}{685.03}
\DefMacro{corpus-bits-num-docs}{10}
\DefMacro{corpus-bits-spec-loc}{1,578}
\DefMacro{corpus-bits-k-spec-loc}{1}
\DefMacro{corpus-bits-proof-loc}{2,463}
\DefMacro{corpus-bits-k-proof-loc}{2}
\DefMacro{corpus-bits-comment-loc}{554}
\DefMacro{corpus-bits-k-comment-loc}{0}
\DefMacro{corpus-bits-total-loc}{4,595}
\DefMacro{corpus-bits-k-total-loc}{4}
\DefMacro{corpus-bits-code-loc}{4,041}
\DefMacro{corpus-bits-k-code-loc}{4}
\DefMacro{corpus-bits-AVG-spec-loc-per-doc}{157.80}
\DefMacro{corpus-bits-SUM-spec-loc-per-doc}{1,578}
\DefMacro{corpus-bits-MAX-spec-loc-per-doc}{407}
\DefMacro{corpus-bits-MIN-spec-loc-per-doc}{3}
\DefMacro{corpus-bits-MEDIAN-spec-loc-per-doc}{178.00}
\DefMacro{corpus-bits-STDEV-spec-loc-per-doc}{118.24}
\DefMacro{corpus-bits-AVG-proof-loc-per-doc}{246.30}
\DefMacro{corpus-bits-SUM-proof-loc-per-doc}{2,463}
\DefMacro{corpus-bits-MAX-proof-loc-per-doc}{1,204}
\DefMacro{corpus-bits-MIN-proof-loc-per-doc}{0}
\DefMacro{corpus-bits-MEDIAN-proof-loc-per-doc}{167.50}
\DefMacro{corpus-bits-STDEV-proof-loc-per-doc}{346.70}
\DefMacro{corpus-bits-AVG-comment-loc-per-doc}{55.40}
\DefMacro{corpus-bits-SUM-comment-loc-per-doc}{554}
\DefMacro{corpus-bits-MAX-comment-loc-per-doc}{188}
\DefMacro{corpus-bits-MIN-comment-loc-per-doc}{0}
\DefMacro{corpus-bits-MEDIAN-comment-loc-per-doc}{58.50}
\DefMacro{corpus-bits-STDEV-comment-loc-per-doc}{52.37}
\DefMacro{corpus-bits-AVG-total-loc-per-doc}{459.50}
\DefMacro{corpus-bits-SUM-total-loc-per-doc}{4,595}
\DefMacro{corpus-bits-MAX-total-loc-per-doc}{1,799}
\DefMacro{corpus-bits-MIN-total-loc-per-doc}{3}
\DefMacro{corpus-bits-MEDIAN-total-loc-per-doc}{411.00}
\DefMacro{corpus-bits-STDEV-total-loc-per-doc}{497.97}
\DefMacro{corpus-bits-AVG-code-loc-per-doc}{404.10}
\DefMacro{corpus-bits-SUM-code-loc-per-doc}{4,041}
\DefMacro{corpus-bits-MAX-code-loc-per-doc}{1,611}
\DefMacro{corpus-bits-MIN-code-loc-per-doc}{3}
\DefMacro{corpus-bits-MEDIAN-code-loc-per-doc}{354.50}
\DefMacro{corpus-bits-STDEV-code-loc-per-doc}{448.48}
\DefMacro{corpus-twosquare-num-docs}{2}
\DefMacro{corpus-twosquare-spec-loc}{413}
\DefMacro{corpus-twosquare-k-spec-loc}{0}
\DefMacro{corpus-twosquare-proof-loc}{1,308}
\DefMacro{corpus-twosquare-k-proof-loc}{1}
\DefMacro{corpus-twosquare-comment-loc}{78}
\DefMacro{corpus-twosquare-k-comment-loc}{0}
\DefMacro{corpus-twosquare-total-loc}{1,799}
\DefMacro{corpus-twosquare-k-total-loc}{1}
\DefMacro{corpus-twosquare-code-loc}{1,721}
\DefMacro{corpus-twosquare-k-code-loc}{1}
\DefMacro{corpus-twosquare-AVG-spec-loc-per-doc}{206.50}
\DefMacro{corpus-twosquare-SUM-spec-loc-per-doc}{413}
\DefMacro{corpus-twosquare-MAX-spec-loc-per-doc}{369}
\DefMacro{corpus-twosquare-MIN-spec-loc-per-doc}{44}
\DefMacro{corpus-twosquare-MEDIAN-spec-loc-per-doc}{206.50}
\DefMacro{corpus-twosquare-STDEV-spec-loc-per-doc}{162.50}
\DefMacro{corpus-twosquare-AVG-proof-loc-per-doc}{654.00}
\DefMacro{corpus-twosquare-SUM-proof-loc-per-doc}{1,308}
\DefMacro{corpus-twosquare-MAX-proof-loc-per-doc}{1,126}
\DefMacro{corpus-twosquare-MIN-proof-loc-per-doc}{182}
\DefMacro{corpus-twosquare-MEDIAN-proof-loc-per-doc}{654.00}
\DefMacro{corpus-twosquare-STDEV-proof-loc-per-doc}{472.00}
\DefMacro{corpus-twosquare-AVG-comment-loc-per-doc}{39.00}
\DefMacro{corpus-twosquare-SUM-comment-loc-per-doc}{78}
\DefMacro{corpus-twosquare-MAX-comment-loc-per-doc}{77}
\DefMacro{corpus-twosquare-MIN-comment-loc-per-doc}{1}
\DefMacro{corpus-twosquare-MEDIAN-comment-loc-per-doc}{39.00}
\DefMacro{corpus-twosquare-STDEV-comment-loc-per-doc}{38.00}
\DefMacro{corpus-twosquare-AVG-total-loc-per-doc}{899.50}
\DefMacro{corpus-twosquare-SUM-total-loc-per-doc}{1,799}
\DefMacro{corpus-twosquare-MAX-total-loc-per-doc}{1,572}
\DefMacro{corpus-twosquare-MIN-total-loc-per-doc}{227}
\DefMacro{corpus-twosquare-MEDIAN-total-loc-per-doc}{899.50}
\DefMacro{corpus-twosquare-STDEV-total-loc-per-doc}{672.50}
\DefMacro{corpus-twosquare-AVG-code-loc-per-doc}{860.50}
\DefMacro{corpus-twosquare-SUM-code-loc-per-doc}{1,721}
\DefMacro{corpus-twosquare-MAX-code-loc-per-doc}{1,495}
\DefMacro{corpus-twosquare-MIN-code-loc-per-doc}{226}
\DefMacro{corpus-twosquare-MEDIAN-code-loc-per-doc}{860.50}
\DefMacro{corpus-twosquare-STDEV-code-loc-per-doc}{634.50}
\DefMacro{corpus-grobner-num-docs}{1}
\DefMacro{corpus-grobner-spec-loc}{312}
\DefMacro{corpus-grobner-k-spec-loc}{0}
\DefMacro{corpus-grobner-proof-loc}{1,018}
\DefMacro{corpus-grobner-k-proof-loc}{1}
\DefMacro{corpus-grobner-comment-loc}{48}
\DefMacro{corpus-grobner-k-comment-loc}{0}
\DefMacro{corpus-grobner-total-loc}{1,378}
\DefMacro{corpus-grobner-k-total-loc}{1}
\DefMacro{corpus-grobner-code-loc}{1,330}
\DefMacro{corpus-grobner-k-code-loc}{1}
\DefMacro{corpus-grobner-AVG-spec-loc-per-doc}{312.00}
\DefMacro{corpus-grobner-SUM-spec-loc-per-doc}{312}
\DefMacro{corpus-grobner-MAX-spec-loc-per-doc}{312}
\DefMacro{corpus-grobner-MIN-spec-loc-per-doc}{312}
\DefMacro{corpus-grobner-MEDIAN-spec-loc-per-doc}{312.00}
\DefMacro{corpus-grobner-STDEV-spec-loc-per-doc}{0.00}
\DefMacro{corpus-grobner-AVG-proof-loc-per-doc}{1,018.00}
\DefMacro{corpus-grobner-SUM-proof-loc-per-doc}{1,018}
\DefMacro{corpus-grobner-MAX-proof-loc-per-doc}{1,018}
\DefMacro{corpus-grobner-MIN-proof-loc-per-doc}{1,018}
\DefMacro{corpus-grobner-MEDIAN-proof-loc-per-doc}{1,018.00}
\DefMacro{corpus-grobner-STDEV-proof-loc-per-doc}{0.00}
\DefMacro{corpus-grobner-AVG-comment-loc-per-doc}{48.00}
\DefMacro{corpus-grobner-SUM-comment-loc-per-doc}{48}
\DefMacro{corpus-grobner-MAX-comment-loc-per-doc}{48}
\DefMacro{corpus-grobner-MIN-comment-loc-per-doc}{48}
\DefMacro{corpus-grobner-MEDIAN-comment-loc-per-doc}{48.00}
\DefMacro{corpus-grobner-STDEV-comment-loc-per-doc}{0.00}
\DefMacro{corpus-grobner-AVG-total-loc-per-doc}{1,378.00}
\DefMacro{corpus-grobner-SUM-total-loc-per-doc}{1,378}
\DefMacro{corpus-grobner-MAX-total-loc-per-doc}{1,378}
\DefMacro{corpus-grobner-MIN-total-loc-per-doc}{1,378}
\DefMacro{corpus-grobner-MEDIAN-total-loc-per-doc}{1,378.00}
\DefMacro{corpus-grobner-STDEV-total-loc-per-doc}{0.00}
\DefMacro{corpus-grobner-AVG-code-loc-per-doc}{1,330.00}
\DefMacro{corpus-grobner-SUM-code-loc-per-doc}{1,330}
\DefMacro{corpus-grobner-MAX-code-loc-per-doc}{1,330}
\DefMacro{corpus-grobner-MIN-code-loc-per-doc}{1,330}
\DefMacro{corpus-grobner-MEDIAN-code-loc-per-doc}{1,330.00}
\DefMacro{corpus-grobner-STDEV-code-loc-per-doc}{0.00}
\DefMacro{corpus-infotheo-num-docs}{81}
\DefMacro{corpus-infotheo-spec-loc}{12,517}
\DefMacro{corpus-infotheo-k-spec-loc}{12}
\DefMacro{corpus-infotheo-proof-loc}{29,778}
\DefMacro{corpus-infotheo-k-proof-loc}{29}
\DefMacro{corpus-infotheo-comment-loc}{1,733}
\DefMacro{corpus-infotheo-k-comment-loc}{1}
\DefMacro{corpus-infotheo-total-loc}{44,028}
\DefMacro{corpus-infotheo-k-total-loc}{44}
\DefMacro{corpus-infotheo-code-loc}{42,295}
\DefMacro{corpus-infotheo-k-code-loc}{42}
\DefMacro{corpus-infotheo-AVG-spec-loc-per-doc}{154.53}
\DefMacro{corpus-infotheo-SUM-spec-loc-per-doc}{12,517}
\DefMacro{corpus-infotheo-MAX-spec-loc-per-doc}{1,265}
\DefMacro{corpus-infotheo-MIN-spec-loc-per-doc}{15}
\DefMacro{corpus-infotheo-MEDIAN-spec-loc-per-doc}{104.00}
\DefMacro{corpus-infotheo-STDEV-spec-loc-per-doc}{181.69}
\DefMacro{corpus-infotheo-AVG-proof-loc-per-doc}{367.63}
\DefMacro{corpus-infotheo-SUM-proof-loc-per-doc}{29,778}
\DefMacro{corpus-infotheo-MAX-proof-loc-per-doc}{2,963}
\DefMacro{corpus-infotheo-MIN-proof-loc-per-doc}{0}
\DefMacro{corpus-infotheo-MEDIAN-proof-loc-per-doc}{235.00}
\DefMacro{corpus-infotheo-STDEV-proof-loc-per-doc}{430.38}
\DefMacro{corpus-infotheo-AVG-comment-loc-per-doc}{21.40}
\DefMacro{corpus-infotheo-SUM-comment-loc-per-doc}{1,733}
\DefMacro{corpus-infotheo-MAX-comment-loc-per-doc}{320}
\DefMacro{corpus-infotheo-MIN-comment-loc-per-doc}{0}
\DefMacro{corpus-infotheo-MEDIAN-comment-loc-per-doc}{13.00}
\DefMacro{corpus-infotheo-STDEV-comment-loc-per-doc}{39.29}
\DefMacro{corpus-infotheo-AVG-total-loc-per-doc}{543.56}
\DefMacro{corpus-infotheo-SUM-total-loc-per-doc}{44,028}
\DefMacro{corpus-infotheo-MAX-total-loc-per-doc}{4,548}
\DefMacro{corpus-infotheo-MIN-total-loc-per-doc}{33}
\DefMacro{corpus-infotheo-MEDIAN-total-loc-per-doc}{342.00}
\DefMacro{corpus-infotheo-STDEV-total-loc-per-doc}{625.20}
\DefMacro{corpus-infotheo-AVG-code-loc-per-doc}{522.16}
\DefMacro{corpus-infotheo-SUM-code-loc-per-doc}{42,295}
\DefMacro{corpus-infotheo-MAX-code-loc-per-doc}{4,228}
\DefMacro{corpus-infotheo-MIN-code-loc-per-doc}{31}
\DefMacro{corpus-infotheo-MEDIAN-code-loc-per-doc}{327.00}
\DefMacro{corpus-infotheo-STDEV-code-loc-per-doc}{591.23}
\DefMacro{corpus-monae-num-docs}{18}
\DefMacro{corpus-monae-spec-loc}{3,422}
\DefMacro{corpus-monae-k-spec-loc}{3}
\DefMacro{corpus-monae-proof-loc}{3,233}
\DefMacro{corpus-monae-k-proof-loc}{3}
\DefMacro{corpus-monae-comment-loc}{414}
\DefMacro{corpus-monae-k-comment-loc}{0}
\DefMacro{corpus-monae-total-loc}{7,069}
\DefMacro{corpus-monae-k-total-loc}{7}
\DefMacro{corpus-monae-code-loc}{6,655}
\DefMacro{corpus-monae-k-code-loc}{6}
\DefMacro{corpus-monae-AVG-spec-loc-per-doc}{190.11}
\DefMacro{corpus-monae-SUM-spec-loc-per-doc}{3,422}
\DefMacro{corpus-monae-MAX-spec-loc-per-doc}{844}
\DefMacro{corpus-monae-MIN-spec-loc-per-doc}{9}
\DefMacro{corpus-monae-MEDIAN-spec-loc-per-doc}{117.00}
\DefMacro{corpus-monae-STDEV-spec-loc-per-doc}{200.22}
\DefMacro{corpus-monae-AVG-proof-loc-per-doc}{179.61}
\DefMacro{corpus-monae-SUM-proof-loc-per-doc}{3,233}
\DefMacro{corpus-monae-MAX-proof-loc-per-doc}{451}
\DefMacro{corpus-monae-MIN-proof-loc-per-doc}{0}
\DefMacro{corpus-monae-MEDIAN-proof-loc-per-doc}{185.00}
\DefMacro{corpus-monae-STDEV-proof-loc-per-doc}{144.72}
\DefMacro{corpus-monae-AVG-comment-loc-per-doc}{23.00}
\DefMacro{corpus-monae-SUM-comment-loc-per-doc}{414}
\DefMacro{corpus-monae-MAX-comment-loc-per-doc}{161}
\DefMacro{corpus-monae-MIN-comment-loc-per-doc}{0}
\DefMacro{corpus-monae-MEDIAN-comment-loc-per-doc}{10.00}
\DefMacro{corpus-monae-STDEV-comment-loc-per-doc}{37.49}
\DefMacro{corpus-monae-AVG-total-loc-per-doc}{392.72}
\DefMacro{corpus-monae-SUM-total-loc-per-doc}{7,069}
\DefMacro{corpus-monae-MAX-total-loc-per-doc}{1,456}
\DefMacro{corpus-monae-MIN-total-loc-per-doc}{9}
\DefMacro{corpus-monae-MEDIAN-total-loc-per-doc}{368.50}
\DefMacro{corpus-monae-STDEV-total-loc-per-doc}{344.59}
\DefMacro{corpus-monae-AVG-code-loc-per-doc}{369.72}
\DefMacro{corpus-monae-SUM-code-loc-per-doc}{6,655}
\DefMacro{corpus-monae-MAX-code-loc-per-doc}{1,295}
\DefMacro{corpus-monae-MIN-code-loc-per-doc}{9}
\DefMacro{corpus-monae-MEDIAN-code-loc-per-doc}{366.00}
\DefMacro{corpus-monae-STDEV-code-loc-per-doc}{311.19}
\DefMacro{corpus-games-num-docs}{12}
\DefMacro{corpus-games-spec-loc}{1,450}
\DefMacro{corpus-games-k-spec-loc}{1}
\DefMacro{corpus-games-proof-loc}{3,503}
\DefMacro{corpus-games-k-proof-loc}{3}
\DefMacro{corpus-games-comment-loc}{258}
\DefMacro{corpus-games-k-comment-loc}{0}
\DefMacro{corpus-games-total-loc}{5,211}
\DefMacro{corpus-games-k-total-loc}{5}
\DefMacro{corpus-games-code-loc}{4,953}
\DefMacro{corpus-games-k-code-loc}{4}
\DefMacro{corpus-games-AVG-spec-loc-per-doc}{120.83}
\DefMacro{corpus-games-SUM-spec-loc-per-doc}{1,450}
\DefMacro{corpus-games-MAX-spec-loc-per-doc}{300}
\DefMacro{corpus-games-MIN-spec-loc-per-doc}{28}
\DefMacro{corpus-games-MEDIAN-spec-loc-per-doc}{99.00}
\DefMacro{corpus-games-STDEV-spec-loc-per-doc}{78.32}
\DefMacro{corpus-games-AVG-proof-loc-per-doc}{291.92}
\DefMacro{corpus-games-SUM-proof-loc-per-doc}{3,503}
\DefMacro{corpus-games-MAX-proof-loc-per-doc}{1,163}
\DefMacro{corpus-games-MIN-proof-loc-per-doc}{37}
\DefMacro{corpus-games-MEDIAN-proof-loc-per-doc}{184.00}
\DefMacro{corpus-games-STDEV-proof-loc-per-doc}{301.35}
\DefMacro{corpus-games-AVG-comment-loc-per-doc}{21.50}
\DefMacro{corpus-games-SUM-comment-loc-per-doc}{258}
\DefMacro{corpus-games-MAX-comment-loc-per-doc}{108}
\DefMacro{corpus-games-MIN-comment-loc-per-doc}{1}
\DefMacro{corpus-games-MEDIAN-comment-loc-per-doc}{14.00}
\DefMacro{corpus-games-STDEV-comment-loc-per-doc}{27.74}
\DefMacro{corpus-games-AVG-total-loc-per-doc}{434.25}
\DefMacro{corpus-games-SUM-total-loc-per-doc}{5,211}
\DefMacro{corpus-games-MAX-total-loc-per-doc}{1,483}
\DefMacro{corpus-games-MIN-total-loc-per-doc}{66}
\DefMacro{corpus-games-MEDIAN-total-loc-per-doc}{303.50}
\DefMacro{corpus-games-STDEV-total-loc-per-doc}{380.89}
\DefMacro{corpus-games-AVG-code-loc-per-doc}{412.75}
\DefMacro{corpus-games-SUM-code-loc-per-doc}{4,953}
\DefMacro{corpus-games-MAX-code-loc-per-doc}{1,463}
\DefMacro{corpus-games-MIN-code-loc-per-doc}{65}
\DefMacro{corpus-games-MEDIAN-code-loc-per-doc}{281.00}
\DefMacro{corpus-games-STDEV-code-loc-per-doc}{370.26}
\DefMacro{corpus-t1-num-projects}{4}
\DefMacro{corpus-t1-AVG-num-docs}{46.75}
\DefMacro{corpus-t1-AVG-spec-loc}{15,890.00}
\DefMacro{corpus-t1-AVG-proof-loc}{25,216.75}
\DefMacro{corpus-t1-AVG-comment-loc}{2,874.50}
\DefMacro{corpus-t1-AVG-total-loc}{43,981.25}
\DefMacro{corpus-t1-AVG-code-loc}{41,106.75}
\DefMacro{corpus-t1-SUM-num-docs}{187}
\DefMacro{corpus-t1-SUM-spec-loc}{63,560}
\DefMacro{corpus-t1-SUM-k-spec-loc}{63}
\DefMacro{corpus-t1-SUM-proof-loc}{100,867}
\DefMacro{corpus-t1-SUM-k-proof-loc}{100}
\DefMacro{corpus-t1-SUM-comment-loc}{11,498}
\DefMacro{corpus-t1-SUM-k-comment-loc}{11}
\DefMacro{corpus-t1-SUM-total-loc}{175,925}
\DefMacro{corpus-t1-SUM-k-total-loc}{175}
\DefMacro{corpus-t1-SUM-code-loc}{164,427}
\DefMacro{corpus-t1-SUM-k-code-loc}{164}
\DefMacro{corpus-t1-MAX-num-docs}{89}
\DefMacro{corpus-t1-MAX-spec-loc}{38,243}
\DefMacro{corpus-t1-MAX-k-spec-loc}{38}
\DefMacro{corpus-t1-MAX-proof-loc}{46,470}
\DefMacro{corpus-t1-MAX-k-proof-loc}{46}
\DefMacro{corpus-t1-MAX-comment-loc}{6,131}
\DefMacro{corpus-t1-MAX-k-comment-loc}{6}
\DefMacro{corpus-t1-MAX-total-loc}{90,844}
\DefMacro{corpus-t1-MAX-k-total-loc}{90}
\DefMacro{corpus-t1-MAX-code-loc}{84,713}
\DefMacro{corpus-t1-MAX-k-code-loc}{84}
\DefMacro{corpus-t1-MIN-num-docs}{4}
\DefMacro{corpus-t1-MIN-spec-loc}{4,260}
\DefMacro{corpus-t1-MIN-k-spec-loc}{4}
\DefMacro{corpus-t1-MIN-proof-loc}{2,191}
\DefMacro{corpus-t1-MIN-k-proof-loc}{2}
\DefMacro{corpus-t1-MIN-comment-loc}{238}
\DefMacro{corpus-t1-MIN-k-comment-loc}{0}
\DefMacro{corpus-t1-MIN-total-loc}{6,689}
\DefMacro{corpus-t1-MIN-k-total-loc}{6}
\DefMacro{corpus-t1-MIN-code-loc}{6,451}
\DefMacro{corpus-t1-MIN-k-code-loc}{6}
\DefMacro{corpus-t1-MEDIAN-num-docs}{47.00}
\DefMacro{corpus-t1-MEDIAN-spec-loc}{10,528.50}
\DefMacro{corpus-t1-MEDIAN-proof-loc}{26,103.00}
\DefMacro{corpus-t1-MEDIAN-comment-loc}{2,564.50}
\DefMacro{corpus-t1-MEDIAN-total-loc}{39,196.00}
\DefMacro{corpus-t1-MEDIAN-code-loc}{36,631.50}
\DefMacro{corpus-t1-STDEV-num-docs}{31.43}
\DefMacro{corpus-t1-STDEV-spec-loc}{13,191.56}
\DefMacro{corpus-t1-STDEV-proof-loc}{15,735.12}
\DefMacro{corpus-t1-STDEV-comment-loc}{2,113.92}
\DefMacro{corpus-t1-STDEV-total-loc}{30,140.40}
\DefMacro{corpus-t1-STDEV-code-loc}{28,031.66}
\DefMacro{corpus-t1-AVG-spec-loc-per-doc}{339.89}
\DefMacro{corpus-t1-SUM-spec-loc-per-doc}{63,560}
\DefMacro{corpus-t1-MAX-spec-loc-per-doc}{4,331}
\DefMacro{corpus-t1-MIN-spec-loc-per-doc}{1}
\DefMacro{corpus-t1-MEDIAN-spec-loc-per-doc}{185.00}
\DefMacro{corpus-t1-STDEV-spec-loc-per-doc}{462.97}
\DefMacro{corpus-t1-AVG-proof-loc-per-doc}{539.40}
\DefMacro{corpus-t1-SUM-proof-loc-per-doc}{100,867}
\DefMacro{corpus-t1-MAX-proof-loc-per-doc}{5,551}
\DefMacro{corpus-t1-MIN-proof-loc-per-doc}{0}
\DefMacro{corpus-t1-MEDIAN-proof-loc-per-doc}{381.00}
\DefMacro{corpus-t1-STDEV-proof-loc-per-doc}{704.12}
\DefMacro{corpus-t1-AVG-comment-loc-per-doc}{61.49}
\DefMacro{corpus-t1-SUM-comment-loc-per-doc}{11,498}
\DefMacro{corpus-t1-MAX-comment-loc-per-doc}{588}
\DefMacro{corpus-t1-MIN-comment-loc-per-doc}{0}
\DefMacro{corpus-t1-MEDIAN-comment-loc-per-doc}{43.00}
\DefMacro{corpus-t1-STDEV-comment-loc-per-doc}{66.75}
\DefMacro{corpus-t1-AVG-total-loc-per-doc}{940.78}
\DefMacro{corpus-t1-SUM-total-loc-per-doc}{175,925}
\DefMacro{corpus-t1-MAX-total-loc-per-doc}{5,619}
\DefMacro{corpus-t1-MIN-total-loc-per-doc}{1}
\DefMacro{corpus-t1-MEDIAN-total-loc-per-doc}{654.00}
\DefMacro{corpus-t1-STDEV-total-loc-per-doc}{964.69}
\DefMacro{corpus-t1-AVG-code-loc-per-doc}{879.29}
\DefMacro{corpus-t1-SUM-code-loc-per-doc}{164,427}
\DefMacro{corpus-t1-MAX-code-loc-per-doc}{5,560}
\DefMacro{corpus-t1-MIN-code-loc-per-doc}{1}
\DefMacro{corpus-t1-MEDIAN-code-loc-per-doc}{622.00}
\DefMacro{corpus-t1-STDEV-code-loc-per-doc}{925.44}
\DefMacro{corpus-t2-num-projects}{8}
\DefMacro{corpus-t2-AVG-num-docs}{8.38}
\DefMacro{corpus-t2-AVG-spec-loc}{2,696.75}
\DefMacro{corpus-t2-AVG-proof-loc}{3,788.50}
\DefMacro{corpus-t2-AVG-comment-loc}{385.62}
\DefMacro{corpus-t2-AVG-total-loc}{6,870.88}
\DefMacro{corpus-t2-AVG-code-loc}{6,485.25}
\DefMacro{corpus-t2-SUM-num-docs}{67}
\DefMacro{corpus-t2-SUM-spec-loc}{21,574}
\DefMacro{corpus-t2-SUM-k-spec-loc}{21}
\DefMacro{corpus-t2-SUM-proof-loc}{30,308}
\DefMacro{corpus-t2-SUM-k-proof-loc}{30}
\DefMacro{corpus-t2-SUM-comment-loc}{3,085}
\DefMacro{corpus-t2-SUM-k-comment-loc}{3}
\DefMacro{corpus-t2-SUM-total-loc}{54,967}
\DefMacro{corpus-t2-SUM-k-total-loc}{54}
\DefMacro{corpus-t2-SUM-code-loc}{51,882}
\DefMacro{corpus-t2-SUM-k-code-loc}{51}
\DefMacro{corpus-t2-MAX-num-docs}{18}
\DefMacro{corpus-t2-MAX-spec-loc}{5,553}
\DefMacro{corpus-t2-MAX-k-spec-loc}{5}
\DefMacro{corpus-t2-MAX-proof-loc}{7,249}
\DefMacro{corpus-t2-MAX-k-proof-loc}{7}
\DefMacro{corpus-t2-MAX-comment-loc}{1,352}
\DefMacro{corpus-t2-MAX-k-comment-loc}{1}
\DefMacro{corpus-t2-MAX-total-loc}{13,091}
\DefMacro{corpus-t2-MAX-k-total-loc}{13}
\DefMacro{corpus-t2-MAX-code-loc}{11,739}
\DefMacro{corpus-t2-MAX-k-code-loc}{11}
\DefMacro{corpus-t2-MIN-num-docs}{1}
\DefMacro{corpus-t2-MIN-spec-loc}{70}
\DefMacro{corpus-t2-MIN-k-spec-loc}{0}
\DefMacro{corpus-t2-MIN-proof-loc}{8}
\DefMacro{corpus-t2-MIN-k-proof-loc}{0}
\DefMacro{corpus-t2-MIN-comment-loc}{16}
\DefMacro{corpus-t2-MIN-k-comment-loc}{0}
\DefMacro{corpus-t2-MIN-total-loc}{94}
\DefMacro{corpus-t2-MIN-k-total-loc}{0}
\DefMacro{corpus-t2-MIN-code-loc}{78}
\DefMacro{corpus-t2-MIN-k-code-loc}{0}
\DefMacro{corpus-t2-MEDIAN-num-docs}{7.50}
\DefMacro{corpus-t2-MEDIAN-spec-loc}{3,498.50}
\DefMacro{corpus-t2-MEDIAN-proof-loc}{4,120.50}
\DefMacro{corpus-t2-MEDIAN-comment-loc}{293.50}
\DefMacro{corpus-t2-MEDIAN-total-loc}{8,476.00}
\DefMacro{corpus-t2-MEDIAN-code-loc}{8,144.00}
\DefMacro{corpus-t2-STDEV-num-docs}{6.63}
\DefMacro{corpus-t2-STDEV-spec-loc}{1,982.91}
\DefMacro{corpus-t2-STDEV-proof-loc}{2,568.53}
\DefMacro{corpus-t2-STDEV-comment-loc}{413.26}
\DefMacro{corpus-t2-STDEV-total-loc}{4,744.27}
\DefMacro{corpus-t2-STDEV-code-loc}{4,414.50}
\DefMacro{corpus-t2-AVG-spec-loc-per-doc}{322.00}
\DefMacro{corpus-t2-SUM-spec-loc-per-doc}{21,574}
\DefMacro{corpus-t2-MAX-spec-loc-per-doc}{2,069}
\DefMacro{corpus-t2-MIN-spec-loc-per-doc}{8}
\DefMacro{corpus-t2-MEDIAN-spec-loc-per-doc}{226.00}
\DefMacro{corpus-t2-STDEV-spec-loc-per-doc}{334.27}
\DefMacro{corpus-t2-AVG-proof-loc-per-doc}{452.36}
\DefMacro{corpus-t2-SUM-proof-loc-per-doc}{30,308}
\DefMacro{corpus-t2-MAX-proof-loc-per-doc}{2,387}
\DefMacro{corpus-t2-MIN-proof-loc-per-doc}{0}
\DefMacro{corpus-t2-MEDIAN-proof-loc-per-doc}{327.00}
\DefMacro{corpus-t2-STDEV-proof-loc-per-doc}{422.40}
\DefMacro{corpus-t2-AVG-comment-loc-per-doc}{46.04}
\DefMacro{corpus-t2-SUM-comment-loc-per-doc}{3,085}
\DefMacro{corpus-t2-MAX-comment-loc-per-doc}{311}
\DefMacro{corpus-t2-MIN-comment-loc-per-doc}{0}
\DefMacro{corpus-t2-MEDIAN-comment-loc-per-doc}{27.00}
\DefMacro{corpus-t2-STDEV-comment-loc-per-doc}{58.59}
\DefMacro{corpus-t2-AVG-total-loc-per-doc}{820.40}
\DefMacro{corpus-t2-SUM-total-loc-per-doc}{54,967}
\DefMacro{corpus-t2-MAX-total-loc-per-doc}{4,611}
\DefMacro{corpus-t2-MIN-total-loc-per-doc}{8}
\DefMacro{corpus-t2-MEDIAN-total-loc-per-doc}{747.00}
\DefMacro{corpus-t2-STDEV-total-loc-per-doc}{717.19}
\DefMacro{corpus-t2-AVG-code-loc-per-doc}{774.36}
\DefMacro{corpus-t2-SUM-code-loc-per-doc}{51,882}
\DefMacro{corpus-t2-MAX-code-loc-per-doc}{4,456}
\DefMacro{corpus-t2-MIN-code-loc-per-doc}{8}
\DefMacro{corpus-t2-MEDIAN-code-loc-per-doc}{720.00}
\DefMacro{corpus-t2-STDEV-code-loc-per-doc}{682.33}
\DefMacro{corpus-t3-num-projects}{8}
\DefMacro{corpus-t3-AVG-num-docs}{14.25}
\DefMacro{corpus-t3-AVG-spec-loc}{2,164.12}
\DefMacro{corpus-t3-AVG-proof-loc}{2,665.62}
\DefMacro{corpus-t3-AVG-comment-loc}{388.00}
\DefMacro{corpus-t3-AVG-total-loc}{5,217.75}
\DefMacro{corpus-t3-AVG-code-loc}{4,829.75}
\DefMacro{corpus-t3-SUM-num-docs}{114}
\DefMacro{corpus-t3-SUM-spec-loc}{17,313}
\DefMacro{corpus-t3-SUM-k-spec-loc}{17}
\DefMacro{corpus-t3-SUM-proof-loc}{21,325}
\DefMacro{corpus-t3-SUM-k-proof-loc}{21}
\DefMacro{corpus-t3-SUM-comment-loc}{3,104}
\DefMacro{corpus-t3-SUM-k-comment-loc}{3}
\DefMacro{corpus-t3-SUM-total-loc}{41,742}
\DefMacro{corpus-t3-SUM-k-total-loc}{41}
\DefMacro{corpus-t3-SUM-code-loc}{38,638}
\DefMacro{corpus-t3-SUM-k-code-loc}{38}
\DefMacro{corpus-t3-MAX-num-docs}{20}
\DefMacro{corpus-t3-MAX-spec-loc}{3,422}
\DefMacro{corpus-t3-MAX-k-spec-loc}{3}
\DefMacro{corpus-t3-MAX-proof-loc}{3,528}
\DefMacro{corpus-t3-MAX-k-proof-loc}{3}
\DefMacro{corpus-t3-MAX-comment-loc}{554}
\DefMacro{corpus-t3-MAX-k-comment-loc}{0}
\DefMacro{corpus-t3-MAX-total-loc}{7,069}
\DefMacro{corpus-t3-MAX-k-total-loc}{7}
\DefMacro{corpus-t3-MAX-code-loc}{6,655}
\DefMacro{corpus-t3-MAX-k-code-loc}{6}
\DefMacro{corpus-t3-MIN-num-docs}{10}
\DefMacro{corpus-t3-MIN-spec-loc}{1,299}
\DefMacro{corpus-t3-MIN-k-spec-loc}{1}
\DefMacro{corpus-t3-MIN-proof-loc}{1,734}
\DefMacro{corpus-t3-MIN-k-proof-loc}{1}
\DefMacro{corpus-t3-MIN-comment-loc}{250}
\DefMacro{corpus-t3-MIN-k-comment-loc}{0}
\DefMacro{corpus-t3-MIN-total-loc}{3,283}
\DefMacro{corpus-t3-MIN-k-total-loc}{3}
\DefMacro{corpus-t3-MIN-code-loc}{3,033}
\DefMacro{corpus-t3-MIN-k-code-loc}{3}
\DefMacro{corpus-t3-MEDIAN-num-docs}{13.00}
\DefMacro{corpus-t3-MEDIAN-spec-loc}{2,026.00}
\DefMacro{corpus-t3-MEDIAN-proof-loc}{2,657.50}
\DefMacro{corpus-t3-MEDIAN-comment-loc}{404.00}
\DefMacro{corpus-t3-MEDIAN-total-loc}{5,094.50}
\DefMacro{corpus-t3-MEDIAN-code-loc}{4,713.00}
\DefMacro{corpus-t3-STDEV-num-docs}{3.23}
\DefMacro{corpus-t3-STDEV-spec-loc}{719.74}
\DefMacro{corpus-t3-STDEV-proof-loc}{671.20}
\DefMacro{corpus-t3-STDEV-comment-loc}{111.59}
\DefMacro{corpus-t3-STDEV-total-loc}{1,058.18}
\DefMacro{corpus-t3-STDEV-code-loc}{1,037.52}
\DefMacro{corpus-t3-AVG-spec-loc-per-doc}{151.87}
\DefMacro{corpus-t3-SUM-spec-loc-per-doc}{17,313}
\DefMacro{corpus-t3-MAX-spec-loc-per-doc}{844}
\DefMacro{corpus-t3-MIN-spec-loc-per-doc}{3}
\DefMacro{corpus-t3-MEDIAN-spec-loc-per-doc}{103.00}
\DefMacro{corpus-t3-STDEV-spec-loc-per-doc}{148.08}
\DefMacro{corpus-t3-AVG-proof-loc-per-doc}{187.06}
\DefMacro{corpus-t3-SUM-proof-loc-per-doc}{21,325}
\DefMacro{corpus-t3-MAX-proof-loc-per-doc}{2,176}
\DefMacro{corpus-t3-MIN-proof-loc-per-doc}{0}
\DefMacro{corpus-t3-MEDIAN-proof-loc-per-doc}{103.50}
\DefMacro{corpus-t3-STDEV-proof-loc-per-doc}{280.20}
\DefMacro{corpus-t3-AVG-comment-loc-per-doc}{27.23}
\DefMacro{corpus-t3-SUM-comment-loc-per-doc}{3,104}
\DefMacro{corpus-t3-MAX-comment-loc-per-doc}{188}
\DefMacro{corpus-t3-MIN-comment-loc-per-doc}{0}
\DefMacro{corpus-t3-MEDIAN-comment-loc-per-doc}{17.00}
\DefMacro{corpus-t3-STDEV-comment-loc-per-doc}{32.96}
\DefMacro{corpus-t3-AVG-total-loc-per-doc}{366.16}
\DefMacro{corpus-t3-SUM-total-loc-per-doc}{41,742}
\DefMacro{corpus-t3-MAX-total-loc-per-doc}{2,821}
\DefMacro{corpus-t3-MIN-total-loc-per-doc}{3}
\DefMacro{corpus-t3-MEDIAN-total-loc-per-doc}{243.50}
\DefMacro{corpus-t3-STDEV-total-loc-per-doc}{413.48}
\DefMacro{corpus-t3-AVG-code-loc-per-doc}{338.93}
\DefMacro{corpus-t3-SUM-code-loc-per-doc}{38,638}
\DefMacro{corpus-t3-MAX-code-loc-per-doc}{2,743}
\DefMacro{corpus-t3-MIN-code-loc-per-doc}{3}
\DefMacro{corpus-t3-MEDIAN-code-loc-per-doc}{216.00}
\DefMacro{corpus-t3-STDEV-code-loc-per-doc}{390.96}
\DefMacro{corpus-lo-num-projects}{1}
\DefMacro{corpus-lo-AVG-num-docs}{81.00}
\DefMacro{corpus-lo-AVG-spec-loc}{12,517.00}
\DefMacro{corpus-lo-AVG-proof-loc}{29,778.00}
\DefMacro{corpus-lo-AVG-comment-loc}{1,733.00}
\DefMacro{corpus-lo-AVG-total-loc}{44,028.00}
\DefMacro{corpus-lo-AVG-code-loc}{42,295.00}
\DefMacro{corpus-lo-SUM-num-docs}{81}
\DefMacro{corpus-lo-SUM-spec-loc}{12,517}
\DefMacro{corpus-lo-SUM-k-spec-loc}{12}
\DefMacro{corpus-lo-SUM-proof-loc}{29,778}
\DefMacro{corpus-lo-SUM-k-proof-loc}{29}
\DefMacro{corpus-lo-SUM-comment-loc}{1,733}
\DefMacro{corpus-lo-SUM-k-comment-loc}{1}
\DefMacro{corpus-lo-SUM-total-loc}{44,028}
\DefMacro{corpus-lo-SUM-k-total-loc}{44}
\DefMacro{corpus-lo-SUM-code-loc}{42,295}
\DefMacro{corpus-lo-SUM-k-code-loc}{42}
\DefMacro{corpus-lo-MAX-num-docs}{81}
\DefMacro{corpus-lo-MAX-spec-loc}{12,517}
\DefMacro{corpus-lo-MAX-k-spec-loc}{12}
\DefMacro{corpus-lo-MAX-proof-loc}{29,778}
\DefMacro{corpus-lo-MAX-k-proof-loc}{29}
\DefMacro{corpus-lo-MAX-comment-loc}{1,733}
\DefMacro{corpus-lo-MAX-k-comment-loc}{1}
\DefMacro{corpus-lo-MAX-total-loc}{44,028}
\DefMacro{corpus-lo-MAX-k-total-loc}{44}
\DefMacro{corpus-lo-MAX-code-loc}{42,295}
\DefMacro{corpus-lo-MAX-k-code-loc}{42}
\DefMacro{corpus-lo-MIN-num-docs}{81}
\DefMacro{corpus-lo-MIN-spec-loc}{12,517}
\DefMacro{corpus-lo-MIN-k-spec-loc}{12}
\DefMacro{corpus-lo-MIN-proof-loc}{29,778}
\DefMacro{corpus-lo-MIN-k-proof-loc}{29}
\DefMacro{corpus-lo-MIN-comment-loc}{1,733}
\DefMacro{corpus-lo-MIN-k-comment-loc}{1}
\DefMacro{corpus-lo-MIN-total-loc}{44,028}
\DefMacro{corpus-lo-MIN-k-total-loc}{44}
\DefMacro{corpus-lo-MIN-code-loc}{42,295}
\DefMacro{corpus-lo-MIN-k-code-loc}{42}
\DefMacro{corpus-lo-MEDIAN-num-docs}{81.00}
\DefMacro{corpus-lo-MEDIAN-spec-loc}{12,517.00}
\DefMacro{corpus-lo-MEDIAN-proof-loc}{29,778.00}
\DefMacro{corpus-lo-MEDIAN-comment-loc}{1,733.00}
\DefMacro{corpus-lo-MEDIAN-total-loc}{44,028.00}
\DefMacro{corpus-lo-MEDIAN-code-loc}{42,295.00}
\DefMacro{corpus-lo-STDEV-num-docs}{0.00}
\DefMacro{corpus-lo-STDEV-spec-loc}{0.00}
\DefMacro{corpus-lo-STDEV-proof-loc}{0.00}
\DefMacro{corpus-lo-STDEV-comment-loc}{0.00}
\DefMacro{corpus-lo-STDEV-total-loc}{0.00}
\DefMacro{corpus-lo-STDEV-code-loc}{0.00}
\DefMacro{corpus-lo-AVG-spec-loc-per-doc}{154.53}
\DefMacro{corpus-lo-SUM-spec-loc-per-doc}{12,517}
\DefMacro{corpus-lo-MAX-spec-loc-per-doc}{1,265}
\DefMacro{corpus-lo-MIN-spec-loc-per-doc}{15}
\DefMacro{corpus-lo-MEDIAN-spec-loc-per-doc}{104.00}
\DefMacro{corpus-lo-STDEV-spec-loc-per-doc}{181.69}
\DefMacro{corpus-lo-AVG-proof-loc-per-doc}{367.63}
\DefMacro{corpus-lo-SUM-proof-loc-per-doc}{29,778}
\DefMacro{corpus-lo-MAX-proof-loc-per-doc}{2,963}
\DefMacro{corpus-lo-MIN-proof-loc-per-doc}{0}
\DefMacro{corpus-lo-MEDIAN-proof-loc-per-doc}{235.00}
\DefMacro{corpus-lo-STDEV-proof-loc-per-doc}{430.38}
\DefMacro{corpus-lo-AVG-comment-loc-per-doc}{21.40}
\DefMacro{corpus-lo-SUM-comment-loc-per-doc}{1,733}
\DefMacro{corpus-lo-MAX-comment-loc-per-doc}{320}
\DefMacro{corpus-lo-MIN-comment-loc-per-doc}{0}
\DefMacro{corpus-lo-MEDIAN-comment-loc-per-doc}{13.00}
\DefMacro{corpus-lo-STDEV-comment-loc-per-doc}{39.29}
\DefMacro{corpus-lo-AVG-total-loc-per-doc}{543.56}
\DefMacro{corpus-lo-SUM-total-loc-per-doc}{44,028}
\DefMacro{corpus-lo-MAX-total-loc-per-doc}{4,548}
\DefMacro{corpus-lo-MIN-total-loc-per-doc}{33}
\DefMacro{corpus-lo-MEDIAN-total-loc-per-doc}{342.00}
\DefMacro{corpus-lo-STDEV-total-loc-per-doc}{625.20}
\DefMacro{corpus-lo-AVG-code-loc-per-doc}{522.16}
\DefMacro{corpus-lo-SUM-code-loc-per-doc}{42,295}
\DefMacro{corpus-lo-MAX-code-loc-per-doc}{4,228}
\DefMacro{corpus-lo-MIN-code-loc-per-doc}{31}
\DefMacro{corpus-lo-MEDIAN-code-loc-per-doc}{327.00}
\DefMacro{corpus-lo-STDEV-code-loc-per-doc}{591.23}
\DefMacro{corpus-ta-num-projects}{20}
\DefMacro{corpus-ta-AVG-num-docs}{18.40}
\DefMacro{corpus-ta-AVG-spec-loc}{5,122.35}
\DefMacro{corpus-ta-AVG-proof-loc}{7,625.00}
\DefMacro{corpus-ta-AVG-comment-loc}{884.35}
\DefMacro{corpus-ta-AVG-total-loc}{13,631.70}
\DefMacro{corpus-ta-AVG-code-loc}{12,747.35}
\DefMacro{corpus-ta-SUM-num-docs}{368}
\DefMacro{corpus-ta-SUM-spec-loc}{102,447}
\DefMacro{corpus-ta-SUM-k-spec-loc}{102}
\DefMacro{corpus-ta-SUM-proof-loc}{152,500}
\DefMacro{corpus-ta-SUM-k-proof-loc}{152}
\DefMacro{corpus-ta-SUM-comment-loc}{17,687}
\DefMacro{corpus-ta-SUM-k-comment-loc}{17}
\DefMacro{corpus-ta-SUM-total-loc}{272,634}
\DefMacro{corpus-ta-SUM-k-total-loc}{272}
\DefMacro{corpus-ta-SUM-code-loc}{254,947}
\DefMacro{corpus-ta-SUM-k-code-loc}{254}
\DefMacro{corpus-ta-MAX-num-docs}{89}
\DefMacro{corpus-ta-MAX-spec-loc}{38,243}
\DefMacro{corpus-ta-MAX-k-spec-loc}{38}
\DefMacro{corpus-ta-MAX-proof-loc}{46,470}
\DefMacro{corpus-ta-MAX-k-proof-loc}{46}
\DefMacro{corpus-ta-MAX-comment-loc}{6,131}
\DefMacro{corpus-ta-MAX-k-comment-loc}{6}
\DefMacro{corpus-ta-MAX-total-loc}{90,844}
\DefMacro{corpus-ta-MAX-k-total-loc}{90}
\DefMacro{corpus-ta-MAX-code-loc}{84,713}
\DefMacro{corpus-ta-MAX-k-code-loc}{84}
\DefMacro{corpus-ta-MIN-num-docs}{1}
\DefMacro{corpus-ta-MIN-spec-loc}{70}
\DefMacro{corpus-ta-MIN-k-spec-loc}{0}
\DefMacro{corpus-ta-MIN-proof-loc}{8}
\DefMacro{corpus-ta-MIN-k-proof-loc}{0}
\DefMacro{corpus-ta-MIN-comment-loc}{16}
\DefMacro{corpus-ta-MIN-k-comment-loc}{0}
\DefMacro{corpus-ta-MIN-total-loc}{94}
\DefMacro{corpus-ta-MIN-k-total-loc}{0}
\DefMacro{corpus-ta-MIN-code-loc}{78}
\DefMacro{corpus-ta-MIN-k-code-loc}{0}
\DefMacro{corpus-ta-MEDIAN-num-docs}{12.50}
\DefMacro{corpus-ta-MEDIAN-spec-loc}{3,117.50}
\DefMacro{corpus-ta-MEDIAN-proof-loc}{3,368.00}
\DefMacro{corpus-ta-MEDIAN-comment-loc}{375.00}
\DefMacro{corpus-ta-MEDIAN-total-loc}{6,472.00}
\DefMacro{corpus-ta-MEDIAN-code-loc}{6,120.00}
\DefMacro{corpus-ta-STDEV-num-docs}{20.67}
\DefMacro{corpus-ta-STDEV-spec-loc}{8,100.97}
\DefMacro{corpus-ta-STDEV-proof-loc}{11,399.89}
\DefMacro{corpus-ta-STDEV-comment-loc}{1,399.00}
\DefMacro{corpus-ta-STDEV-total-loc}{20,541.67}
\DefMacro{corpus-ta-STDEV-code-loc}{19,157.05}
\DefMacro{corpus-ta-AVG-spec-loc-per-doc}{278.39}
\DefMacro{corpus-ta-SUM-spec-loc-per-doc}{102,447}
\DefMacro{corpus-ta-MAX-spec-loc-per-doc}{4,331}
\DefMacro{corpus-ta-MIN-spec-loc-per-doc}{1}
\DefMacro{corpus-ta-MEDIAN-spec-loc-per-doc}{159.00}
\DefMacro{corpus-ta-STDEV-spec-loc-per-doc}{378.53}
\DefMacro{corpus-ta-AVG-proof-loc-per-doc}{414.40}
\DefMacro{corpus-ta-SUM-proof-loc-per-doc}{152,500}
\DefMacro{corpus-ta-MAX-proof-loc-per-doc}{5,551}
\DefMacro{corpus-ta-MIN-proof-loc-per-doc}{0}
\DefMacro{corpus-ta-MEDIAN-proof-loc-per-doc}{251.00}
\DefMacro{corpus-ta-STDEV-proof-loc-per-doc}{577.02}
\DefMacro{corpus-ta-AVG-comment-loc-per-doc}{48.06}
\DefMacro{corpus-ta-SUM-comment-loc-per-doc}{17,687}
\DefMacro{corpus-ta-MAX-comment-loc-per-doc}{588}
\DefMacro{corpus-ta-MIN-comment-loc-per-doc}{0}
\DefMacro{corpus-ta-MEDIAN-comment-loc-per-doc}{32.00}
\DefMacro{corpus-ta-STDEV-comment-loc-per-doc}{58.76}
\DefMacro{corpus-ta-AVG-total-loc-per-doc}{740.85}
\DefMacro{corpus-ta-SUM-total-loc-per-doc}{272,634}
\DefMacro{corpus-ta-MAX-total-loc-per-doc}{5,619}
\DefMacro{corpus-ta-MIN-total-loc-per-doc}{1}
\DefMacro{corpus-ta-MEDIAN-total-loc-per-doc}{481.50}
\DefMacro{corpus-ta-STDEV-total-loc-per-doc}{827.33}
\DefMacro{corpus-ta-AVG-code-loc-per-doc}{692.79}
\DefMacro{corpus-ta-SUM-code-loc-per-doc}{254,947}
\DefMacro{corpus-ta-MAX-code-loc-per-doc}{5,560}
\DefMacro{corpus-ta-MIN-code-loc-per-doc}{1}
\DefMacro{corpus-ta-MEDIAN-code-loc-per-doc}{445.50}
\DefMacro{corpus-ta-STDEV-code-loc-per-doc}{790.57}
\DefMacro{corpus-allgroup-num-projects}{21}
\DefMacro{corpus-allgroup-AVG-num-docs}{21.38}
\DefMacro{corpus-allgroup-AVG-spec-loc}{5,474.48}
\DefMacro{corpus-allgroup-AVG-proof-loc}{8,679.90}
\DefMacro{corpus-allgroup-AVG-comment-loc}{924.76}
\DefMacro{corpus-allgroup-AVG-total-loc}{15,079.14}
\DefMacro{corpus-allgroup-AVG-code-loc}{14,154.38}
\DefMacro{corpus-allgroup-SUM-num-docs}{449}
\DefMacro{corpus-allgroup-SUM-spec-loc}{114,964}
\DefMacro{corpus-allgroup-SUM-k-spec-loc}{114}
\DefMacro{corpus-allgroup-SUM-proof-loc}{182,278}
\DefMacro{corpus-allgroup-SUM-k-proof-loc}{182}
\DefMacro{corpus-allgroup-SUM-comment-loc}{19,420}
\DefMacro{corpus-allgroup-SUM-k-comment-loc}{19}
\DefMacro{corpus-allgroup-SUM-total-loc}{316,662}
\DefMacro{corpus-allgroup-SUM-k-total-loc}{316}
\DefMacro{corpus-allgroup-SUM-code-loc}{297,242}
\DefMacro{corpus-allgroup-SUM-k-code-loc}{297}
\DefMacro{corpus-allgroup-MAX-num-docs}{89}
\DefMacro{corpus-allgroup-MAX-spec-loc}{38,243}
\DefMacro{corpus-allgroup-MAX-k-spec-loc}{38}
\DefMacro{corpus-allgroup-MAX-proof-loc}{46,470}
\DefMacro{corpus-allgroup-MAX-k-proof-loc}{46}
\DefMacro{corpus-allgroup-MAX-comment-loc}{6,131}
\DefMacro{corpus-allgroup-MAX-k-comment-loc}{6}
\DefMacro{corpus-allgroup-MAX-total-loc}{90,844}
\DefMacro{corpus-allgroup-MAX-k-total-loc}{90}
\DefMacro{corpus-allgroup-MAX-code-loc}{84,713}
\DefMacro{corpus-allgroup-MAX-k-code-loc}{84}
\DefMacro{corpus-allgroup-MIN-num-docs}{1}
\DefMacro{corpus-allgroup-MIN-spec-loc}{70}
\DefMacro{corpus-allgroup-MIN-k-spec-loc}{0}
\DefMacro{corpus-allgroup-MIN-proof-loc}{8}
\DefMacro{corpus-allgroup-MIN-k-proof-loc}{0}
\DefMacro{corpus-allgroup-MIN-comment-loc}{16}
\DefMacro{corpus-allgroup-MIN-k-comment-loc}{0}
\DefMacro{corpus-allgroup-MIN-total-loc}{94}
\DefMacro{corpus-allgroup-MIN-k-total-loc}{0}
\DefMacro{corpus-allgroup-MIN-code-loc}{78}
\DefMacro{corpus-allgroup-MIN-k-code-loc}{0}
\DefMacro{corpus-allgroup-MEDIAN-num-docs}{13.00}
\DefMacro{corpus-allgroup-MEDIAN-spec-loc}{3,298.00}
\DefMacro{corpus-allgroup-MEDIAN-proof-loc}{3,503.00}
\DefMacro{corpus-allgroup-MEDIAN-comment-loc}{394.00}
\DefMacro{corpus-allgroup-MEDIAN-total-loc}{6,689.00}
\DefMacro{corpus-allgroup-MEDIAN-code-loc}{6,451.00}
\DefMacro{corpus-allgroup-STDEV-num-docs}{24.18}
\DefMacro{corpus-allgroup-STDEV-spec-loc}{8,061.05}
\DefMacro{corpus-allgroup-STDEV-proof-loc}{12,084.10}
\DefMacro{corpus-allgroup-STDEV-comment-loc}{1,377.19}
\DefMacro{corpus-allgroup-STDEV-total-loc}{21,065.82}
\DefMacro{corpus-allgroup-STDEV-code-loc}{19,725.91}
\DefMacro{corpus-allgroup-AVG-spec-loc-per-doc}{256.04}
\DefMacro{corpus-allgroup-SUM-spec-loc-per-doc}{114,964}
\DefMacro{corpus-allgroup-MAX-spec-loc-per-doc}{4,331}
\DefMacro{corpus-allgroup-MIN-spec-loc-per-doc}{1}
\DefMacro{corpus-allgroup-MEDIAN-spec-loc-per-doc}{151.00}
\DefMacro{corpus-allgroup-STDEV-spec-loc-per-doc}{354.48}
\DefMacro{corpus-allgroup-AVG-proof-loc-per-doc}{405.96}
\DefMacro{corpus-allgroup-SUM-proof-loc-per-doc}{182,278}
\DefMacro{corpus-allgroup-MAX-proof-loc-per-doc}{5,551}
\DefMacro{corpus-allgroup-MIN-proof-loc-per-doc}{0}
\DefMacro{corpus-allgroup-MEDIAN-proof-loc-per-doc}{247.00}
\DefMacro{corpus-allgroup-STDEV-proof-loc-per-doc}{553.74}
\DefMacro{corpus-allgroup-AVG-comment-loc-per-doc}{43.25}
\DefMacro{corpus-allgroup-SUM-comment-loc-per-doc}{19,420}
\DefMacro{corpus-allgroup-MAX-comment-loc-per-doc}{588}
\DefMacro{corpus-allgroup-MIN-comment-loc-per-doc}{0}
\DefMacro{corpus-allgroup-MEDIAN-comment-loc-per-doc}{26.00}
\DefMacro{corpus-allgroup-STDEV-comment-loc-per-doc}{56.69}
\DefMacro{corpus-allgroup-AVG-total-loc-per-doc}{705.26}
\DefMacro{corpus-allgroup-SUM-total-loc-per-doc}{316,662}
\DefMacro{corpus-allgroup-MAX-total-loc-per-doc}{5,619}
\DefMacro{corpus-allgroup-MIN-total-loc-per-doc}{1}
\DefMacro{corpus-allgroup-MEDIAN-total-loc-per-doc}{462.00}
\DefMacro{corpus-allgroup-STDEV-total-loc-per-doc}{798.29}
\DefMacro{corpus-allgroup-AVG-code-loc-per-doc}{662.01}
\DefMacro{corpus-allgroup-SUM-code-loc-per-doc}{297,242}
\DefMacro{corpus-allgroup-MAX-code-loc-per-doc}{5,560}
\DefMacro{corpus-allgroup-MIN-code-loc-per-doc}{1}
\DefMacro{corpus-allgroup-MEDIAN-code-loc-per-doc}{428.00}
\DefMacro{corpus-allgroup-STDEV-code-loc-per-doc}{761.32}

%% file: tables/numbers-results-ln-lo.tex
\DefMacro{lo-train-lo-25}{105}
\DefMacro{lo-train-lo-5}{223}
\DefMacro{lo-train-lo-75}{505}
\DefMacro{lo-train-lo}{580}

%% file: tables/numbers-results-ln-lo-t1.tex
\DefMacro{results-ln-lo-t1-ln-s+bsexpl1+attn+copy_t1-val-AVG-BLEU-4}{40.8}
\DefMacro{results-ln-lo-t1-ln-s+bsexpl1+attn+copy_t1-val-AVG-char-acc}{30.9\%}
\DefMacro{results-ln-lo-t1-ln-s+bsexpl1+attn+copy_t1-val-AVG-count-top-1}{108500.0\%}
\DefMacro{results-ln-lo-t1-ln-s+bsexpl1+attn+copy_t1-val-AVG-count-top-2}{108500.0\%}
\DefMacro{results-ln-lo-t1-ln-s+bsexpl1+attn+copy_t1-val-AVG-count-top-3}{108500.0\%}
\DefMacro{results-ln-lo-t1-ln-s+bsexpl1+attn+copy_t1-val-AVG-count-top-5}{108500.0\%}
\DefMacro{results-ln-lo-t1-ln-s+bsexpl1+attn+copy_t1-val-AVG-frag-acc}{16.6\%}
\DefMacro{results-ln-lo-t1-ln-s+bsexpl1+attn+copy_t1-val-AVG-full-acc-top-1}{4.4\%}
\DefMacro{results-ln-lo-t1-ln-s+bsexpl1+attn+copy_t1-val-AVG-full-acc-top-2}{7.4\%}
\DefMacro{results-ln-lo-t1-ln-s+bsexpl1+attn+copy_t1-val-AVG-full-acc-top-3}{8.9\%}
\DefMacro{results-ln-lo-t1-ln-s+bsexpl1+attn+copy_t1-val-AVG-full-acc-top-5}{10.6\%}
\DefMacro{results-ln-lo-t1-ln-s+bsexpl1+attn+copy_t1-val-AVG-full-correct-top-1}{4733.3\%}
\DefMacro{results-ln-lo-t1-ln-s+bsexpl1+attn+copy_t1-val-AVG-full-correct-top-2}{8066.7\%}
\DefMacro{results-ln-lo-t1-ln-s+bsexpl1+attn+copy_t1-val-AVG-full-correct-top-3}{9666.7\%}
\DefMacro{results-ln-lo-t1-ln-s+bsexpl1+attn+copy_t1-val-AVG-full-correct-top-5}{11466.7\%}
\DefMacro{results-ln-lo-t1-ln-s+bsexpl1+attn+copy_t1-val-MAX-BLEU-4}{42.9}
\DefMacro{results-ln-lo-t1-ln-s+bsexpl1+attn+copy_t1-val-MAX-char-acc}{33.5\%}
\DefMacro{results-ln-lo-t1-ln-s+bsexpl1+attn+copy_t1-val-MAX-count-top-1}{1085}
\DefMacro{results-ln-lo-t1-ln-s+bsexpl1+attn+copy_t1-val-MAX-count-top-2}{1085}
\DefMacro{results-ln-lo-t1-ln-s+bsexpl1+attn+copy_t1-val-MAX-count-top-3}{1085}
\DefMacro{results-ln-lo-t1-ln-s+bsexpl1+attn+copy_t1-val-MAX-count-top-5}{1085}
\DefMacro{results-ln-lo-t1-ln-s+bsexpl1+attn+copy_t1-val-MAX-frag-acc}{19.5\%}
\DefMacro{results-ln-lo-t1-ln-s+bsexpl1+attn+copy_t1-val-MAX-full-acc-top-1}{6.7\%}
\DefMacro{results-ln-lo-t1-ln-s+bsexpl1+attn+copy_t1-val-MAX-full-acc-top-2}{10.0\%}
\DefMacro{results-ln-lo-t1-ln-s+bsexpl1+attn+copy_t1-val-MAX-full-acc-top-3}{10.9\%}
\DefMacro{results-ln-lo-t1-ln-s+bsexpl1+attn+copy_t1-val-MAX-full-acc-top-5}{11.8\%}
\DefMacro{results-ln-lo-t1-ln-s+bsexpl1+attn+copy_t1-val-MAX-full-correct-top-1}{73}
\DefMacro{results-ln-lo-t1-ln-s+bsexpl1+attn+copy_t1-val-MAX-full-correct-top-2}{108}
\DefMacro{results-ln-lo-t1-ln-s+bsexpl1+attn+copy_t1-val-MAX-full-correct-top-3}{118}
\DefMacro{results-ln-lo-t1-ln-s+bsexpl1+attn+copy_t1-val-MAX-full-correct-top-5}{128}
\DefMacro{results-ln-lo-t1-ln-s+bsexpl1+attn+copy_t1-val-MEDIAN-BLEU-4}{40.1}
\DefMacro{results-ln-lo-t1-ln-s+bsexpl1+attn+copy_t1-val-MEDIAN-char-acc}{29.7\%}
\DefMacro{results-ln-lo-t1-ln-s+bsexpl1+attn+copy_t1-val-MEDIAN-count-top-1}{108500.0\%}
\DefMacro{results-ln-lo-t1-ln-s+bsexpl1+attn+copy_t1-val-MEDIAN-count-top-2}{108500.0\%}
\DefMacro{results-ln-lo-t1-ln-s+bsexpl1+attn+copy_t1-val-MEDIAN-count-top-3}{108500.0\%}
\DefMacro{results-ln-lo-t1-ln-s+bsexpl1+attn+copy_t1-val-MEDIAN-count-top-5}{108500.0\%}
\DefMacro{results-ln-lo-t1-ln-s+bsexpl1+attn+copy_t1-val-MEDIAN-frag-acc}{15.4\%}
\DefMacro{results-ln-lo-t1-ln-s+bsexpl1+attn+copy_t1-val-MEDIAN-full-acc-top-1}{3.2\%}
\DefMacro{results-ln-lo-t1-ln-s+bsexpl1+attn+copy_t1-val-MEDIAN-full-acc-top-2}{6.5\%}
\DefMacro{results-ln-lo-t1-ln-s+bsexpl1+attn+copy_t1-val-MEDIAN-full-acc-top-3}{8.1\%}
\DefMacro{results-ln-lo-t1-ln-s+bsexpl1+attn+copy_t1-val-MEDIAN-full-acc-top-5}{10.5\%}
\DefMacro{results-ln-lo-t1-ln-s+bsexpl1+attn+copy_t1-val-MEDIAN-full-correct-top-1}{3500.0\%}
\DefMacro{results-ln-lo-t1-ln-s+bsexpl1+attn+copy_t1-val-MEDIAN-full-correct-top-2}{7100.0\%}
\DefMacro{results-ln-lo-t1-ln-s+bsexpl1+attn+copy_t1-val-MEDIAN-full-correct-top-3}{8800.0\%}
\DefMacro{results-ln-lo-t1-ln-s+bsexpl1+attn+copy_t1-val-MEDIAN-full-correct-top-5}{11400.0\%}
\DefMacro{results-ln-lo-t1-ln-s+bsexpl1+attn+copy_t1-val-MIN-BLEU-4}{39.4}
\DefMacro{results-ln-lo-t1-ln-s+bsexpl1+attn+copy_t1-val-MIN-char-acc}{29.3\%}
\DefMacro{results-ln-lo-t1-ln-s+bsexpl1+attn+copy_t1-val-MIN-count-top-1}{1085}
\DefMacro{results-ln-lo-t1-ln-s+bsexpl1+attn+copy_t1-val-MIN-count-top-2}{1085}
\DefMacro{results-ln-lo-t1-ln-s+bsexpl1+attn+copy_t1-val-MIN-count-top-3}{1085}
\DefMacro{results-ln-lo-t1-ln-s+bsexpl1+attn+copy_t1-val-MIN-count-top-5}{1085}
\DefMacro{results-ln-lo-t1-ln-s+bsexpl1+attn+copy_t1-val-MIN-frag-acc}{14.8\%}
\DefMacro{results-ln-lo-t1-ln-s+bsexpl1+attn+copy_t1-val-MIN-full-acc-top-1}{3.1\%}
\DefMacro{results-ln-lo-t1-ln-s+bsexpl1+attn+copy_t1-val-MIN-full-acc-top-2}{5.8\%}
\DefMacro{results-ln-lo-t1-ln-s+bsexpl1+attn+copy_t1-val-MIN-full-acc-top-3}{7.7\%}
\DefMacro{results-ln-lo-t1-ln-s+bsexpl1+attn+copy_t1-val-MIN-full-acc-top-5}{9.4\%}
\DefMacro{results-ln-lo-t1-ln-s+bsexpl1+attn+copy_t1-val-MIN-full-correct-top-1}{34}
\DefMacro{results-ln-lo-t1-ln-s+bsexpl1+attn+copy_t1-val-MIN-full-correct-top-2}{63}
\DefMacro{results-ln-lo-t1-ln-s+bsexpl1+attn+copy_t1-val-MIN-full-correct-top-3}{84}
\DefMacro{results-ln-lo-t1-ln-s+bsexpl1+attn+copy_t1-val-MIN-full-correct-top-5}{102}
\DefMacro{results-ln-lo-t1-ln-s+bsexpl1+attn+copy_t1-val-STDEV-BLEU-4}{1.5}
\DefMacro{results-ln-lo-t1-ln-s+bsexpl1+attn+copy_t1-val-STDEV-char-acc}{1.9\%}
\DefMacro{results-ln-lo-t1-ln-s+bsexpl1+attn+copy_t1-val-STDEV-count-top-1}{0.0\%}
\DefMacro{results-ln-lo-t1-ln-s+bsexpl1+attn+copy_t1-val-STDEV-count-top-2}{0.0\%}
\DefMacro{results-ln-lo-t1-ln-s+bsexpl1+attn+copy_t1-val-STDEV-count-top-3}{0.0\%}
\DefMacro{results-ln-lo-t1-ln-s+bsexpl1+attn+copy_t1-val-STDEV-count-top-5}{0.0\%}
\DefMacro{results-ln-lo-t1-ln-s+bsexpl1+attn+copy_t1-val-STDEV-frag-acc}{2.1\%}
\DefMacro{results-ln-lo-t1-ln-s+bsexpl1+attn+copy_t1-val-STDEV-full-acc-top-1}{1.7\%}
\DefMacro{results-ln-lo-t1-ln-s+bsexpl1+attn+copy_t1-val-STDEV-full-acc-top-2}{1.8\%}
\DefMacro{results-ln-lo-t1-ln-s+bsexpl1+attn+copy_t1-val-STDEV-full-acc-top-3}{1.4\%}
\DefMacro{results-ln-lo-t1-ln-s+bsexpl1+attn+copy_t1-val-STDEV-full-acc-top-5}{1.0\%}
\DefMacro{results-ln-lo-t1-ln-s+bsexpl1+attn+copy_t1-val-STDEV-full-correct-top-1}{1815.4\%}
\DefMacro{results-ln-lo-t1-ln-s+bsexpl1+attn+copy_t1-val-STDEV-full-correct-top-2}{1960.2\%}
\DefMacro{results-ln-lo-t1-ln-s+bsexpl1+attn+copy_t1-val-STDEV-full-correct-top-3}{1517.3\%}
\DefMacro{results-ln-lo-t1-ln-s+bsexpl1+attn+copy_t1-val-STDEV-full-correct-top-5}{1062.5\%}
\DefMacro{results-ln-lo-t1-ln-s+bsexpl1+attn+copy_t1-val-SUM-BLEU-4}{122.5}
\DefMacro{results-ln-lo-t1-ln-s+bsexpl1+attn+copy_t1-val-SUM-char-acc}{92.6\%}
\DefMacro{results-ln-lo-t1-ln-s+bsexpl1+attn+copy_t1-val-SUM-count-top-1}{3255}
\DefMacro{results-ln-lo-t1-ln-s+bsexpl1+attn+copy_t1-val-SUM-count-top-2}{3255}
\DefMacro{results-ln-lo-t1-ln-s+bsexpl1+attn+copy_t1-val-SUM-count-top-3}{3255}
\DefMacro{results-ln-lo-t1-ln-s+bsexpl1+attn+copy_t1-val-SUM-count-top-5}{3255}
\DefMacro{results-ln-lo-t1-ln-s+bsexpl1+attn+copy_t1-val-SUM-frag-acc}{49.7\%}
\DefMacro{results-ln-lo-t1-ln-s+bsexpl1+attn+copy_t1-val-SUM-full-acc-top-1}{13.1\%}
\DefMacro{results-ln-lo-t1-ln-s+bsexpl1+attn+copy_t1-val-SUM-full-acc-top-2}{22.3\%}
\DefMacro{results-ln-lo-t1-ln-s+bsexpl1+attn+copy_t1-val-SUM-full-acc-top-3}{26.7\%}
\DefMacro{results-ln-lo-t1-ln-s+bsexpl1+attn+copy_t1-val-SUM-full-acc-top-5}{31.7\%}
\DefMacro{results-ln-lo-t1-ln-s+bsexpl1+attn+copy_t1-val-SUM-full-correct-top-1}{142}
\DefMacro{results-ln-lo-t1-ln-s+bsexpl1+attn+copy_t1-val-SUM-full-correct-top-2}{242}
\DefMacro{results-ln-lo-t1-ln-s+bsexpl1+attn+copy_t1-val-SUM-full-correct-top-3}{290}
\DefMacro{results-ln-lo-t1-ln-s+bsexpl1+attn+copy_t1-val-SUM-full-correct-top-5}{344}
\DefMacro{results-ln-lo-t1-ln-s+bsexpl1+attn+copy_t1-test-AVG-BLEU-4}{31.3}
\DefMacro{results-ln-lo-t1-ln-s+bsexpl1+attn+copy_t1-test-AVG-char-acc}{18.4\%}
\DefMacro{results-ln-lo-t1-ln-s+bsexpl1+attn+copy_t1-test-AVG-count-top-1}{116700.0\%}
\DefMacro{results-ln-lo-t1-ln-s+bsexpl1+attn+copy_t1-test-AVG-count-top-2}{116700.0\%}
\DefMacro{results-ln-lo-t1-ln-s+bsexpl1+attn+copy_t1-test-AVG-count-top-3}{116700.0\%}
\DefMacro{results-ln-lo-t1-ln-s+bsexpl1+attn+copy_t1-test-AVG-count-top-5}{116700.0\%}
\DefMacro{results-ln-lo-t1-ln-s+bsexpl1+attn+copy_t1-test-AVG-frag-acc}{18.8\%}
\DefMacro{results-ln-lo-t1-ln-s+bsexpl1+attn+copy_t1-test-AVG-full-acc-top-1}{3.1\%}
\DefMacro{results-ln-lo-t1-ln-s+bsexpl1+attn+copy_t1-test-AVG-full-acc-top-2}{4.8\%}
\DefMacro{results-ln-lo-t1-ln-s+bsexpl1+attn+copy_t1-test-AVG-full-acc-top-3}{6.0\%}
\DefMacro{results-ln-lo-t1-ln-s+bsexpl1+attn+copy_t1-test-AVG-full-acc-top-5}{7.1\%}
\DefMacro{results-ln-lo-t1-ln-s+bsexpl1+attn+copy_t1-test-AVG-full-correct-top-1}{3633.3\%}
\DefMacro{results-ln-lo-t1-ln-s+bsexpl1+attn+copy_t1-test-AVG-full-correct-top-2}{5600.0\%}
\DefMacro{results-ln-lo-t1-ln-s+bsexpl1+attn+copy_t1-test-AVG-full-correct-top-3}{6966.7\%}
\DefMacro{results-ln-lo-t1-ln-s+bsexpl1+attn+copy_t1-test-AVG-full-correct-top-5}{8266.7\%}
\DefMacro{results-ln-lo-t1-ln-s+bsexpl1+attn+copy_t1-test-MAX-BLEU-4}{33.9}
\DefMacro{results-ln-lo-t1-ln-s+bsexpl1+attn+copy_t1-test-MAX-char-acc}{19.7\%}
\DefMacro{results-ln-lo-t1-ln-s+bsexpl1+attn+copy_t1-test-MAX-count-top-1}{1167}
\DefMacro{results-ln-lo-t1-ln-s+bsexpl1+attn+copy_t1-test-MAX-count-top-2}{1167}
\DefMacro{results-ln-lo-t1-ln-s+bsexpl1+attn+copy_t1-test-MAX-count-top-3}{1167}
\DefMacro{results-ln-lo-t1-ln-s+bsexpl1+attn+copy_t1-test-MAX-count-top-5}{1167}
\DefMacro{results-ln-lo-t1-ln-s+bsexpl1+attn+copy_t1-test-MAX-frag-acc}{19.5\%}
\DefMacro{results-ln-lo-t1-ln-s+bsexpl1+attn+copy_t1-test-MAX-full-acc-top-1}{3.4\%}
\DefMacro{results-ln-lo-t1-ln-s+bsexpl1+attn+copy_t1-test-MAX-full-acc-top-2}{6.0\%}
\DefMacro{results-ln-lo-t1-ln-s+bsexpl1+attn+copy_t1-test-MAX-full-acc-top-3}{7.1\%}
\DefMacro{results-ln-lo-t1-ln-s+bsexpl1+attn+copy_t1-test-MAX-full-acc-top-5}{7.8\%}
\DefMacro{results-ln-lo-t1-ln-s+bsexpl1+attn+copy_t1-test-MAX-full-correct-top-1}{40}
\DefMacro{results-ln-lo-t1-ln-s+bsexpl1+attn+copy_t1-test-MAX-full-correct-top-2}{70}
\DefMacro{results-ln-lo-t1-ln-s+bsexpl1+attn+copy_t1-test-MAX-full-correct-top-3}{83}
\DefMacro{results-ln-lo-t1-ln-s+bsexpl1+attn+copy_t1-test-MAX-full-correct-top-5}{91}
\DefMacro{results-ln-lo-t1-ln-s+bsexpl1+attn+copy_t1-test-MEDIAN-BLEU-4}{30.8}
\DefMacro{results-ln-lo-t1-ln-s+bsexpl1+attn+copy_t1-test-MEDIAN-char-acc}{18.3\%}
\DefMacro{results-ln-lo-t1-ln-s+bsexpl1+attn+copy_t1-test-MEDIAN-count-top-1}{116700.0\%}
\DefMacro{results-ln-lo-t1-ln-s+bsexpl1+attn+copy_t1-test-MEDIAN-count-top-2}{116700.0\%}
\DefMacro{results-ln-lo-t1-ln-s+bsexpl1+attn+copy_t1-test-MEDIAN-count-top-3}{116700.0\%}
\DefMacro{results-ln-lo-t1-ln-s+bsexpl1+attn+copy_t1-test-MEDIAN-count-top-5}{116700.0\%}
\DefMacro{results-ln-lo-t1-ln-s+bsexpl1+attn+copy_t1-test-MEDIAN-frag-acc}{19.1\%}
\DefMacro{results-ln-lo-t1-ln-s+bsexpl1+attn+copy_t1-test-MEDIAN-full-acc-top-1}{3.0\%}
\DefMacro{results-ln-lo-t1-ln-s+bsexpl1+attn+copy_t1-test-MEDIAN-full-acc-top-2}{4.3\%}
\DefMacro{results-ln-lo-t1-ln-s+bsexpl1+attn+copy_t1-test-MEDIAN-full-acc-top-3}{5.7\%}
\DefMacro{results-ln-lo-t1-ln-s+bsexpl1+attn+copy_t1-test-MEDIAN-full-acc-top-5}{6.9\%}
\DefMacro{results-ln-lo-t1-ln-s+bsexpl1+attn+copy_t1-test-MEDIAN-full-correct-top-1}{3500.0\%}
\DefMacro{results-ln-lo-t1-ln-s+bsexpl1+attn+copy_t1-test-MEDIAN-full-correct-top-2}{5000.0\%}
\DefMacro{results-ln-lo-t1-ln-s+bsexpl1+attn+copy_t1-test-MEDIAN-full-correct-top-3}{6600.0\%}
\DefMacro{results-ln-lo-t1-ln-s+bsexpl1+attn+copy_t1-test-MEDIAN-full-correct-top-5}{8100.0\%}
\DefMacro{results-ln-lo-t1-ln-s+bsexpl1+attn+copy_t1-test-MIN-BLEU-4}{29.2}
\DefMacro{results-ln-lo-t1-ln-s+bsexpl1+attn+copy_t1-test-MIN-char-acc}{17.2\%}
\DefMacro{results-ln-lo-t1-ln-s+bsexpl1+attn+copy_t1-test-MIN-count-top-1}{1167}
\DefMacro{results-ln-lo-t1-ln-s+bsexpl1+attn+copy_t1-test-MIN-count-top-2}{1167}
\DefMacro{results-ln-lo-t1-ln-s+bsexpl1+attn+copy_t1-test-MIN-count-top-3}{1167}
\DefMacro{results-ln-lo-t1-ln-s+bsexpl1+attn+copy_t1-test-MIN-count-top-5}{1167}
\DefMacro{results-ln-lo-t1-ln-s+bsexpl1+attn+copy_t1-test-MIN-frag-acc}{18.0\%}
\DefMacro{results-ln-lo-t1-ln-s+bsexpl1+attn+copy_t1-test-MIN-full-acc-top-1}{2.9\%}
\DefMacro{results-ln-lo-t1-ln-s+bsexpl1+attn+copy_t1-test-MIN-full-acc-top-2}{4.1\%}
\DefMacro{results-ln-lo-t1-ln-s+bsexpl1+attn+copy_t1-test-MIN-full-acc-top-3}{5.1\%}
\DefMacro{results-ln-lo-t1-ln-s+bsexpl1+attn+copy_t1-test-MIN-full-acc-top-5}{6.5\%}
\DefMacro{results-ln-lo-t1-ln-s+bsexpl1+attn+copy_t1-test-MIN-full-correct-top-1}{34}
\DefMacro{results-ln-lo-t1-ln-s+bsexpl1+attn+copy_t1-test-MIN-full-correct-top-2}{48}
\DefMacro{results-ln-lo-t1-ln-s+bsexpl1+attn+copy_t1-test-MIN-full-correct-top-3}{60}
\DefMacro{results-ln-lo-t1-ln-s+bsexpl1+attn+copy_t1-test-MIN-full-correct-top-5}{76}
\DefMacro{results-ln-lo-t1-ln-s+bsexpl1+attn+copy_t1-test-STDEV-BLEU-4}{2.0}
\DefMacro{results-ln-lo-t1-ln-s+bsexpl1+attn+copy_t1-test-STDEV-char-acc}{1.0\%}
\DefMacro{results-ln-lo-t1-ln-s+bsexpl1+attn+copy_t1-test-STDEV-count-top-1}{0.0\%}
\DefMacro{results-ln-lo-t1-ln-s+bsexpl1+attn+copy_t1-test-STDEV-count-top-2}{0.0\%}
\DefMacro{results-ln-lo-t1-ln-s+bsexpl1+attn+copy_t1-test-STDEV-count-top-3}{0.0\%}
\DefMacro{results-ln-lo-t1-ln-s+bsexpl1+attn+copy_t1-test-STDEV-count-top-5}{0.0\%}
\DefMacro{results-ln-lo-t1-ln-s+bsexpl1+attn+copy_t1-test-STDEV-frag-acc}{0.6\%}
\DefMacro{results-ln-lo-t1-ln-s+bsexpl1+attn+copy_t1-test-STDEV-full-acc-top-1}{0.2\%}
\DefMacro{results-ln-lo-t1-ln-s+bsexpl1+attn+copy_t1-test-STDEV-full-acc-top-2}{0.9\%}
\DefMacro{results-ln-lo-t1-ln-s+bsexpl1+attn+copy_t1-test-STDEV-full-acc-top-3}{0.8\%}
\DefMacro{results-ln-lo-t1-ln-s+bsexpl1+attn+copy_t1-test-STDEV-full-acc-top-5}{0.5\%}
\DefMacro{results-ln-lo-t1-ln-s+bsexpl1+attn+copy_t1-test-STDEV-full-correct-top-1}{262.5\%}
\DefMacro{results-ln-lo-t1-ln-s+bsexpl1+attn+copy_t1-test-STDEV-full-correct-top-2}{993.3\%}
\DefMacro{results-ln-lo-t1-ln-s+bsexpl1+attn+copy_t1-test-STDEV-full-correct-top-3}{974.1\%}
\DefMacro{results-ln-lo-t1-ln-s+bsexpl1+attn+copy_t1-test-STDEV-full-correct-top-5}{623.6\%}
\DefMacro{results-ln-lo-t1-ln-s+bsexpl1+attn+copy_t1-test-SUM-BLEU-4}{93.8}
\DefMacro{results-ln-lo-t1-ln-s+bsexpl1+attn+copy_t1-test-SUM-char-acc}{55.1\%}
\DefMacro{results-ln-lo-t1-ln-s+bsexpl1+attn+copy_t1-test-SUM-count-top-1}{3501}
\DefMacro{results-ln-lo-t1-ln-s+bsexpl1+attn+copy_t1-test-SUM-count-top-2}{3501}
\DefMacro{results-ln-lo-t1-ln-s+bsexpl1+attn+copy_t1-test-SUM-count-top-3}{3501}
\DefMacro{results-ln-lo-t1-ln-s+bsexpl1+attn+copy_t1-test-SUM-count-top-5}{3501}
\DefMacro{results-ln-lo-t1-ln-s+bsexpl1+attn+copy_t1-test-SUM-frag-acc}{56.5\%}
\DefMacro{results-ln-lo-t1-ln-s+bsexpl1+attn+copy_t1-test-SUM-full-acc-top-1}{9.3\%}
\DefMacro{results-ln-lo-t1-ln-s+bsexpl1+attn+copy_t1-test-SUM-full-acc-top-2}{14.4\%}
\DefMacro{results-ln-lo-t1-ln-s+bsexpl1+attn+copy_t1-test-SUM-full-acc-top-3}{17.9\%}
\DefMacro{results-ln-lo-t1-ln-s+bsexpl1+attn+copy_t1-test-SUM-full-acc-top-5}{21.3\%}
\DefMacro{results-ln-lo-t1-ln-s+bsexpl1+attn+copy_t1-test-SUM-full-correct-top-1}{109}
\DefMacro{results-ln-lo-t1-ln-s+bsexpl1+attn+copy_t1-test-SUM-full-correct-top-2}{168}
\DefMacro{results-ln-lo-t1-ln-s+bsexpl1+attn+copy_t1-test-SUM-full-correct-top-3}{209}
\DefMacro{results-ln-lo-t1-ln-s+bsexpl1+attn+copy_t1-test-SUM-full-correct-top-5}{248}
\DefMacro{results-ln-lo-lo-25-ln-s+bsexpl1+attn+copy_t1-val-AVG-BLEU-4}{47.3}
\DefMacro{results-ln-lo-lo-25-ln-s+bsexpl1+attn+copy_t1-val-AVG-char-acc}{31.2\%}
\DefMacro{results-ln-lo-lo-25-ln-s+bsexpl1+attn+copy_t1-val-AVG-count-top-1}{14400.0\%}
\DefMacro{results-ln-lo-lo-25-ln-s+bsexpl1+attn+copy_t1-val-AVG-count-top-2}{14400.0\%}
\DefMacro{results-ln-lo-lo-25-ln-s+bsexpl1+attn+copy_t1-val-AVG-count-top-3}{14400.0\%}
\DefMacro{results-ln-lo-lo-25-ln-s+bsexpl1+attn+copy_t1-val-AVG-count-top-5}{14400.0\%}
\DefMacro{results-ln-lo-lo-25-ln-s+bsexpl1+attn+copy_t1-val-AVG-frag-acc}{35.8\%}
\DefMacro{results-ln-lo-lo-25-ln-s+bsexpl1+attn+copy_t1-val-AVG-full-acc-top-1}{3.2\%}
\DefMacro{results-ln-lo-lo-25-ln-s+bsexpl1+attn+copy_t1-val-AVG-full-acc-top-2}{5.3\%}
\DefMacro{results-ln-lo-lo-25-ln-s+bsexpl1+attn+copy_t1-val-AVG-full-acc-top-3}{6.9\%}
\DefMacro{results-ln-lo-lo-25-ln-s+bsexpl1+attn+copy_t1-val-AVG-full-acc-top-5}{7.9\%}
\DefMacro{results-ln-lo-lo-25-ln-s+bsexpl1+attn+copy_t1-val-AVG-full-correct-top-1}{466.7\%}
\DefMacro{results-ln-lo-lo-25-ln-s+bsexpl1+attn+copy_t1-val-AVG-full-correct-top-2}{766.7\%}
\DefMacro{results-ln-lo-lo-25-ln-s+bsexpl1+attn+copy_t1-val-AVG-full-correct-top-3}{1000.0\%}
\DefMacro{results-ln-lo-lo-25-ln-s+bsexpl1+attn+copy_t1-val-AVG-full-correct-top-5}{1133.3\%}
\DefMacro{results-ln-lo-lo-25-ln-s+bsexpl1+attn+copy_t1-val-MAX-BLEU-4}{49.2}
\DefMacro{results-ln-lo-lo-25-ln-s+bsexpl1+attn+copy_t1-val-MAX-char-acc}{33.3\%}
\DefMacro{results-ln-lo-lo-25-ln-s+bsexpl1+attn+copy_t1-val-MAX-count-top-1}{144}
\DefMacro{results-ln-lo-lo-25-ln-s+bsexpl1+attn+copy_t1-val-MAX-count-top-2}{144}
\DefMacro{results-ln-lo-lo-25-ln-s+bsexpl1+attn+copy_t1-val-MAX-count-top-3}{144}
\DefMacro{results-ln-lo-lo-25-ln-s+bsexpl1+attn+copy_t1-val-MAX-count-top-5}{144}
\DefMacro{results-ln-lo-lo-25-ln-s+bsexpl1+attn+copy_t1-val-MAX-frag-acc}{36.0\%}
\DefMacro{results-ln-lo-lo-25-ln-s+bsexpl1+attn+copy_t1-val-MAX-full-acc-top-1}{5.6\%}
\DefMacro{results-ln-lo-lo-25-ln-s+bsexpl1+attn+copy_t1-val-MAX-full-acc-top-2}{7.6\%}
\DefMacro{results-ln-lo-lo-25-ln-s+bsexpl1+attn+copy_t1-val-MAX-full-acc-top-3}{8.3\%}
\DefMacro{results-ln-lo-lo-25-ln-s+bsexpl1+attn+copy_t1-val-MAX-full-acc-top-5}{8.3\%}
\DefMacro{results-ln-lo-lo-25-ln-s+bsexpl1+attn+copy_t1-val-MAX-full-correct-top-1}{8}
\DefMacro{results-ln-lo-lo-25-ln-s+bsexpl1+attn+copy_t1-val-MAX-full-correct-top-2}{11}
\DefMacro{results-ln-lo-lo-25-ln-s+bsexpl1+attn+copy_t1-val-MAX-full-correct-top-3}{12}
\DefMacro{results-ln-lo-lo-25-ln-s+bsexpl1+attn+copy_t1-val-MAX-full-correct-top-5}{12}
\DefMacro{results-ln-lo-lo-25-ln-s+bsexpl1+attn+copy_t1-val-MEDIAN-BLEU-4}{47.0}
\DefMacro{results-ln-lo-lo-25-ln-s+bsexpl1+attn+copy_t1-val-MEDIAN-char-acc}{33.0\%}
\DefMacro{results-ln-lo-lo-25-ln-s+bsexpl1+attn+copy_t1-val-MEDIAN-count-top-1}{14400.0\%}
\DefMacro{results-ln-lo-lo-25-ln-s+bsexpl1+attn+copy_t1-val-MEDIAN-count-top-2}{14400.0\%}
\DefMacro{results-ln-lo-lo-25-ln-s+bsexpl1+attn+copy_t1-val-MEDIAN-count-top-3}{14400.0\%}
\DefMacro{results-ln-lo-lo-25-ln-s+bsexpl1+attn+copy_t1-val-MEDIAN-count-top-5}{14400.0\%}
\DefMacro{results-ln-lo-lo-25-ln-s+bsexpl1+attn+copy_t1-val-MEDIAN-frag-acc}{35.7\%}
\DefMacro{results-ln-lo-lo-25-ln-s+bsexpl1+attn+copy_t1-val-MEDIAN-full-acc-top-1}{2.1\%}
\DefMacro{results-ln-lo-lo-25-ln-s+bsexpl1+attn+copy_t1-val-MEDIAN-full-acc-top-2}{4.2\%}
\DefMacro{results-ln-lo-lo-25-ln-s+bsexpl1+attn+copy_t1-val-MEDIAN-full-acc-top-3}{6.9\%}
\DefMacro{results-ln-lo-lo-25-ln-s+bsexpl1+attn+copy_t1-val-MEDIAN-full-acc-top-5}{7.6\%}
\DefMacro{results-ln-lo-lo-25-ln-s+bsexpl1+attn+copy_t1-val-MEDIAN-full-correct-top-1}{300.0\%}
\DefMacro{results-ln-lo-lo-25-ln-s+bsexpl1+attn+copy_t1-val-MEDIAN-full-correct-top-2}{600.0\%}
\DefMacro{results-ln-lo-lo-25-ln-s+bsexpl1+attn+copy_t1-val-MEDIAN-full-correct-top-3}{1000.0\%}
\DefMacro{results-ln-lo-lo-25-ln-s+bsexpl1+attn+copy_t1-val-MEDIAN-full-correct-top-5}{1100.0\%}
\DefMacro{results-ln-lo-lo-25-ln-s+bsexpl1+attn+copy_t1-val-MIN-BLEU-4}{45.6}
\DefMacro{results-ln-lo-lo-25-ln-s+bsexpl1+attn+copy_t1-val-MIN-char-acc}{27.3\%}
\DefMacro{results-ln-lo-lo-25-ln-s+bsexpl1+attn+copy_t1-val-MIN-count-top-1}{144}
\DefMacro{results-ln-lo-lo-25-ln-s+bsexpl1+attn+copy_t1-val-MIN-count-top-2}{144}
\DefMacro{results-ln-lo-lo-25-ln-s+bsexpl1+attn+copy_t1-val-MIN-count-top-3}{144}
\DefMacro{results-ln-lo-lo-25-ln-s+bsexpl1+attn+copy_t1-val-MIN-count-top-5}{144}
\DefMacro{results-ln-lo-lo-25-ln-s+bsexpl1+attn+copy_t1-val-MIN-frag-acc}{35.6\%}
\DefMacro{results-ln-lo-lo-25-ln-s+bsexpl1+attn+copy_t1-val-MIN-full-acc-top-1}{2.1\%}
\DefMacro{results-ln-lo-lo-25-ln-s+bsexpl1+attn+copy_t1-val-MIN-full-acc-top-2}{4.2\%}
\DefMacro{results-ln-lo-lo-25-ln-s+bsexpl1+attn+copy_t1-val-MIN-full-acc-top-3}{5.6\%}
\DefMacro{results-ln-lo-lo-25-ln-s+bsexpl1+attn+copy_t1-val-MIN-full-acc-top-5}{7.6\%}
\DefMacro{results-ln-lo-lo-25-ln-s+bsexpl1+attn+copy_t1-val-MIN-full-correct-top-1}{3}
\DefMacro{results-ln-lo-lo-25-ln-s+bsexpl1+attn+copy_t1-val-MIN-full-correct-top-2}{6}
\DefMacro{results-ln-lo-lo-25-ln-s+bsexpl1+attn+copy_t1-val-MIN-full-correct-top-3}{8}
\DefMacro{results-ln-lo-lo-25-ln-s+bsexpl1+attn+copy_t1-val-MIN-full-correct-top-5}{11}
\DefMacro{results-ln-lo-lo-25-ln-s+bsexpl1+attn+copy_t1-val-STDEV-BLEU-4}{1.5}
\DefMacro{results-ln-lo-lo-25-ln-s+bsexpl1+attn+copy_t1-val-STDEV-char-acc}{2.8\%}
\DefMacro{results-ln-lo-lo-25-ln-s+bsexpl1+attn+copy_t1-val-STDEV-count-top-1}{0.0\%}
\DefMacro{results-ln-lo-lo-25-ln-s+bsexpl1+attn+copy_t1-val-STDEV-count-top-2}{0.0\%}
\DefMacro{results-ln-lo-lo-25-ln-s+bsexpl1+attn+copy_t1-val-STDEV-count-top-3}{0.0\%}
\DefMacro{results-ln-lo-lo-25-ln-s+bsexpl1+attn+copy_t1-val-STDEV-count-top-5}{0.0\%}
\DefMacro{results-ln-lo-lo-25-ln-s+bsexpl1+attn+copy_t1-val-STDEV-frag-acc}{0.2\%}
\DefMacro{results-ln-lo-lo-25-ln-s+bsexpl1+attn+copy_t1-val-STDEV-full-acc-top-1}{1.6\%}
\DefMacro{results-ln-lo-lo-25-ln-s+bsexpl1+attn+copy_t1-val-STDEV-full-acc-top-2}{1.6\%}
\DefMacro{results-ln-lo-lo-25-ln-s+bsexpl1+attn+copy_t1-val-STDEV-full-acc-top-3}{1.1\%}
\DefMacro{results-ln-lo-lo-25-ln-s+bsexpl1+attn+copy_t1-val-STDEV-full-acc-top-5}{0.3\%}
\DefMacro{results-ln-lo-lo-25-ln-s+bsexpl1+attn+copy_t1-val-STDEV-full-correct-top-1}{235.7\%}
\DefMacro{results-ln-lo-lo-25-ln-s+bsexpl1+attn+copy_t1-val-STDEV-full-correct-top-2}{235.7\%}
\DefMacro{results-ln-lo-lo-25-ln-s+bsexpl1+attn+copy_t1-val-STDEV-full-correct-top-3}{163.3\%}
\DefMacro{results-ln-lo-lo-25-ln-s+bsexpl1+attn+copy_t1-val-STDEV-full-correct-top-5}{47.1\%}
\DefMacro{results-ln-lo-lo-25-ln-s+bsexpl1+attn+copy_t1-val-SUM-BLEU-4}{141.8}
\DefMacro{results-ln-lo-lo-25-ln-s+bsexpl1+attn+copy_t1-val-SUM-char-acc}{93.7\%}
\DefMacro{results-ln-lo-lo-25-ln-s+bsexpl1+attn+copy_t1-val-SUM-count-top-1}{432}
\DefMacro{results-ln-lo-lo-25-ln-s+bsexpl1+attn+copy_t1-val-SUM-count-top-2}{432}
\DefMacro{results-ln-lo-lo-25-ln-s+bsexpl1+attn+copy_t1-val-SUM-count-top-3}{432}
\DefMacro{results-ln-lo-lo-25-ln-s+bsexpl1+attn+copy_t1-val-SUM-count-top-5}{432}
\DefMacro{results-ln-lo-lo-25-ln-s+bsexpl1+attn+copy_t1-val-SUM-frag-acc}{107.3\%}
\DefMacro{results-ln-lo-lo-25-ln-s+bsexpl1+attn+copy_t1-val-SUM-full-acc-top-1}{9.7\%}
\DefMacro{results-ln-lo-lo-25-ln-s+bsexpl1+attn+copy_t1-val-SUM-full-acc-top-2}{16.0\%}
\DefMacro{results-ln-lo-lo-25-ln-s+bsexpl1+attn+copy_t1-val-SUM-full-acc-top-3}{20.8\%}
\DefMacro{results-ln-lo-lo-25-ln-s+bsexpl1+attn+copy_t1-val-SUM-full-acc-top-5}{23.6\%}
\DefMacro{results-ln-lo-lo-25-ln-s+bsexpl1+attn+copy_t1-val-SUM-full-correct-top-1}{14}
\DefMacro{results-ln-lo-lo-25-ln-s+bsexpl1+attn+copy_t1-val-SUM-full-correct-top-2}{23}
\DefMacro{results-ln-lo-lo-25-ln-s+bsexpl1+attn+copy_t1-val-SUM-full-correct-top-3}{30}
\DefMacro{results-ln-lo-lo-25-ln-s+bsexpl1+attn+copy_t1-val-SUM-full-correct-top-5}{34}
\DefMacro{results-ln-lo-lo-25-ln-s+bsexpl1+attn+copy_t1-test-AVG-BLEU-4}{31.2}
\DefMacro{results-ln-lo-lo-25-ln-s+bsexpl1+attn+copy_t1-test-AVG-char-acc}{19.5\%}
\DefMacro{results-ln-lo-lo-25-ln-s+bsexpl1+attn+copy_t1-test-AVG-count-top-1}{116700.0\%}
\DefMacro{results-ln-lo-lo-25-ln-s+bsexpl1+attn+copy_t1-test-AVG-count-top-2}{116700.0\%}
\DefMacro{results-ln-lo-lo-25-ln-s+bsexpl1+attn+copy_t1-test-AVG-count-top-3}{116700.0\%}
\DefMacro{results-ln-lo-lo-25-ln-s+bsexpl1+attn+copy_t1-test-AVG-count-top-5}{116700.0\%}
\DefMacro{results-ln-lo-lo-25-ln-s+bsexpl1+attn+copy_t1-test-AVG-frag-acc}{19.9\%}
\DefMacro{results-ln-lo-lo-25-ln-s+bsexpl1+attn+copy_t1-test-AVG-full-acc-top-1}{2.4\%}
\DefMacro{results-ln-lo-lo-25-ln-s+bsexpl1+attn+copy_t1-test-AVG-full-acc-top-2}{3.2\%}
\DefMacro{results-ln-lo-lo-25-ln-s+bsexpl1+attn+copy_t1-test-AVG-full-acc-top-3}{3.7\%}
\DefMacro{results-ln-lo-lo-25-ln-s+bsexpl1+attn+copy_t1-test-AVG-full-acc-top-5}{4.5\%}
\DefMacro{results-ln-lo-lo-25-ln-s+bsexpl1+attn+copy_t1-test-AVG-full-correct-top-1}{2833.3\%}
\DefMacro{results-ln-lo-lo-25-ln-s+bsexpl1+attn+copy_t1-test-AVG-full-correct-top-2}{3733.3\%}
\DefMacro{results-ln-lo-lo-25-ln-s+bsexpl1+attn+copy_t1-test-AVG-full-correct-top-3}{4333.3\%}
\DefMacro{results-ln-lo-lo-25-ln-s+bsexpl1+attn+copy_t1-test-AVG-full-correct-top-5}{5200.0\%}
\DefMacro{results-ln-lo-lo-25-ln-s+bsexpl1+attn+copy_t1-test-MAX-BLEU-4}{32.6}
\DefMacro{results-ln-lo-lo-25-ln-s+bsexpl1+attn+copy_t1-test-MAX-char-acc}{20.4\%}
\DefMacro{results-ln-lo-lo-25-ln-s+bsexpl1+attn+copy_t1-test-MAX-count-top-1}{1167}
\DefMacro{results-ln-lo-lo-25-ln-s+bsexpl1+attn+copy_t1-test-MAX-count-top-2}{1167}
\DefMacro{results-ln-lo-lo-25-ln-s+bsexpl1+attn+copy_t1-test-MAX-count-top-3}{1167}
\DefMacro{results-ln-lo-lo-25-ln-s+bsexpl1+attn+copy_t1-test-MAX-count-top-5}{1167}
\DefMacro{results-ln-lo-lo-25-ln-s+bsexpl1+attn+copy_t1-test-MAX-frag-acc}{20.7\%}
\DefMacro{results-ln-lo-lo-25-ln-s+bsexpl1+attn+copy_t1-test-MAX-full-acc-top-1}{3.0\%}
\DefMacro{results-ln-lo-lo-25-ln-s+bsexpl1+attn+copy_t1-test-MAX-full-acc-top-2}{3.9\%}
\DefMacro{results-ln-lo-lo-25-ln-s+bsexpl1+attn+copy_t1-test-MAX-full-acc-top-3}{4.5\%}
\DefMacro{results-ln-lo-lo-25-ln-s+bsexpl1+attn+copy_t1-test-MAX-full-acc-top-5}{5.3\%}
\DefMacro{results-ln-lo-lo-25-ln-s+bsexpl1+attn+copy_t1-test-MAX-full-correct-top-1}{35}
\DefMacro{results-ln-lo-lo-25-ln-s+bsexpl1+attn+copy_t1-test-MAX-full-correct-top-2}{46}
\DefMacro{results-ln-lo-lo-25-ln-s+bsexpl1+attn+copy_t1-test-MAX-full-correct-top-3}{53}
\DefMacro{results-ln-lo-lo-25-ln-s+bsexpl1+attn+copy_t1-test-MAX-full-correct-top-5}{62}
\DefMacro{results-ln-lo-lo-25-ln-s+bsexpl1+attn+copy_t1-test-MEDIAN-BLEU-4}{31.0}
\DefMacro{results-ln-lo-lo-25-ln-s+bsexpl1+attn+copy_t1-test-MEDIAN-char-acc}{19.5\%}
\DefMacro{results-ln-lo-lo-25-ln-s+bsexpl1+attn+copy_t1-test-MEDIAN-count-top-1}{116700.0\%}
\DefMacro{results-ln-lo-lo-25-ln-s+bsexpl1+attn+copy_t1-test-MEDIAN-count-top-2}{116700.0\%}
\DefMacro{results-ln-lo-lo-25-ln-s+bsexpl1+attn+copy_t1-test-MEDIAN-count-top-3}{116700.0\%}
\DefMacro{results-ln-lo-lo-25-ln-s+bsexpl1+attn+copy_t1-test-MEDIAN-count-top-5}{116700.0\%}
\DefMacro{results-ln-lo-lo-25-ln-s+bsexpl1+attn+copy_t1-test-MEDIAN-frag-acc}{19.7\%}
\DefMacro{results-ln-lo-lo-25-ln-s+bsexpl1+attn+copy_t1-test-MEDIAN-full-acc-top-1}{2.2\%}
\DefMacro{results-ln-lo-lo-25-ln-s+bsexpl1+attn+copy_t1-test-MEDIAN-full-acc-top-2}{3.0\%}
\DefMacro{results-ln-lo-lo-25-ln-s+bsexpl1+attn+copy_t1-test-MEDIAN-full-acc-top-3}{3.9\%}
\DefMacro{results-ln-lo-lo-25-ln-s+bsexpl1+attn+copy_t1-test-MEDIAN-full-acc-top-5}{4.5\%}
\DefMacro{results-ln-lo-lo-25-ln-s+bsexpl1+attn+copy_t1-test-MEDIAN-full-correct-top-1}{2600.0\%}
\DefMacro{results-ln-lo-lo-25-ln-s+bsexpl1+attn+copy_t1-test-MEDIAN-full-correct-top-2}{3500.0\%}
\DefMacro{results-ln-lo-lo-25-ln-s+bsexpl1+attn+copy_t1-test-MEDIAN-full-correct-top-3}{4500.0\%}
\DefMacro{results-ln-lo-lo-25-ln-s+bsexpl1+attn+copy_t1-test-MEDIAN-full-correct-top-5}{5200.0\%}
\DefMacro{results-ln-lo-lo-25-ln-s+bsexpl1+attn+copy_t1-test-MIN-BLEU-4}{29.9}
\DefMacro{results-ln-lo-lo-25-ln-s+bsexpl1+attn+copy_t1-test-MIN-char-acc}{18.8\%}
\DefMacro{results-ln-lo-lo-25-ln-s+bsexpl1+attn+copy_t1-test-MIN-count-top-1}{1167}
\DefMacro{results-ln-lo-lo-25-ln-s+bsexpl1+attn+copy_t1-test-MIN-count-top-2}{1167}
\DefMacro{results-ln-lo-lo-25-ln-s+bsexpl1+attn+copy_t1-test-MIN-count-top-3}{1167}
\DefMacro{results-ln-lo-lo-25-ln-s+bsexpl1+attn+copy_t1-test-MIN-count-top-5}{1167}
\DefMacro{results-ln-lo-lo-25-ln-s+bsexpl1+attn+copy_t1-test-MIN-frag-acc}{19.2\%}
\DefMacro{results-ln-lo-lo-25-ln-s+bsexpl1+attn+copy_t1-test-MIN-full-acc-top-1}{2.1\%}
\DefMacro{results-ln-lo-lo-25-ln-s+bsexpl1+attn+copy_t1-test-MIN-full-acc-top-2}{2.7\%}
\DefMacro{results-ln-lo-lo-25-ln-s+bsexpl1+attn+copy_t1-test-MIN-full-acc-top-3}{2.7\%}
\DefMacro{results-ln-lo-lo-25-ln-s+bsexpl1+attn+copy_t1-test-MIN-full-acc-top-5}{3.6\%}
\DefMacro{results-ln-lo-lo-25-ln-s+bsexpl1+attn+copy_t1-test-MIN-full-correct-top-1}{24}
\DefMacro{results-ln-lo-lo-25-ln-s+bsexpl1+attn+copy_t1-test-MIN-full-correct-top-2}{31}
\DefMacro{results-ln-lo-lo-25-ln-s+bsexpl1+attn+copy_t1-test-MIN-full-correct-top-3}{32}
\DefMacro{results-ln-lo-lo-25-ln-s+bsexpl1+attn+copy_t1-test-MIN-full-correct-top-5}{42}
\DefMacro{results-ln-lo-lo-25-ln-s+bsexpl1+attn+copy_t1-test-STDEV-BLEU-4}{1.1}
\DefMacro{results-ln-lo-lo-25-ln-s+bsexpl1+attn+copy_t1-test-STDEV-char-acc}{0.7\%}
\DefMacro{results-ln-lo-lo-25-ln-s+bsexpl1+attn+copy_t1-test-STDEV-count-top-1}{0.0\%}
\DefMacro{results-ln-lo-lo-25-ln-s+bsexpl1+attn+copy_t1-test-STDEV-count-top-2}{0.0\%}
\DefMacro{results-ln-lo-lo-25-ln-s+bsexpl1+attn+copy_t1-test-STDEV-count-top-3}{0.0\%}
\DefMacro{results-ln-lo-lo-25-ln-s+bsexpl1+attn+copy_t1-test-STDEV-count-top-5}{0.0\%}
\DefMacro{results-ln-lo-lo-25-ln-s+bsexpl1+attn+copy_t1-test-STDEV-frag-acc}{0.6\%}
\DefMacro{results-ln-lo-lo-25-ln-s+bsexpl1+attn+copy_t1-test-STDEV-full-acc-top-1}{0.4\%}
\DefMacro{results-ln-lo-lo-25-ln-s+bsexpl1+attn+copy_t1-test-STDEV-full-acc-top-2}{0.5\%}
\DefMacro{results-ln-lo-lo-25-ln-s+bsexpl1+attn+copy_t1-test-STDEV-full-acc-top-3}{0.7\%}
\DefMacro{results-ln-lo-lo-25-ln-s+bsexpl1+attn+copy_t1-test-STDEV-full-acc-top-5}{0.7\%}
\DefMacro{results-ln-lo-lo-25-ln-s+bsexpl1+attn+copy_t1-test-STDEV-full-correct-top-1}{478.4\%}
\DefMacro{results-ln-lo-lo-25-ln-s+bsexpl1+attn+copy_t1-test-STDEV-full-correct-top-2}{634.2\%}
\DefMacro{results-ln-lo-lo-25-ln-s+bsexpl1+attn+copy_t1-test-STDEV-full-correct-top-3}{865.4\%}
\DefMacro{results-ln-lo-lo-25-ln-s+bsexpl1+attn+copy_t1-test-STDEV-full-correct-top-5}{816.5\%}
\DefMacro{results-ln-lo-lo-25-ln-s+bsexpl1+attn+copy_t1-test-SUM-BLEU-4}{93.6}
\DefMacro{results-ln-lo-lo-25-ln-s+bsexpl1+attn+copy_t1-test-SUM-char-acc}{58.6\%}
\DefMacro{results-ln-lo-lo-25-ln-s+bsexpl1+attn+copy_t1-test-SUM-count-top-1}{3501}
\DefMacro{results-ln-lo-lo-25-ln-s+bsexpl1+attn+copy_t1-test-SUM-count-top-2}{3501}
\DefMacro{results-ln-lo-lo-25-ln-s+bsexpl1+attn+copy_t1-test-SUM-count-top-3}{3501}
\DefMacro{results-ln-lo-lo-25-ln-s+bsexpl1+attn+copy_t1-test-SUM-count-top-5}{3501}
\DefMacro{results-ln-lo-lo-25-ln-s+bsexpl1+attn+copy_t1-test-SUM-frag-acc}{59.6\%}
\DefMacro{results-ln-lo-lo-25-ln-s+bsexpl1+attn+copy_t1-test-SUM-full-acc-top-1}{7.3\%}
\DefMacro{results-ln-lo-lo-25-ln-s+bsexpl1+attn+copy_t1-test-SUM-full-acc-top-2}{9.6\%}
\DefMacro{results-ln-lo-lo-25-ln-s+bsexpl1+attn+copy_t1-test-SUM-full-acc-top-3}{11.1\%}
\DefMacro{results-ln-lo-lo-25-ln-s+bsexpl1+attn+copy_t1-test-SUM-full-acc-top-5}{13.4\%}
\DefMacro{results-ln-lo-lo-25-ln-s+bsexpl1+attn+copy_t1-test-SUM-full-correct-top-1}{85}
\DefMacro{results-ln-lo-lo-25-ln-s+bsexpl1+attn+copy_t1-test-SUM-full-correct-top-2}{112}
\DefMacro{results-ln-lo-lo-25-ln-s+bsexpl1+attn+copy_t1-test-SUM-full-correct-top-3}{130}
\DefMacro{results-ln-lo-lo-25-ln-s+bsexpl1+attn+copy_t1-test-SUM-full-correct-top-5}{156}
\DefMacro{results-ln-lo-lo-5-ln-s+bsexpl1+attn+copy_t1-val-AVG-BLEU-4}{43.3}
\DefMacro{results-ln-lo-lo-5-ln-s+bsexpl1+attn+copy_t1-val-AVG-char-acc}{29.6\%}
\DefMacro{results-ln-lo-lo-5-ln-s+bsexpl1+attn+copy_t1-val-AVG-count-top-1}{14400.0\%}
\DefMacro{results-ln-lo-lo-5-ln-s+bsexpl1+attn+copy_t1-val-AVG-count-top-2}{14400.0\%}
\DefMacro{results-ln-lo-lo-5-ln-s+bsexpl1+attn+copy_t1-val-AVG-count-top-3}{14400.0\%}
\DefMacro{results-ln-lo-lo-5-ln-s+bsexpl1+attn+copy_t1-val-AVG-count-top-5}{14400.0\%}
\DefMacro{results-ln-lo-lo-5-ln-s+bsexpl1+attn+copy_t1-val-AVG-frag-acc}{33.5\%}
\DefMacro{results-ln-lo-lo-5-ln-s+bsexpl1+attn+copy_t1-val-AVG-full-acc-top-1}{4.9\%}
\DefMacro{results-ln-lo-lo-5-ln-s+bsexpl1+attn+copy_t1-val-AVG-full-acc-top-2}{6.2\%}
\DefMacro{results-ln-lo-lo-5-ln-s+bsexpl1+attn+copy_t1-val-AVG-full-acc-top-3}{6.7\%}
\DefMacro{results-ln-lo-lo-5-ln-s+bsexpl1+attn+copy_t1-val-AVG-full-acc-top-5}{7.4\%}
\DefMacro{results-ln-lo-lo-5-ln-s+bsexpl1+attn+copy_t1-val-AVG-full-correct-top-1}{700.0\%}
\DefMacro{results-ln-lo-lo-5-ln-s+bsexpl1+attn+copy_t1-val-AVG-full-correct-top-2}{900.0\%}
\DefMacro{results-ln-lo-lo-5-ln-s+bsexpl1+attn+copy_t1-val-AVG-full-correct-top-3}{966.7\%}
\DefMacro{results-ln-lo-lo-5-ln-s+bsexpl1+attn+copy_t1-val-AVG-full-correct-top-5}{1066.7\%}
\DefMacro{results-ln-lo-lo-5-ln-s+bsexpl1+attn+copy_t1-val-MAX-BLEU-4}{45.4}
\DefMacro{results-ln-lo-lo-5-ln-s+bsexpl1+attn+copy_t1-val-MAX-char-acc}{31.3\%}
\DefMacro{results-ln-lo-lo-5-ln-s+bsexpl1+attn+copy_t1-val-MAX-count-top-1}{144}
\DefMacro{results-ln-lo-lo-5-ln-s+bsexpl1+attn+copy_t1-val-MAX-count-top-2}{144}
\DefMacro{results-ln-lo-lo-5-ln-s+bsexpl1+attn+copy_t1-val-MAX-count-top-3}{144}
\DefMacro{results-ln-lo-lo-5-ln-s+bsexpl1+attn+copy_t1-val-MAX-count-top-5}{144}
\DefMacro{results-ln-lo-lo-5-ln-s+bsexpl1+attn+copy_t1-val-MAX-frag-acc}{35.2\%}
\DefMacro{results-ln-lo-lo-5-ln-s+bsexpl1+attn+copy_t1-val-MAX-full-acc-top-1}{6.9\%}
\DefMacro{results-ln-lo-lo-5-ln-s+bsexpl1+attn+copy_t1-val-MAX-full-acc-top-2}{9.0\%}
\DefMacro{results-ln-lo-lo-5-ln-s+bsexpl1+attn+copy_t1-val-MAX-full-acc-top-3}{9.7\%}
\DefMacro{results-ln-lo-lo-5-ln-s+bsexpl1+attn+copy_t1-val-MAX-full-acc-top-5}{10.4\%}
\DefMacro{results-ln-lo-lo-5-ln-s+bsexpl1+attn+copy_t1-val-MAX-full-correct-top-1}{10}
\DefMacro{results-ln-lo-lo-5-ln-s+bsexpl1+attn+copy_t1-val-MAX-full-correct-top-2}{13}
\DefMacro{results-ln-lo-lo-5-ln-s+bsexpl1+attn+copy_t1-val-MAX-full-correct-top-3}{14}
\DefMacro{results-ln-lo-lo-5-ln-s+bsexpl1+attn+copy_t1-val-MAX-full-correct-top-5}{15}
\DefMacro{results-ln-lo-lo-5-ln-s+bsexpl1+attn+copy_t1-val-MEDIAN-BLEU-4}{43.3}
\DefMacro{results-ln-lo-lo-5-ln-s+bsexpl1+attn+copy_t1-val-MEDIAN-char-acc}{29.6\%}
\DefMacro{results-ln-lo-lo-5-ln-s+bsexpl1+attn+copy_t1-val-MEDIAN-count-top-1}{14400.0\%}
\DefMacro{results-ln-lo-lo-5-ln-s+bsexpl1+attn+copy_t1-val-MEDIAN-count-top-2}{14400.0\%}
\DefMacro{results-ln-lo-lo-5-ln-s+bsexpl1+attn+copy_t1-val-MEDIAN-count-top-3}{14400.0\%}
\DefMacro{results-ln-lo-lo-5-ln-s+bsexpl1+attn+copy_t1-val-MEDIAN-count-top-5}{14400.0\%}
\DefMacro{results-ln-lo-lo-5-ln-s+bsexpl1+attn+copy_t1-val-MEDIAN-frag-acc}{32.8\%}
\DefMacro{results-ln-lo-lo-5-ln-s+bsexpl1+attn+copy_t1-val-MEDIAN-full-acc-top-1}{4.2\%}
\DefMacro{results-ln-lo-lo-5-ln-s+bsexpl1+attn+copy_t1-val-MEDIAN-full-acc-top-2}{5.6\%}
\DefMacro{results-ln-lo-lo-5-ln-s+bsexpl1+attn+copy_t1-val-MEDIAN-full-acc-top-3}{5.6\%}
\DefMacro{results-ln-lo-lo-5-ln-s+bsexpl1+attn+copy_t1-val-MEDIAN-full-acc-top-5}{6.2\%}
\DefMacro{results-ln-lo-lo-5-ln-s+bsexpl1+attn+copy_t1-val-MEDIAN-full-correct-top-1}{600.0\%}
\DefMacro{results-ln-lo-lo-5-ln-s+bsexpl1+attn+copy_t1-val-MEDIAN-full-correct-top-2}{800.0\%}
\DefMacro{results-ln-lo-lo-5-ln-s+bsexpl1+attn+copy_t1-val-MEDIAN-full-correct-top-3}{800.0\%}
\DefMacro{results-ln-lo-lo-5-ln-s+bsexpl1+attn+copy_t1-val-MEDIAN-full-correct-top-5}{900.0\%}
\DefMacro{results-ln-lo-lo-5-ln-s+bsexpl1+attn+copy_t1-val-MIN-BLEU-4}{41.3}
\DefMacro{results-ln-lo-lo-5-ln-s+bsexpl1+attn+copy_t1-val-MIN-char-acc}{27.9\%}
\DefMacro{results-ln-lo-lo-5-ln-s+bsexpl1+attn+copy_t1-val-MIN-count-top-1}{144}
\DefMacro{results-ln-lo-lo-5-ln-s+bsexpl1+attn+copy_t1-val-MIN-count-top-2}{144}
\DefMacro{results-ln-lo-lo-5-ln-s+bsexpl1+attn+copy_t1-val-MIN-count-top-3}{144}
\DefMacro{results-ln-lo-lo-5-ln-s+bsexpl1+attn+copy_t1-val-MIN-count-top-5}{144}
\DefMacro{results-ln-lo-lo-5-ln-s+bsexpl1+attn+copy_t1-val-MIN-frag-acc}{32.5\%}
\DefMacro{results-ln-lo-lo-5-ln-s+bsexpl1+attn+copy_t1-val-MIN-full-acc-top-1}{3.5\%}
\DefMacro{results-ln-lo-lo-5-ln-s+bsexpl1+attn+copy_t1-val-MIN-full-acc-top-2}{4.2\%}
\DefMacro{results-ln-lo-lo-5-ln-s+bsexpl1+attn+copy_t1-val-MIN-full-acc-top-3}{4.9\%}
\DefMacro{results-ln-lo-lo-5-ln-s+bsexpl1+attn+copy_t1-val-MIN-full-acc-top-5}{5.6\%}
\DefMacro{results-ln-lo-lo-5-ln-s+bsexpl1+attn+copy_t1-val-MIN-full-correct-top-1}{5}
\DefMacro{results-ln-lo-lo-5-ln-s+bsexpl1+attn+copy_t1-val-MIN-full-correct-top-2}{6}
\DefMacro{results-ln-lo-lo-5-ln-s+bsexpl1+attn+copy_t1-val-MIN-full-correct-top-3}{7}
\DefMacro{results-ln-lo-lo-5-ln-s+bsexpl1+attn+copy_t1-val-MIN-full-correct-top-5}{8}
\DefMacro{results-ln-lo-lo-5-ln-s+bsexpl1+attn+copy_t1-val-STDEV-BLEU-4}{1.7}
\DefMacro{results-ln-lo-lo-5-ln-s+bsexpl1+attn+copy_t1-val-STDEV-char-acc}{1.4\%}
\DefMacro{results-ln-lo-lo-5-ln-s+bsexpl1+attn+copy_t1-val-STDEV-count-top-1}{0.0\%}
\DefMacro{results-ln-lo-lo-5-ln-s+bsexpl1+attn+copy_t1-val-STDEV-count-top-2}{0.0\%}
\DefMacro{results-ln-lo-lo-5-ln-s+bsexpl1+attn+copy_t1-val-STDEV-count-top-3}{0.0\%}
\DefMacro{results-ln-lo-lo-5-ln-s+bsexpl1+attn+copy_t1-val-STDEV-count-top-5}{0.0\%}
\DefMacro{results-ln-lo-lo-5-ln-s+bsexpl1+attn+copy_t1-val-STDEV-frag-acc}{1.2\%}
\DefMacro{results-ln-lo-lo-5-ln-s+bsexpl1+attn+copy_t1-val-STDEV-full-acc-top-1}{1.5\%}
\DefMacro{results-ln-lo-lo-5-ln-s+bsexpl1+attn+copy_t1-val-STDEV-full-acc-top-2}{2.0\%}
\DefMacro{results-ln-lo-lo-5-ln-s+bsexpl1+attn+copy_t1-val-STDEV-full-acc-top-3}{2.1\%}
\DefMacro{results-ln-lo-lo-5-ln-s+bsexpl1+attn+copy_t1-val-STDEV-full-acc-top-5}{2.1\%}
\DefMacro{results-ln-lo-lo-5-ln-s+bsexpl1+attn+copy_t1-val-STDEV-full-correct-top-1}{216.0\%}
\DefMacro{results-ln-lo-lo-5-ln-s+bsexpl1+attn+copy_t1-val-STDEV-full-correct-top-2}{294.4\%}
\DefMacro{results-ln-lo-lo-5-ln-s+bsexpl1+attn+copy_t1-val-STDEV-full-correct-top-3}{309.1\%}
\DefMacro{results-ln-lo-lo-5-ln-s+bsexpl1+attn+copy_t1-val-STDEV-full-correct-top-5}{309.1\%}
\DefMacro{results-ln-lo-lo-5-ln-s+bsexpl1+attn+copy_t1-val-SUM-BLEU-4}{130.0}
\DefMacro{results-ln-lo-lo-5-ln-s+bsexpl1+attn+copy_t1-val-SUM-char-acc}{88.9\%}
\DefMacro{results-ln-lo-lo-5-ln-s+bsexpl1+attn+copy_t1-val-SUM-count-top-1}{432}
\DefMacro{results-ln-lo-lo-5-ln-s+bsexpl1+attn+copy_t1-val-SUM-count-top-2}{432}
\DefMacro{results-ln-lo-lo-5-ln-s+bsexpl1+attn+copy_t1-val-SUM-count-top-3}{432}
\DefMacro{results-ln-lo-lo-5-ln-s+bsexpl1+attn+copy_t1-val-SUM-count-top-5}{432}
\DefMacro{results-ln-lo-lo-5-ln-s+bsexpl1+attn+copy_t1-val-SUM-frag-acc}{100.5\%}
\DefMacro{results-ln-lo-lo-5-ln-s+bsexpl1+attn+copy_t1-val-SUM-full-acc-top-1}{14.6\%}
\DefMacro{results-ln-lo-lo-5-ln-s+bsexpl1+attn+copy_t1-val-SUM-full-acc-top-2}{18.8\%}
\DefMacro{results-ln-lo-lo-5-ln-s+bsexpl1+attn+copy_t1-val-SUM-full-acc-top-3}{20.1\%}
\DefMacro{results-ln-lo-lo-5-ln-s+bsexpl1+attn+copy_t1-val-SUM-full-acc-top-5}{22.2\%}
\DefMacro{results-ln-lo-lo-5-ln-s+bsexpl1+attn+copy_t1-val-SUM-full-correct-top-1}{21}
\DefMacro{results-ln-lo-lo-5-ln-s+bsexpl1+attn+copy_t1-val-SUM-full-correct-top-2}{27}
\DefMacro{results-ln-lo-lo-5-ln-s+bsexpl1+attn+copy_t1-val-SUM-full-correct-top-3}{29}
\DefMacro{results-ln-lo-lo-5-ln-s+bsexpl1+attn+copy_t1-val-SUM-full-correct-top-5}{32}
\DefMacro{results-ln-lo-lo-5-ln-s+bsexpl1+attn+copy_t1-test-AVG-BLEU-4}{31.9}
\DefMacro{results-ln-lo-lo-5-ln-s+bsexpl1+attn+copy_t1-test-AVG-char-acc}{19.8\%}
\DefMacro{results-ln-lo-lo-5-ln-s+bsexpl1+attn+copy_t1-test-AVG-count-top-1}{116700.0\%}
\DefMacro{results-ln-lo-lo-5-ln-s+bsexpl1+attn+copy_t1-test-AVG-count-top-2}{116700.0\%}
\DefMacro{results-ln-lo-lo-5-ln-s+bsexpl1+attn+copy_t1-test-AVG-count-top-3}{116700.0\%}
\DefMacro{results-ln-lo-lo-5-ln-s+bsexpl1+attn+copy_t1-test-AVG-count-top-5}{116700.0\%}
\DefMacro{results-ln-lo-lo-5-ln-s+bsexpl1+attn+copy_t1-test-AVG-frag-acc}{20.2\%}
\DefMacro{results-ln-lo-lo-5-ln-s+bsexpl1+attn+copy_t1-test-AVG-full-acc-top-1}{2.9\%}
\DefMacro{results-ln-lo-lo-5-ln-s+bsexpl1+attn+copy_t1-test-AVG-full-acc-top-2}{3.7\%}
\DefMacro{results-ln-lo-lo-5-ln-s+bsexpl1+attn+copy_t1-test-AVG-full-acc-top-3}{4.4\%}
\DefMacro{results-ln-lo-lo-5-ln-s+bsexpl1+attn+copy_t1-test-AVG-full-acc-top-5}{5.1\%}
\DefMacro{results-ln-lo-lo-5-ln-s+bsexpl1+attn+copy_t1-test-AVG-full-correct-top-1}{3366.7\%}
\DefMacro{results-ln-lo-lo-5-ln-s+bsexpl1+attn+copy_t1-test-AVG-full-correct-top-2}{4333.3\%}
\DefMacro{results-ln-lo-lo-5-ln-s+bsexpl1+attn+copy_t1-test-AVG-full-correct-top-3}{5100.0\%}
\DefMacro{results-ln-lo-lo-5-ln-s+bsexpl1+attn+copy_t1-test-AVG-full-correct-top-5}{5900.0\%}
\DefMacro{results-ln-lo-lo-5-ln-s+bsexpl1+attn+copy_t1-test-MAX-BLEU-4}{32.5}
\DefMacro{results-ln-lo-lo-5-ln-s+bsexpl1+attn+copy_t1-test-MAX-char-acc}{20.7\%}
\DefMacro{results-ln-lo-lo-5-ln-s+bsexpl1+attn+copy_t1-test-MAX-count-top-1}{1167}
\DefMacro{results-ln-lo-lo-5-ln-s+bsexpl1+attn+copy_t1-test-MAX-count-top-2}{1167}
\DefMacro{results-ln-lo-lo-5-ln-s+bsexpl1+attn+copy_t1-test-MAX-count-top-3}{1167}
\DefMacro{results-ln-lo-lo-5-ln-s+bsexpl1+attn+copy_t1-test-MAX-count-top-5}{1167}
\DefMacro{results-ln-lo-lo-5-ln-s+bsexpl1+attn+copy_t1-test-MAX-frag-acc}{20.7\%}
\DefMacro{results-ln-lo-lo-5-ln-s+bsexpl1+attn+copy_t1-test-MAX-full-acc-top-1}{3.1\%}
\DefMacro{results-ln-lo-lo-5-ln-s+bsexpl1+attn+copy_t1-test-MAX-full-acc-top-2}{4.3\%}
\DefMacro{results-ln-lo-lo-5-ln-s+bsexpl1+attn+copy_t1-test-MAX-full-acc-top-3}{4.7\%}
\DefMacro{results-ln-lo-lo-5-ln-s+bsexpl1+attn+copy_t1-test-MAX-full-acc-top-5}{5.6\%}
\DefMacro{results-ln-lo-lo-5-ln-s+bsexpl1+attn+copy_t1-test-MAX-full-correct-top-1}{36}
\DefMacro{results-ln-lo-lo-5-ln-s+bsexpl1+attn+copy_t1-test-MAX-full-correct-top-2}{50}
\DefMacro{results-ln-lo-lo-5-ln-s+bsexpl1+attn+copy_t1-test-MAX-full-correct-top-3}{55}
\DefMacro{results-ln-lo-lo-5-ln-s+bsexpl1+attn+copy_t1-test-MAX-full-correct-top-5}{65}
\DefMacro{results-ln-lo-lo-5-ln-s+bsexpl1+attn+copy_t1-test-MEDIAN-BLEU-4}{32.0}
\DefMacro{results-ln-lo-lo-5-ln-s+bsexpl1+attn+copy_t1-test-MEDIAN-char-acc}{19.7\%}
\DefMacro{results-ln-lo-lo-5-ln-s+bsexpl1+attn+copy_t1-test-MEDIAN-count-top-1}{116700.0\%}
\DefMacro{results-ln-lo-lo-5-ln-s+bsexpl1+attn+copy_t1-test-MEDIAN-count-top-2}{116700.0\%}
\DefMacro{results-ln-lo-lo-5-ln-s+bsexpl1+attn+copy_t1-test-MEDIAN-count-top-3}{116700.0\%}
\DefMacro{results-ln-lo-lo-5-ln-s+bsexpl1+attn+copy_t1-test-MEDIAN-count-top-5}{116700.0\%}
\DefMacro{results-ln-lo-lo-5-ln-s+bsexpl1+attn+copy_t1-test-MEDIAN-frag-acc}{20.6\%}
\DefMacro{results-ln-lo-lo-5-ln-s+bsexpl1+attn+copy_t1-test-MEDIAN-full-acc-top-1}{2.9\%}
\DefMacro{results-ln-lo-lo-5-ln-s+bsexpl1+attn+copy_t1-test-MEDIAN-full-acc-top-2}{3.5\%}
\DefMacro{results-ln-lo-lo-5-ln-s+bsexpl1+attn+copy_t1-test-MEDIAN-full-acc-top-3}{4.3\%}
\DefMacro{results-ln-lo-lo-5-ln-s+bsexpl1+attn+copy_t1-test-MEDIAN-full-acc-top-5}{5.2\%}
\DefMacro{results-ln-lo-lo-5-ln-s+bsexpl1+attn+copy_t1-test-MEDIAN-full-correct-top-1}{3400.0\%}
\DefMacro{results-ln-lo-lo-5-ln-s+bsexpl1+attn+copy_t1-test-MEDIAN-full-correct-top-2}{4100.0\%}
\DefMacro{results-ln-lo-lo-5-ln-s+bsexpl1+attn+copy_t1-test-MEDIAN-full-correct-top-3}{5000.0\%}
\DefMacro{results-ln-lo-lo-5-ln-s+bsexpl1+attn+copy_t1-test-MEDIAN-full-correct-top-5}{6100.0\%}
\DefMacro{results-ln-lo-lo-5-ln-s+bsexpl1+attn+copy_t1-test-MIN-BLEU-4}{31.1}
\DefMacro{results-ln-lo-lo-5-ln-s+bsexpl1+attn+copy_t1-test-MIN-char-acc}{19.0\%}
\DefMacro{results-ln-lo-lo-5-ln-s+bsexpl1+attn+copy_t1-test-MIN-count-top-1}{1167}
\DefMacro{results-ln-lo-lo-5-ln-s+bsexpl1+attn+copy_t1-test-MIN-count-top-2}{1167}
\DefMacro{results-ln-lo-lo-5-ln-s+bsexpl1+attn+copy_t1-test-MIN-count-top-3}{1167}
\DefMacro{results-ln-lo-lo-5-ln-s+bsexpl1+attn+copy_t1-test-MIN-count-top-5}{1167}
\DefMacro{results-ln-lo-lo-5-ln-s+bsexpl1+attn+copy_t1-test-MIN-frag-acc}{19.2\%}
\DefMacro{results-ln-lo-lo-5-ln-s+bsexpl1+attn+copy_t1-test-MIN-full-acc-top-1}{2.7\%}
\DefMacro{results-ln-lo-lo-5-ln-s+bsexpl1+attn+copy_t1-test-MIN-full-acc-top-2}{3.3\%}
\DefMacro{results-ln-lo-lo-5-ln-s+bsexpl1+attn+copy_t1-test-MIN-full-acc-top-3}{4.1\%}
\DefMacro{results-ln-lo-lo-5-ln-s+bsexpl1+attn+copy_t1-test-MIN-full-acc-top-5}{4.4\%}
\DefMacro{results-ln-lo-lo-5-ln-s+bsexpl1+attn+copy_t1-test-MIN-full-correct-top-1}{31}
\DefMacro{results-ln-lo-lo-5-ln-s+bsexpl1+attn+copy_t1-test-MIN-full-correct-top-2}{39}
\DefMacro{results-ln-lo-lo-5-ln-s+bsexpl1+attn+copy_t1-test-MIN-full-correct-top-3}{48}
\DefMacro{results-ln-lo-lo-5-ln-s+bsexpl1+attn+copy_t1-test-MIN-full-correct-top-5}{51}
\DefMacro{results-ln-lo-lo-5-ln-s+bsexpl1+attn+copy_t1-test-STDEV-BLEU-4}{0.6}
\DefMacro{results-ln-lo-lo-5-ln-s+bsexpl1+attn+copy_t1-test-STDEV-char-acc}{0.7\%}
\DefMacro{results-ln-lo-lo-5-ln-s+bsexpl1+attn+copy_t1-test-STDEV-count-top-1}{0.0\%}
\DefMacro{results-ln-lo-lo-5-ln-s+bsexpl1+attn+copy_t1-test-STDEV-count-top-2}{0.0\%}
\DefMacro{results-ln-lo-lo-5-ln-s+bsexpl1+attn+copy_t1-test-STDEV-count-top-3}{0.0\%}
\DefMacro{results-ln-lo-lo-5-ln-s+bsexpl1+attn+copy_t1-test-STDEV-count-top-5}{0.0\%}
\DefMacro{results-ln-lo-lo-5-ln-s+bsexpl1+attn+copy_t1-test-STDEV-frag-acc}{0.7\%}
\DefMacro{results-ln-lo-lo-5-ln-s+bsexpl1+attn+copy_t1-test-STDEV-full-acc-top-1}{0.2\%}
\DefMacro{results-ln-lo-lo-5-ln-s+bsexpl1+attn+copy_t1-test-STDEV-full-acc-top-2}{0.4\%}
\DefMacro{results-ln-lo-lo-5-ln-s+bsexpl1+attn+copy_t1-test-STDEV-full-acc-top-3}{0.3\%}
\DefMacro{results-ln-lo-lo-5-ln-s+bsexpl1+attn+copy_t1-test-STDEV-full-acc-top-5}{0.5\%}
\DefMacro{results-ln-lo-lo-5-ln-s+bsexpl1+attn+copy_t1-test-STDEV-full-correct-top-1}{205.5\%}
\DefMacro{results-ln-lo-lo-5-ln-s+bsexpl1+attn+copy_t1-test-STDEV-full-correct-top-2}{478.4\%}
\DefMacro{results-ln-lo-lo-5-ln-s+bsexpl1+attn+copy_t1-test-STDEV-full-correct-top-3}{294.4\%}
\DefMacro{results-ln-lo-lo-5-ln-s+bsexpl1+attn+copy_t1-test-STDEV-full-correct-top-5}{588.8\%}
\DefMacro{results-ln-lo-lo-5-ln-s+bsexpl1+attn+copy_t1-test-SUM-BLEU-4}{95.6}
\DefMacro{results-ln-lo-lo-5-ln-s+bsexpl1+attn+copy_t1-test-SUM-char-acc}{59.4\%}
\DefMacro{results-ln-lo-lo-5-ln-s+bsexpl1+attn+copy_t1-test-SUM-count-top-1}{3501}
\DefMacro{results-ln-lo-lo-5-ln-s+bsexpl1+attn+copy_t1-test-SUM-count-top-2}{3501}
\DefMacro{results-ln-lo-lo-5-ln-s+bsexpl1+attn+copy_t1-test-SUM-count-top-3}{3501}
\DefMacro{results-ln-lo-lo-5-ln-s+bsexpl1+attn+copy_t1-test-SUM-count-top-5}{3501}
\DefMacro{results-ln-lo-lo-5-ln-s+bsexpl1+attn+copy_t1-test-SUM-frag-acc}{60.5\%}
\DefMacro{results-ln-lo-lo-5-ln-s+bsexpl1+attn+copy_t1-test-SUM-full-acc-top-1}{8.7\%}
\DefMacro{results-ln-lo-lo-5-ln-s+bsexpl1+attn+copy_t1-test-SUM-full-acc-top-2}{11.1\%}
\DefMacro{results-ln-lo-lo-5-ln-s+bsexpl1+attn+copy_t1-test-SUM-full-acc-top-3}{13.1\%}
\DefMacro{results-ln-lo-lo-5-ln-s+bsexpl1+attn+copy_t1-test-SUM-full-acc-top-5}{15.2\%}
\DefMacro{results-ln-lo-lo-5-ln-s+bsexpl1+attn+copy_t1-test-SUM-full-correct-top-1}{101}
\DefMacro{results-ln-lo-lo-5-ln-s+bsexpl1+attn+copy_t1-test-SUM-full-correct-top-2}{130}
\DefMacro{results-ln-lo-lo-5-ln-s+bsexpl1+attn+copy_t1-test-SUM-full-correct-top-3}{153}
\DefMacro{results-ln-lo-lo-5-ln-s+bsexpl1+attn+copy_t1-test-SUM-full-correct-top-5}{177}
\DefMacro{results-ln-lo-lo-75-ln-s+bsexpl1+attn+copy_t1-val-AVG-BLEU-4}{46.5}
\DefMacro{results-ln-lo-lo-75-ln-s+bsexpl1+attn+copy_t1-val-AVG-char-acc}{32.1\%}
\DefMacro{results-ln-lo-lo-75-ln-s+bsexpl1+attn+copy_t1-val-AVG-count-top-1}{14400.0\%}
\DefMacro{results-ln-lo-lo-75-ln-s+bsexpl1+attn+copy_t1-val-AVG-count-top-2}{14400.0\%}
\DefMacro{results-ln-lo-lo-75-ln-s+bsexpl1+attn+copy_t1-val-AVG-count-top-3}{14400.0\%}
\DefMacro{results-ln-lo-lo-75-ln-s+bsexpl1+attn+copy_t1-val-AVG-count-top-5}{14400.0\%}
\DefMacro{results-ln-lo-lo-75-ln-s+bsexpl1+attn+copy_t1-val-AVG-frag-acc}{35.6\%}
\DefMacro{results-ln-lo-lo-75-ln-s+bsexpl1+attn+copy_t1-val-AVG-full-acc-top-1}{4.4\%}
\DefMacro{results-ln-lo-lo-75-ln-s+bsexpl1+attn+copy_t1-val-AVG-full-acc-top-2}{6.9\%}
\DefMacro{results-ln-lo-lo-75-ln-s+bsexpl1+attn+copy_t1-val-AVG-full-acc-top-3}{7.6\%}
\DefMacro{results-ln-lo-lo-75-ln-s+bsexpl1+attn+copy_t1-val-AVG-full-acc-top-5}{9.0\%}
\DefMacro{results-ln-lo-lo-75-ln-s+bsexpl1+attn+copy_t1-val-AVG-full-correct-top-1}{633.3\%}
\DefMacro{results-ln-lo-lo-75-ln-s+bsexpl1+attn+copy_t1-val-AVG-full-correct-top-2}{1000.0\%}
\DefMacro{results-ln-lo-lo-75-ln-s+bsexpl1+attn+copy_t1-val-AVG-full-correct-top-3}{1100.0\%}
\DefMacro{results-ln-lo-lo-75-ln-s+bsexpl1+attn+copy_t1-val-AVG-full-correct-top-5}{1300.0\%}
\DefMacro{results-ln-lo-lo-75-ln-s+bsexpl1+attn+copy_t1-val-MAX-BLEU-4}{48.8}
\DefMacro{results-ln-lo-lo-75-ln-s+bsexpl1+attn+copy_t1-val-MAX-char-acc}{33.1\%}
\DefMacro{results-ln-lo-lo-75-ln-s+bsexpl1+attn+copy_t1-val-MAX-count-top-1}{144}
\DefMacro{results-ln-lo-lo-75-ln-s+bsexpl1+attn+copy_t1-val-MAX-count-top-2}{144}
\DefMacro{results-ln-lo-lo-75-ln-s+bsexpl1+attn+copy_t1-val-MAX-count-top-3}{144}
\DefMacro{results-ln-lo-lo-75-ln-s+bsexpl1+attn+copy_t1-val-MAX-count-top-5}{144}
\DefMacro{results-ln-lo-lo-75-ln-s+bsexpl1+attn+copy_t1-val-MAX-frag-acc}{38.9\%}
\DefMacro{results-ln-lo-lo-75-ln-s+bsexpl1+attn+copy_t1-val-MAX-full-acc-top-1}{5.6\%}
\DefMacro{results-ln-lo-lo-75-ln-s+bsexpl1+attn+copy_t1-val-MAX-full-acc-top-2}{9.0\%}
\DefMacro{results-ln-lo-lo-75-ln-s+bsexpl1+attn+copy_t1-val-MAX-full-acc-top-3}{10.4\%}
\DefMacro{results-ln-lo-lo-75-ln-s+bsexpl1+attn+copy_t1-val-MAX-full-acc-top-5}{11.1\%}
\DefMacro{results-ln-lo-lo-75-ln-s+bsexpl1+attn+copy_t1-val-MAX-full-correct-top-1}{8}
\DefMacro{results-ln-lo-lo-75-ln-s+bsexpl1+attn+copy_t1-val-MAX-full-correct-top-2}{13}
\DefMacro{results-ln-lo-lo-75-ln-s+bsexpl1+attn+copy_t1-val-MAX-full-correct-top-3}{15}
\DefMacro{results-ln-lo-lo-75-ln-s+bsexpl1+attn+copy_t1-val-MAX-full-correct-top-5}{16}
\DefMacro{results-ln-lo-lo-75-ln-s+bsexpl1+attn+copy_t1-val-MEDIAN-BLEU-4}{45.6}
\DefMacro{results-ln-lo-lo-75-ln-s+bsexpl1+attn+copy_t1-val-MEDIAN-char-acc}{33.1\%}
\DefMacro{results-ln-lo-lo-75-ln-s+bsexpl1+attn+copy_t1-val-MEDIAN-count-top-1}{14400.0\%}
\DefMacro{results-ln-lo-lo-75-ln-s+bsexpl1+attn+copy_t1-val-MEDIAN-count-top-2}{14400.0\%}
\DefMacro{results-ln-lo-lo-75-ln-s+bsexpl1+attn+copy_t1-val-MEDIAN-count-top-3}{14400.0\%}
\DefMacro{results-ln-lo-lo-75-ln-s+bsexpl1+attn+copy_t1-val-MEDIAN-count-top-5}{14400.0\%}
\DefMacro{results-ln-lo-lo-75-ln-s+bsexpl1+attn+copy_t1-val-MEDIAN-frag-acc}{33.9\%}
\DefMacro{results-ln-lo-lo-75-ln-s+bsexpl1+attn+copy_t1-val-MEDIAN-full-acc-top-1}{4.2\%}
\DefMacro{results-ln-lo-lo-75-ln-s+bsexpl1+attn+copy_t1-val-MEDIAN-full-acc-top-2}{6.9\%}
\DefMacro{results-ln-lo-lo-75-ln-s+bsexpl1+attn+copy_t1-val-MEDIAN-full-acc-top-3}{6.9\%}
\DefMacro{results-ln-lo-lo-75-ln-s+bsexpl1+attn+copy_t1-val-MEDIAN-full-acc-top-5}{8.3\%}
\DefMacro{results-ln-lo-lo-75-ln-s+bsexpl1+attn+copy_t1-val-MEDIAN-full-correct-top-1}{600.0\%}
\DefMacro{results-ln-lo-lo-75-ln-s+bsexpl1+attn+copy_t1-val-MEDIAN-full-correct-top-2}{1000.0\%}
\DefMacro{results-ln-lo-lo-75-ln-s+bsexpl1+attn+copy_t1-val-MEDIAN-full-correct-top-3}{1000.0\%}
\DefMacro{results-ln-lo-lo-75-ln-s+bsexpl1+attn+copy_t1-val-MEDIAN-full-correct-top-5}{1200.0\%}
\DefMacro{results-ln-lo-lo-75-ln-s+bsexpl1+attn+copy_t1-val-MIN-BLEU-4}{45.0}
\DefMacro{results-ln-lo-lo-75-ln-s+bsexpl1+attn+copy_t1-val-MIN-char-acc}{30.0\%}
\DefMacro{results-ln-lo-lo-75-ln-s+bsexpl1+attn+copy_t1-val-MIN-count-top-1}{144}
\DefMacro{results-ln-lo-lo-75-ln-s+bsexpl1+attn+copy_t1-val-MIN-count-top-2}{144}
\DefMacro{results-ln-lo-lo-75-ln-s+bsexpl1+attn+copy_t1-val-MIN-count-top-3}{144}
\DefMacro{results-ln-lo-lo-75-ln-s+bsexpl1+attn+copy_t1-val-MIN-count-top-5}{144}
\DefMacro{results-ln-lo-lo-75-ln-s+bsexpl1+attn+copy_t1-val-MIN-frag-acc}{33.9\%}
\DefMacro{results-ln-lo-lo-75-ln-s+bsexpl1+attn+copy_t1-val-MIN-full-acc-top-1}{3.5\%}
\DefMacro{results-ln-lo-lo-75-ln-s+bsexpl1+attn+copy_t1-val-MIN-full-acc-top-2}{4.9\%}
\DefMacro{results-ln-lo-lo-75-ln-s+bsexpl1+attn+copy_t1-val-MIN-full-acc-top-3}{5.6\%}
\DefMacro{results-ln-lo-lo-75-ln-s+bsexpl1+attn+copy_t1-val-MIN-full-acc-top-5}{7.6\%}
\DefMacro{results-ln-lo-lo-75-ln-s+bsexpl1+attn+copy_t1-val-MIN-full-correct-top-1}{5}
\DefMacro{results-ln-lo-lo-75-ln-s+bsexpl1+attn+copy_t1-val-MIN-full-correct-top-2}{7}
\DefMacro{results-ln-lo-lo-75-ln-s+bsexpl1+attn+copy_t1-val-MIN-full-correct-top-3}{8}
\DefMacro{results-ln-lo-lo-75-ln-s+bsexpl1+attn+copy_t1-val-MIN-full-correct-top-5}{11}
\DefMacro{results-ln-lo-lo-75-ln-s+bsexpl1+attn+copy_t1-val-STDEV-BLEU-4}{1.7}
\DefMacro{results-ln-lo-lo-75-ln-s+bsexpl1+attn+copy_t1-val-STDEV-char-acc}{1.4\%}
\DefMacro{results-ln-lo-lo-75-ln-s+bsexpl1+attn+copy_t1-val-STDEV-count-top-1}{0.0\%}
\DefMacro{results-ln-lo-lo-75-ln-s+bsexpl1+attn+copy_t1-val-STDEV-count-top-2}{0.0\%}
\DefMacro{results-ln-lo-lo-75-ln-s+bsexpl1+attn+copy_t1-val-STDEV-count-top-3}{0.0\%}
\DefMacro{results-ln-lo-lo-75-ln-s+bsexpl1+attn+copy_t1-val-STDEV-count-top-5}{0.0\%}
\DefMacro{results-ln-lo-lo-75-ln-s+bsexpl1+attn+copy_t1-val-STDEV-frag-acc}{2.3\%}
\DefMacro{results-ln-lo-lo-75-ln-s+bsexpl1+attn+copy_t1-val-STDEV-full-acc-top-1}{0.9\%}
\DefMacro{results-ln-lo-lo-75-ln-s+bsexpl1+attn+copy_t1-val-STDEV-full-acc-top-2}{1.7\%}
\DefMacro{results-ln-lo-lo-75-ln-s+bsexpl1+attn+copy_t1-val-STDEV-full-acc-top-3}{2.0\%}
\DefMacro{results-ln-lo-lo-75-ln-s+bsexpl1+attn+copy_t1-val-STDEV-full-acc-top-5}{1.5\%}
\DefMacro{results-ln-lo-lo-75-ln-s+bsexpl1+attn+copy_t1-val-STDEV-full-correct-top-1}{124.7\%}
\DefMacro{results-ln-lo-lo-75-ln-s+bsexpl1+attn+copy_t1-val-STDEV-full-correct-top-2}{244.9\%}
\DefMacro{results-ln-lo-lo-75-ln-s+bsexpl1+attn+copy_t1-val-STDEV-full-correct-top-3}{294.4\%}
\DefMacro{results-ln-lo-lo-75-ln-s+bsexpl1+attn+copy_t1-val-STDEV-full-correct-top-5}{216.0\%}
\DefMacro{results-ln-lo-lo-75-ln-s+bsexpl1+attn+copy_t1-val-SUM-BLEU-4}{139.4}
\DefMacro{results-ln-lo-lo-75-ln-s+bsexpl1+attn+copy_t1-val-SUM-char-acc}{96.3\%}
\DefMacro{results-ln-lo-lo-75-ln-s+bsexpl1+attn+copy_t1-val-SUM-count-top-1}{432}
\DefMacro{results-ln-lo-lo-75-ln-s+bsexpl1+attn+copy_t1-val-SUM-count-top-2}{432}
\DefMacro{results-ln-lo-lo-75-ln-s+bsexpl1+attn+copy_t1-val-SUM-count-top-3}{432}
\DefMacro{results-ln-lo-lo-75-ln-s+bsexpl1+attn+copy_t1-val-SUM-count-top-5}{432}
\DefMacro{results-ln-lo-lo-75-ln-s+bsexpl1+attn+copy_t1-val-SUM-frag-acc}{106.7\%}
\DefMacro{results-ln-lo-lo-75-ln-s+bsexpl1+attn+copy_t1-val-SUM-full-acc-top-1}{13.2\%}
\DefMacro{results-ln-lo-lo-75-ln-s+bsexpl1+attn+copy_t1-val-SUM-full-acc-top-2}{20.8\%}
\DefMacro{results-ln-lo-lo-75-ln-s+bsexpl1+attn+copy_t1-val-SUM-full-acc-top-3}{22.9\%}
\DefMacro{results-ln-lo-lo-75-ln-s+bsexpl1+attn+copy_t1-val-SUM-full-acc-top-5}{27.1\%}
\DefMacro{results-ln-lo-lo-75-ln-s+bsexpl1+attn+copy_t1-val-SUM-full-correct-top-1}{19}
\DefMacro{results-ln-lo-lo-75-ln-s+bsexpl1+attn+copy_t1-val-SUM-full-correct-top-2}{30}
\DefMacro{results-ln-lo-lo-75-ln-s+bsexpl1+attn+copy_t1-val-SUM-full-correct-top-3}{33}
\DefMacro{results-ln-lo-lo-75-ln-s+bsexpl1+attn+copy_t1-val-SUM-full-correct-top-5}{39}
\DefMacro{results-ln-lo-lo-75-ln-s+bsexpl1+attn+copy_t1-test-AVG-BLEU-4}{36.6}
\DefMacro{results-ln-lo-lo-75-ln-s+bsexpl1+attn+copy_t1-test-AVG-char-acc}{23.9\%}
\DefMacro{results-ln-lo-lo-75-ln-s+bsexpl1+attn+copy_t1-test-AVG-count-top-1}{116700.0\%}
\DefMacro{results-ln-lo-lo-75-ln-s+bsexpl1+attn+copy_t1-test-AVG-count-top-2}{116700.0\%}
\DefMacro{results-ln-lo-lo-75-ln-s+bsexpl1+attn+copy_t1-test-AVG-count-top-3}{116700.0\%}
\DefMacro{results-ln-lo-lo-75-ln-s+bsexpl1+attn+copy_t1-test-AVG-count-top-5}{116700.0\%}
\DefMacro{results-ln-lo-lo-75-ln-s+bsexpl1+attn+copy_t1-test-AVG-frag-acc}{24.2\%}
\DefMacro{results-ln-lo-lo-75-ln-s+bsexpl1+attn+copy_t1-test-AVG-full-acc-top-1}{4.5\%}
\DefMacro{results-ln-lo-lo-75-ln-s+bsexpl1+attn+copy_t1-test-AVG-full-acc-top-2}{6.3\%}
\DefMacro{results-ln-lo-lo-75-ln-s+bsexpl1+attn+copy_t1-test-AVG-full-acc-top-3}{7.4\%}
\DefMacro{results-ln-lo-lo-75-ln-s+bsexpl1+attn+copy_t1-test-AVG-full-acc-top-5}{8.4\%}
\DefMacro{results-ln-lo-lo-75-ln-s+bsexpl1+attn+copy_t1-test-AVG-full-correct-top-1}{5266.7\%}
\DefMacro{results-ln-lo-lo-75-ln-s+bsexpl1+attn+copy_t1-test-AVG-full-correct-top-2}{7333.3\%}
\DefMacro{results-ln-lo-lo-75-ln-s+bsexpl1+attn+copy_t1-test-AVG-full-correct-top-3}{8600.0\%}
\DefMacro{results-ln-lo-lo-75-ln-s+bsexpl1+attn+copy_t1-test-AVG-full-correct-top-5}{9800.0\%}
\DefMacro{results-ln-lo-lo-75-ln-s+bsexpl1+attn+copy_t1-test-MAX-BLEU-4}{38.3}
\DefMacro{results-ln-lo-lo-75-ln-s+bsexpl1+attn+copy_t1-test-MAX-char-acc}{25.2\%}
\DefMacro{results-ln-lo-lo-75-ln-s+bsexpl1+attn+copy_t1-test-MAX-count-top-1}{1167}
\DefMacro{results-ln-lo-lo-75-ln-s+bsexpl1+attn+copy_t1-test-MAX-count-top-2}{1167}
\DefMacro{results-ln-lo-lo-75-ln-s+bsexpl1+attn+copy_t1-test-MAX-count-top-3}{1167}
\DefMacro{results-ln-lo-lo-75-ln-s+bsexpl1+attn+copy_t1-test-MAX-count-top-5}{1167}
\DefMacro{results-ln-lo-lo-75-ln-s+bsexpl1+attn+copy_t1-test-MAX-frag-acc}{25.2\%}
\DefMacro{results-ln-lo-lo-75-ln-s+bsexpl1+attn+copy_t1-test-MAX-full-acc-top-1}{4.6\%}
\DefMacro{results-ln-lo-lo-75-ln-s+bsexpl1+attn+copy_t1-test-MAX-full-acc-top-2}{6.3\%}
\DefMacro{results-ln-lo-lo-75-ln-s+bsexpl1+attn+copy_t1-test-MAX-full-acc-top-3}{7.7\%}
\DefMacro{results-ln-lo-lo-75-ln-s+bsexpl1+attn+copy_t1-test-MAX-full-acc-top-5}{8.7\%}
\DefMacro{results-ln-lo-lo-75-ln-s+bsexpl1+attn+copy_t1-test-MAX-full-correct-top-1}{54}
\DefMacro{results-ln-lo-lo-75-ln-s+bsexpl1+attn+copy_t1-test-MAX-full-correct-top-2}{74}
\DefMacro{results-ln-lo-lo-75-ln-s+bsexpl1+attn+copy_t1-test-MAX-full-correct-top-3}{90}
\DefMacro{results-ln-lo-lo-75-ln-s+bsexpl1+attn+copy_t1-test-MAX-full-correct-top-5}{101}
\DefMacro{results-ln-lo-lo-75-ln-s+bsexpl1+attn+copy_t1-test-MEDIAN-BLEU-4}{36.4}
\DefMacro{results-ln-lo-lo-75-ln-s+bsexpl1+attn+copy_t1-test-MEDIAN-char-acc}{23.9\%}
\DefMacro{results-ln-lo-lo-75-ln-s+bsexpl1+attn+copy_t1-test-MEDIAN-count-top-1}{116700.0\%}
\DefMacro{results-ln-lo-lo-75-ln-s+bsexpl1+attn+copy_t1-test-MEDIAN-count-top-2}{116700.0\%}
\DefMacro{results-ln-lo-lo-75-ln-s+bsexpl1+attn+copy_t1-test-MEDIAN-count-top-3}{116700.0\%}
\DefMacro{results-ln-lo-lo-75-ln-s+bsexpl1+attn+copy_t1-test-MEDIAN-count-top-5}{116700.0\%}
\DefMacro{results-ln-lo-lo-75-ln-s+bsexpl1+attn+copy_t1-test-MEDIAN-frag-acc}{24.2\%}
\DefMacro{results-ln-lo-lo-75-ln-s+bsexpl1+attn+copy_t1-test-MEDIAN-full-acc-top-1}{4.5\%}
\DefMacro{results-ln-lo-lo-75-ln-s+bsexpl1+attn+copy_t1-test-MEDIAN-full-acc-top-2}{6.3\%}
\DefMacro{results-ln-lo-lo-75-ln-s+bsexpl1+attn+copy_t1-test-MEDIAN-full-acc-top-3}{7.3\%}
\DefMacro{results-ln-lo-lo-75-ln-s+bsexpl1+attn+copy_t1-test-MEDIAN-full-acc-top-5}{8.6\%}
\DefMacro{results-ln-lo-lo-75-ln-s+bsexpl1+attn+copy_t1-test-MEDIAN-full-correct-top-1}{5300.0\%}
\DefMacro{results-ln-lo-lo-75-ln-s+bsexpl1+attn+copy_t1-test-MEDIAN-full-correct-top-2}{7400.0\%}
\DefMacro{results-ln-lo-lo-75-ln-s+bsexpl1+attn+copy_t1-test-MEDIAN-full-correct-top-3}{8500.0\%}
\DefMacro{results-ln-lo-lo-75-ln-s+bsexpl1+attn+copy_t1-test-MEDIAN-full-correct-top-5}{10000.0\%}
\DefMacro{results-ln-lo-lo-75-ln-s+bsexpl1+attn+copy_t1-test-MIN-BLEU-4}{35.0}
\DefMacro{results-ln-lo-lo-75-ln-s+bsexpl1+attn+copy_t1-test-MIN-char-acc}{22.5\%}
\DefMacro{results-ln-lo-lo-75-ln-s+bsexpl1+attn+copy_t1-test-MIN-count-top-1}{1167}
\DefMacro{results-ln-lo-lo-75-ln-s+bsexpl1+attn+copy_t1-test-MIN-count-top-2}{1167}
\DefMacro{results-ln-lo-lo-75-ln-s+bsexpl1+attn+copy_t1-test-MIN-count-top-3}{1167}
\DefMacro{results-ln-lo-lo-75-ln-s+bsexpl1+attn+copy_t1-test-MIN-count-top-5}{1167}
\DefMacro{results-ln-lo-lo-75-ln-s+bsexpl1+attn+copy_t1-test-MIN-frag-acc}{23.1\%}
\DefMacro{results-ln-lo-lo-75-ln-s+bsexpl1+attn+copy_t1-test-MIN-full-acc-top-1}{4.4\%}
\DefMacro{results-ln-lo-lo-75-ln-s+bsexpl1+attn+copy_t1-test-MIN-full-acc-top-2}{6.2\%}
\DefMacro{results-ln-lo-lo-75-ln-s+bsexpl1+attn+copy_t1-test-MIN-full-acc-top-3}{7.1\%}
\DefMacro{results-ln-lo-lo-75-ln-s+bsexpl1+attn+copy_t1-test-MIN-full-acc-top-5}{8.0\%}
\DefMacro{results-ln-lo-lo-75-ln-s+bsexpl1+attn+copy_t1-test-MIN-full-correct-top-1}{51}
\DefMacro{results-ln-lo-lo-75-ln-s+bsexpl1+attn+copy_t1-test-MIN-full-correct-top-2}{72}
\DefMacro{results-ln-lo-lo-75-ln-s+bsexpl1+attn+copy_t1-test-MIN-full-correct-top-3}{83}
\DefMacro{results-ln-lo-lo-75-ln-s+bsexpl1+attn+copy_t1-test-MIN-full-correct-top-5}{93}
\DefMacro{results-ln-lo-lo-75-ln-s+bsexpl1+attn+copy_t1-test-STDEV-BLEU-4}{1.4}
\DefMacro{results-ln-lo-lo-75-ln-s+bsexpl1+attn+copy_t1-test-STDEV-char-acc}{1.1\%}
\DefMacro{results-ln-lo-lo-75-ln-s+bsexpl1+attn+copy_t1-test-STDEV-count-top-1}{0.0\%}
\DefMacro{results-ln-lo-lo-75-ln-s+bsexpl1+attn+copy_t1-test-STDEV-count-top-2}{0.0\%}
\DefMacro{results-ln-lo-lo-75-ln-s+bsexpl1+attn+copy_t1-test-STDEV-count-top-3}{0.0\%}
\DefMacro{results-ln-lo-lo-75-ln-s+bsexpl1+attn+copy_t1-test-STDEV-count-top-5}{0.0\%}
\DefMacro{results-ln-lo-lo-75-ln-s+bsexpl1+attn+copy_t1-test-STDEV-frag-acc}{0.9\%}
\DefMacro{results-ln-lo-lo-75-ln-s+bsexpl1+attn+copy_t1-test-STDEV-full-acc-top-1}{0.1\%}
\DefMacro{results-ln-lo-lo-75-ln-s+bsexpl1+attn+copy_t1-test-STDEV-full-acc-top-2}{0.1\%}
\DefMacro{results-ln-lo-lo-75-ln-s+bsexpl1+attn+copy_t1-test-STDEV-full-acc-top-3}{0.3\%}
\DefMacro{results-ln-lo-lo-75-ln-s+bsexpl1+attn+copy_t1-test-STDEV-full-acc-top-5}{0.3\%}
\DefMacro{results-ln-lo-lo-75-ln-s+bsexpl1+attn+copy_t1-test-STDEV-full-correct-top-1}{124.7\%}
\DefMacro{results-ln-lo-lo-75-ln-s+bsexpl1+attn+copy_t1-test-STDEV-full-correct-top-2}{94.3\%}
\DefMacro{results-ln-lo-lo-75-ln-s+bsexpl1+attn+copy_t1-test-STDEV-full-correct-top-3}{294.4\%}
\DefMacro{results-ln-lo-lo-75-ln-s+bsexpl1+attn+copy_t1-test-STDEV-full-correct-top-5}{355.9\%}
\DefMacro{results-ln-lo-lo-75-ln-s+bsexpl1+attn+copy_t1-test-SUM-BLEU-4}{109.8}
\DefMacro{results-ln-lo-lo-75-ln-s+bsexpl1+attn+copy_t1-test-SUM-char-acc}{71.6\%}
\DefMacro{results-ln-lo-lo-75-ln-s+bsexpl1+attn+copy_t1-test-SUM-count-top-1}{3501}
\DefMacro{results-ln-lo-lo-75-ln-s+bsexpl1+attn+copy_t1-test-SUM-count-top-2}{3501}
\DefMacro{results-ln-lo-lo-75-ln-s+bsexpl1+attn+copy_t1-test-SUM-count-top-3}{3501}
\DefMacro{results-ln-lo-lo-75-ln-s+bsexpl1+attn+copy_t1-test-SUM-count-top-5}{3501}
\DefMacro{results-ln-lo-lo-75-ln-s+bsexpl1+attn+copy_t1-test-SUM-frag-acc}{72.6\%}
\DefMacro{results-ln-lo-lo-75-ln-s+bsexpl1+attn+copy_t1-test-SUM-full-acc-top-1}{13.5\%}
\DefMacro{results-ln-lo-lo-75-ln-s+bsexpl1+attn+copy_t1-test-SUM-full-acc-top-2}{18.9\%}
\DefMacro{results-ln-lo-lo-75-ln-s+bsexpl1+attn+copy_t1-test-SUM-full-acc-top-3}{22.1\%}
\DefMacro{results-ln-lo-lo-75-ln-s+bsexpl1+attn+copy_t1-test-SUM-full-acc-top-5}{25.2\%}
\DefMacro{results-ln-lo-lo-75-ln-s+bsexpl1+attn+copy_t1-test-SUM-full-correct-top-1}{158}
\DefMacro{results-ln-lo-lo-75-ln-s+bsexpl1+attn+copy_t1-test-SUM-full-correct-top-2}{220}
\DefMacro{results-ln-lo-lo-75-ln-s+bsexpl1+attn+copy_t1-test-SUM-full-correct-top-3}{258}
\DefMacro{results-ln-lo-lo-75-ln-s+bsexpl1+attn+copy_t1-test-SUM-full-correct-top-5}{294}
\DefMacro{results-ln-lo-lo-ln-s+bsexpl1+attn+copy_t1-val-AVG-BLEU-4}{46.3}
\DefMacro{results-ln-lo-lo-ln-s+bsexpl1+attn+copy_t1-val-AVG-char-acc}{32.2\%}
\DefMacro{results-ln-lo-lo-ln-s+bsexpl1+attn+copy_t1-val-AVG-count-top-1}{14400.0\%}
\DefMacro{results-ln-lo-lo-ln-s+bsexpl1+attn+copy_t1-val-AVG-count-top-2}{14400.0\%}
\DefMacro{results-ln-lo-lo-ln-s+bsexpl1+attn+copy_t1-val-AVG-count-top-3}{14400.0\%}
\DefMacro{results-ln-lo-lo-ln-s+bsexpl1+attn+copy_t1-val-AVG-count-top-5}{14400.0\%}
\DefMacro{results-ln-lo-lo-ln-s+bsexpl1+attn+copy_t1-val-AVG-frag-acc}{35.7\%}
\DefMacro{results-ln-lo-lo-ln-s+bsexpl1+attn+copy_t1-val-AVG-full-acc-top-1}{4.6\%}
\DefMacro{results-ln-lo-lo-ln-s+bsexpl1+attn+copy_t1-val-AVG-full-acc-top-2}{6.9\%}
\DefMacro{results-ln-lo-lo-ln-s+bsexpl1+attn+copy_t1-val-AVG-full-acc-top-3}{9.5\%}
\DefMacro{results-ln-lo-lo-ln-s+bsexpl1+attn+copy_t1-val-AVG-full-acc-top-5}{9.7\%}
\DefMacro{results-ln-lo-lo-ln-s+bsexpl1+attn+copy_t1-val-AVG-full-correct-top-1}{666.7\%}
\DefMacro{results-ln-lo-lo-ln-s+bsexpl1+attn+copy_t1-val-AVG-full-correct-top-2}{1000.0\%}
\DefMacro{results-ln-lo-lo-ln-s+bsexpl1+attn+copy_t1-val-AVG-full-correct-top-3}{1366.7\%}
\DefMacro{results-ln-lo-lo-ln-s+bsexpl1+attn+copy_t1-val-AVG-full-correct-top-5}{1400.0\%}
\DefMacro{results-ln-lo-lo-ln-s+bsexpl1+attn+copy_t1-val-MAX-BLEU-4}{46.6}
\DefMacro{results-ln-lo-lo-ln-s+bsexpl1+attn+copy_t1-val-MAX-char-acc}{33.6\%}
\DefMacro{results-ln-lo-lo-ln-s+bsexpl1+attn+copy_t1-val-MAX-count-top-1}{144}
\DefMacro{results-ln-lo-lo-ln-s+bsexpl1+attn+copy_t1-val-MAX-count-top-2}{144}
\DefMacro{results-ln-lo-lo-ln-s+bsexpl1+attn+copy_t1-val-MAX-count-top-3}{144}
\DefMacro{results-ln-lo-lo-ln-s+bsexpl1+attn+copy_t1-val-MAX-count-top-5}{144}
\DefMacro{results-ln-lo-lo-ln-s+bsexpl1+attn+copy_t1-val-MAX-frag-acc}{37.0\%}
\DefMacro{results-ln-lo-lo-ln-s+bsexpl1+attn+copy_t1-val-MAX-full-acc-top-1}{5.6\%}
\DefMacro{results-ln-lo-lo-ln-s+bsexpl1+attn+copy_t1-val-MAX-full-acc-top-2}{9.0\%}
\DefMacro{results-ln-lo-lo-ln-s+bsexpl1+attn+copy_t1-val-MAX-full-acc-top-3}{10.4\%}
\DefMacro{results-ln-lo-lo-ln-s+bsexpl1+attn+copy_t1-val-MAX-full-acc-top-5}{10.4\%}
\DefMacro{results-ln-lo-lo-ln-s+bsexpl1+attn+copy_t1-val-MAX-full-correct-top-1}{8}
\DefMacro{results-ln-lo-lo-ln-s+bsexpl1+attn+copy_t1-val-MAX-full-correct-top-2}{13}
\DefMacro{results-ln-lo-lo-ln-s+bsexpl1+attn+copy_t1-val-MAX-full-correct-top-3}{15}
\DefMacro{results-ln-lo-lo-ln-s+bsexpl1+attn+copy_t1-val-MAX-full-correct-top-5}{15}
\DefMacro{results-ln-lo-lo-ln-s+bsexpl1+attn+copy_t1-val-MEDIAN-BLEU-4}{46.4}
\DefMacro{results-ln-lo-lo-ln-s+bsexpl1+attn+copy_t1-val-MEDIAN-char-acc}{32.3\%}
\DefMacro{results-ln-lo-lo-ln-s+bsexpl1+attn+copy_t1-val-MEDIAN-count-top-1}{14400.0\%}
\DefMacro{results-ln-lo-lo-ln-s+bsexpl1+attn+copy_t1-val-MEDIAN-count-top-2}{14400.0\%}
\DefMacro{results-ln-lo-lo-ln-s+bsexpl1+attn+copy_t1-val-MEDIAN-count-top-3}{14400.0\%}
\DefMacro{results-ln-lo-lo-ln-s+bsexpl1+attn+copy_t1-val-MEDIAN-count-top-5}{14400.0\%}
\DefMacro{results-ln-lo-lo-ln-s+bsexpl1+attn+copy_t1-val-MEDIAN-frag-acc}{35.4\%}
\DefMacro{results-ln-lo-lo-ln-s+bsexpl1+attn+copy_t1-val-MEDIAN-full-acc-top-1}{4.9\%}
\DefMacro{results-ln-lo-lo-ln-s+bsexpl1+attn+copy_t1-val-MEDIAN-full-acc-top-2}{6.2\%}
\DefMacro{results-ln-lo-lo-ln-s+bsexpl1+attn+copy_t1-val-MEDIAN-full-acc-top-3}{9.7\%}
\DefMacro{results-ln-lo-lo-ln-s+bsexpl1+attn+copy_t1-val-MEDIAN-full-acc-top-5}{9.7\%}
\DefMacro{results-ln-lo-lo-ln-s+bsexpl1+attn+copy_t1-val-MEDIAN-full-correct-top-1}{700.0\%}
\DefMacro{results-ln-lo-lo-ln-s+bsexpl1+attn+copy_t1-val-MEDIAN-full-correct-top-2}{900.0\%}
\DefMacro{results-ln-lo-lo-ln-s+bsexpl1+attn+copy_t1-val-MEDIAN-full-correct-top-3}{1400.0\%}
\DefMacro{results-ln-lo-lo-ln-s+bsexpl1+attn+copy_t1-val-MEDIAN-full-correct-top-5}{1400.0\%}
\DefMacro{results-ln-lo-lo-ln-s+bsexpl1+attn+copy_t1-val-MIN-BLEU-4}{46.1}
\DefMacro{results-ln-lo-lo-ln-s+bsexpl1+attn+copy_t1-val-MIN-char-acc}{30.6\%}
\DefMacro{results-ln-lo-lo-ln-s+bsexpl1+attn+copy_t1-val-MIN-count-top-1}{144}
\DefMacro{results-ln-lo-lo-ln-s+bsexpl1+attn+copy_t1-val-MIN-count-top-2}{144}
\DefMacro{results-ln-lo-lo-ln-s+bsexpl1+attn+copy_t1-val-MIN-count-top-3}{144}
\DefMacro{results-ln-lo-lo-ln-s+bsexpl1+attn+copy_t1-val-MIN-count-top-5}{144}
\DefMacro{results-ln-lo-lo-ln-s+bsexpl1+attn+copy_t1-val-MIN-frag-acc}{34.8\%}
\DefMacro{results-ln-lo-lo-ln-s+bsexpl1+attn+copy_t1-val-MIN-full-acc-top-1}{3.5\%}
\DefMacro{results-ln-lo-lo-ln-s+bsexpl1+attn+copy_t1-val-MIN-full-acc-top-2}{5.6\%}
\DefMacro{results-ln-lo-lo-ln-s+bsexpl1+attn+copy_t1-val-MIN-full-acc-top-3}{8.3\%}
\DefMacro{results-ln-lo-lo-ln-s+bsexpl1+attn+copy_t1-val-MIN-full-acc-top-5}{9.0\%}
\DefMacro{results-ln-lo-lo-ln-s+bsexpl1+attn+copy_t1-val-MIN-full-correct-top-1}{5}
\DefMacro{results-ln-lo-lo-ln-s+bsexpl1+attn+copy_t1-val-MIN-full-correct-top-2}{8}
\DefMacro{results-ln-lo-lo-ln-s+bsexpl1+attn+copy_t1-val-MIN-full-correct-top-3}{12}
\DefMacro{results-ln-lo-lo-ln-s+bsexpl1+attn+copy_t1-val-MIN-full-correct-top-5}{13}
\DefMacro{results-ln-lo-lo-ln-s+bsexpl1+attn+copy_t1-val-STDEV-BLEU-4}{0.2}
\DefMacro{results-ln-lo-lo-ln-s+bsexpl1+attn+copy_t1-val-STDEV-char-acc}{1.3\%}
\DefMacro{results-ln-lo-lo-ln-s+bsexpl1+attn+copy_t1-val-STDEV-count-top-1}{0.0\%}
\DefMacro{results-ln-lo-lo-ln-s+bsexpl1+attn+copy_t1-val-STDEV-count-top-2}{0.0\%}
\DefMacro{results-ln-lo-lo-ln-s+bsexpl1+attn+copy_t1-val-STDEV-count-top-3}{0.0\%}
\DefMacro{results-ln-lo-lo-ln-s+bsexpl1+attn+copy_t1-val-STDEV-count-top-5}{0.0\%}
\DefMacro{results-ln-lo-lo-ln-s+bsexpl1+attn+copy_t1-val-STDEV-frag-acc}{0.9\%}
\DefMacro{results-ln-lo-lo-ln-s+bsexpl1+attn+copy_t1-val-STDEV-full-acc-top-1}{0.9\%}
\DefMacro{results-ln-lo-lo-ln-s+bsexpl1+attn+copy_t1-val-STDEV-full-acc-top-2}{1.5\%}
\DefMacro{results-ln-lo-lo-ln-s+bsexpl1+attn+copy_t1-val-STDEV-full-acc-top-3}{0.9\%}
\DefMacro{results-ln-lo-lo-ln-s+bsexpl1+attn+copy_t1-val-STDEV-full-acc-top-5}{0.6\%}
\DefMacro{results-ln-lo-lo-ln-s+bsexpl1+attn+copy_t1-val-STDEV-full-correct-top-1}{124.7\%}
\DefMacro{results-ln-lo-lo-ln-s+bsexpl1+attn+copy_t1-val-STDEV-full-correct-top-2}{216.0\%}
\DefMacro{results-ln-lo-lo-ln-s+bsexpl1+attn+copy_t1-val-STDEV-full-correct-top-3}{124.7\%}
\DefMacro{results-ln-lo-lo-ln-s+bsexpl1+attn+copy_t1-val-STDEV-full-correct-top-5}{81.6\%}
\DefMacro{results-ln-lo-lo-ln-s+bsexpl1+attn+copy_t1-val-SUM-BLEU-4}{139.0}
\DefMacro{results-ln-lo-lo-ln-s+bsexpl1+attn+copy_t1-val-SUM-char-acc}{96.5\%}
\DefMacro{results-ln-lo-lo-ln-s+bsexpl1+attn+copy_t1-val-SUM-count-top-1}{432}
\DefMacro{results-ln-lo-lo-ln-s+bsexpl1+attn+copy_t1-val-SUM-count-top-2}{432}
\DefMacro{results-ln-lo-lo-ln-s+bsexpl1+attn+copy_t1-val-SUM-count-top-3}{432}
\DefMacro{results-ln-lo-lo-ln-s+bsexpl1+attn+copy_t1-val-SUM-count-top-5}{432}
\DefMacro{results-ln-lo-lo-ln-s+bsexpl1+attn+copy_t1-val-SUM-frag-acc}{107.1\%}
\DefMacro{results-ln-lo-lo-ln-s+bsexpl1+attn+copy_t1-val-SUM-full-acc-top-1}{13.9\%}
\DefMacro{results-ln-lo-lo-ln-s+bsexpl1+attn+copy_t1-val-SUM-full-acc-top-2}{20.8\%}
\DefMacro{results-ln-lo-lo-ln-s+bsexpl1+attn+copy_t1-val-SUM-full-acc-top-3}{28.5\%}
\DefMacro{results-ln-lo-lo-ln-s+bsexpl1+attn+copy_t1-val-SUM-full-acc-top-5}{29.2\%}
\DefMacro{results-ln-lo-lo-ln-s+bsexpl1+attn+copy_t1-val-SUM-full-correct-top-1}{20}
\DefMacro{results-ln-lo-lo-ln-s+bsexpl1+attn+copy_t1-val-SUM-full-correct-top-2}{30}
\DefMacro{results-ln-lo-lo-ln-s+bsexpl1+attn+copy_t1-val-SUM-full-correct-top-3}{41}
\DefMacro{results-ln-lo-lo-ln-s+bsexpl1+attn+copy_t1-val-SUM-full-correct-top-5}{42}
\DefMacro{results-ln-lo-lo-ln-s+bsexpl1+attn+copy_t1-test-AVG-BLEU-4}{35.9}
\DefMacro{results-ln-lo-lo-ln-s+bsexpl1+attn+copy_t1-test-AVG-char-acc}{25.3\%}
\DefMacro{results-ln-lo-lo-ln-s+bsexpl1+attn+copy_t1-test-AVG-count-top-1}{116700.0\%}
\DefMacro{results-ln-lo-lo-ln-s+bsexpl1+attn+copy_t1-test-AVG-count-top-2}{116700.0\%}
\DefMacro{results-ln-lo-lo-ln-s+bsexpl1+attn+copy_t1-test-AVG-count-top-3}{116700.0\%}
\DefMacro{results-ln-lo-lo-ln-s+bsexpl1+attn+copy_t1-test-AVG-count-top-5}{116700.0\%}
\DefMacro{results-ln-lo-lo-ln-s+bsexpl1+attn+copy_t1-test-AVG-frag-acc}{24.2\%}
\DefMacro{results-ln-lo-lo-ln-s+bsexpl1+attn+copy_t1-test-AVG-full-acc-top-1}{6.6\%}
\DefMacro{results-ln-lo-lo-ln-s+bsexpl1+attn+copy_t1-test-AVG-full-acc-top-2}{8.7\%}
\DefMacro{results-ln-lo-lo-ln-s+bsexpl1+attn+copy_t1-test-AVG-full-acc-top-3}{10.0\%}
\DefMacro{results-ln-lo-lo-ln-s+bsexpl1+attn+copy_t1-test-AVG-full-acc-top-5}{10.8\%}
\DefMacro{results-ln-lo-lo-ln-s+bsexpl1+attn+copy_t1-test-AVG-full-correct-top-1}{7733.3\%}
\DefMacro{results-ln-lo-lo-ln-s+bsexpl1+attn+copy_t1-test-AVG-full-correct-top-2}{10200.0\%}
\DefMacro{results-ln-lo-lo-ln-s+bsexpl1+attn+copy_t1-test-AVG-full-correct-top-3}{11633.3\%}
\DefMacro{results-ln-lo-lo-ln-s+bsexpl1+attn+copy_t1-test-AVG-full-correct-top-5}{12633.3\%}
\DefMacro{results-ln-lo-lo-ln-s+bsexpl1+attn+copy_t1-test-MAX-BLEU-4}{37.8}
\DefMacro{results-ln-lo-lo-ln-s+bsexpl1+attn+copy_t1-test-MAX-char-acc}{26.6\%}
\DefMacro{results-ln-lo-lo-ln-s+bsexpl1+attn+copy_t1-test-MAX-count-top-1}{1167}
\DefMacro{results-ln-lo-lo-ln-s+bsexpl1+attn+copy_t1-test-MAX-count-top-2}{1167}
\DefMacro{results-ln-lo-lo-ln-s+bsexpl1+attn+copy_t1-test-MAX-count-top-3}{1167}
\DefMacro{results-ln-lo-lo-ln-s+bsexpl1+attn+copy_t1-test-MAX-count-top-5}{1167}
\DefMacro{results-ln-lo-lo-ln-s+bsexpl1+attn+copy_t1-test-MAX-frag-acc}{24.7\%}
\DefMacro{results-ln-lo-lo-ln-s+bsexpl1+attn+copy_t1-test-MAX-full-acc-top-1}{6.9\%}
\DefMacro{results-ln-lo-lo-ln-s+bsexpl1+attn+copy_t1-test-MAX-full-acc-top-2}{9.4\%}
\DefMacro{results-ln-lo-lo-ln-s+bsexpl1+attn+copy_t1-test-MAX-full-acc-top-3}{10.5\%}
\DefMacro{results-ln-lo-lo-ln-s+bsexpl1+attn+copy_t1-test-MAX-full-acc-top-5}{11.3\%}
\DefMacro{results-ln-lo-lo-ln-s+bsexpl1+attn+copy_t1-test-MAX-full-correct-top-1}{80}
\DefMacro{results-ln-lo-lo-ln-s+bsexpl1+attn+copy_t1-test-MAX-full-correct-top-2}{110}
\DefMacro{results-ln-lo-lo-ln-s+bsexpl1+attn+copy_t1-test-MAX-full-correct-top-3}{122}
\DefMacro{results-ln-lo-lo-ln-s+bsexpl1+attn+copy_t1-test-MAX-full-correct-top-5}{132}
\DefMacro{results-ln-lo-lo-ln-s+bsexpl1+attn+copy_t1-test-MEDIAN-BLEU-4}{35.0}
\DefMacro{results-ln-lo-lo-ln-s+bsexpl1+attn+copy_t1-test-MEDIAN-char-acc}{25.5\%}
\DefMacro{results-ln-lo-lo-ln-s+bsexpl1+attn+copy_t1-test-MEDIAN-count-top-1}{116700.0\%}
\DefMacro{results-ln-lo-lo-ln-s+bsexpl1+attn+copy_t1-test-MEDIAN-count-top-2}{116700.0\%}
\DefMacro{results-ln-lo-lo-ln-s+bsexpl1+attn+copy_t1-test-MEDIAN-count-top-3}{116700.0\%}
\DefMacro{results-ln-lo-lo-ln-s+bsexpl1+attn+copy_t1-test-MEDIAN-count-top-5}{116700.0\%}
\DefMacro{results-ln-lo-lo-ln-s+bsexpl1+attn+copy_t1-test-MEDIAN-frag-acc}{24.4\%}
\DefMacro{results-ln-lo-lo-ln-s+bsexpl1+attn+copy_t1-test-MEDIAN-full-acc-top-1}{6.5\%}
\DefMacro{results-ln-lo-lo-ln-s+bsexpl1+attn+copy_t1-test-MEDIAN-full-acc-top-2}{8.7\%}
\DefMacro{results-ln-lo-lo-ln-s+bsexpl1+attn+copy_t1-test-MEDIAN-full-acc-top-3}{9.9\%}
\DefMacro{results-ln-lo-lo-ln-s+bsexpl1+attn+copy_t1-test-MEDIAN-full-acc-top-5}{10.6\%}
\DefMacro{results-ln-lo-lo-ln-s+bsexpl1+attn+copy_t1-test-MEDIAN-full-correct-top-1}{7600.0\%}
\DefMacro{results-ln-lo-lo-ln-s+bsexpl1+attn+copy_t1-test-MEDIAN-full-correct-top-2}{10200.0\%}
\DefMacro{results-ln-lo-lo-ln-s+bsexpl1+attn+copy_t1-test-MEDIAN-full-correct-top-3}{11500.0\%}
\DefMacro{results-ln-lo-lo-ln-s+bsexpl1+attn+copy_t1-test-MEDIAN-full-correct-top-5}{12400.0\%}
\DefMacro{results-ln-lo-lo-ln-s+bsexpl1+attn+copy_t1-test-MIN-BLEU-4}{34.9}
\DefMacro{results-ln-lo-lo-ln-s+bsexpl1+attn+copy_t1-test-MIN-char-acc}{23.8\%}
\DefMacro{results-ln-lo-lo-ln-s+bsexpl1+attn+copy_t1-test-MIN-count-top-1}{1167}
\DefMacro{results-ln-lo-lo-ln-s+bsexpl1+attn+copy_t1-test-MIN-count-top-2}{1167}
\DefMacro{results-ln-lo-lo-ln-s+bsexpl1+attn+copy_t1-test-MIN-count-top-3}{1167}
\DefMacro{results-ln-lo-lo-ln-s+bsexpl1+attn+copy_t1-test-MIN-count-top-5}{1167}
\DefMacro{results-ln-lo-lo-ln-s+bsexpl1+attn+copy_t1-test-MIN-frag-acc}{23.6\%}
\DefMacro{results-ln-lo-lo-ln-s+bsexpl1+attn+copy_t1-test-MIN-full-acc-top-1}{6.5\%}
\DefMacro{results-ln-lo-lo-ln-s+bsexpl1+attn+copy_t1-test-MIN-full-acc-top-2}{8.1\%}
\DefMacro{results-ln-lo-lo-ln-s+bsexpl1+attn+copy_t1-test-MIN-full-acc-top-3}{9.6\%}
\DefMacro{results-ln-lo-lo-ln-s+bsexpl1+attn+copy_t1-test-MIN-full-acc-top-5}{10.5\%}
\DefMacro{results-ln-lo-lo-ln-s+bsexpl1+attn+copy_t1-test-MIN-full-correct-top-1}{76}
\DefMacro{results-ln-lo-lo-ln-s+bsexpl1+attn+copy_t1-test-MIN-full-correct-top-2}{94}
\DefMacro{results-ln-lo-lo-ln-s+bsexpl1+attn+copy_t1-test-MIN-full-correct-top-3}{112}
\DefMacro{results-ln-lo-lo-ln-s+bsexpl1+attn+copy_t1-test-MIN-full-correct-top-5}{123}
\DefMacro{results-ln-lo-lo-ln-s+bsexpl1+attn+copy_t1-test-STDEV-BLEU-4}{1.3}
\DefMacro{results-ln-lo-lo-ln-s+bsexpl1+attn+copy_t1-test-STDEV-char-acc}{1.1\%}
\DefMacro{results-ln-lo-lo-ln-s+bsexpl1+attn+copy_t1-test-STDEV-count-top-1}{0.0\%}
\DefMacro{results-ln-lo-lo-ln-s+bsexpl1+attn+copy_t1-test-STDEV-count-top-2}{0.0\%}
\DefMacro{results-ln-lo-lo-ln-s+bsexpl1+attn+copy_t1-test-STDEV-count-top-3}{0.0\%}
\DefMacro{results-ln-lo-lo-ln-s+bsexpl1+attn+copy_t1-test-STDEV-count-top-5}{0.0\%}
\DefMacro{results-ln-lo-lo-ln-s+bsexpl1+attn+copy_t1-test-STDEV-frag-acc}{0.5\%}
\DefMacro{results-ln-lo-lo-ln-s+bsexpl1+attn+copy_t1-test-STDEV-full-acc-top-1}{0.2\%}
\DefMacro{results-ln-lo-lo-ln-s+bsexpl1+attn+copy_t1-test-STDEV-full-acc-top-2}{0.6\%}
\DefMacro{results-ln-lo-lo-ln-s+bsexpl1+attn+copy_t1-test-STDEV-full-acc-top-3}{0.4\%}
\DefMacro{results-ln-lo-lo-ln-s+bsexpl1+attn+copy_t1-test-STDEV-full-acc-top-5}{0.3\%}
\DefMacro{results-ln-lo-lo-ln-s+bsexpl1+attn+copy_t1-test-STDEV-full-correct-top-1}{188.6\%}
\DefMacro{results-ln-lo-lo-ln-s+bsexpl1+attn+copy_t1-test-STDEV-full-correct-top-2}{653.2\%}
\DefMacro{results-ln-lo-lo-ln-s+bsexpl1+attn+copy_t1-test-STDEV-full-correct-top-3}{419.0\%}
\DefMacro{results-ln-lo-lo-ln-s+bsexpl1+attn+copy_t1-test-STDEV-full-correct-top-5}{402.8\%}
\DefMacro{results-ln-lo-lo-ln-s+bsexpl1+attn+copy_t1-test-SUM-BLEU-4}{107.7}
\DefMacro{results-ln-lo-lo-ln-s+bsexpl1+attn+copy_t1-test-SUM-char-acc}{75.9\%}
\DefMacro{results-ln-lo-lo-ln-s+bsexpl1+attn+copy_t1-test-SUM-count-top-1}{3501}
\DefMacro{results-ln-lo-lo-ln-s+bsexpl1+attn+copy_t1-test-SUM-count-top-2}{3501}
\DefMacro{results-ln-lo-lo-ln-s+bsexpl1+attn+copy_t1-test-SUM-count-top-3}{3501}
\DefMacro{results-ln-lo-lo-ln-s+bsexpl1+attn+copy_t1-test-SUM-count-top-5}{3501}
\DefMacro{results-ln-lo-lo-ln-s+bsexpl1+attn+copy_t1-test-SUM-frag-acc}{72.7\%}
\DefMacro{results-ln-lo-lo-ln-s+bsexpl1+attn+copy_t1-test-SUM-full-acc-top-1}{19.9\%}
\DefMacro{results-ln-lo-lo-ln-s+bsexpl1+attn+copy_t1-test-SUM-full-acc-top-2}{26.2\%}
\DefMacro{results-ln-lo-lo-ln-s+bsexpl1+attn+copy_t1-test-SUM-full-acc-top-3}{29.9\%}
\DefMacro{results-ln-lo-lo-ln-s+bsexpl1+attn+copy_t1-test-SUM-full-acc-top-5}{32.5\%}
\DefMacro{results-ln-lo-lo-ln-s+bsexpl1+attn+copy_t1-test-SUM-full-correct-top-1}{232}
\DefMacro{results-ln-lo-lo-ln-s+bsexpl1+attn+copy_t1-test-SUM-full-correct-top-2}{306}
\DefMacro{results-ln-lo-lo-ln-s+bsexpl1+attn+copy_t1-test-SUM-full-correct-top-3}{349}
\DefMacro{results-ln-lo-lo-ln-s+bsexpl1+attn+copy_t1-test-SUM-full-correct-top-5}{379}

%% file: tables/numbers-results-ln-lo-ta.tex
\DefMacro{results-ln-lo-ta-ln-s+bsexpl1+attn+copy_ta-val-AVG-BLEU-4}{43.7}
\DefMacro{results-ln-lo-ta-ln-s+bsexpl1+attn+copy_ta-val-AVG-char-acc}{34.0\%}
\DefMacro{results-ln-lo-ta-ln-s+bsexpl1+attn+copy_ta-val-AVG-count-top-1}{155600.0\%}
\DefMacro{results-ln-lo-ta-ln-s+bsexpl1+attn+copy_ta-val-AVG-count-top-2}{155600.0\%}
\DefMacro{results-ln-lo-ta-ln-s+bsexpl1+attn+copy_ta-val-AVG-count-top-3}{155600.0\%}
\DefMacro{results-ln-lo-ta-ln-s+bsexpl1+attn+copy_ta-val-AVG-count-top-5}{155600.0\%}
\DefMacro{results-ln-lo-ta-ln-s+bsexpl1+attn+copy_ta-val-AVG-frag-acc}{19.9\%}
\DefMacro{results-ln-lo-ta-ln-s+bsexpl1+attn+copy_ta-val-AVG-full-acc-top-1}{7.7\%}
\DefMacro{results-ln-lo-ta-ln-s+bsexpl1+attn+copy_ta-val-AVG-full-acc-top-2}{11.0\%}
\DefMacro{results-ln-lo-ta-ln-s+bsexpl1+attn+copy_ta-val-AVG-full-acc-top-3}{13.2\%}
\DefMacro{results-ln-lo-ta-ln-s+bsexpl1+attn+copy_ta-val-AVG-full-acc-top-5}{15.6\%}
\DefMacro{results-ln-lo-ta-ln-s+bsexpl1+attn+copy_ta-val-AVG-full-correct-top-1}{11933.3\%}
\DefMacro{results-ln-lo-ta-ln-s+bsexpl1+attn+copy_ta-val-AVG-full-correct-top-2}{17166.7\%}
\DefMacro{results-ln-lo-ta-ln-s+bsexpl1+attn+copy_ta-val-AVG-full-correct-top-3}{20533.3\%}
\DefMacro{results-ln-lo-ta-ln-s+bsexpl1+attn+copy_ta-val-AVG-full-correct-top-5}{24333.3\%}
\DefMacro{results-ln-lo-ta-ln-s+bsexpl1+attn+copy_ta-val-MAX-BLEU-4}{46.2}
\DefMacro{results-ln-lo-ta-ln-s+bsexpl1+attn+copy_ta-val-MAX-char-acc}{35.7\%}
\DefMacro{results-ln-lo-ta-ln-s+bsexpl1+attn+copy_ta-val-MAX-count-top-1}{1556}
\DefMacro{results-ln-lo-ta-ln-s+bsexpl1+attn+copy_ta-val-MAX-count-top-2}{1556}
\DefMacro{results-ln-lo-ta-ln-s+bsexpl1+attn+copy_ta-val-MAX-count-top-3}{1556}
\DefMacro{results-ln-lo-ta-ln-s+bsexpl1+attn+copy_ta-val-MAX-count-top-5}{1556}
\DefMacro{results-ln-lo-ta-ln-s+bsexpl1+attn+copy_ta-val-MAX-frag-acc}{22.5\%}
\DefMacro{results-ln-lo-ta-ln-s+bsexpl1+attn+copy_ta-val-MAX-full-acc-top-1}{10.0\%}
\DefMacro{results-ln-lo-ta-ln-s+bsexpl1+attn+copy_ta-val-MAX-full-acc-top-2}{13.4\%}
\DefMacro{results-ln-lo-ta-ln-s+bsexpl1+attn+copy_ta-val-MAX-full-acc-top-3}{15.9\%}
\DefMacro{results-ln-lo-ta-ln-s+bsexpl1+attn+copy_ta-val-MAX-full-acc-top-5}{18.7\%}
\DefMacro{results-ln-lo-ta-ln-s+bsexpl1+attn+copy_ta-val-MAX-full-correct-top-1}{156}
\DefMacro{results-ln-lo-ta-ln-s+bsexpl1+attn+copy_ta-val-MAX-full-correct-top-2}{209}
\DefMacro{results-ln-lo-ta-ln-s+bsexpl1+attn+copy_ta-val-MAX-full-correct-top-3}{248}
\DefMacro{results-ln-lo-ta-ln-s+bsexpl1+attn+copy_ta-val-MAX-full-correct-top-5}{291}
\DefMacro{results-ln-lo-ta-ln-s+bsexpl1+attn+copy_ta-val-MEDIAN-BLEU-4}{43.7}
\DefMacro{results-ln-lo-ta-ln-s+bsexpl1+attn+copy_ta-val-MEDIAN-char-acc}{33.6\%}
\DefMacro{results-ln-lo-ta-ln-s+bsexpl1+attn+copy_ta-val-MEDIAN-count-top-1}{155600.0\%}
\DefMacro{results-ln-lo-ta-ln-s+bsexpl1+attn+copy_ta-val-MEDIAN-count-top-2}{155600.0\%}
\DefMacro{results-ln-lo-ta-ln-s+bsexpl1+attn+copy_ta-val-MEDIAN-count-top-3}{155600.0\%}
\DefMacro{results-ln-lo-ta-ln-s+bsexpl1+attn+copy_ta-val-MEDIAN-count-top-5}{155600.0\%}
\DefMacro{results-ln-lo-ta-ln-s+bsexpl1+attn+copy_ta-val-MEDIAN-frag-acc}{20.5\%}
\DefMacro{results-ln-lo-ta-ln-s+bsexpl1+attn+copy_ta-val-MEDIAN-full-acc-top-1}{7.3\%}
\DefMacro{results-ln-lo-ta-ln-s+bsexpl1+attn+copy_ta-val-MEDIAN-full-acc-top-2}{10.0\%}
\DefMacro{results-ln-lo-ta-ln-s+bsexpl1+attn+copy_ta-val-MEDIAN-full-acc-top-3}{12.4\%}
\DefMacro{results-ln-lo-ta-ln-s+bsexpl1+attn+copy_ta-val-MEDIAN-full-acc-top-5}{14.6\%}
\DefMacro{results-ln-lo-ta-ln-s+bsexpl1+attn+copy_ta-val-MEDIAN-full-correct-top-1}{11400.0\%}
\DefMacro{results-ln-lo-ta-ln-s+bsexpl1+attn+copy_ta-val-MEDIAN-full-correct-top-2}{15500.0\%}
\DefMacro{results-ln-lo-ta-ln-s+bsexpl1+attn+copy_ta-val-MEDIAN-full-correct-top-3}{19300.0\%}
\DefMacro{results-ln-lo-ta-ln-s+bsexpl1+attn+copy_ta-val-MEDIAN-full-correct-top-5}{22700.0\%}
\DefMacro{results-ln-lo-ta-ln-s+bsexpl1+attn+copy_ta-val-MIN-BLEU-4}{41.1}
\DefMacro{results-ln-lo-ta-ln-s+bsexpl1+attn+copy_ta-val-MIN-char-acc}{32.7\%}
\DefMacro{results-ln-lo-ta-ln-s+bsexpl1+attn+copy_ta-val-MIN-count-top-1}{1556}
\DefMacro{results-ln-lo-ta-ln-s+bsexpl1+attn+copy_ta-val-MIN-count-top-2}{1556}
\DefMacro{results-ln-lo-ta-ln-s+bsexpl1+attn+copy_ta-val-MIN-count-top-3}{1556}
\DefMacro{results-ln-lo-ta-ln-s+bsexpl1+attn+copy_ta-val-MIN-count-top-5}{1556}
\DefMacro{results-ln-lo-ta-ln-s+bsexpl1+attn+copy_ta-val-MIN-frag-acc}{16.7\%}
\DefMacro{results-ln-lo-ta-ln-s+bsexpl1+attn+copy_ta-val-MIN-full-acc-top-1}{5.7\%}
\DefMacro{results-ln-lo-ta-ln-s+bsexpl1+attn+copy_ta-val-MIN-full-acc-top-2}{9.7\%}
\DefMacro{results-ln-lo-ta-ln-s+bsexpl1+attn+copy_ta-val-MIN-full-acc-top-3}{11.2\%}
\DefMacro{results-ln-lo-ta-ln-s+bsexpl1+attn+copy_ta-val-MIN-full-acc-top-5}{13.6\%}
\DefMacro{results-ln-lo-ta-ln-s+bsexpl1+attn+copy_ta-val-MIN-full-correct-top-1}{88}
\DefMacro{results-ln-lo-ta-ln-s+bsexpl1+attn+copy_ta-val-MIN-full-correct-top-2}{151}
\DefMacro{results-ln-lo-ta-ln-s+bsexpl1+attn+copy_ta-val-MIN-full-correct-top-3}{175}
\DefMacro{results-ln-lo-ta-ln-s+bsexpl1+attn+copy_ta-val-MIN-full-correct-top-5}{212}
\DefMacro{results-ln-lo-ta-ln-s+bsexpl1+attn+copy_ta-val-STDEV-BLEU-4}{2.1}
\DefMacro{results-ln-lo-ta-ln-s+bsexpl1+attn+copy_ta-val-STDEV-char-acc}{1.2\%}
\DefMacro{results-ln-lo-ta-ln-s+bsexpl1+attn+copy_ta-val-STDEV-count-top-1}{0.0\%}
\DefMacro{results-ln-lo-ta-ln-s+bsexpl1+attn+copy_ta-val-STDEV-count-top-2}{0.0\%}
\DefMacro{results-ln-lo-ta-ln-s+bsexpl1+attn+copy_ta-val-STDEV-count-top-3}{0.0\%}
\DefMacro{results-ln-lo-ta-ln-s+bsexpl1+attn+copy_ta-val-STDEV-count-top-5}{0.0\%}
\DefMacro{results-ln-lo-ta-ln-s+bsexpl1+attn+copy_ta-val-STDEV-frag-acc}{2.4\%}
\DefMacro{results-ln-lo-ta-ln-s+bsexpl1+attn+copy_ta-val-STDEV-full-acc-top-1}{1.8\%}
\DefMacro{results-ln-lo-ta-ln-s+bsexpl1+attn+copy_ta-val-STDEV-full-acc-top-2}{1.7\%}
\DefMacro{results-ln-lo-ta-ln-s+bsexpl1+attn+copy_ta-val-STDEV-full-acc-top-3}{2.0\%}
\DefMacro{results-ln-lo-ta-ln-s+bsexpl1+attn+copy_ta-val-STDEV-full-acc-top-5}{2.2\%}
\DefMacro{results-ln-lo-ta-ln-s+bsexpl1+attn+copy_ta-val-STDEV-full-correct-top-1}{2801.6\%}
\DefMacro{results-ln-lo-ta-ln-s+bsexpl1+attn+copy_ta-val-STDEV-full-correct-top-2}{2644.9\%}
\DefMacro{results-ln-lo-ta-ln-s+bsexpl1+attn+copy_ta-val-STDEV-full-correct-top-3}{3105.2\%}
\DefMacro{results-ln-lo-ta-ln-s+bsexpl1+attn+copy_ta-val-STDEV-full-correct-top-5}{3425.7\%}
\DefMacro{results-ln-lo-ta-ln-s+bsexpl1+attn+copy_ta-val-SUM-BLEU-4}{131.1}
\DefMacro{results-ln-lo-ta-ln-s+bsexpl1+attn+copy_ta-val-SUM-char-acc}{102.0\%}
\DefMacro{results-ln-lo-ta-ln-s+bsexpl1+attn+copy_ta-val-SUM-count-top-1}{4668}
\DefMacro{results-ln-lo-ta-ln-s+bsexpl1+attn+copy_ta-val-SUM-count-top-2}{4668}
\DefMacro{results-ln-lo-ta-ln-s+bsexpl1+attn+copy_ta-val-SUM-count-top-3}{4668}
\DefMacro{results-ln-lo-ta-ln-s+bsexpl1+attn+copy_ta-val-SUM-count-top-5}{4668}
\DefMacro{results-ln-lo-ta-ln-s+bsexpl1+attn+copy_ta-val-SUM-frag-acc}{59.6\%}
\DefMacro{results-ln-lo-ta-ln-s+bsexpl1+attn+copy_ta-val-SUM-full-acc-top-1}{23.0\%}
\DefMacro{results-ln-lo-ta-ln-s+bsexpl1+attn+copy_ta-val-SUM-full-acc-top-2}{33.1\%}
\DefMacro{results-ln-lo-ta-ln-s+bsexpl1+attn+copy_ta-val-SUM-full-acc-top-3}{39.6\%}
\DefMacro{results-ln-lo-ta-ln-s+bsexpl1+attn+copy_ta-val-SUM-full-acc-top-5}{46.9\%}
\DefMacro{results-ln-lo-ta-ln-s+bsexpl1+attn+copy_ta-val-SUM-full-correct-top-1}{358}
\DefMacro{results-ln-lo-ta-ln-s+bsexpl1+attn+copy_ta-val-SUM-full-correct-top-2}{515}
\DefMacro{results-ln-lo-ta-ln-s+bsexpl1+attn+copy_ta-val-SUM-full-correct-top-3}{616}
\DefMacro{results-ln-lo-ta-ln-s+bsexpl1+attn+copy_ta-val-SUM-full-correct-top-5}{730}
\DefMacro{results-ln-lo-ta-ln-s+bsexpl1+attn+copy_ta-test-AVG-BLEU-4}{33.9}
\DefMacro{results-ln-lo-ta-ln-s+bsexpl1+attn+copy_ta-test-AVG-char-acc}{22.1\%}
\DefMacro{results-ln-lo-ta-ln-s+bsexpl1+attn+copy_ta-test-AVG-count-top-1}{116700.0\%}
\DefMacro{results-ln-lo-ta-ln-s+bsexpl1+attn+copy_ta-test-AVG-count-top-2}{116700.0\%}
\DefMacro{results-ln-lo-ta-ln-s+bsexpl1+attn+copy_ta-test-AVG-count-top-3}{116700.0\%}
\DefMacro{results-ln-lo-ta-ln-s+bsexpl1+attn+copy_ta-test-AVG-count-top-5}{116700.0\%}
\DefMacro{results-ln-lo-ta-ln-s+bsexpl1+attn+copy_ta-test-AVG-frag-acc}{21.3\%}
\DefMacro{results-ln-lo-ta-ln-s+bsexpl1+attn+copy_ta-test-AVG-full-acc-top-1}{4.4\%}
\DefMacro{results-ln-lo-ta-ln-s+bsexpl1+attn+copy_ta-test-AVG-full-acc-top-2}{6.4\%}
\DefMacro{results-ln-lo-ta-ln-s+bsexpl1+attn+copy_ta-test-AVG-full-acc-top-3}{7.7\%}
\DefMacro{results-ln-lo-ta-ln-s+bsexpl1+attn+copy_ta-test-AVG-full-acc-top-5}{8.9\%}
\DefMacro{results-ln-lo-ta-ln-s+bsexpl1+attn+copy_ta-test-AVG-full-correct-top-1}{5100.0\%}
\DefMacro{results-ln-lo-ta-ln-s+bsexpl1+attn+copy_ta-test-AVG-full-correct-top-2}{7466.7\%}
\DefMacro{results-ln-lo-ta-ln-s+bsexpl1+attn+copy_ta-test-AVG-full-correct-top-3}{8933.3\%}
\DefMacro{results-ln-lo-ta-ln-s+bsexpl1+attn+copy_ta-test-AVG-full-correct-top-5}{10433.3\%}
\DefMacro{results-ln-lo-ta-ln-s+bsexpl1+attn+copy_ta-test-MAX-BLEU-4}{34.5}
\DefMacro{results-ln-lo-ta-ln-s+bsexpl1+attn+copy_ta-test-MAX-char-acc}{23.0\%}
\DefMacro{results-ln-lo-ta-ln-s+bsexpl1+attn+copy_ta-test-MAX-count-top-1}{1167}
\DefMacro{results-ln-lo-ta-ln-s+bsexpl1+attn+copy_ta-test-MAX-count-top-2}{1167}
\DefMacro{results-ln-lo-ta-ln-s+bsexpl1+attn+copy_ta-test-MAX-count-top-3}{1167}
\DefMacro{results-ln-lo-ta-ln-s+bsexpl1+attn+copy_ta-test-MAX-count-top-5}{1167}
\DefMacro{results-ln-lo-ta-ln-s+bsexpl1+attn+copy_ta-test-MAX-frag-acc}{22.2\%}
\DefMacro{results-ln-lo-ta-ln-s+bsexpl1+attn+copy_ta-test-MAX-full-acc-top-1}{5.4\%}
\DefMacro{results-ln-lo-ta-ln-s+bsexpl1+attn+copy_ta-test-MAX-full-acc-top-2}{7.3\%}
\DefMacro{results-ln-lo-ta-ln-s+bsexpl1+attn+copy_ta-test-MAX-full-acc-top-3}{8.8\%}
\DefMacro{results-ln-lo-ta-ln-s+bsexpl1+attn+copy_ta-test-MAX-full-acc-top-5}{10.1\%}
\DefMacro{results-ln-lo-ta-ln-s+bsexpl1+attn+copy_ta-test-MAX-full-correct-top-1}{63}
\DefMacro{results-ln-lo-ta-ln-s+bsexpl1+attn+copy_ta-test-MAX-full-correct-top-2}{85}
\DefMacro{results-ln-lo-ta-ln-s+bsexpl1+attn+copy_ta-test-MAX-full-correct-top-3}{103}
\DefMacro{results-ln-lo-ta-ln-s+bsexpl1+attn+copy_ta-test-MAX-full-correct-top-5}{118}
\DefMacro{results-ln-lo-ta-ln-s+bsexpl1+attn+copy_ta-test-MEDIAN-BLEU-4}{34.3}
\DefMacro{results-ln-lo-ta-ln-s+bsexpl1+attn+copy_ta-test-MEDIAN-char-acc}{21.8\%}
\DefMacro{results-ln-lo-ta-ln-s+bsexpl1+attn+copy_ta-test-MEDIAN-count-top-1}{116700.0\%}
\DefMacro{results-ln-lo-ta-ln-s+bsexpl1+attn+copy_ta-test-MEDIAN-count-top-2}{116700.0\%}
\DefMacro{results-ln-lo-ta-ln-s+bsexpl1+attn+copy_ta-test-MEDIAN-count-top-3}{116700.0\%}
\DefMacro{results-ln-lo-ta-ln-s+bsexpl1+attn+copy_ta-test-MEDIAN-count-top-5}{116700.0\%}
\DefMacro{results-ln-lo-ta-ln-s+bsexpl1+attn+copy_ta-test-MEDIAN-frag-acc}{21.9\%}
\DefMacro{results-ln-lo-ta-ln-s+bsexpl1+attn+copy_ta-test-MEDIAN-full-acc-top-1}{3.9\%}
\DefMacro{results-ln-lo-ta-ln-s+bsexpl1+attn+copy_ta-test-MEDIAN-full-acc-top-2}{6.2\%}
\DefMacro{results-ln-lo-ta-ln-s+bsexpl1+attn+copy_ta-test-MEDIAN-full-acc-top-3}{7.5\%}
\DefMacro{results-ln-lo-ta-ln-s+bsexpl1+attn+copy_ta-test-MEDIAN-full-acc-top-5}{8.9\%}
\DefMacro{results-ln-lo-ta-ln-s+bsexpl1+attn+copy_ta-test-MEDIAN-full-correct-top-1}{4600.0\%}
\DefMacro{results-ln-lo-ta-ln-s+bsexpl1+attn+copy_ta-test-MEDIAN-full-correct-top-2}{7200.0\%}
\DefMacro{results-ln-lo-ta-ln-s+bsexpl1+attn+copy_ta-test-MEDIAN-full-correct-top-3}{8700.0\%}
\DefMacro{results-ln-lo-ta-ln-s+bsexpl1+attn+copy_ta-test-MEDIAN-full-correct-top-5}{10400.0\%}
\DefMacro{results-ln-lo-ta-ln-s+bsexpl1+attn+copy_ta-test-MIN-BLEU-4}{32.9}
\DefMacro{results-ln-lo-ta-ln-s+bsexpl1+attn+copy_ta-test-MIN-char-acc}{21.4\%}
\DefMacro{results-ln-lo-ta-ln-s+bsexpl1+attn+copy_ta-test-MIN-count-top-1}{1167}
\DefMacro{results-ln-lo-ta-ln-s+bsexpl1+attn+copy_ta-test-MIN-count-top-2}{1167}
\DefMacro{results-ln-lo-ta-ln-s+bsexpl1+attn+copy_ta-test-MIN-count-top-3}{1167}
\DefMacro{results-ln-lo-ta-ln-s+bsexpl1+attn+copy_ta-test-MIN-count-top-5}{1167}
\DefMacro{results-ln-lo-ta-ln-s+bsexpl1+attn+copy_ta-test-MIN-frag-acc}{19.8\%}
\DefMacro{results-ln-lo-ta-ln-s+bsexpl1+attn+copy_ta-test-MIN-full-acc-top-1}{3.8\%}
\DefMacro{results-ln-lo-ta-ln-s+bsexpl1+attn+copy_ta-test-MIN-full-acc-top-2}{5.7\%}
\DefMacro{results-ln-lo-ta-ln-s+bsexpl1+attn+copy_ta-test-MIN-full-acc-top-3}{6.7\%}
\DefMacro{results-ln-lo-ta-ln-s+bsexpl1+attn+copy_ta-test-MIN-full-acc-top-5}{7.8\%}
\DefMacro{results-ln-lo-ta-ln-s+bsexpl1+attn+copy_ta-test-MIN-full-correct-top-1}{44}
\DefMacro{results-ln-lo-ta-ln-s+bsexpl1+attn+copy_ta-test-MIN-full-correct-top-2}{67}
\DefMacro{results-ln-lo-ta-ln-s+bsexpl1+attn+copy_ta-test-MIN-full-correct-top-3}{78}
\DefMacro{results-ln-lo-ta-ln-s+bsexpl1+attn+copy_ta-test-MIN-full-correct-top-5}{91}
\DefMacro{results-ln-lo-ta-ln-s+bsexpl1+attn+copy_ta-test-STDEV-BLEU-4}{0.7}
\DefMacro{results-ln-lo-ta-ln-s+bsexpl1+attn+copy_ta-test-STDEV-char-acc}{0.7\%}
\DefMacro{results-ln-lo-ta-ln-s+bsexpl1+attn+copy_ta-test-STDEV-count-top-1}{0.0\%}
\DefMacro{results-ln-lo-ta-ln-s+bsexpl1+attn+copy_ta-test-STDEV-count-top-2}{0.0\%}
\DefMacro{results-ln-lo-ta-ln-s+bsexpl1+attn+copy_ta-test-STDEV-count-top-3}{0.0\%}
\DefMacro{results-ln-lo-ta-ln-s+bsexpl1+attn+copy_ta-test-STDEV-count-top-5}{0.0\%}
\DefMacro{results-ln-lo-ta-ln-s+bsexpl1+attn+copy_ta-test-STDEV-frag-acc}{1.1\%}
\DefMacro{results-ln-lo-ta-ln-s+bsexpl1+attn+copy_ta-test-STDEV-full-acc-top-1}{0.7\%}
\DefMacro{results-ln-lo-ta-ln-s+bsexpl1+attn+copy_ta-test-STDEV-full-acc-top-2}{0.7\%}
\DefMacro{results-ln-lo-ta-ln-s+bsexpl1+attn+copy_ta-test-STDEV-full-acc-top-3}{0.9\%}
\DefMacro{results-ln-lo-ta-ln-s+bsexpl1+attn+copy_ta-test-STDEV-full-acc-top-5}{0.9\%}
\DefMacro{results-ln-lo-ta-ln-s+bsexpl1+attn+copy_ta-test-STDEV-full-correct-top-1}{852.4\%}
\DefMacro{results-ln-lo-ta-ln-s+bsexpl1+attn+copy_ta-test-STDEV-full-correct-top-2}{758.7\%}
\DefMacro{results-ln-lo-ta-ln-s+bsexpl1+attn+copy_ta-test-STDEV-full-correct-top-3}{1033.9\%}
\DefMacro{results-ln-lo-ta-ln-s+bsexpl1+attn+copy_ta-test-STDEV-full-correct-top-5}{1102.5\%}
\DefMacro{results-ln-lo-ta-ln-s+bsexpl1+attn+copy_ta-test-SUM-BLEU-4}{101.7}
\DefMacro{results-ln-lo-ta-ln-s+bsexpl1+attn+copy_ta-test-SUM-char-acc}{66.2\%}
\DefMacro{results-ln-lo-ta-ln-s+bsexpl1+attn+copy_ta-test-SUM-count-top-1}{3501}
\DefMacro{results-ln-lo-ta-ln-s+bsexpl1+attn+copy_ta-test-SUM-count-top-2}{3501}
\DefMacro{results-ln-lo-ta-ln-s+bsexpl1+attn+copy_ta-test-SUM-count-top-3}{3501}
\DefMacro{results-ln-lo-ta-ln-s+bsexpl1+attn+copy_ta-test-SUM-count-top-5}{3501}
\DefMacro{results-ln-lo-ta-ln-s+bsexpl1+attn+copy_ta-test-SUM-frag-acc}{63.8\%}
\DefMacro{results-ln-lo-ta-ln-s+bsexpl1+attn+copy_ta-test-SUM-full-acc-top-1}{13.1\%}
\DefMacro{results-ln-lo-ta-ln-s+bsexpl1+attn+copy_ta-test-SUM-full-acc-top-2}{19.2\%}
\DefMacro{results-ln-lo-ta-ln-s+bsexpl1+attn+copy_ta-test-SUM-full-acc-top-3}{23.0\%}
\DefMacro{results-ln-lo-ta-ln-s+bsexpl1+attn+copy_ta-test-SUM-full-acc-top-5}{26.8\%}
\DefMacro{results-ln-lo-ta-ln-s+bsexpl1+attn+copy_ta-test-SUM-full-correct-top-1}{153}
\DefMacro{results-ln-lo-ta-ln-s+bsexpl1+attn+copy_ta-test-SUM-full-correct-top-2}{224}
\DefMacro{results-ln-lo-ta-ln-s+bsexpl1+attn+copy_ta-test-SUM-full-correct-top-3}{268}
\DefMacro{results-ln-lo-ta-ln-s+bsexpl1+attn+copy_ta-test-SUM-full-correct-top-5}{313}
\DefMacro{results-ln-lo-lo-25-ln-s+bsexpl1+attn+copy_ta-val-AVG-BLEU-4}{49.3}
\DefMacro{results-ln-lo-lo-25-ln-s+bsexpl1+attn+copy_ta-val-AVG-char-acc}{34.3\%}
\DefMacro{results-ln-lo-lo-25-ln-s+bsexpl1+attn+copy_ta-val-AVG-count-top-1}{14400.0\%}
\DefMacro{results-ln-lo-lo-25-ln-s+bsexpl1+attn+copy_ta-val-AVG-count-top-2}{14400.0\%}
\DefMacro{results-ln-lo-lo-25-ln-s+bsexpl1+attn+copy_ta-val-AVG-count-top-3}{14400.0\%}
\DefMacro{results-ln-lo-lo-25-ln-s+bsexpl1+attn+copy_ta-val-AVG-count-top-5}{14400.0\%}
\DefMacro{results-ln-lo-lo-25-ln-s+bsexpl1+attn+copy_ta-val-AVG-frag-acc}{40.3\%}
\DefMacro{results-ln-lo-lo-25-ln-s+bsexpl1+attn+copy_ta-val-AVG-full-acc-top-1}{5.6\%}
\DefMacro{results-ln-lo-lo-25-ln-s+bsexpl1+attn+copy_ta-val-AVG-full-acc-top-2}{8.1\%}
\DefMacro{results-ln-lo-lo-25-ln-s+bsexpl1+attn+copy_ta-val-AVG-full-acc-top-3}{9.3\%}
\DefMacro{results-ln-lo-lo-25-ln-s+bsexpl1+attn+copy_ta-val-AVG-full-acc-top-5}{10.2\%}
\DefMacro{results-ln-lo-lo-25-ln-s+bsexpl1+attn+copy_ta-val-AVG-full-correct-top-1}{800.0\%}
\DefMacro{results-ln-lo-lo-25-ln-s+bsexpl1+attn+copy_ta-val-AVG-full-correct-top-2}{1166.7\%}
\DefMacro{results-ln-lo-lo-25-ln-s+bsexpl1+attn+copy_ta-val-AVG-full-correct-top-3}{1333.3\%}
\DefMacro{results-ln-lo-lo-25-ln-s+bsexpl1+attn+copy_ta-val-AVG-full-correct-top-5}{1466.7\%}
\DefMacro{results-ln-lo-lo-25-ln-s+bsexpl1+attn+copy_ta-val-MAX-BLEU-4}{51.2}
\DefMacro{results-ln-lo-lo-25-ln-s+bsexpl1+attn+copy_ta-val-MAX-char-acc}{37.0\%}
\DefMacro{results-ln-lo-lo-25-ln-s+bsexpl1+attn+copy_ta-val-MAX-count-top-1}{144}
\DefMacro{results-ln-lo-lo-25-ln-s+bsexpl1+attn+copy_ta-val-MAX-count-top-2}{144}
\DefMacro{results-ln-lo-lo-25-ln-s+bsexpl1+attn+copy_ta-val-MAX-count-top-3}{144}
\DefMacro{results-ln-lo-lo-25-ln-s+bsexpl1+attn+copy_ta-val-MAX-count-top-5}{144}
\DefMacro{results-ln-lo-lo-25-ln-s+bsexpl1+attn+copy_ta-val-MAX-frag-acc}{43.5\%}
\DefMacro{results-ln-lo-lo-25-ln-s+bsexpl1+attn+copy_ta-val-MAX-full-acc-top-1}{8.3\%}
\DefMacro{results-ln-lo-lo-25-ln-s+bsexpl1+attn+copy_ta-val-MAX-full-acc-top-2}{9.7\%}
\DefMacro{results-ln-lo-lo-25-ln-s+bsexpl1+attn+copy_ta-val-MAX-full-acc-top-3}{10.4\%}
\DefMacro{results-ln-lo-lo-25-ln-s+bsexpl1+attn+copy_ta-val-MAX-full-acc-top-5}{12.5\%}
\DefMacro{results-ln-lo-lo-25-ln-s+bsexpl1+attn+copy_ta-val-MAX-full-correct-top-1}{12}
\DefMacro{results-ln-lo-lo-25-ln-s+bsexpl1+attn+copy_ta-val-MAX-full-correct-top-2}{14}
\DefMacro{results-ln-lo-lo-25-ln-s+bsexpl1+attn+copy_ta-val-MAX-full-correct-top-3}{15}
\DefMacro{results-ln-lo-lo-25-ln-s+bsexpl1+attn+copy_ta-val-MAX-full-correct-top-5}{18}
\DefMacro{results-ln-lo-lo-25-ln-s+bsexpl1+attn+copy_ta-val-MEDIAN-BLEU-4}{49.7}
\DefMacro{results-ln-lo-lo-25-ln-s+bsexpl1+attn+copy_ta-val-MEDIAN-char-acc}{35.2\%}
\DefMacro{results-ln-lo-lo-25-ln-s+bsexpl1+attn+copy_ta-val-MEDIAN-count-top-1}{14400.0\%}
\DefMacro{results-ln-lo-lo-25-ln-s+bsexpl1+attn+copy_ta-val-MEDIAN-count-top-2}{14400.0\%}
\DefMacro{results-ln-lo-lo-25-ln-s+bsexpl1+attn+copy_ta-val-MEDIAN-count-top-3}{14400.0\%}
\DefMacro{results-ln-lo-lo-25-ln-s+bsexpl1+attn+copy_ta-val-MEDIAN-count-top-5}{14400.0\%}
\DefMacro{results-ln-lo-lo-25-ln-s+bsexpl1+attn+copy_ta-val-MEDIAN-frag-acc}{40.1\%}
\DefMacro{results-ln-lo-lo-25-ln-s+bsexpl1+attn+copy_ta-val-MEDIAN-full-acc-top-1}{4.2\%}
\DefMacro{results-ln-lo-lo-25-ln-s+bsexpl1+attn+copy_ta-val-MEDIAN-full-acc-top-2}{7.6\%}
\DefMacro{results-ln-lo-lo-25-ln-s+bsexpl1+attn+copy_ta-val-MEDIAN-full-acc-top-3}{9.0\%}
\DefMacro{results-ln-lo-lo-25-ln-s+bsexpl1+attn+copy_ta-val-MEDIAN-full-acc-top-5}{9.0\%}
\DefMacro{results-ln-lo-lo-25-ln-s+bsexpl1+attn+copy_ta-val-MEDIAN-full-correct-top-1}{600.0\%}
\DefMacro{results-ln-lo-lo-25-ln-s+bsexpl1+attn+copy_ta-val-MEDIAN-full-correct-top-2}{1100.0\%}
\DefMacro{results-ln-lo-lo-25-ln-s+bsexpl1+attn+copy_ta-val-MEDIAN-full-correct-top-3}{1300.0\%}
\DefMacro{results-ln-lo-lo-25-ln-s+bsexpl1+attn+copy_ta-val-MEDIAN-full-correct-top-5}{1300.0\%}
\DefMacro{results-ln-lo-lo-25-ln-s+bsexpl1+attn+copy_ta-val-MIN-BLEU-4}{47.1}
\DefMacro{results-ln-lo-lo-25-ln-s+bsexpl1+attn+copy_ta-val-MIN-char-acc}{30.7\%}
\DefMacro{results-ln-lo-lo-25-ln-s+bsexpl1+attn+copy_ta-val-MIN-count-top-1}{144}
\DefMacro{results-ln-lo-lo-25-ln-s+bsexpl1+attn+copy_ta-val-MIN-count-top-2}{144}
\DefMacro{results-ln-lo-lo-25-ln-s+bsexpl1+attn+copy_ta-val-MIN-count-top-3}{144}
\DefMacro{results-ln-lo-lo-25-ln-s+bsexpl1+attn+copy_ta-val-MIN-count-top-5}{144}
\DefMacro{results-ln-lo-lo-25-ln-s+bsexpl1+attn+copy_ta-val-MIN-frag-acc}{37.4\%}
\DefMacro{results-ln-lo-lo-25-ln-s+bsexpl1+attn+copy_ta-val-MIN-full-acc-top-1}{4.2\%}
\DefMacro{results-ln-lo-lo-25-ln-s+bsexpl1+attn+copy_ta-val-MIN-full-acc-top-2}{6.9\%}
\DefMacro{results-ln-lo-lo-25-ln-s+bsexpl1+attn+copy_ta-val-MIN-full-acc-top-3}{8.3\%}
\DefMacro{results-ln-lo-lo-25-ln-s+bsexpl1+attn+copy_ta-val-MIN-full-acc-top-5}{9.0\%}
\DefMacro{results-ln-lo-lo-25-ln-s+bsexpl1+attn+copy_ta-val-MIN-full-correct-top-1}{6}
\DefMacro{results-ln-lo-lo-25-ln-s+bsexpl1+attn+copy_ta-val-MIN-full-correct-top-2}{10}
\DefMacro{results-ln-lo-lo-25-ln-s+bsexpl1+attn+copy_ta-val-MIN-full-correct-top-3}{12}
\DefMacro{results-ln-lo-lo-25-ln-s+bsexpl1+attn+copy_ta-val-MIN-full-correct-top-5}{13}
\DefMacro{results-ln-lo-lo-25-ln-s+bsexpl1+attn+copy_ta-val-STDEV-BLEU-4}{1.7}
\DefMacro{results-ln-lo-lo-25-ln-s+bsexpl1+attn+copy_ta-val-STDEV-char-acc}{2.7\%}
\DefMacro{results-ln-lo-lo-25-ln-s+bsexpl1+attn+copy_ta-val-STDEV-count-top-1}{0.0\%}
\DefMacro{results-ln-lo-lo-25-ln-s+bsexpl1+attn+copy_ta-val-STDEV-count-top-2}{0.0\%}
\DefMacro{results-ln-lo-lo-25-ln-s+bsexpl1+attn+copy_ta-val-STDEV-count-top-3}{0.0\%}
\DefMacro{results-ln-lo-lo-25-ln-s+bsexpl1+attn+copy_ta-val-STDEV-count-top-5}{0.0\%}
\DefMacro{results-ln-lo-lo-25-ln-s+bsexpl1+attn+copy_ta-val-STDEV-frag-acc}{2.5\%}
\DefMacro{results-ln-lo-lo-25-ln-s+bsexpl1+attn+copy_ta-val-STDEV-full-acc-top-1}{2.0\%}
\DefMacro{results-ln-lo-lo-25-ln-s+bsexpl1+attn+copy_ta-val-STDEV-full-acc-top-2}{1.2\%}
\DefMacro{results-ln-lo-lo-25-ln-s+bsexpl1+attn+copy_ta-val-STDEV-full-acc-top-3}{0.9\%}
\DefMacro{results-ln-lo-lo-25-ln-s+bsexpl1+attn+copy_ta-val-STDEV-full-acc-top-5}{1.6\%}
\DefMacro{results-ln-lo-lo-25-ln-s+bsexpl1+attn+copy_ta-val-STDEV-full-correct-top-1}{282.8\%}
\DefMacro{results-ln-lo-lo-25-ln-s+bsexpl1+attn+copy_ta-val-STDEV-full-correct-top-2}{170.0\%}
\DefMacro{results-ln-lo-lo-25-ln-s+bsexpl1+attn+copy_ta-val-STDEV-full-correct-top-3}{124.7\%}
\DefMacro{results-ln-lo-lo-25-ln-s+bsexpl1+attn+copy_ta-val-STDEV-full-correct-top-5}{235.7\%}
\DefMacro{results-ln-lo-lo-25-ln-s+bsexpl1+attn+copy_ta-val-SUM-BLEU-4}{148.0}
\DefMacro{results-ln-lo-lo-25-ln-s+bsexpl1+attn+copy_ta-val-SUM-char-acc}{102.8\%}
\DefMacro{results-ln-lo-lo-25-ln-s+bsexpl1+attn+copy_ta-val-SUM-count-top-1}{432}
\DefMacro{results-ln-lo-lo-25-ln-s+bsexpl1+attn+copy_ta-val-SUM-count-top-2}{432}
\DefMacro{results-ln-lo-lo-25-ln-s+bsexpl1+attn+copy_ta-val-SUM-count-top-3}{432}
\DefMacro{results-ln-lo-lo-25-ln-s+bsexpl1+attn+copy_ta-val-SUM-count-top-5}{432}
\DefMacro{results-ln-lo-lo-25-ln-s+bsexpl1+attn+copy_ta-val-SUM-frag-acc}{120.9\%}
\DefMacro{results-ln-lo-lo-25-ln-s+bsexpl1+attn+copy_ta-val-SUM-full-acc-top-1}{16.7\%}
\DefMacro{results-ln-lo-lo-25-ln-s+bsexpl1+attn+copy_ta-val-SUM-full-acc-top-2}{24.3\%}
\DefMacro{results-ln-lo-lo-25-ln-s+bsexpl1+attn+copy_ta-val-SUM-full-acc-top-3}{27.8\%}
\DefMacro{results-ln-lo-lo-25-ln-s+bsexpl1+attn+copy_ta-val-SUM-full-acc-top-5}{30.6\%}
\DefMacro{results-ln-lo-lo-25-ln-s+bsexpl1+attn+copy_ta-val-SUM-full-correct-top-1}{24}
\DefMacro{results-ln-lo-lo-25-ln-s+bsexpl1+attn+copy_ta-val-SUM-full-correct-top-2}{35}
\DefMacro{results-ln-lo-lo-25-ln-s+bsexpl1+attn+copy_ta-val-SUM-full-correct-top-3}{40}
\DefMacro{results-ln-lo-lo-25-ln-s+bsexpl1+attn+copy_ta-val-SUM-full-correct-top-5}{44}
\DefMacro{results-ln-lo-lo-25-ln-s+bsexpl1+attn+copy_ta-test-AVG-BLEU-4}{32.6}
\DefMacro{results-ln-lo-lo-25-ln-s+bsexpl1+attn+copy_ta-test-AVG-char-acc}{21.3\%}
\DefMacro{results-ln-lo-lo-25-ln-s+bsexpl1+attn+copy_ta-test-AVG-count-top-1}{116700.0\%}
\DefMacro{results-ln-lo-lo-25-ln-s+bsexpl1+attn+copy_ta-test-AVG-count-top-2}{116700.0\%}
\DefMacro{results-ln-lo-lo-25-ln-s+bsexpl1+attn+copy_ta-test-AVG-count-top-3}{116700.0\%}
\DefMacro{results-ln-lo-lo-25-ln-s+bsexpl1+attn+copy_ta-test-AVG-count-top-5}{116700.0\%}
\DefMacro{results-ln-lo-lo-25-ln-s+bsexpl1+attn+copy_ta-test-AVG-frag-acc}{21.5\%}
\DefMacro{results-ln-lo-lo-25-ln-s+bsexpl1+attn+copy_ta-test-AVG-full-acc-top-1}{3.3\%}
\DefMacro{results-ln-lo-lo-25-ln-s+bsexpl1+attn+copy_ta-test-AVG-full-acc-top-2}{4.2\%}
\DefMacro{results-ln-lo-lo-25-ln-s+bsexpl1+attn+copy_ta-test-AVG-full-acc-top-3}{4.6\%}
\DefMacro{results-ln-lo-lo-25-ln-s+bsexpl1+attn+copy_ta-test-AVG-full-acc-top-5}{5.3\%}
\DefMacro{results-ln-lo-lo-25-ln-s+bsexpl1+attn+copy_ta-test-AVG-full-correct-top-1}{3866.7\%}
\DefMacro{results-ln-lo-lo-25-ln-s+bsexpl1+attn+copy_ta-test-AVG-full-correct-top-2}{4900.0\%}
\DefMacro{results-ln-lo-lo-25-ln-s+bsexpl1+attn+copy_ta-test-AVG-full-correct-top-3}{5366.7\%}
\DefMacro{results-ln-lo-lo-25-ln-s+bsexpl1+attn+copy_ta-test-AVG-full-correct-top-5}{6200.0\%}
\DefMacro{results-ln-lo-lo-25-ln-s+bsexpl1+attn+copy_ta-test-MAX-BLEU-4}{33.4}
\DefMacro{results-ln-lo-lo-25-ln-s+bsexpl1+attn+copy_ta-test-MAX-char-acc}{22.0\%}
\DefMacro{results-ln-lo-lo-25-ln-s+bsexpl1+attn+copy_ta-test-MAX-count-top-1}{1167}
\DefMacro{results-ln-lo-lo-25-ln-s+bsexpl1+attn+copy_ta-test-MAX-count-top-2}{1167}
\DefMacro{results-ln-lo-lo-25-ln-s+bsexpl1+attn+copy_ta-test-MAX-count-top-3}{1167}
\DefMacro{results-ln-lo-lo-25-ln-s+bsexpl1+attn+copy_ta-test-MAX-count-top-5}{1167}
\DefMacro{results-ln-lo-lo-25-ln-s+bsexpl1+attn+copy_ta-test-MAX-frag-acc}{22.1\%}
\DefMacro{results-ln-lo-lo-25-ln-s+bsexpl1+attn+copy_ta-test-MAX-full-acc-top-1}{3.9\%}
\DefMacro{results-ln-lo-lo-25-ln-s+bsexpl1+attn+copy_ta-test-MAX-full-acc-top-2}{5.1\%}
\DefMacro{results-ln-lo-lo-25-ln-s+bsexpl1+attn+copy_ta-test-MAX-full-acc-top-3}{5.6\%}
\DefMacro{results-ln-lo-lo-25-ln-s+bsexpl1+attn+copy_ta-test-MAX-full-acc-top-5}{6.1\%}
\DefMacro{results-ln-lo-lo-25-ln-s+bsexpl1+attn+copy_ta-test-MAX-full-correct-top-1}{46}
\DefMacro{results-ln-lo-lo-25-ln-s+bsexpl1+attn+copy_ta-test-MAX-full-correct-top-2}{60}
\DefMacro{results-ln-lo-lo-25-ln-s+bsexpl1+attn+copy_ta-test-MAX-full-correct-top-3}{65}
\DefMacro{results-ln-lo-lo-25-ln-s+bsexpl1+attn+copy_ta-test-MAX-full-correct-top-5}{71}
\DefMacro{results-ln-lo-lo-25-ln-s+bsexpl1+attn+copy_ta-test-MEDIAN-BLEU-4}{32.2}
\DefMacro{results-ln-lo-lo-25-ln-s+bsexpl1+attn+copy_ta-test-MEDIAN-char-acc}{21.7\%}
\DefMacro{results-ln-lo-lo-25-ln-s+bsexpl1+attn+copy_ta-test-MEDIAN-count-top-1}{116700.0\%}
\DefMacro{results-ln-lo-lo-25-ln-s+bsexpl1+attn+copy_ta-test-MEDIAN-count-top-2}{116700.0\%}
\DefMacro{results-ln-lo-lo-25-ln-s+bsexpl1+attn+copy_ta-test-MEDIAN-count-top-3}{116700.0\%}
\DefMacro{results-ln-lo-lo-25-ln-s+bsexpl1+attn+copy_ta-test-MEDIAN-count-top-5}{116700.0\%}
\DefMacro{results-ln-lo-lo-25-ln-s+bsexpl1+attn+copy_ta-test-MEDIAN-frag-acc}{21.3\%}
\DefMacro{results-ln-lo-lo-25-ln-s+bsexpl1+attn+copy_ta-test-MEDIAN-full-acc-top-1}{3.3\%}
\DefMacro{results-ln-lo-lo-25-ln-s+bsexpl1+attn+copy_ta-test-MEDIAN-full-acc-top-2}{3.9\%}
\DefMacro{results-ln-lo-lo-25-ln-s+bsexpl1+attn+copy_ta-test-MEDIAN-full-acc-top-3}{4.5\%}
\DefMacro{results-ln-lo-lo-25-ln-s+bsexpl1+attn+copy_ta-test-MEDIAN-full-acc-top-5}{5.2\%}
\DefMacro{results-ln-lo-lo-25-ln-s+bsexpl1+attn+copy_ta-test-MEDIAN-full-correct-top-1}{3800.0\%}
\DefMacro{results-ln-lo-lo-25-ln-s+bsexpl1+attn+copy_ta-test-MEDIAN-full-correct-top-2}{4600.0\%}
\DefMacro{results-ln-lo-lo-25-ln-s+bsexpl1+attn+copy_ta-test-MEDIAN-full-correct-top-3}{5300.0\%}
\DefMacro{results-ln-lo-lo-25-ln-s+bsexpl1+attn+copy_ta-test-MEDIAN-full-correct-top-5}{6100.0\%}
\DefMacro{results-ln-lo-lo-25-ln-s+bsexpl1+attn+copy_ta-test-MIN-BLEU-4}{32.2}
\DefMacro{results-ln-lo-lo-25-ln-s+bsexpl1+attn+copy_ta-test-MIN-char-acc}{20.3\%}
\DefMacro{results-ln-lo-lo-25-ln-s+bsexpl1+attn+copy_ta-test-MIN-count-top-1}{1167}
\DefMacro{results-ln-lo-lo-25-ln-s+bsexpl1+attn+copy_ta-test-MIN-count-top-2}{1167}
\DefMacro{results-ln-lo-lo-25-ln-s+bsexpl1+attn+copy_ta-test-MIN-count-top-3}{1167}
\DefMacro{results-ln-lo-lo-25-ln-s+bsexpl1+attn+copy_ta-test-MIN-count-top-5}{1167}
\DefMacro{results-ln-lo-lo-25-ln-s+bsexpl1+attn+copy_ta-test-MIN-frag-acc}{21.1\%}
\DefMacro{results-ln-lo-lo-25-ln-s+bsexpl1+attn+copy_ta-test-MIN-full-acc-top-1}{2.7\%}
\DefMacro{results-ln-lo-lo-25-ln-s+bsexpl1+attn+copy_ta-test-MIN-full-acc-top-2}{3.5\%}
\DefMacro{results-ln-lo-lo-25-ln-s+bsexpl1+attn+copy_ta-test-MIN-full-acc-top-3}{3.7\%}
\DefMacro{results-ln-lo-lo-25-ln-s+bsexpl1+attn+copy_ta-test-MIN-full-acc-top-5}{4.6\%}
\DefMacro{results-ln-lo-lo-25-ln-s+bsexpl1+attn+copy_ta-test-MIN-full-correct-top-1}{32}
\DefMacro{results-ln-lo-lo-25-ln-s+bsexpl1+attn+copy_ta-test-MIN-full-correct-top-2}{41}
\DefMacro{results-ln-lo-lo-25-ln-s+bsexpl1+attn+copy_ta-test-MIN-full-correct-top-3}{43}
\DefMacro{results-ln-lo-lo-25-ln-s+bsexpl1+attn+copy_ta-test-MIN-full-correct-top-5}{54}
\DefMacro{results-ln-lo-lo-25-ln-s+bsexpl1+attn+copy_ta-test-STDEV-BLEU-4}{0.6}
\DefMacro{results-ln-lo-lo-25-ln-s+bsexpl1+attn+copy_ta-test-STDEV-char-acc}{0.8\%}
\DefMacro{results-ln-lo-lo-25-ln-s+bsexpl1+attn+copy_ta-test-STDEV-count-top-1}{0.0\%}
\DefMacro{results-ln-lo-lo-25-ln-s+bsexpl1+attn+copy_ta-test-STDEV-count-top-2}{0.0\%}
\DefMacro{results-ln-lo-lo-25-ln-s+bsexpl1+attn+copy_ta-test-STDEV-count-top-3}{0.0\%}
\DefMacro{results-ln-lo-lo-25-ln-s+bsexpl1+attn+copy_ta-test-STDEV-count-top-5}{0.0\%}
\DefMacro{results-ln-lo-lo-25-ln-s+bsexpl1+attn+copy_ta-test-STDEV-frag-acc}{0.5\%}
\DefMacro{results-ln-lo-lo-25-ln-s+bsexpl1+attn+copy_ta-test-STDEV-full-acc-top-1}{0.5\%}
\DefMacro{results-ln-lo-lo-25-ln-s+bsexpl1+attn+copy_ta-test-STDEV-full-acc-top-2}{0.7\%}
\DefMacro{results-ln-lo-lo-25-ln-s+bsexpl1+attn+copy_ta-test-STDEV-full-acc-top-3}{0.8\%}
\DefMacro{results-ln-lo-lo-25-ln-s+bsexpl1+attn+copy_ta-test-STDEV-full-acc-top-5}{0.6\%}
\DefMacro{results-ln-lo-lo-25-ln-s+bsexpl1+attn+copy_ta-test-STDEV-full-correct-top-1}{573.5\%}
\DefMacro{results-ln-lo-lo-25-ln-s+bsexpl1+attn+copy_ta-test-STDEV-full-correct-top-2}{804.2\%}
\DefMacro{results-ln-lo-lo-25-ln-s+bsexpl1+attn+copy_ta-test-STDEV-full-correct-top-3}{899.4\%}
\DefMacro{results-ln-lo-lo-25-ln-s+bsexpl1+attn+copy_ta-test-STDEV-full-correct-top-5}{697.6\%}
\DefMacro{results-ln-lo-lo-25-ln-s+bsexpl1+attn+copy_ta-test-SUM-BLEU-4}{97.8}
\DefMacro{results-ln-lo-lo-25-ln-s+bsexpl1+attn+copy_ta-test-SUM-char-acc}{64.0\%}
\DefMacro{results-ln-lo-lo-25-ln-s+bsexpl1+attn+copy_ta-test-SUM-count-top-1}{3501}
\DefMacro{results-ln-lo-lo-25-ln-s+bsexpl1+attn+copy_ta-test-SUM-count-top-2}{3501}
\DefMacro{results-ln-lo-lo-25-ln-s+bsexpl1+attn+copy_ta-test-SUM-count-top-3}{3501}
\DefMacro{results-ln-lo-lo-25-ln-s+bsexpl1+attn+copy_ta-test-SUM-count-top-5}{3501}
\DefMacro{results-ln-lo-lo-25-ln-s+bsexpl1+attn+copy_ta-test-SUM-frag-acc}{64.5\%}
\DefMacro{results-ln-lo-lo-25-ln-s+bsexpl1+attn+copy_ta-test-SUM-full-acc-top-1}{9.9\%}
\DefMacro{results-ln-lo-lo-25-ln-s+bsexpl1+attn+copy_ta-test-SUM-full-acc-top-2}{12.6\%}
\DefMacro{results-ln-lo-lo-25-ln-s+bsexpl1+attn+copy_ta-test-SUM-full-acc-top-3}{13.8\%}
\DefMacro{results-ln-lo-lo-25-ln-s+bsexpl1+attn+copy_ta-test-SUM-full-acc-top-5}{15.9\%}
\DefMacro{results-ln-lo-lo-25-ln-s+bsexpl1+attn+copy_ta-test-SUM-full-correct-top-1}{116}
\DefMacro{results-ln-lo-lo-25-ln-s+bsexpl1+attn+copy_ta-test-SUM-full-correct-top-2}{147}
\DefMacro{results-ln-lo-lo-25-ln-s+bsexpl1+attn+copy_ta-test-SUM-full-correct-top-3}{161}
\DefMacro{results-ln-lo-lo-25-ln-s+bsexpl1+attn+copy_ta-test-SUM-full-correct-top-5}{186}
\DefMacro{results-ln-lo-lo-5-ln-s+bsexpl1+attn+copy_ta-val-AVG-BLEU-4}{46.1}
\DefMacro{results-ln-lo-lo-5-ln-s+bsexpl1+attn+copy_ta-val-AVG-char-acc}{31.9\%}
\DefMacro{results-ln-lo-lo-5-ln-s+bsexpl1+attn+copy_ta-val-AVG-count-top-1}{14400.0\%}
\DefMacro{results-ln-lo-lo-5-ln-s+bsexpl1+attn+copy_ta-val-AVG-count-top-2}{14400.0\%}
\DefMacro{results-ln-lo-lo-5-ln-s+bsexpl1+attn+copy_ta-val-AVG-count-top-3}{14400.0\%}
\DefMacro{results-ln-lo-lo-5-ln-s+bsexpl1+attn+copy_ta-val-AVG-count-top-5}{14400.0\%}
\DefMacro{results-ln-lo-lo-5-ln-s+bsexpl1+attn+copy_ta-val-AVG-frag-acc}{37.3\%}
\DefMacro{results-ln-lo-lo-5-ln-s+bsexpl1+attn+copy_ta-val-AVG-full-acc-top-1}{4.2\%}
\DefMacro{results-ln-lo-lo-5-ln-s+bsexpl1+attn+copy_ta-val-AVG-full-acc-top-2}{6.5\%}
\DefMacro{results-ln-lo-lo-5-ln-s+bsexpl1+attn+copy_ta-val-AVG-full-acc-top-3}{7.9\%}
\DefMacro{results-ln-lo-lo-5-ln-s+bsexpl1+attn+copy_ta-val-AVG-full-acc-top-5}{8.8\%}
\DefMacro{results-ln-lo-lo-5-ln-s+bsexpl1+attn+copy_ta-val-AVG-full-correct-top-1}{600.0\%}
\DefMacro{results-ln-lo-lo-5-ln-s+bsexpl1+attn+copy_ta-val-AVG-full-correct-top-2}{933.3\%}
\DefMacro{results-ln-lo-lo-5-ln-s+bsexpl1+attn+copy_ta-val-AVG-full-correct-top-3}{1133.3\%}
\DefMacro{results-ln-lo-lo-5-ln-s+bsexpl1+attn+copy_ta-val-AVG-full-correct-top-5}{1266.7\%}
\DefMacro{results-ln-lo-lo-5-ln-s+bsexpl1+attn+copy_ta-val-MAX-BLEU-4}{48.1}
\DefMacro{results-ln-lo-lo-5-ln-s+bsexpl1+attn+copy_ta-val-MAX-char-acc}{34.8\%}
\DefMacro{results-ln-lo-lo-5-ln-s+bsexpl1+attn+copy_ta-val-MAX-count-top-1}{144}
\DefMacro{results-ln-lo-lo-5-ln-s+bsexpl1+attn+copy_ta-val-MAX-count-top-2}{144}
\DefMacro{results-ln-lo-lo-5-ln-s+bsexpl1+attn+copy_ta-val-MAX-count-top-3}{144}
\DefMacro{results-ln-lo-lo-5-ln-s+bsexpl1+attn+copy_ta-val-MAX-count-top-5}{144}
\DefMacro{results-ln-lo-lo-5-ln-s+bsexpl1+attn+copy_ta-val-MAX-frag-acc}{39.8\%}
\DefMacro{results-ln-lo-lo-5-ln-s+bsexpl1+attn+copy_ta-val-MAX-full-acc-top-1}{5.6\%}
\DefMacro{results-ln-lo-lo-5-ln-s+bsexpl1+attn+copy_ta-val-MAX-full-acc-top-2}{7.6\%}
\DefMacro{results-ln-lo-lo-5-ln-s+bsexpl1+attn+copy_ta-val-MAX-full-acc-top-3}{10.4\%}
\DefMacro{results-ln-lo-lo-5-ln-s+bsexpl1+attn+copy_ta-val-MAX-full-acc-top-5}{11.8\%}
\DefMacro{results-ln-lo-lo-5-ln-s+bsexpl1+attn+copy_ta-val-MAX-full-correct-top-1}{8}
\DefMacro{results-ln-lo-lo-5-ln-s+bsexpl1+attn+copy_ta-val-MAX-full-correct-top-2}{11}
\DefMacro{results-ln-lo-lo-5-ln-s+bsexpl1+attn+copy_ta-val-MAX-full-correct-top-3}{15}
\DefMacro{results-ln-lo-lo-5-ln-s+bsexpl1+attn+copy_ta-val-MAX-full-correct-top-5}{17}
\DefMacro{results-ln-lo-lo-5-ln-s+bsexpl1+attn+copy_ta-val-MEDIAN-BLEU-4}{46.2}
\DefMacro{results-ln-lo-lo-5-ln-s+bsexpl1+attn+copy_ta-val-MEDIAN-char-acc}{31.8\%}
\DefMacro{results-ln-lo-lo-5-ln-s+bsexpl1+attn+copy_ta-val-MEDIAN-count-top-1}{14400.0\%}
\DefMacro{results-ln-lo-lo-5-ln-s+bsexpl1+attn+copy_ta-val-MEDIAN-count-top-2}{14400.0\%}
\DefMacro{results-ln-lo-lo-5-ln-s+bsexpl1+attn+copy_ta-val-MEDIAN-count-top-3}{14400.0\%}
\DefMacro{results-ln-lo-lo-5-ln-s+bsexpl1+attn+copy_ta-val-MEDIAN-count-top-5}{14400.0\%}
\DefMacro{results-ln-lo-lo-5-ln-s+bsexpl1+attn+copy_ta-val-MEDIAN-frag-acc}{37.8\%}
\DefMacro{results-ln-lo-lo-5-ln-s+bsexpl1+attn+copy_ta-val-MEDIAN-full-acc-top-1}{3.5\%}
\DefMacro{results-ln-lo-lo-5-ln-s+bsexpl1+attn+copy_ta-val-MEDIAN-full-acc-top-2}{6.9\%}
\DefMacro{results-ln-lo-lo-5-ln-s+bsexpl1+attn+copy_ta-val-MEDIAN-full-acc-top-3}{7.6\%}
\DefMacro{results-ln-lo-lo-5-ln-s+bsexpl1+attn+copy_ta-val-MEDIAN-full-acc-top-5}{7.6\%}
\DefMacro{results-ln-lo-lo-5-ln-s+bsexpl1+attn+copy_ta-val-MEDIAN-full-correct-top-1}{500.0\%}
\DefMacro{results-ln-lo-lo-5-ln-s+bsexpl1+attn+copy_ta-val-MEDIAN-full-correct-top-2}{1000.0\%}
\DefMacro{results-ln-lo-lo-5-ln-s+bsexpl1+attn+copy_ta-val-MEDIAN-full-correct-top-3}{1100.0\%}
\DefMacro{results-ln-lo-lo-5-ln-s+bsexpl1+attn+copy_ta-val-MEDIAN-full-correct-top-5}{1100.0\%}
\DefMacro{results-ln-lo-lo-5-ln-s+bsexpl1+attn+copy_ta-val-MIN-BLEU-4}{44.1}
\DefMacro{results-ln-lo-lo-5-ln-s+bsexpl1+attn+copy_ta-val-MIN-char-acc}{29.2\%}
\DefMacro{results-ln-lo-lo-5-ln-s+bsexpl1+attn+copy_ta-val-MIN-count-top-1}{144}
\DefMacro{results-ln-lo-lo-5-ln-s+bsexpl1+attn+copy_ta-val-MIN-count-top-2}{144}
\DefMacro{results-ln-lo-lo-5-ln-s+bsexpl1+attn+copy_ta-val-MIN-count-top-3}{144}
\DefMacro{results-ln-lo-lo-5-ln-s+bsexpl1+attn+copy_ta-val-MIN-count-top-5}{144}
\DefMacro{results-ln-lo-lo-5-ln-s+bsexpl1+attn+copy_ta-val-MIN-frag-acc}{34.3\%}
\DefMacro{results-ln-lo-lo-5-ln-s+bsexpl1+attn+copy_ta-val-MIN-full-acc-top-1}{3.5\%}
\DefMacro{results-ln-lo-lo-5-ln-s+bsexpl1+attn+copy_ta-val-MIN-full-acc-top-2}{4.9\%}
\DefMacro{results-ln-lo-lo-5-ln-s+bsexpl1+attn+copy_ta-val-MIN-full-acc-top-3}{5.6\%}
\DefMacro{results-ln-lo-lo-5-ln-s+bsexpl1+attn+copy_ta-val-MIN-full-acc-top-5}{6.9\%}
\DefMacro{results-ln-lo-lo-5-ln-s+bsexpl1+attn+copy_ta-val-MIN-full-correct-top-1}{5}
\DefMacro{results-ln-lo-lo-5-ln-s+bsexpl1+attn+copy_ta-val-MIN-full-correct-top-2}{7}
\DefMacro{results-ln-lo-lo-5-ln-s+bsexpl1+attn+copy_ta-val-MIN-full-correct-top-3}{8}
\DefMacro{results-ln-lo-lo-5-ln-s+bsexpl1+attn+copy_ta-val-MIN-full-correct-top-5}{10}
\DefMacro{results-ln-lo-lo-5-ln-s+bsexpl1+attn+copy_ta-val-STDEV-BLEU-4}{1.6}
\DefMacro{results-ln-lo-lo-5-ln-s+bsexpl1+attn+copy_ta-val-STDEV-char-acc}{2.3\%}
\DefMacro{results-ln-lo-lo-5-ln-s+bsexpl1+attn+copy_ta-val-STDEV-count-top-1}{0.0\%}
\DefMacro{results-ln-lo-lo-5-ln-s+bsexpl1+attn+copy_ta-val-STDEV-count-top-2}{0.0\%}
\DefMacro{results-ln-lo-lo-5-ln-s+bsexpl1+attn+copy_ta-val-STDEV-count-top-3}{0.0\%}
\DefMacro{results-ln-lo-lo-5-ln-s+bsexpl1+attn+copy_ta-val-STDEV-count-top-5}{0.0\%}
\DefMacro{results-ln-lo-lo-5-ln-s+bsexpl1+attn+copy_ta-val-STDEV-frag-acc}{2.3\%}
\DefMacro{results-ln-lo-lo-5-ln-s+bsexpl1+attn+copy_ta-val-STDEV-full-acc-top-1}{1.0\%}
\DefMacro{results-ln-lo-lo-5-ln-s+bsexpl1+attn+copy_ta-val-STDEV-full-acc-top-2}{1.2\%}
\DefMacro{results-ln-lo-lo-5-ln-s+bsexpl1+attn+copy_ta-val-STDEV-full-acc-top-3}{2.0\%}
\DefMacro{results-ln-lo-lo-5-ln-s+bsexpl1+attn+copy_ta-val-STDEV-full-acc-top-5}{2.1\%}
\DefMacro{results-ln-lo-lo-5-ln-s+bsexpl1+attn+copy_ta-val-STDEV-full-correct-top-1}{141.4\%}
\DefMacro{results-ln-lo-lo-5-ln-s+bsexpl1+attn+copy_ta-val-STDEV-full-correct-top-2}{170.0\%}
\DefMacro{results-ln-lo-lo-5-ln-s+bsexpl1+attn+copy_ta-val-STDEV-full-correct-top-3}{286.7\%}
\DefMacro{results-ln-lo-lo-5-ln-s+bsexpl1+attn+copy_ta-val-STDEV-full-correct-top-5}{309.1\%}
\DefMacro{results-ln-lo-lo-5-ln-s+bsexpl1+attn+copy_ta-val-SUM-BLEU-4}{138.3}
\DefMacro{results-ln-lo-lo-5-ln-s+bsexpl1+attn+copy_ta-val-SUM-char-acc}{95.7\%}
\DefMacro{results-ln-lo-lo-5-ln-s+bsexpl1+attn+copy_ta-val-SUM-count-top-1}{432}
\DefMacro{results-ln-lo-lo-5-ln-s+bsexpl1+attn+copy_ta-val-SUM-count-top-2}{432}
\DefMacro{results-ln-lo-lo-5-ln-s+bsexpl1+attn+copy_ta-val-SUM-count-top-3}{432}
\DefMacro{results-ln-lo-lo-5-ln-s+bsexpl1+attn+copy_ta-val-SUM-count-top-5}{432}
\DefMacro{results-ln-lo-lo-5-ln-s+bsexpl1+attn+copy_ta-val-SUM-frag-acc}{111.9\%}
\DefMacro{results-ln-lo-lo-5-ln-s+bsexpl1+attn+copy_ta-val-SUM-full-acc-top-1}{12.5\%}
\DefMacro{results-ln-lo-lo-5-ln-s+bsexpl1+attn+copy_ta-val-SUM-full-acc-top-2}{19.4\%}
\DefMacro{results-ln-lo-lo-5-ln-s+bsexpl1+attn+copy_ta-val-SUM-full-acc-top-3}{23.6\%}
\DefMacro{results-ln-lo-lo-5-ln-s+bsexpl1+attn+copy_ta-val-SUM-full-acc-top-5}{26.4\%}
\DefMacro{results-ln-lo-lo-5-ln-s+bsexpl1+attn+copy_ta-val-SUM-full-correct-top-1}{18}
\DefMacro{results-ln-lo-lo-5-ln-s+bsexpl1+attn+copy_ta-val-SUM-full-correct-top-2}{28}
\DefMacro{results-ln-lo-lo-5-ln-s+bsexpl1+attn+copy_ta-val-SUM-full-correct-top-3}{34}
\DefMacro{results-ln-lo-lo-5-ln-s+bsexpl1+attn+copy_ta-val-SUM-full-correct-top-5}{38}
\DefMacro{results-ln-lo-lo-5-ln-s+bsexpl1+attn+copy_ta-test-AVG-BLEU-4}{34.1}
\DefMacro{results-ln-lo-lo-5-ln-s+bsexpl1+attn+copy_ta-test-AVG-char-acc}{22.6\%}
\DefMacro{results-ln-lo-lo-5-ln-s+bsexpl1+attn+copy_ta-test-AVG-count-top-1}{116700.0\%}
\DefMacro{results-ln-lo-lo-5-ln-s+bsexpl1+attn+copy_ta-test-AVG-count-top-2}{116700.0\%}
\DefMacro{results-ln-lo-lo-5-ln-s+bsexpl1+attn+copy_ta-test-AVG-count-top-3}{116700.0\%}
\DefMacro{results-ln-lo-lo-5-ln-s+bsexpl1+attn+copy_ta-test-AVG-count-top-5}{116700.0\%}
\DefMacro{results-ln-lo-lo-5-ln-s+bsexpl1+attn+copy_ta-test-AVG-frag-acc}{22.7\%}
\DefMacro{results-ln-lo-lo-5-ln-s+bsexpl1+attn+copy_ta-test-AVG-full-acc-top-1}{3.8\%}
\DefMacro{results-ln-lo-lo-5-ln-s+bsexpl1+attn+copy_ta-test-AVG-full-acc-top-2}{5.2\%}
\DefMacro{results-ln-lo-lo-5-ln-s+bsexpl1+attn+copy_ta-test-AVG-full-acc-top-3}{6.1\%}
\DefMacro{results-ln-lo-lo-5-ln-s+bsexpl1+attn+copy_ta-test-AVG-full-acc-top-5}{6.9\%}
\DefMacro{results-ln-lo-lo-5-ln-s+bsexpl1+attn+copy_ta-test-AVG-full-correct-top-1}{4400.0\%}
\DefMacro{results-ln-lo-lo-5-ln-s+bsexpl1+attn+copy_ta-test-AVG-full-correct-top-2}{6033.3\%}
\DefMacro{results-ln-lo-lo-5-ln-s+bsexpl1+attn+copy_ta-test-AVG-full-correct-top-3}{7100.0\%}
\DefMacro{results-ln-lo-lo-5-ln-s+bsexpl1+attn+copy_ta-test-AVG-full-correct-top-5}{8033.3\%}
\DefMacro{results-ln-lo-lo-5-ln-s+bsexpl1+attn+copy_ta-test-MAX-BLEU-4}{36.0}
\DefMacro{results-ln-lo-lo-5-ln-s+bsexpl1+attn+copy_ta-test-MAX-char-acc}{24.1\%}
\DefMacro{results-ln-lo-lo-5-ln-s+bsexpl1+attn+copy_ta-test-MAX-count-top-1}{1167}
\DefMacro{results-ln-lo-lo-5-ln-s+bsexpl1+attn+copy_ta-test-MAX-count-top-2}{1167}
\DefMacro{results-ln-lo-lo-5-ln-s+bsexpl1+attn+copy_ta-test-MAX-count-top-3}{1167}
\DefMacro{results-ln-lo-lo-5-ln-s+bsexpl1+attn+copy_ta-test-MAX-count-top-5}{1167}
\DefMacro{results-ln-lo-lo-5-ln-s+bsexpl1+attn+copy_ta-test-MAX-frag-acc}{25.6\%}
\DefMacro{results-ln-lo-lo-5-ln-s+bsexpl1+attn+copy_ta-test-MAX-full-acc-top-1}{4.5\%}
\DefMacro{results-ln-lo-lo-5-ln-s+bsexpl1+attn+copy_ta-test-MAX-full-acc-top-2}{6.2\%}
\DefMacro{results-ln-lo-lo-5-ln-s+bsexpl1+attn+copy_ta-test-MAX-full-acc-top-3}{7.0\%}
\DefMacro{results-ln-lo-lo-5-ln-s+bsexpl1+attn+copy_ta-test-MAX-full-acc-top-5}{7.7\%}
\DefMacro{results-ln-lo-lo-5-ln-s+bsexpl1+attn+copy_ta-test-MAX-full-correct-top-1}{53}
\DefMacro{results-ln-lo-lo-5-ln-s+bsexpl1+attn+copy_ta-test-MAX-full-correct-top-2}{72}
\DefMacro{results-ln-lo-lo-5-ln-s+bsexpl1+attn+copy_ta-test-MAX-full-correct-top-3}{82}
\DefMacro{results-ln-lo-lo-5-ln-s+bsexpl1+attn+copy_ta-test-MAX-full-correct-top-5}{90}
\DefMacro{results-ln-lo-lo-5-ln-s+bsexpl1+attn+copy_ta-test-MEDIAN-BLEU-4}{33.6}
\DefMacro{results-ln-lo-lo-5-ln-s+bsexpl1+attn+copy_ta-test-MEDIAN-char-acc}{22.3\%}
\DefMacro{results-ln-lo-lo-5-ln-s+bsexpl1+attn+copy_ta-test-MEDIAN-count-top-1}{116700.0\%}
\DefMacro{results-ln-lo-lo-5-ln-s+bsexpl1+attn+copy_ta-test-MEDIAN-count-top-2}{116700.0\%}
\DefMacro{results-ln-lo-lo-5-ln-s+bsexpl1+attn+copy_ta-test-MEDIAN-count-top-3}{116700.0\%}
\DefMacro{results-ln-lo-lo-5-ln-s+bsexpl1+attn+copy_ta-test-MEDIAN-count-top-5}{116700.0\%}
\DefMacro{results-ln-lo-lo-5-ln-s+bsexpl1+attn+copy_ta-test-MEDIAN-frag-acc}{22.2\%}
\DefMacro{results-ln-lo-lo-5-ln-s+bsexpl1+attn+copy_ta-test-MEDIAN-full-acc-top-1}{3.4\%}
\DefMacro{results-ln-lo-lo-5-ln-s+bsexpl1+attn+copy_ta-test-MEDIAN-full-acc-top-2}{5.0\%}
\DefMacro{results-ln-lo-lo-5-ln-s+bsexpl1+attn+copy_ta-test-MEDIAN-full-acc-top-3}{5.7\%}
\DefMacro{results-ln-lo-lo-5-ln-s+bsexpl1+attn+copy_ta-test-MEDIAN-full-acc-top-5}{6.6\%}
\DefMacro{results-ln-lo-lo-5-ln-s+bsexpl1+attn+copy_ta-test-MEDIAN-full-correct-top-1}{4000.0\%}
\DefMacro{results-ln-lo-lo-5-ln-s+bsexpl1+attn+copy_ta-test-MEDIAN-full-correct-top-2}{5800.0\%}
\DefMacro{results-ln-lo-lo-5-ln-s+bsexpl1+attn+copy_ta-test-MEDIAN-full-correct-top-3}{6600.0\%}
\DefMacro{results-ln-lo-lo-5-ln-s+bsexpl1+attn+copy_ta-test-MEDIAN-full-correct-top-5}{7700.0\%}
\DefMacro{results-ln-lo-lo-5-ln-s+bsexpl1+attn+copy_ta-test-MIN-BLEU-4}{32.7}
\DefMacro{results-ln-lo-lo-5-ln-s+bsexpl1+attn+copy_ta-test-MIN-char-acc}{21.4\%}
\DefMacro{results-ln-lo-lo-5-ln-s+bsexpl1+attn+copy_ta-test-MIN-count-top-1}{1167}
\DefMacro{results-ln-lo-lo-5-ln-s+bsexpl1+attn+copy_ta-test-MIN-count-top-2}{1167}
\DefMacro{results-ln-lo-lo-5-ln-s+bsexpl1+attn+copy_ta-test-MIN-count-top-3}{1167}
\DefMacro{results-ln-lo-lo-5-ln-s+bsexpl1+attn+copy_ta-test-MIN-count-top-5}{1167}
\DefMacro{results-ln-lo-lo-5-ln-s+bsexpl1+attn+copy_ta-test-MIN-frag-acc}{20.4\%}
\DefMacro{results-ln-lo-lo-5-ln-s+bsexpl1+attn+copy_ta-test-MIN-full-acc-top-1}{3.3\%}
\DefMacro{results-ln-lo-lo-5-ln-s+bsexpl1+attn+copy_ta-test-MIN-full-acc-top-2}{4.4\%}
\DefMacro{results-ln-lo-lo-5-ln-s+bsexpl1+attn+copy_ta-test-MIN-full-acc-top-3}{5.6\%}
\DefMacro{results-ln-lo-lo-5-ln-s+bsexpl1+attn+copy_ta-test-MIN-full-acc-top-5}{6.3\%}
\DefMacro{results-ln-lo-lo-5-ln-s+bsexpl1+attn+copy_ta-test-MIN-full-correct-top-1}{39}
\DefMacro{results-ln-lo-lo-5-ln-s+bsexpl1+attn+copy_ta-test-MIN-full-correct-top-2}{51}
\DefMacro{results-ln-lo-lo-5-ln-s+bsexpl1+attn+copy_ta-test-MIN-full-correct-top-3}{65}
\DefMacro{results-ln-lo-lo-5-ln-s+bsexpl1+attn+copy_ta-test-MIN-full-correct-top-5}{74}
\DefMacro{results-ln-lo-lo-5-ln-s+bsexpl1+attn+copy_ta-test-STDEV-BLEU-4}{1.4}
\DefMacro{results-ln-lo-lo-5-ln-s+bsexpl1+attn+copy_ta-test-STDEV-char-acc}{1.1\%}
\DefMacro{results-ln-lo-lo-5-ln-s+bsexpl1+attn+copy_ta-test-STDEV-count-top-1}{0.0\%}
\DefMacro{results-ln-lo-lo-5-ln-s+bsexpl1+attn+copy_ta-test-STDEV-count-top-2}{0.0\%}
\DefMacro{results-ln-lo-lo-5-ln-s+bsexpl1+attn+copy_ta-test-STDEV-count-top-3}{0.0\%}
\DefMacro{results-ln-lo-lo-5-ln-s+bsexpl1+attn+copy_ta-test-STDEV-count-top-5}{0.0\%}
\DefMacro{results-ln-lo-lo-5-ln-s+bsexpl1+attn+copy_ta-test-STDEV-frag-acc}{2.2\%}
\DefMacro{results-ln-lo-lo-5-ln-s+bsexpl1+attn+copy_ta-test-STDEV-full-acc-top-1}{0.5\%}
\DefMacro{results-ln-lo-lo-5-ln-s+bsexpl1+attn+copy_ta-test-STDEV-full-acc-top-2}{0.7\%}
\DefMacro{results-ln-lo-lo-5-ln-s+bsexpl1+attn+copy_ta-test-STDEV-full-acc-top-3}{0.7\%}
\DefMacro{results-ln-lo-lo-5-ln-s+bsexpl1+attn+copy_ta-test-STDEV-full-acc-top-5}{0.6\%}
\DefMacro{results-ln-lo-lo-5-ln-s+bsexpl1+attn+copy_ta-test-STDEV-full-correct-top-1}{637.7\%}
\DefMacro{results-ln-lo-lo-5-ln-s+bsexpl1+attn+copy_ta-test-STDEV-full-correct-top-2}{873.1\%}
\DefMacro{results-ln-lo-lo-5-ln-s+bsexpl1+attn+copy_ta-test-STDEV-full-correct-top-3}{778.9\%}
\DefMacro{results-ln-lo-lo-5-ln-s+bsexpl1+attn+copy_ta-test-STDEV-full-correct-top-5}{694.4\%}
\DefMacro{results-ln-lo-lo-5-ln-s+bsexpl1+attn+copy_ta-test-SUM-BLEU-4}{102.4}
\DefMacro{results-ln-lo-lo-5-ln-s+bsexpl1+attn+copy_ta-test-SUM-char-acc}{67.8\%}
\DefMacro{results-ln-lo-lo-5-ln-s+bsexpl1+attn+copy_ta-test-SUM-count-top-1}{3501}
\DefMacro{results-ln-lo-lo-5-ln-s+bsexpl1+attn+copy_ta-test-SUM-count-top-2}{3501}
\DefMacro{results-ln-lo-lo-5-ln-s+bsexpl1+attn+copy_ta-test-SUM-count-top-3}{3501}
\DefMacro{results-ln-lo-lo-5-ln-s+bsexpl1+attn+copy_ta-test-SUM-count-top-5}{3501}
\DefMacro{results-ln-lo-lo-5-ln-s+bsexpl1+attn+copy_ta-test-SUM-frag-acc}{68.2\%}
\DefMacro{results-ln-lo-lo-5-ln-s+bsexpl1+attn+copy_ta-test-SUM-full-acc-top-1}{11.3\%}
\DefMacro{results-ln-lo-lo-5-ln-s+bsexpl1+attn+copy_ta-test-SUM-full-acc-top-2}{15.5\%}
\DefMacro{results-ln-lo-lo-5-ln-s+bsexpl1+attn+copy_ta-test-SUM-full-acc-top-3}{18.3\%}
\DefMacro{results-ln-lo-lo-5-ln-s+bsexpl1+attn+copy_ta-test-SUM-full-acc-top-5}{20.7\%}
\DefMacro{results-ln-lo-lo-5-ln-s+bsexpl1+attn+copy_ta-test-SUM-full-correct-top-1}{132}
\DefMacro{results-ln-lo-lo-5-ln-s+bsexpl1+attn+copy_ta-test-SUM-full-correct-top-2}{181}
\DefMacro{results-ln-lo-lo-5-ln-s+bsexpl1+attn+copy_ta-test-SUM-full-correct-top-3}{213}
\DefMacro{results-ln-lo-lo-5-ln-s+bsexpl1+attn+copy_ta-test-SUM-full-correct-top-5}{241}
\DefMacro{results-ln-lo-lo-75-ln-s+bsexpl1+attn+copy_ta-val-AVG-BLEU-4}{47.6}
\DefMacro{results-ln-lo-lo-75-ln-s+bsexpl1+attn+copy_ta-val-AVG-char-acc}{34.1\%}
\DefMacro{results-ln-lo-lo-75-ln-s+bsexpl1+attn+copy_ta-val-AVG-count-top-1}{14400.0\%}
\DefMacro{results-ln-lo-lo-75-ln-s+bsexpl1+attn+copy_ta-val-AVG-count-top-2}{14400.0\%}
\DefMacro{results-ln-lo-lo-75-ln-s+bsexpl1+attn+copy_ta-val-AVG-count-top-3}{14400.0\%}
\DefMacro{results-ln-lo-lo-75-ln-s+bsexpl1+attn+copy_ta-val-AVG-count-top-5}{14400.0\%}
\DefMacro{results-ln-lo-lo-75-ln-s+bsexpl1+attn+copy_ta-val-AVG-frag-acc}{38.2\%}
\DefMacro{results-ln-lo-lo-75-ln-s+bsexpl1+attn+copy_ta-val-AVG-full-acc-top-1}{4.9\%}
\DefMacro{results-ln-lo-lo-75-ln-s+bsexpl1+attn+copy_ta-val-AVG-full-acc-top-2}{7.4\%}
\DefMacro{results-ln-lo-lo-75-ln-s+bsexpl1+attn+copy_ta-val-AVG-full-acc-top-3}{9.0\%}
\DefMacro{results-ln-lo-lo-75-ln-s+bsexpl1+attn+copy_ta-val-AVG-full-acc-top-5}{11.3\%}
\DefMacro{results-ln-lo-lo-75-ln-s+bsexpl1+attn+copy_ta-val-AVG-full-correct-top-1}{700.0\%}
\DefMacro{results-ln-lo-lo-75-ln-s+bsexpl1+attn+copy_ta-val-AVG-full-correct-top-2}{1066.7\%}
\DefMacro{results-ln-lo-lo-75-ln-s+bsexpl1+attn+copy_ta-val-AVG-full-correct-top-3}{1300.0\%}
\DefMacro{results-ln-lo-lo-75-ln-s+bsexpl1+attn+copy_ta-val-AVG-full-correct-top-5}{1633.3\%}
\DefMacro{results-ln-lo-lo-75-ln-s+bsexpl1+attn+copy_ta-val-MAX-BLEU-4}{49.1}
\DefMacro{results-ln-lo-lo-75-ln-s+bsexpl1+attn+copy_ta-val-MAX-char-acc}{37.2\%}
\DefMacro{results-ln-lo-lo-75-ln-s+bsexpl1+attn+copy_ta-val-MAX-count-top-1}{144}
\DefMacro{results-ln-lo-lo-75-ln-s+bsexpl1+attn+copy_ta-val-MAX-count-top-2}{144}
\DefMacro{results-ln-lo-lo-75-ln-s+bsexpl1+attn+copy_ta-val-MAX-count-top-3}{144}
\DefMacro{results-ln-lo-lo-75-ln-s+bsexpl1+attn+copy_ta-val-MAX-count-top-5}{144}
\DefMacro{results-ln-lo-lo-75-ln-s+bsexpl1+attn+copy_ta-val-MAX-frag-acc}{38.8\%}
\DefMacro{results-ln-lo-lo-75-ln-s+bsexpl1+attn+copy_ta-val-MAX-full-acc-top-1}{6.2\%}
\DefMacro{results-ln-lo-lo-75-ln-s+bsexpl1+attn+copy_ta-val-MAX-full-acc-top-2}{9.0\%}
\DefMacro{results-ln-lo-lo-75-ln-s+bsexpl1+attn+copy_ta-val-MAX-full-acc-top-3}{10.4\%}
\DefMacro{results-ln-lo-lo-75-ln-s+bsexpl1+attn+copy_ta-val-MAX-full-acc-top-5}{13.2\%}
\DefMacro{results-ln-lo-lo-75-ln-s+bsexpl1+attn+copy_ta-val-MAX-full-correct-top-1}{9}
\DefMacro{results-ln-lo-lo-75-ln-s+bsexpl1+attn+copy_ta-val-MAX-full-correct-top-2}{13}
\DefMacro{results-ln-lo-lo-75-ln-s+bsexpl1+attn+copy_ta-val-MAX-full-correct-top-3}{15}
\DefMacro{results-ln-lo-lo-75-ln-s+bsexpl1+attn+copy_ta-val-MAX-full-correct-top-5}{19}
\DefMacro{results-ln-lo-lo-75-ln-s+bsexpl1+attn+copy_ta-val-MEDIAN-BLEU-4}{47.2}
\DefMacro{results-ln-lo-lo-75-ln-s+bsexpl1+attn+copy_ta-val-MEDIAN-char-acc}{33.9\%}
\DefMacro{results-ln-lo-lo-75-ln-s+bsexpl1+attn+copy_ta-val-MEDIAN-count-top-1}{14400.0\%}
\DefMacro{results-ln-lo-lo-75-ln-s+bsexpl1+attn+copy_ta-val-MEDIAN-count-top-2}{14400.0\%}
\DefMacro{results-ln-lo-lo-75-ln-s+bsexpl1+attn+copy_ta-val-MEDIAN-count-top-3}{14400.0\%}
\DefMacro{results-ln-lo-lo-75-ln-s+bsexpl1+attn+copy_ta-val-MEDIAN-count-top-5}{14400.0\%}
\DefMacro{results-ln-lo-lo-75-ln-s+bsexpl1+attn+copy_ta-val-MEDIAN-frag-acc}{38.7\%}
\DefMacro{results-ln-lo-lo-75-ln-s+bsexpl1+attn+copy_ta-val-MEDIAN-full-acc-top-1}{4.2\%}
\DefMacro{results-ln-lo-lo-75-ln-s+bsexpl1+attn+copy_ta-val-MEDIAN-full-acc-top-2}{8.3\%}
\DefMacro{results-ln-lo-lo-75-ln-s+bsexpl1+attn+copy_ta-val-MEDIAN-full-acc-top-3}{10.4\%}
\DefMacro{results-ln-lo-lo-75-ln-s+bsexpl1+attn+copy_ta-val-MEDIAN-full-acc-top-5}{11.8\%}
\DefMacro{results-ln-lo-lo-75-ln-s+bsexpl1+attn+copy_ta-val-MEDIAN-full-correct-top-1}{600.0\%}
\DefMacro{results-ln-lo-lo-75-ln-s+bsexpl1+attn+copy_ta-val-MEDIAN-full-correct-top-2}{1200.0\%}
\DefMacro{results-ln-lo-lo-75-ln-s+bsexpl1+attn+copy_ta-val-MEDIAN-full-correct-top-3}{1500.0\%}
\DefMacro{results-ln-lo-lo-75-ln-s+bsexpl1+attn+copy_ta-val-MEDIAN-full-correct-top-5}{1700.0\%}
\DefMacro{results-ln-lo-lo-75-ln-s+bsexpl1+attn+copy_ta-val-MIN-BLEU-4}{46.4}
\DefMacro{results-ln-lo-lo-75-ln-s+bsexpl1+attn+copy_ta-val-MIN-char-acc}{31.2\%}
\DefMacro{results-ln-lo-lo-75-ln-s+bsexpl1+attn+copy_ta-val-MIN-count-top-1}{144}
\DefMacro{results-ln-lo-lo-75-ln-s+bsexpl1+attn+copy_ta-val-MIN-count-top-2}{144}
\DefMacro{results-ln-lo-lo-75-ln-s+bsexpl1+attn+copy_ta-val-MIN-count-top-3}{144}
\DefMacro{results-ln-lo-lo-75-ln-s+bsexpl1+attn+copy_ta-val-MIN-count-top-5}{144}
\DefMacro{results-ln-lo-lo-75-ln-s+bsexpl1+attn+copy_ta-val-MIN-frag-acc}{37.0\%}
\DefMacro{results-ln-lo-lo-75-ln-s+bsexpl1+attn+copy_ta-val-MIN-full-acc-top-1}{4.2\%}
\DefMacro{results-ln-lo-lo-75-ln-s+bsexpl1+attn+copy_ta-val-MIN-full-acc-top-2}{4.9\%}
\DefMacro{results-ln-lo-lo-75-ln-s+bsexpl1+attn+copy_ta-val-MIN-full-acc-top-3}{6.2\%}
\DefMacro{results-ln-lo-lo-75-ln-s+bsexpl1+attn+copy_ta-val-MIN-full-acc-top-5}{9.0\%}
\DefMacro{results-ln-lo-lo-75-ln-s+bsexpl1+attn+copy_ta-val-MIN-full-correct-top-1}{6}
\DefMacro{results-ln-lo-lo-75-ln-s+bsexpl1+attn+copy_ta-val-MIN-full-correct-top-2}{7}
\DefMacro{results-ln-lo-lo-75-ln-s+bsexpl1+attn+copy_ta-val-MIN-full-correct-top-3}{9}
\DefMacro{results-ln-lo-lo-75-ln-s+bsexpl1+attn+copy_ta-val-MIN-full-correct-top-5}{13}
\DefMacro{results-ln-lo-lo-75-ln-s+bsexpl1+attn+copy_ta-val-STDEV-BLEU-4}{1.1}
\DefMacro{results-ln-lo-lo-75-ln-s+bsexpl1+attn+copy_ta-val-STDEV-char-acc}{2.5\%}
\DefMacro{results-ln-lo-lo-75-ln-s+bsexpl1+attn+copy_ta-val-STDEV-count-top-1}{0.0\%}
\DefMacro{results-ln-lo-lo-75-ln-s+bsexpl1+attn+copy_ta-val-STDEV-count-top-2}{0.0\%}
\DefMacro{results-ln-lo-lo-75-ln-s+bsexpl1+attn+copy_ta-val-STDEV-count-top-3}{0.0\%}
\DefMacro{results-ln-lo-lo-75-ln-s+bsexpl1+attn+copy_ta-val-STDEV-count-top-5}{0.0\%}
\DefMacro{results-ln-lo-lo-75-ln-s+bsexpl1+attn+copy_ta-val-STDEV-frag-acc}{0.8\%}
\DefMacro{results-ln-lo-lo-75-ln-s+bsexpl1+attn+copy_ta-val-STDEV-full-acc-top-1}{1.0\%}
\DefMacro{results-ln-lo-lo-75-ln-s+bsexpl1+attn+copy_ta-val-STDEV-full-acc-top-2}{1.8\%}
\DefMacro{results-ln-lo-lo-75-ln-s+bsexpl1+attn+copy_ta-val-STDEV-full-acc-top-3}{2.0\%}
\DefMacro{results-ln-lo-lo-75-ln-s+bsexpl1+attn+copy_ta-val-STDEV-full-acc-top-5}{1.7\%}
\DefMacro{results-ln-lo-lo-75-ln-s+bsexpl1+attn+copy_ta-val-STDEV-full-correct-top-1}{141.4\%}
\DefMacro{results-ln-lo-lo-75-ln-s+bsexpl1+attn+copy_ta-val-STDEV-full-correct-top-2}{262.5\%}
\DefMacro{results-ln-lo-lo-75-ln-s+bsexpl1+attn+copy_ta-val-STDEV-full-correct-top-3}{282.8\%}
\DefMacro{results-ln-lo-lo-75-ln-s+bsexpl1+attn+copy_ta-val-STDEV-full-correct-top-5}{249.4\%}
\DefMacro{results-ln-lo-lo-75-ln-s+bsexpl1+attn+copy_ta-val-SUM-BLEU-4}{142.8}
\DefMacro{results-ln-lo-lo-75-ln-s+bsexpl1+attn+copy_ta-val-SUM-char-acc}{102.3\%}
\DefMacro{results-ln-lo-lo-75-ln-s+bsexpl1+attn+copy_ta-val-SUM-count-top-1}{432}
\DefMacro{results-ln-lo-lo-75-ln-s+bsexpl1+attn+copy_ta-val-SUM-count-top-2}{432}
\DefMacro{results-ln-lo-lo-75-ln-s+bsexpl1+attn+copy_ta-val-SUM-count-top-3}{432}
\DefMacro{results-ln-lo-lo-75-ln-s+bsexpl1+attn+copy_ta-val-SUM-count-top-5}{432}
\DefMacro{results-ln-lo-lo-75-ln-s+bsexpl1+attn+copy_ta-val-SUM-frag-acc}{114.6\%}
\DefMacro{results-ln-lo-lo-75-ln-s+bsexpl1+attn+copy_ta-val-SUM-full-acc-top-1}{14.6\%}
\DefMacro{results-ln-lo-lo-75-ln-s+bsexpl1+attn+copy_ta-val-SUM-full-acc-top-2}{22.2\%}
\DefMacro{results-ln-lo-lo-75-ln-s+bsexpl1+attn+copy_ta-val-SUM-full-acc-top-3}{27.1\%}
\DefMacro{results-ln-lo-lo-75-ln-s+bsexpl1+attn+copy_ta-val-SUM-full-acc-top-5}{34.0\%}
\DefMacro{results-ln-lo-lo-75-ln-s+bsexpl1+attn+copy_ta-val-SUM-full-correct-top-1}{21}
\DefMacro{results-ln-lo-lo-75-ln-s+bsexpl1+attn+copy_ta-val-SUM-full-correct-top-2}{32}
\DefMacro{results-ln-lo-lo-75-ln-s+bsexpl1+attn+copy_ta-val-SUM-full-correct-top-3}{39}
\DefMacro{results-ln-lo-lo-75-ln-s+bsexpl1+attn+copy_ta-val-SUM-full-correct-top-5}{49}
\DefMacro{results-ln-lo-lo-75-ln-s+bsexpl1+attn+copy_ta-test-AVG-BLEU-4}{35.7}
\DefMacro{results-ln-lo-lo-75-ln-s+bsexpl1+attn+copy_ta-test-AVG-char-acc}{24.5\%}
\DefMacro{results-ln-lo-lo-75-ln-s+bsexpl1+attn+copy_ta-test-AVG-count-top-1}{116700.0\%}
\DefMacro{results-ln-lo-lo-75-ln-s+bsexpl1+attn+copy_ta-test-AVG-count-top-2}{116700.0\%}
\DefMacro{results-ln-lo-lo-75-ln-s+bsexpl1+attn+copy_ta-test-AVG-count-top-3}{116700.0\%}
\DefMacro{results-ln-lo-lo-75-ln-s+bsexpl1+attn+copy_ta-test-AVG-count-top-5}{116700.0\%}
\DefMacro{results-ln-lo-lo-75-ln-s+bsexpl1+attn+copy_ta-test-AVG-frag-acc}{24.3\%}
\DefMacro{results-ln-lo-lo-75-ln-s+bsexpl1+attn+copy_ta-test-AVG-full-acc-top-1}{5.0\%}
\DefMacro{results-ln-lo-lo-75-ln-s+bsexpl1+attn+copy_ta-test-AVG-full-acc-top-2}{6.7\%}
\DefMacro{results-ln-lo-lo-75-ln-s+bsexpl1+attn+copy_ta-test-AVG-full-acc-top-3}{7.5\%}
\DefMacro{results-ln-lo-lo-75-ln-s+bsexpl1+attn+copy_ta-test-AVG-full-acc-top-5}{8.7\%}
\DefMacro{results-ln-lo-lo-75-ln-s+bsexpl1+attn+copy_ta-test-AVG-full-correct-top-1}{5800.0\%}
\DefMacro{results-ln-lo-lo-75-ln-s+bsexpl1+attn+copy_ta-test-AVG-full-correct-top-2}{7766.7\%}
\DefMacro{results-ln-lo-lo-75-ln-s+bsexpl1+attn+copy_ta-test-AVG-full-correct-top-3}{8733.3\%}
\DefMacro{results-ln-lo-lo-75-ln-s+bsexpl1+attn+copy_ta-test-AVG-full-correct-top-5}{10100.0\%}
\DefMacro{results-ln-lo-lo-75-ln-s+bsexpl1+attn+copy_ta-test-MAX-BLEU-4}{37.2}
\DefMacro{results-ln-lo-lo-75-ln-s+bsexpl1+attn+copy_ta-test-MAX-char-acc}{26.3\%}
\DefMacro{results-ln-lo-lo-75-ln-s+bsexpl1+attn+copy_ta-test-MAX-count-top-1}{1167}
\DefMacro{results-ln-lo-lo-75-ln-s+bsexpl1+attn+copy_ta-test-MAX-count-top-2}{1167}
\DefMacro{results-ln-lo-lo-75-ln-s+bsexpl1+attn+copy_ta-test-MAX-count-top-3}{1167}
\DefMacro{results-ln-lo-lo-75-ln-s+bsexpl1+attn+copy_ta-test-MAX-count-top-5}{1167}
\DefMacro{results-ln-lo-lo-75-ln-s+bsexpl1+attn+copy_ta-test-MAX-frag-acc}{26.5\%}
\DefMacro{results-ln-lo-lo-75-ln-s+bsexpl1+attn+copy_ta-test-MAX-full-acc-top-1}{6.1\%}
\DefMacro{results-ln-lo-lo-75-ln-s+bsexpl1+attn+copy_ta-test-MAX-full-acc-top-2}{8.6\%}
\DefMacro{results-ln-lo-lo-75-ln-s+bsexpl1+attn+copy_ta-test-MAX-full-acc-top-3}{9.4\%}
\DefMacro{results-ln-lo-lo-75-ln-s+bsexpl1+attn+copy_ta-test-MAX-full-acc-top-5}{10.8\%}
\DefMacro{results-ln-lo-lo-75-ln-s+bsexpl1+attn+copy_ta-test-MAX-full-correct-top-1}{71}
\DefMacro{results-ln-lo-lo-75-ln-s+bsexpl1+attn+copy_ta-test-MAX-full-correct-top-2}{100}
\DefMacro{results-ln-lo-lo-75-ln-s+bsexpl1+attn+copy_ta-test-MAX-full-correct-top-3}{110}
\DefMacro{results-ln-lo-lo-75-ln-s+bsexpl1+attn+copy_ta-test-MAX-full-correct-top-5}{126}
\DefMacro{results-ln-lo-lo-75-ln-s+bsexpl1+attn+copy_ta-test-MEDIAN-BLEU-4}{35.2}
\DefMacro{results-ln-lo-lo-75-ln-s+bsexpl1+attn+copy_ta-test-MEDIAN-char-acc}{23.7\%}
\DefMacro{results-ln-lo-lo-75-ln-s+bsexpl1+attn+copy_ta-test-MEDIAN-count-top-1}{116700.0\%}
\DefMacro{results-ln-lo-lo-75-ln-s+bsexpl1+attn+copy_ta-test-MEDIAN-count-top-2}{116700.0\%}
\DefMacro{results-ln-lo-lo-75-ln-s+bsexpl1+attn+copy_ta-test-MEDIAN-count-top-3}{116700.0\%}
\DefMacro{results-ln-lo-lo-75-ln-s+bsexpl1+attn+copy_ta-test-MEDIAN-count-top-5}{116700.0\%}
\DefMacro{results-ln-lo-lo-75-ln-s+bsexpl1+attn+copy_ta-test-MEDIAN-frag-acc}{23.6\%}
\DefMacro{results-ln-lo-lo-75-ln-s+bsexpl1+attn+copy_ta-test-MEDIAN-full-acc-top-1}{4.6\%}
\DefMacro{results-ln-lo-lo-75-ln-s+bsexpl1+attn+copy_ta-test-MEDIAN-full-acc-top-2}{5.7\%}
\DefMacro{results-ln-lo-lo-75-ln-s+bsexpl1+attn+copy_ta-test-MEDIAN-full-acc-top-3}{6.6\%}
\DefMacro{results-ln-lo-lo-75-ln-s+bsexpl1+attn+copy_ta-test-MEDIAN-full-acc-top-5}{8.1\%}
\DefMacro{results-ln-lo-lo-75-ln-s+bsexpl1+attn+copy_ta-test-MEDIAN-full-correct-top-1}{5400.0\%}
\DefMacro{results-ln-lo-lo-75-ln-s+bsexpl1+attn+copy_ta-test-MEDIAN-full-correct-top-2}{6700.0\%}
\DefMacro{results-ln-lo-lo-75-ln-s+bsexpl1+attn+copy_ta-test-MEDIAN-full-correct-top-3}{7700.0\%}
\DefMacro{results-ln-lo-lo-75-ln-s+bsexpl1+attn+copy_ta-test-MEDIAN-full-correct-top-5}{9400.0\%}
\DefMacro{results-ln-lo-lo-75-ln-s+bsexpl1+attn+copy_ta-test-MIN-BLEU-4}{34.6}
\DefMacro{results-ln-lo-lo-75-ln-s+bsexpl1+attn+copy_ta-test-MIN-char-acc}{23.5\%}
\DefMacro{results-ln-lo-lo-75-ln-s+bsexpl1+attn+copy_ta-test-MIN-count-top-1}{1167}
\DefMacro{results-ln-lo-lo-75-ln-s+bsexpl1+attn+copy_ta-test-MIN-count-top-2}{1167}
\DefMacro{results-ln-lo-lo-75-ln-s+bsexpl1+attn+copy_ta-test-MIN-count-top-3}{1167}
\DefMacro{results-ln-lo-lo-75-ln-s+bsexpl1+attn+copy_ta-test-MIN-count-top-5}{1167}
\DefMacro{results-ln-lo-lo-75-ln-s+bsexpl1+attn+copy_ta-test-MIN-frag-acc}{22.8\%}
\DefMacro{results-ln-lo-lo-75-ln-s+bsexpl1+attn+copy_ta-test-MIN-full-acc-top-1}{4.2\%}
\DefMacro{results-ln-lo-lo-75-ln-s+bsexpl1+attn+copy_ta-test-MIN-full-acc-top-2}{5.7\%}
\DefMacro{results-ln-lo-lo-75-ln-s+bsexpl1+attn+copy_ta-test-MIN-full-acc-top-3}{6.4\%}
\DefMacro{results-ln-lo-lo-75-ln-s+bsexpl1+attn+copy_ta-test-MIN-full-acc-top-5}{7.1\%}
\DefMacro{results-ln-lo-lo-75-ln-s+bsexpl1+attn+copy_ta-test-MIN-full-correct-top-1}{49}
\DefMacro{results-ln-lo-lo-75-ln-s+bsexpl1+attn+copy_ta-test-MIN-full-correct-top-2}{66}
\DefMacro{results-ln-lo-lo-75-ln-s+bsexpl1+attn+copy_ta-test-MIN-full-correct-top-3}{75}
\DefMacro{results-ln-lo-lo-75-ln-s+bsexpl1+attn+copy_ta-test-MIN-full-correct-top-5}{83}
\DefMacro{results-ln-lo-lo-75-ln-s+bsexpl1+attn+copy_ta-test-STDEV-BLEU-4}{1.1}
\DefMacro{results-ln-lo-lo-75-ln-s+bsexpl1+attn+copy_ta-test-STDEV-char-acc}{1.3\%}
\DefMacro{results-ln-lo-lo-75-ln-s+bsexpl1+attn+copy_ta-test-STDEV-count-top-1}{0.0\%}
\DefMacro{results-ln-lo-lo-75-ln-s+bsexpl1+attn+copy_ta-test-STDEV-count-top-2}{0.0\%}
\DefMacro{results-ln-lo-lo-75-ln-s+bsexpl1+attn+copy_ta-test-STDEV-count-top-3}{0.0\%}
\DefMacro{results-ln-lo-lo-75-ln-s+bsexpl1+attn+copy_ta-test-STDEV-count-top-5}{0.0\%}
\DefMacro{results-ln-lo-lo-75-ln-s+bsexpl1+attn+copy_ta-test-STDEV-frag-acc}{1.6\%}
\DefMacro{results-ln-lo-lo-75-ln-s+bsexpl1+attn+copy_ta-test-STDEV-full-acc-top-1}{0.8\%}
\DefMacro{results-ln-lo-lo-75-ln-s+bsexpl1+attn+copy_ta-test-STDEV-full-acc-top-2}{1.4\%}
\DefMacro{results-ln-lo-lo-75-ln-s+bsexpl1+attn+copy_ta-test-STDEV-full-acc-top-3}{1.4\%}
\DefMacro{results-ln-lo-lo-75-ln-s+bsexpl1+attn+copy_ta-test-STDEV-full-acc-top-5}{1.6\%}
\DefMacro{results-ln-lo-lo-75-ln-s+bsexpl1+attn+copy_ta-test-STDEV-full-correct-top-1}{941.6\%}
\DefMacro{results-ln-lo-lo-75-ln-s+bsexpl1+attn+copy_ta-test-STDEV-full-correct-top-2}{1579.7\%}
\DefMacro{results-ln-lo-lo-75-ln-s+bsexpl1+attn+copy_ta-test-STDEV-full-correct-top-3}{1604.9\%}
\DefMacro{results-ln-lo-lo-75-ln-s+bsexpl1+attn+copy_ta-test-STDEV-full-correct-top-5}{1823.9\%}
\DefMacro{results-ln-lo-lo-75-ln-s+bsexpl1+attn+copy_ta-test-SUM-BLEU-4}{107.0}
\DefMacro{results-ln-lo-lo-75-ln-s+bsexpl1+attn+copy_ta-test-SUM-char-acc}{73.5\%}
\DefMacro{results-ln-lo-lo-75-ln-s+bsexpl1+attn+copy_ta-test-SUM-count-top-1}{3501}
\DefMacro{results-ln-lo-lo-75-ln-s+bsexpl1+attn+copy_ta-test-SUM-count-top-2}{3501}
\DefMacro{results-ln-lo-lo-75-ln-s+bsexpl1+attn+copy_ta-test-SUM-count-top-3}{3501}
\DefMacro{results-ln-lo-lo-75-ln-s+bsexpl1+attn+copy_ta-test-SUM-count-top-5}{3501}
\DefMacro{results-ln-lo-lo-75-ln-s+bsexpl1+attn+copy_ta-test-SUM-frag-acc}{72.8\%}
\DefMacro{results-ln-lo-lo-75-ln-s+bsexpl1+attn+copy_ta-test-SUM-full-acc-top-1}{14.9\%}
\DefMacro{results-ln-lo-lo-75-ln-s+bsexpl1+attn+copy_ta-test-SUM-full-acc-top-2}{20.0\%}
\DefMacro{results-ln-lo-lo-75-ln-s+bsexpl1+attn+copy_ta-test-SUM-full-acc-top-3}{22.5\%}
\DefMacro{results-ln-lo-lo-75-ln-s+bsexpl1+attn+copy_ta-test-SUM-full-acc-top-5}{26.0\%}
\DefMacro{results-ln-lo-lo-75-ln-s+bsexpl1+attn+copy_ta-test-SUM-full-correct-top-1}{174}
\DefMacro{results-ln-lo-lo-75-ln-s+bsexpl1+attn+copy_ta-test-SUM-full-correct-top-2}{233}
\DefMacro{results-ln-lo-lo-75-ln-s+bsexpl1+attn+copy_ta-test-SUM-full-correct-top-3}{262}
\DefMacro{results-ln-lo-lo-75-ln-s+bsexpl1+attn+copy_ta-test-SUM-full-correct-top-5}{303}
\DefMacro{results-ln-lo-lo-ln-s+bsexpl1+attn+copy_ta-val-AVG-BLEU-4}{49.7}
\DefMacro{results-ln-lo-lo-ln-s+bsexpl1+attn+copy_ta-val-AVG-char-acc}{36.8\%}
\DefMacro{results-ln-lo-lo-ln-s+bsexpl1+attn+copy_ta-val-AVG-count-top-1}{14400.0\%}
\DefMacro{results-ln-lo-lo-ln-s+bsexpl1+attn+copy_ta-val-AVG-count-top-2}{14400.0\%}
\DefMacro{results-ln-lo-lo-ln-s+bsexpl1+attn+copy_ta-val-AVG-count-top-3}{14400.0\%}
\DefMacro{results-ln-lo-lo-ln-s+bsexpl1+attn+copy_ta-val-AVG-count-top-5}{14400.0\%}
\DefMacro{results-ln-lo-lo-ln-s+bsexpl1+attn+copy_ta-val-AVG-frag-acc}{40.5\%}
\DefMacro{results-ln-lo-lo-ln-s+bsexpl1+attn+copy_ta-val-AVG-full-acc-top-1}{6.9\%}
\DefMacro{results-ln-lo-lo-ln-s+bsexpl1+attn+copy_ta-val-AVG-full-acc-top-2}{9.7\%}
\DefMacro{results-ln-lo-lo-ln-s+bsexpl1+attn+copy_ta-val-AVG-full-acc-top-3}{11.6\%}
\DefMacro{results-ln-lo-lo-ln-s+bsexpl1+attn+copy_ta-val-AVG-full-acc-top-5}{13.2\%}
\DefMacro{results-ln-lo-lo-ln-s+bsexpl1+attn+copy_ta-val-AVG-full-correct-top-1}{1000.0\%}
\DefMacro{results-ln-lo-lo-ln-s+bsexpl1+attn+copy_ta-val-AVG-full-correct-top-2}{1400.0\%}
\DefMacro{results-ln-lo-lo-ln-s+bsexpl1+attn+copy_ta-val-AVG-full-correct-top-3}{1666.7\%}
\DefMacro{results-ln-lo-lo-ln-s+bsexpl1+attn+copy_ta-val-AVG-full-correct-top-5}{1900.0\%}
\DefMacro{results-ln-lo-lo-ln-s+bsexpl1+attn+copy_ta-val-MAX-BLEU-4}{51.1}
\DefMacro{results-ln-lo-lo-ln-s+bsexpl1+attn+copy_ta-val-MAX-char-acc}{40.7\%}
\DefMacro{results-ln-lo-lo-ln-s+bsexpl1+attn+copy_ta-val-MAX-count-top-1}{144}
\DefMacro{results-ln-lo-lo-ln-s+bsexpl1+attn+copy_ta-val-MAX-count-top-2}{144}
\DefMacro{results-ln-lo-lo-ln-s+bsexpl1+attn+copy_ta-val-MAX-count-top-3}{144}
\DefMacro{results-ln-lo-lo-ln-s+bsexpl1+attn+copy_ta-val-MAX-count-top-5}{144}
\DefMacro{results-ln-lo-lo-ln-s+bsexpl1+attn+copy_ta-val-MAX-frag-acc}{42.1\%}
\DefMacro{results-ln-lo-lo-ln-s+bsexpl1+attn+copy_ta-val-MAX-full-acc-top-1}{7.6\%}
\DefMacro{results-ln-lo-lo-ln-s+bsexpl1+attn+copy_ta-val-MAX-full-acc-top-2}{11.8\%}
\DefMacro{results-ln-lo-lo-ln-s+bsexpl1+attn+copy_ta-val-MAX-full-acc-top-3}{13.9\%}
\DefMacro{results-ln-lo-lo-ln-s+bsexpl1+attn+copy_ta-val-MAX-full-acc-top-5}{15.3\%}
\DefMacro{results-ln-lo-lo-ln-s+bsexpl1+attn+copy_ta-val-MAX-full-correct-top-1}{11}
\DefMacro{results-ln-lo-lo-ln-s+bsexpl1+attn+copy_ta-val-MAX-full-correct-top-2}{17}
\DefMacro{results-ln-lo-lo-ln-s+bsexpl1+attn+copy_ta-val-MAX-full-correct-top-3}{20}
\DefMacro{results-ln-lo-lo-ln-s+bsexpl1+attn+copy_ta-val-MAX-full-correct-top-5}{22}
\DefMacro{results-ln-lo-lo-ln-s+bsexpl1+attn+copy_ta-val-MEDIAN-BLEU-4}{49.8}
\DefMacro{results-ln-lo-lo-ln-s+bsexpl1+attn+copy_ta-val-MEDIAN-char-acc}{35.4\%}
\DefMacro{results-ln-lo-lo-ln-s+bsexpl1+attn+copy_ta-val-MEDIAN-count-top-1}{14400.0\%}
\DefMacro{results-ln-lo-lo-ln-s+bsexpl1+attn+copy_ta-val-MEDIAN-count-top-2}{14400.0\%}
\DefMacro{results-ln-lo-lo-ln-s+bsexpl1+attn+copy_ta-val-MEDIAN-count-top-3}{14400.0\%}
\DefMacro{results-ln-lo-lo-ln-s+bsexpl1+attn+copy_ta-val-MEDIAN-count-top-5}{14400.0\%}
\DefMacro{results-ln-lo-lo-ln-s+bsexpl1+attn+copy_ta-val-MEDIAN-frag-acc}{40.3\%}
\DefMacro{results-ln-lo-lo-ln-s+bsexpl1+attn+copy_ta-val-MEDIAN-full-acc-top-1}{6.9\%}
\DefMacro{results-ln-lo-lo-ln-s+bsexpl1+attn+copy_ta-val-MEDIAN-full-acc-top-2}{9.0\%}
\DefMacro{results-ln-lo-lo-ln-s+bsexpl1+attn+copy_ta-val-MEDIAN-full-acc-top-3}{11.8\%}
\DefMacro{results-ln-lo-lo-ln-s+bsexpl1+attn+copy_ta-val-MEDIAN-full-acc-top-5}{13.2\%}
\DefMacro{results-ln-lo-lo-ln-s+bsexpl1+attn+copy_ta-val-MEDIAN-full-correct-top-1}{1000.0\%}
\DefMacro{results-ln-lo-lo-ln-s+bsexpl1+attn+copy_ta-val-MEDIAN-full-correct-top-2}{1300.0\%}
\DefMacro{results-ln-lo-lo-ln-s+bsexpl1+attn+copy_ta-val-MEDIAN-full-correct-top-3}{1700.0\%}
\DefMacro{results-ln-lo-lo-ln-s+bsexpl1+attn+copy_ta-val-MEDIAN-full-correct-top-5}{1900.0\%}
\DefMacro{results-ln-lo-lo-ln-s+bsexpl1+attn+copy_ta-val-MIN-BLEU-4}{48.4}
\DefMacro{results-ln-lo-lo-ln-s+bsexpl1+attn+copy_ta-val-MIN-char-acc}{34.1\%}
\DefMacro{results-ln-lo-lo-ln-s+bsexpl1+attn+copy_ta-val-MIN-count-top-1}{144}
\DefMacro{results-ln-lo-lo-ln-s+bsexpl1+attn+copy_ta-val-MIN-count-top-2}{144}
\DefMacro{results-ln-lo-lo-ln-s+bsexpl1+attn+copy_ta-val-MIN-count-top-3}{144}
\DefMacro{results-ln-lo-lo-ln-s+bsexpl1+attn+copy_ta-val-MIN-count-top-5}{144}
\DefMacro{results-ln-lo-lo-ln-s+bsexpl1+attn+copy_ta-val-MIN-frag-acc}{39.2\%}
\DefMacro{results-ln-lo-lo-ln-s+bsexpl1+attn+copy_ta-val-MIN-full-acc-top-1}{6.2\%}
\DefMacro{results-ln-lo-lo-ln-s+bsexpl1+attn+copy_ta-val-MIN-full-acc-top-2}{8.3\%}
\DefMacro{results-ln-lo-lo-ln-s+bsexpl1+attn+copy_ta-val-MIN-full-acc-top-3}{9.0\%}
\DefMacro{results-ln-lo-lo-ln-s+bsexpl1+attn+copy_ta-val-MIN-full-acc-top-5}{11.1\%}
\DefMacro{results-ln-lo-lo-ln-s+bsexpl1+attn+copy_ta-val-MIN-full-correct-top-1}{9}
\DefMacro{results-ln-lo-lo-ln-s+bsexpl1+attn+copy_ta-val-MIN-full-correct-top-2}{12}
\DefMacro{results-ln-lo-lo-ln-s+bsexpl1+attn+copy_ta-val-MIN-full-correct-top-3}{13}
\DefMacro{results-ln-lo-lo-ln-s+bsexpl1+attn+copy_ta-val-MIN-full-correct-top-5}{16}
\DefMacro{results-ln-lo-lo-ln-s+bsexpl1+attn+copy_ta-val-STDEV-BLEU-4}{1.1}
\DefMacro{results-ln-lo-lo-ln-s+bsexpl1+attn+copy_ta-val-STDEV-char-acc}{2.9\%}
\DefMacro{results-ln-lo-lo-ln-s+bsexpl1+attn+copy_ta-val-STDEV-count-top-1}{0.0\%}
\DefMacro{results-ln-lo-lo-ln-s+bsexpl1+attn+copy_ta-val-STDEV-count-top-2}{0.0\%}
\DefMacro{results-ln-lo-lo-ln-s+bsexpl1+attn+copy_ta-val-STDEV-count-top-3}{0.0\%}
\DefMacro{results-ln-lo-lo-ln-s+bsexpl1+attn+copy_ta-val-STDEV-count-top-5}{0.0\%}
\DefMacro{results-ln-lo-lo-ln-s+bsexpl1+attn+copy_ta-val-STDEV-frag-acc}{1.2\%}
\DefMacro{results-ln-lo-lo-ln-s+bsexpl1+attn+copy_ta-val-STDEV-full-acc-top-1}{0.6\%}
\DefMacro{results-ln-lo-lo-ln-s+bsexpl1+attn+copy_ta-val-STDEV-full-acc-top-2}{1.5\%}
\DefMacro{results-ln-lo-lo-ln-s+bsexpl1+attn+copy_ta-val-STDEV-full-acc-top-3}{2.0\%}
\DefMacro{results-ln-lo-lo-ln-s+bsexpl1+attn+copy_ta-val-STDEV-full-acc-top-5}{1.7\%}
\DefMacro{results-ln-lo-lo-ln-s+bsexpl1+attn+copy_ta-val-STDEV-full-correct-top-1}{81.6\%}
\DefMacro{results-ln-lo-lo-ln-s+bsexpl1+attn+copy_ta-val-STDEV-full-correct-top-2}{216.0\%}
\DefMacro{results-ln-lo-lo-ln-s+bsexpl1+attn+copy_ta-val-STDEV-full-correct-top-3}{286.7\%}
\DefMacro{results-ln-lo-lo-ln-s+bsexpl1+attn+copy_ta-val-STDEV-full-correct-top-5}{244.9\%}
\DefMacro{results-ln-lo-lo-ln-s+bsexpl1+attn+copy_ta-val-SUM-BLEU-4}{149.2}
\DefMacro{results-ln-lo-lo-ln-s+bsexpl1+attn+copy_ta-val-SUM-char-acc}{110.3\%}
\DefMacro{results-ln-lo-lo-ln-s+bsexpl1+attn+copy_ta-val-SUM-count-top-1}{432}
\DefMacro{results-ln-lo-lo-ln-s+bsexpl1+attn+copy_ta-val-SUM-count-top-2}{432}
\DefMacro{results-ln-lo-lo-ln-s+bsexpl1+attn+copy_ta-val-SUM-count-top-3}{432}
\DefMacro{results-ln-lo-lo-ln-s+bsexpl1+attn+copy_ta-val-SUM-count-top-5}{432}
\DefMacro{results-ln-lo-lo-ln-s+bsexpl1+attn+copy_ta-val-SUM-frag-acc}{121.6\%}
\DefMacro{results-ln-lo-lo-ln-s+bsexpl1+attn+copy_ta-val-SUM-full-acc-top-1}{20.8\%}
\DefMacro{results-ln-lo-lo-ln-s+bsexpl1+attn+copy_ta-val-SUM-full-acc-top-2}{29.2\%}
\DefMacro{results-ln-lo-lo-ln-s+bsexpl1+attn+copy_ta-val-SUM-full-acc-top-3}{34.7\%}
\DefMacro{results-ln-lo-lo-ln-s+bsexpl1+attn+copy_ta-val-SUM-full-acc-top-5}{39.6\%}
\DefMacro{results-ln-lo-lo-ln-s+bsexpl1+attn+copy_ta-val-SUM-full-correct-top-1}{30}
\DefMacro{results-ln-lo-lo-ln-s+bsexpl1+attn+copy_ta-val-SUM-full-correct-top-2}{42}
\DefMacro{results-ln-lo-lo-ln-s+bsexpl1+attn+copy_ta-val-SUM-full-correct-top-3}{50}
\DefMacro{results-ln-lo-lo-ln-s+bsexpl1+attn+copy_ta-val-SUM-full-correct-top-5}{57}
\DefMacro{results-ln-lo-lo-ln-s+bsexpl1+attn+copy_ta-test-AVG-BLEU-4}{37.4}
\DefMacro{results-ln-lo-lo-ln-s+bsexpl1+attn+copy_ta-test-AVG-char-acc}{28.0\%}
\DefMacro{results-ln-lo-lo-ln-s+bsexpl1+attn+copy_ta-test-AVG-count-top-1}{116700.0\%}
\DefMacro{results-ln-lo-lo-ln-s+bsexpl1+attn+copy_ta-test-AVG-count-top-2}{116700.0\%}
\DefMacro{results-ln-lo-lo-ln-s+bsexpl1+attn+copy_ta-test-AVG-count-top-3}{116700.0\%}
\DefMacro{results-ln-lo-lo-ln-s+bsexpl1+attn+copy_ta-test-AVG-count-top-5}{116700.0\%}
\DefMacro{results-ln-lo-lo-ln-s+bsexpl1+attn+copy_ta-test-AVG-frag-acc}{26.5\%}
\DefMacro{results-ln-lo-lo-ln-s+bsexpl1+attn+copy_ta-test-AVG-full-acc-top-1}{7.4\%}
\DefMacro{results-ln-lo-lo-ln-s+bsexpl1+attn+copy_ta-test-AVG-full-acc-top-2}{10.2\%}
\DefMacro{results-ln-lo-lo-ln-s+bsexpl1+attn+copy_ta-test-AVG-full-acc-top-3}{11.3\%}
\DefMacro{results-ln-lo-lo-ln-s+bsexpl1+attn+copy_ta-test-AVG-full-acc-top-5}{12.5\%}
\DefMacro{results-ln-lo-lo-ln-s+bsexpl1+attn+copy_ta-test-AVG-full-correct-top-1}{8666.7\%}
\DefMacro{results-ln-lo-lo-ln-s+bsexpl1+attn+copy_ta-test-AVG-full-correct-top-2}{11900.0\%}
\DefMacro{results-ln-lo-lo-ln-s+bsexpl1+attn+copy_ta-test-AVG-full-correct-top-3}{13133.3\%}
\DefMacro{results-ln-lo-lo-ln-s+bsexpl1+attn+copy_ta-test-AVG-full-correct-top-5}{14533.3\%}
\DefMacro{results-ln-lo-lo-ln-s+bsexpl1+attn+copy_ta-test-MAX-BLEU-4}{39.8}
\DefMacro{results-ln-lo-lo-ln-s+bsexpl1+attn+copy_ta-test-MAX-char-acc}{29.6\%}
\DefMacro{results-ln-lo-lo-ln-s+bsexpl1+attn+copy_ta-test-MAX-count-top-1}{1167}
\DefMacro{results-ln-lo-lo-ln-s+bsexpl1+attn+copy_ta-test-MAX-count-top-2}{1167}
\DefMacro{results-ln-lo-lo-ln-s+bsexpl1+attn+copy_ta-test-MAX-count-top-3}{1167}
\DefMacro{results-ln-lo-lo-ln-s+bsexpl1+attn+copy_ta-test-MAX-count-top-5}{1167}
\DefMacro{results-ln-lo-lo-ln-s+bsexpl1+attn+copy_ta-test-MAX-frag-acc}{29.5\%}
\DefMacro{results-ln-lo-lo-ln-s+bsexpl1+attn+copy_ta-test-MAX-full-acc-top-1}{8.9\%}
\DefMacro{results-ln-lo-lo-ln-s+bsexpl1+attn+copy_ta-test-MAX-full-acc-top-2}{11.7\%}
\DefMacro{results-ln-lo-lo-ln-s+bsexpl1+attn+copy_ta-test-MAX-full-acc-top-3}{12.7\%}
\DefMacro{results-ln-lo-lo-ln-s+bsexpl1+attn+copy_ta-test-MAX-full-acc-top-5}{14.2\%}
\DefMacro{results-ln-lo-lo-ln-s+bsexpl1+attn+copy_ta-test-MAX-full-correct-top-1}{104}
\DefMacro{results-ln-lo-lo-ln-s+bsexpl1+attn+copy_ta-test-MAX-full-correct-top-2}{137}
\DefMacro{results-ln-lo-lo-ln-s+bsexpl1+attn+copy_ta-test-MAX-full-correct-top-3}{148}
\DefMacro{results-ln-lo-lo-ln-s+bsexpl1+attn+copy_ta-test-MAX-full-correct-top-5}{166}
\DefMacro{results-ln-lo-lo-ln-s+bsexpl1+attn+copy_ta-test-MEDIAN-BLEU-4}{36.7}
\DefMacro{results-ln-lo-lo-ln-s+bsexpl1+attn+copy_ta-test-MEDIAN-char-acc}{27.4\%}
\DefMacro{results-ln-lo-lo-ln-s+bsexpl1+attn+copy_ta-test-MEDIAN-count-top-1}{116700.0\%}
\DefMacro{results-ln-lo-lo-ln-s+bsexpl1+attn+copy_ta-test-MEDIAN-count-top-2}{116700.0\%}
\DefMacro{results-ln-lo-lo-ln-s+bsexpl1+attn+copy_ta-test-MEDIAN-count-top-3}{116700.0\%}
\DefMacro{results-ln-lo-lo-ln-s+bsexpl1+attn+copy_ta-test-MEDIAN-count-top-5}{116700.0\%}
\DefMacro{results-ln-lo-lo-ln-s+bsexpl1+attn+copy_ta-test-MEDIAN-frag-acc}{25.3\%}
\DefMacro{results-ln-lo-lo-ln-s+bsexpl1+attn+copy_ta-test-MEDIAN-full-acc-top-1}{7.0\%}
\DefMacro{results-ln-lo-lo-ln-s+bsexpl1+attn+copy_ta-test-MEDIAN-full-acc-top-2}{9.9\%}
\DefMacro{results-ln-lo-lo-ln-s+bsexpl1+attn+copy_ta-test-MEDIAN-full-acc-top-3}{10.6\%}
\DefMacro{results-ln-lo-lo-ln-s+bsexpl1+attn+copy_ta-test-MEDIAN-full-acc-top-5}{11.8\%}
\DefMacro{results-ln-lo-lo-ln-s+bsexpl1+attn+copy_ta-test-MEDIAN-full-correct-top-1}{8200.0\%}
\DefMacro{results-ln-lo-lo-ln-s+bsexpl1+attn+copy_ta-test-MEDIAN-full-correct-top-2}{11500.0\%}
\DefMacro{results-ln-lo-lo-ln-s+bsexpl1+attn+copy_ta-test-MEDIAN-full-correct-top-3}{12400.0\%}
\DefMacro{results-ln-lo-lo-ln-s+bsexpl1+attn+copy_ta-test-MEDIAN-full-correct-top-5}{13800.0\%}
\DefMacro{results-ln-lo-lo-ln-s+bsexpl1+attn+copy_ta-test-MIN-BLEU-4}{35.8}
\DefMacro{results-ln-lo-lo-ln-s+bsexpl1+attn+copy_ta-test-MIN-char-acc}{27.0\%}
\DefMacro{results-ln-lo-lo-ln-s+bsexpl1+attn+copy_ta-test-MIN-count-top-1}{1167}
\DefMacro{results-ln-lo-lo-ln-s+bsexpl1+attn+copy_ta-test-MIN-count-top-2}{1167}
\DefMacro{results-ln-lo-lo-ln-s+bsexpl1+attn+copy_ta-test-MIN-count-top-3}{1167}
\DefMacro{results-ln-lo-lo-ln-s+bsexpl1+attn+copy_ta-test-MIN-count-top-5}{1167}
\DefMacro{results-ln-lo-lo-ln-s+bsexpl1+attn+copy_ta-test-MIN-frag-acc}{24.7\%}
\DefMacro{results-ln-lo-lo-ln-s+bsexpl1+attn+copy_ta-test-MIN-full-acc-top-1}{6.3\%}
\DefMacro{results-ln-lo-lo-ln-s+bsexpl1+attn+copy_ta-test-MIN-full-acc-top-2}{9.0\%}
\DefMacro{results-ln-lo-lo-ln-s+bsexpl1+attn+copy_ta-test-MIN-full-acc-top-3}{10.5\%}
\DefMacro{results-ln-lo-lo-ln-s+bsexpl1+attn+copy_ta-test-MIN-full-acc-top-5}{11.3\%}
\DefMacro{results-ln-lo-lo-ln-s+bsexpl1+attn+copy_ta-test-MIN-full-correct-top-1}{74}
\DefMacro{results-ln-lo-lo-ln-s+bsexpl1+attn+copy_ta-test-MIN-full-correct-top-2}{105}
\DefMacro{results-ln-lo-lo-ln-s+bsexpl1+attn+copy_ta-test-MIN-full-correct-top-3}{122}
\DefMacro{results-ln-lo-lo-ln-s+bsexpl1+attn+copy_ta-test-MIN-full-correct-top-5}{132}
\DefMacro{results-ln-lo-lo-ln-s+bsexpl1+attn+copy_ta-test-STDEV-BLEU-4}{1.7}
\DefMacro{results-ln-lo-lo-ln-s+bsexpl1+attn+copy_ta-test-STDEV-char-acc}{1.1\%}
\DefMacro{results-ln-lo-lo-ln-s+bsexpl1+attn+copy_ta-test-STDEV-count-top-1}{0.0\%}
\DefMacro{results-ln-lo-lo-ln-s+bsexpl1+attn+copy_ta-test-STDEV-count-top-2}{0.0\%}
\DefMacro{results-ln-lo-lo-ln-s+bsexpl1+attn+copy_ta-test-STDEV-count-top-3}{0.0\%}
\DefMacro{results-ln-lo-lo-ln-s+bsexpl1+attn+copy_ta-test-STDEV-count-top-5}{0.0\%}
\DefMacro{results-ln-lo-lo-ln-s+bsexpl1+attn+copy_ta-test-STDEV-frag-acc}{2.1\%}
\DefMacro{results-ln-lo-lo-ln-s+bsexpl1+attn+copy_ta-test-STDEV-full-acc-top-1}{1.1\%}
\DefMacro{results-ln-lo-lo-ln-s+bsexpl1+attn+copy_ta-test-STDEV-full-acc-top-2}{1.1\%}
\DefMacro{results-ln-lo-lo-ln-s+bsexpl1+attn+copy_ta-test-STDEV-full-acc-top-3}{1.0\%}
\DefMacro{results-ln-lo-lo-ln-s+bsexpl1+attn+copy_ta-test-STDEV-full-acc-top-5}{1.3\%}
\DefMacro{results-ln-lo-lo-ln-s+bsexpl1+attn+copy_ta-test-STDEV-full-correct-top-1}{1268.4\%}
\DefMacro{results-ln-lo-lo-ln-s+bsexpl1+attn+copy_ta-test-STDEV-full-correct-top-2}{1336.7\%}
\DefMacro{results-ln-lo-lo-ln-s+bsexpl1+attn+copy_ta-test-STDEV-full-correct-top-3}{1181.3\%}
\DefMacro{results-ln-lo-lo-ln-s+bsexpl1+attn+copy_ta-test-STDEV-full-correct-top-5}{1481.7\%}
\DefMacro{results-ln-lo-lo-ln-s+bsexpl1+attn+copy_ta-test-SUM-BLEU-4}{112.3}
\DefMacro{results-ln-lo-lo-ln-s+bsexpl1+attn+copy_ta-test-SUM-char-acc}{84.0\%}
\DefMacro{results-ln-lo-lo-ln-s+bsexpl1+attn+copy_ta-test-SUM-count-top-1}{3501}
\DefMacro{results-ln-lo-lo-ln-s+bsexpl1+attn+copy_ta-test-SUM-count-top-2}{3501}
\DefMacro{results-ln-lo-lo-ln-s+bsexpl1+attn+copy_ta-test-SUM-count-top-3}{3501}
\DefMacro{results-ln-lo-lo-ln-s+bsexpl1+attn+copy_ta-test-SUM-count-top-5}{3501}
\DefMacro{results-ln-lo-lo-ln-s+bsexpl1+attn+copy_ta-test-SUM-frag-acc}{79.6\%}
\DefMacro{results-ln-lo-lo-ln-s+bsexpl1+attn+copy_ta-test-SUM-full-acc-top-1}{22.3\%}
\DefMacro{results-ln-lo-lo-ln-s+bsexpl1+attn+copy_ta-test-SUM-full-acc-top-2}{30.6\%}
\DefMacro{results-ln-lo-lo-ln-s+bsexpl1+attn+copy_ta-test-SUM-full-acc-top-3}{33.8\%}
\DefMacro{results-ln-lo-lo-ln-s+bsexpl1+attn+copy_ta-test-SUM-full-acc-top-5}{37.4\%}
\DefMacro{results-ln-lo-lo-ln-s+bsexpl1+attn+copy_ta-test-SUM-full-correct-top-1}{260}
\DefMacro{results-ln-lo-lo-ln-s+bsexpl1+attn+copy_ta-test-SUM-full-correct-top-2}{357}
\DefMacro{results-ln-lo-lo-ln-s+bsexpl1+attn+copy_ta-test-SUM-full-correct-top-3}{394}
\DefMacro{results-ln-lo-lo-ln-s+bsexpl1+attn+copy_ta-test-SUM-full-correct-top-5}{436}

%% file: tables/numbers-misc.tex
\DefMacro{sub-token-repeat-rate}{1.37\%}
\DefMacro{vocab-size-token}{14,299}
\DefMacro{vocab-size-sub-token}{2,328}
\DefMacro{qs-fcslpcm-ln-BLEU-4}{36.3}
\DefMacro{qs-fcslpcm-ln-frag-acc}{17\%}
\DefMacro{qs-fcslpcm-ln-full-acc-top-1}{5\%}
\DefMacro{qs-fcslpcm-ln-full-acc-top-5}{10\%}
\DefMacro{qs-fcslpcm-num-lemmas}{690}
\DefMacro{qs-fcslpcm-ln-full-correct-top-1}{36}
\DefMacro{qs-fcslpcm-ln-full-correct-top-1-percent}{5\%}
\DefMacro{qs-fcslpcm-f-accuracy-all}{0.929}
\DefMacro{qs-fcslpcm-f-accuracy-top-3}{0.988}